\documentclass[a4paper,12pt,times,print,index,custommargin,custombib]{Classes/PhDThesisPSnPDF}
\usepackage[normalem]{ulem}

\input{Preamble/preamble}

\title{The Galactic Millisecond Pulsar Population}

\subtitle{Implications for the Galactic Center Excess}

\author{Harrison Ploeg}

\dept{School of Physical and Chemical Sciences}

\university{University of Canterbury}
\crest{\includegraphics[width=0.3\textwidth]{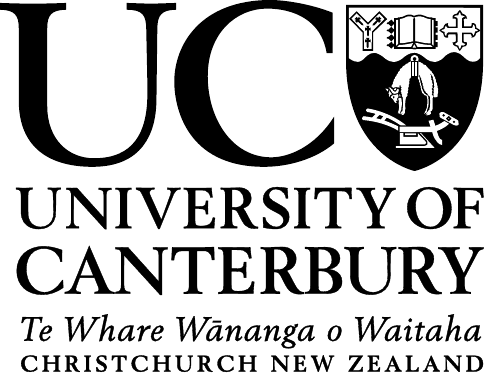}}

\degreetitle{Doctor of Philosophy}

\subject{LaTeX} \keywords{{LaTeX} {PhD Thesis} {Physics} {University of
Canterbury}}

\ifdefineAbstract
\fi

\ifdefineChapter
\fi

\begin{document}

\frontmatter

\maketitle

\begin{acknowledgements}      

I would like to thank my supervisor Chris Gordon for his support and guidance throughout these last few years. His feedback and advice was always useful. I'd also like to thank Roland Crocker and Oscar Macias for their comments and contributions to the work that makes up this thesis.

\end{acknowledgements}

\begin{abstract}

Analysis of Fermi Large Area Telescope (Fermi-LAT) data has uncovered an extended gamma-ray source in the central region of the Milky Way. This Galactic Center Excess (GCE) has a spectral peak at a few GeV and appeared, initially, to have a spherically symmetric profile. These properties suggested that it may be evidence of self-annihilating weakly interacting massive particles (WIMPs) with a Navarro-Frenk-White (NFW) profile. However, the GCE gamma-ray spectrum is also similar to those of millisecond pulsars (MSPs) resolved in the Fermi-LAT data. An alternative possibility would then be that the GCE is produced by a population of MSPs in the Galactic Center too faint to be resolved individually as point sources at the present time. Providing further support for the MSP scenario, in recent years it has become apparent that the GCE may not be spherically symmetric, but may be spatially correlated with the distribution of stellar mass in the Galactic bulge.

In this thesis, we perform detailed modelling of the Galactic MSP population using data from the Fermi Large Area Telescope fourth source catalog data release 2 (4FGL-DR2) and the Australia Telescope National Facility (ATNF) pulsar catalog. Including in our model the spin down between formation and observation, we allow MSP luminosities to depend on intrinsic properties such as period $P$, magnetic field strength $B$ and spectral energy cutoff $E_{\rm cut}$.
We find a model in which luminosity $L \propto E_{\rm cut}^{a_{\gamma}} B^{b_{\gamma}} \dot{E}^{d_{\gamma}}$ provides the best fit to the data, where $a_{\gamma} = 1.2 \pm 0.3$, $b_{\gamma} = 0.1 \pm 0.4$ and $d_{\gamma} = 0.5 \pm 0.1$, and where $\dot{E} \propto B^2 / P^4$ is the spin-down power. This model is significantly better than one in which luminosity is independent of other properties of MSPs, with the luminosity distribution being fitted directly.
The Milky Way disk is expected to be the source of the resolved MSPs, with the GCE potentially produced by MSPs in the Galactic bulge. The Galactic bulge is divided into two structures: the boxy bulge, a significant bar structure extending a few kpc from the Galactic Center; and the nuclear bulge, a less massive component with radius $\sim230$ pc.
Due to differing star formation histories it is expected that the MSPs in the Galactic bulge are older and therefore dimmer than those in the Galactic disk. Additionally, correlations between the spectral parameters of the MSPs and the spin-down rate of the corresponding neutron stars
have been observed. This implies that the bulge MSPs may be spectrally different from the disk MSPs. Although we confirm these correlations, we do not find they are sufficiently large to significantly differentiate the spectra of the bulge MSPs and disk MSPs when the uncertainties are accounted for. We find the age distributions of MSPs cannot be distinguished from a uniform birth rate, based on current data.
Our results demonstrate that the population of MSPs that can explain the gamma-ray signal from the resolved MSPs in the Galactic disk and the unresolved MSPs in the boxy bulge and nuclear bulge can consistently be described as arising from a common evolutionary trajectory for some subset of astrophysical sources common to all these different environments. We do not require that there is anything systematically different about the inner Galaxy MSPs to explain the GCE.
Additionally, we use a more accurate geometry for the distribution of bulge MSPs and incorporate dispersion measure estimates of the MSPs' distances. We find that the elongated boxy bulge morphology means that some bulge MSPs are closer to us and so easier to resolve.  We identify three resolved MSPs that may belong to the bulge population.

In the ``recycling'' channel of MSP formation the neutron star forms from a core collapse supernovae that undergoes a random ``kick'' due to the asymmetry of the explosion. This would imply a smoothing out of the spatial distribution of the MSPs. We use $N$-body simulations to model how the MSP spatial distribution changes. We estimate the probability distribution of natal kick velocities using the resolved gamma-ray MSP proper motions, where MSPs have velocities relative to circular motion of $\sigma_v = 77\pm6$~km/s. The scale of these peculiar velocities are determined as part of our Galactic MSP population model.
We find that, due to the natal kicks,  there is an approximately 10\% increase in each of the bulge MSP spatial distribution dimensions and also the bulge MSP distribution becomes less boxy.
We estimate that natal kicks change the axis ratios of the MSP distribution in the Galactic
boxy bulge from  $\sim 1 : 0.43 : 0.40$ to $\sim 1 : 0.46 : 0.44$. Therefore, the bulge MSP distribution is still far from spherical.
\end{abstract}

\tableofcontents

\listoffigures

\listoftables

\printnomenclature

\mainmatter

\chapter{Introduction} 
\label{ch:introduction}

\graphicspath{{Introduction/Figs/}}

\nomenclature[z-4FGL-DR2]{4FGL-DR2}{Fermi Large Area Telescope fourth source catalog data release 2}
\nomenclature[z-AIC]{AIC}{Accretion Induced Collapse}
\nomenclature[z-ATNF]{ATNF}{Australia Telescope National Facility}
\nomenclature[z-COBE]{COBE}{Cosmic Background Explorer}
\nomenclature[z-DIRBE]{DIRBE}{Diffuse Infrared Background Experiment}
\nomenclature[z-DM]{DM}{Dispersion Measure}
\nomenclature[z-DTD]{DTD}{Delay Time Distribution}
\nomenclature[z-Fermi-LAT]{Fermi-LAT}{Fermi Large Area Telescope}
\nomenclature[z-GCE]{GCE}{Galactic Center Excess}
\nomenclature[z-LMXB]{LMXB}{Low Mass X-ray Binary}
\nomenclature[z-MCMC]{MCMC}{Markov Chain Monte Carlo}
\nomenclature[z-MSP]{MSP}{Millisecond Pulsar}
\nomenclature[z-NFW]{NFW}{Navarro-Frenk-White}
\nomenclature[z-NSC]{NSC}{Nuclear Stellar Cluster}
\nomenclature[z-NSD]{NSD}{Nuclear Stellar Disk}
\nomenclature[z-OGLE]{OGLE}{Optical Gravitational Lensing Experiment}
\nomenclature[z-SFR]{SFR}{Star Formation Rate}
\nomenclature[z-VVV]{VVV}{VISTA Variables in the Via Lactea}
\nomenclature[z-WAIC]{WAIC}{Watanabe Akaike Information Criterion}
\nomenclature[z-WIMP]{WIMP}{Weakly Interacting Massive Particle}
 
The Galactic Center Excess (GCE) is an extended gamma-ray source detected in the Fermi Large Area Telescope (Fermi-LAT) data in the central region of the Galaxy. This source, which has a spectral peak at a few GeV, initially appeared to have a spherically symmetric density profile, suggesting that it may be evidence of dark matter self-annihilating in the form of weakly interacting massive particles (WIMPs) with a Navarro-Frenk-White (NFW) distribution \citep{Goodenough:2009gk,Hooper_2011,Abazajian:2012pn,Gordon:2013vta}. More recently, however, more detailed examination has revealed that the GCE may not be spherically symmetric but exhibits a spatial morphology that is correlated with the distribution of stellar mass in the Galactic bulge \citep{Macias_2018,Bartels2017, Macias19, Abazajian2020,Coleman19}. One recent study, however, using different methods, still argues for a spherically symmetric GCE \cite{DiMauro2021}. If the GCE does trace the stellar mass of the bulge this would disfavor a dark matter origin and would point to a scenario in which it is produced by a population of dim, unresolved, astrophysical point sources such as Millisecond Pulsars (MSPs) \citep{Abazajian:2010zy}. Millisecond pulsars are rapidly spinning neutron stars with millisecond periods which emit gamma radiation with a spectrum that also peaks at a few GeV.
There is some debate about whether the resolved MSPs are consistent with the needed bulge population, see for example refs.~\cite{Hooper:2015jlu,Haggard_2017,Ploeg:2017vai,Bartels2018}.

In this thesis we focus on the Galactic population of MSPs and explore whether observations could be consistent with a GCE produced by an unresolved population of MSPs. In Chapter \ref{ch:msp_pop}, based on Ploeg et al.~\cite{Ploeg2020}, we model the Galactic population of MSPs using the resolved gamma-ray MSPs detected by Fermi-LAT. Extending the model of Ploeg et al.~\cite{Ploeg:2017vai}, where the only intrinsic property of pulsars was their luminosity, we now include other properties such as the period, magnetic field stength, and spectral parameters, upon which the luminosity may depend \cite{Kalapotharakos_2019,Gonthier2018}. In Chapter \ref{ch:msp_pop} the bulge MSPs that could produce the GCE are assumed to differ only in their star formation rate (SFR) and so we can, under the assumption that the GCE is entirely produced by MSPs, estimate the total number of bulge MSPs we may expect to have already resolved. Fitting several models to data, we find the best fit is one in which the luminosities of MSPs depend on their spectral energy cutoff; their magnetic field strength; and their spin-down power, the rate at which the pulsar loses rotational kinetic energy. We find that Galactic bulge MSPs do not need to be systematically different to explain the GCE, which could be produced by an inner Galaxy population of a few tens of thousands. In Chapter \ref{ch:nbody}, based on Ploeg and Gordon \cite{Ploeg2021}, we investigate the effect of birth kicks on the spatial distribution of MSPs, with a particular interest in the effect on the Galactic bulge distribution. We do this by running $N$-body simulations intended to approximate the Milky Way and introduce a distribution of particles which are given a random Maxwell distributed kick of an appropriate scale. We then fit a parametric model to the distributions of particles with no kick, kicks occurring at the beginning of the simulation, and kicks occurring at a uniform rate throughout the simulation. We find that the boxy bulge structure in the center of simulated galaxies is broadened slightly by kicks and becomes less boxy, with the scale parameters increasing by approximately $10\%$, however, it remains far from spherical. In the remainder of this chapter we provide further background on the Galactic bulge, GCE, and MSPs.

\section{The Galactic Bulge}
\label{sec:intro_galactic_bulge}

\begin{figure}
    \centering
    \includegraphics[width=0.99\linewidth]{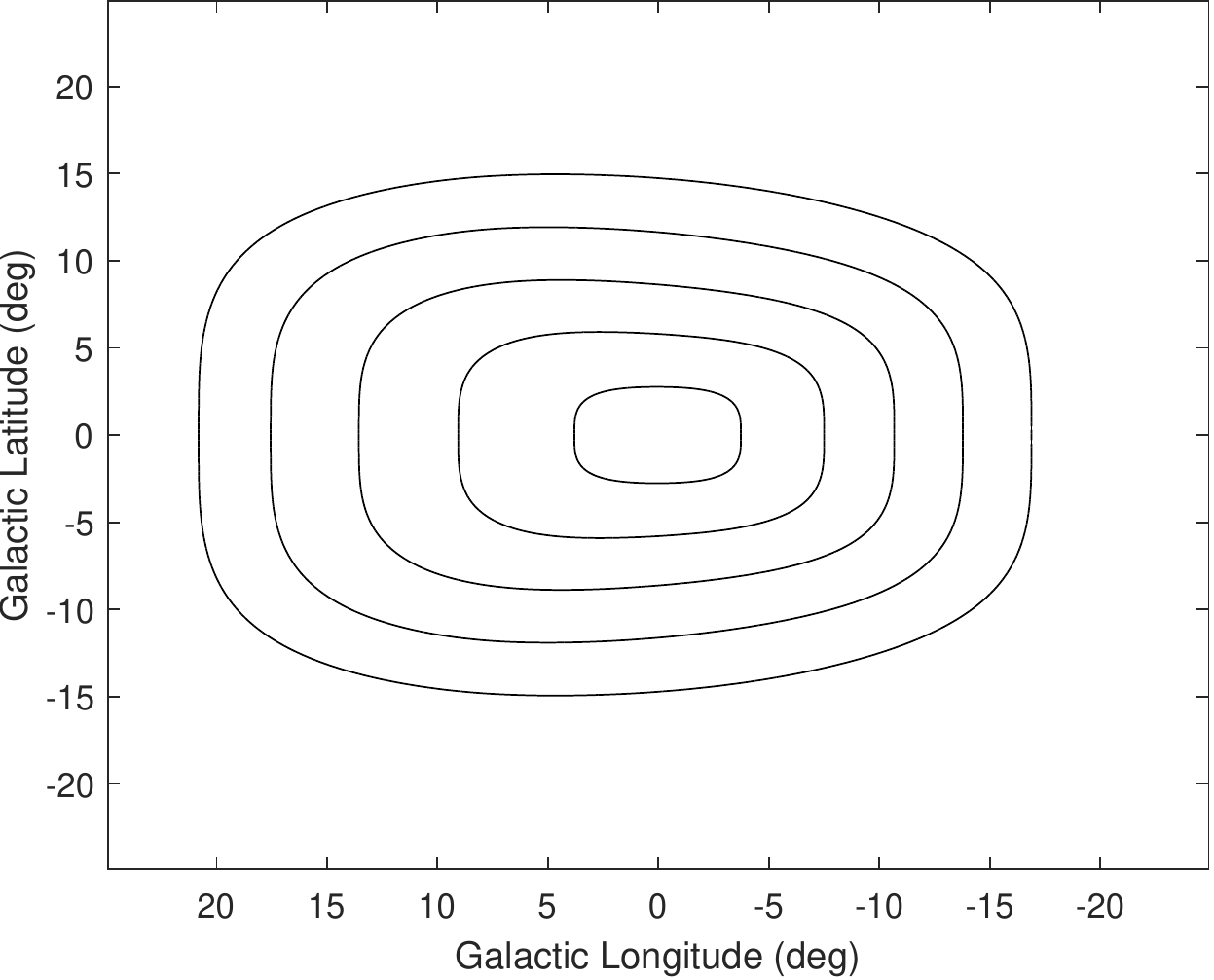}
    \caption[Bar model of Freudenreich \cite{Freudenreich:1997bx}.]{Integrated density along lines of sight for Freudenreich \cite{Freudenreich:1997bx} bar model. The contours are at $1/16$, $1/4$, $1$, $4$ and $16$ times the mean integrated density in this region.}
    \label{fig:intro_F98_bar}
\end{figure}

The inner region of the Milky Way has a significant bar or boxy bulge structure. This structure can be found in near infrared imagery from the Diffuse Infrared Background Experiment (DIRBE) of the Cosmic Background Explorer (COBE) after accounting for emission from material between the Sun and the Galactic Center, and also extinction associated with dust \cite{Weiland1994,Dwek95,Binney1997,Freudenreich:1997bx}. For example, Freudenreich \cite{Freudenreich:1997bx} fitted a $47$ parameter model of the Milky Way to $1.25$, $2.2$, $3.5$ and $4.9$ $\mu$m images. This model included a warped disk with a central hole, a dust density model of a similar form, and three alternative bar models. In addition to contributing to the modelled integrated emission along lines of sight, the dust model was used to calculate an extinction correction factor $\exp(-\tau_\nu(s))$ at a distance $s$. The best bar model was of the form:
\begin{equation}
\label{eq:intro_f98}
    \rho_{\rm F98,~bar} \propto \sech^2(R_s)
\end{equation}
\noindent where:
\begin{equation}
R_\perp^{C_\perp} = \left(\frac{\abs{x}}{a_x} \right)^{C_\perp} + \left(\frac{\abs{y}}{a_y} \right)^{C_\perp}
\end{equation}
\begin{equation}
R_s^{C_\parallel} = R_\perp^{C_\parallel} + \left(\frac{\abs{z}}{a_z} \right)^{C_\parallel}
\end{equation}
\noindent and where $x$, $y$ and $z$ are coordinates in the bar frame, which may be rotated to determine the angle between the location of the Sun and the bar major axis along $x$. The parameters $a_x$, $a_y$ and $a_z$ are the scale lengths along $x$, $y$ and $z$ respectively. The face-on and edge-on shape parameters, $C_\perp$ and $C_\parallel$, lead to a diamond shape when less than $2$ and a boxy shape when greater. Assuming the distance to Galactic Center is $8.5$ kpc, Freudenreich found $a_x=1.7$ kpc, $a_y=0.64$ kpc and $a_z=0.44$ kpc. With ${C_\perp} = 1.6$ and ${C_\parallel}=3.5$, the bar has a boxy shape when viewed edge-on. The angle of the bar relative to the Sun was $13.8 \deg$. The contour plot of Fig.~\ref{fig:intro_F98_bar} shows this bar as viewed from the location of the Sun, where we have integrated along lines of sight through the density model. The boxy appearance is clear.
Using red clump giants, work using observations from the Optical Gravitational Lensing Experiment (OGLE) \cite{Stanek:1997,Rattenbury2007,Cao:2013dwa} and the VISTA Variables in the Via Lactea (VVV) survey \cite{Wegg2013,Simion2017} have come to similar conclusions. As red clump giants have a relatively narrow luminosity distribution, estimates of their distance can be made using their apparent magnitude. Cao et al.~\cite{Cao:2013dwa} use a boxy bulge model with density:
\begin{equation}
\label{eq:intro_cao13}
    \rho_{\rm Cao13,~bar} \propto K_0(R_s)
\end{equation}
\noindent where $K_0$ is the modified Bessel function of the second kind and with fixed parameters ${C_\perp} = 2$ and ${C_\parallel}=4$. They find the bar is at an angle of $29.4 \deg$ and has scale parameters of $0.67$ kpc, $0.29$ kpc and $0.27$ kpc along $x$, $y$ and $z$ respectively. Combining red clump giant data from multiple surveys, Wegg et al.~\cite{Wegg2015} find a total bar half length of $5.0 \pm 0.2$ kpc. As part of this structure, they also find evidence for a two component long bar, a thin extension to the boxy bulge. The scale heights of these two components are $\approx 180$ pc and $\approx 45$ pc, and they appear to be aligned with the boxy bulge with an angle of $28$ to $33$ degrees. Overall, these studies of the Galactic bulge tend to find the sun is located at an angle $\lesssim 30 \deg$ from the major axis of the bar.

A second, much smaller, structure present at the Galactic Center is the nuclear bulge. This consists of two components: the Nuclear Stellar Cluster (NSC), and the Nuclear Stellar Disk (NSD). The NSC, with a mass of $\sim3 \times 10^{7} \msun$, has an $r^{-2}$ density profile in the inner few parsecs, while the far more massive NSD, $M_{\rm NSD} = (1.4 \pm 0.6) \times 10^{9} \msun$, is a disk with radius $\sim230$ pc and scaleheight of $45 \pm 5$ pc \cite{Launhardt2002}.

If the GCE is produced by a population of unresolved astrophysical sources, such as MSPs, we expect that those sources would have a spatial distribution similar to that of the Galactic bulge stellar mass. In Chapter \ref{ch:msp_pop} the Freudenreich \cite{Freudenreich:1997bx} model of the boxy bulge (Eq.~\ref{eq:intro_f98}) is used to model the bar population of MSPs; a parametric model described in that chapter is used for the nuclear bulge population. In Chapter \ref{ch:nbody} a model of the same form as the Cao et al.~\cite{Cao:2013dwa} boxy bulge (Eq.~\ref{eq:intro_cao13}) is used as a component of the parametric model fitted to the spatial distribution of particles at the end of $N$-body simulations.

\section{The Galactic Center Excess}
\label{sec:intro_gce}

The Large Area Telescope (LAT) is a gamma-ray telescope on board the Fermi Gamma-ray Space Telescope (Fermi) \cite{Atwood2009}. Fermi-LAT, which images the entire sky approximately every $3$ hours, detects gamma-rays when they convert to electron-positron pairs upon interacting with a thin layer of tungsten foil. The energy of the gamma-ray is estimated by measuring the energy of the charged particles produced and tracking them allows the incoming direction to be reconstructed. It is sensitive to gamma-rays in the range of $20$ MeV to $300$ GeV, with uncertainty in direction declining as energy increases, decreasing from an angular resolution of a few degrees at $100$ MeV to $\sim 0.1 \deg$ at $100$ GeV \cite{Ackermann2012}.

Based on about a year of gamma-ray data from Fermi, Goodenough and Hooper \cite{Goodenough:2009gk} reported that the radiation from the inner few degrees around the center of the Galaxy could be consistent with annihilating dark matter with an inner $r^{-\gamma}$ profile with $\gamma=1.1$. Further studies \cite{Hooper_2011,Abazajian:2012pn,Gordon:2013vta,Macias:2013vya,Abazajian:2014fta,Daylan:2014rsa,TheFermi-LAT:2015kwa} confirmed the presence of this apparently spherically symmetric extended source of gamma-rays with $\gamma \approx 1.2$ and with a spectrum peaking at a few GeV. The spatial distribution of the dark matter halo can be modelled using:
\begin{equation}
    \rho(r) = \frac{\rho_0}{\left( \frac{r}{r_s} \right)^{\gamma} \left( 1 + \left(\frac{r}{r_s}\right)^{\alpha} \right)^{(\beta - \gamma)/\alpha}}
\end{equation}
\noindent which is a generalized version of the Navarro-Frenk-White (NFW) profile where $\alpha = 1$, $\beta = 3$ and $\gamma = 1$ \cite{Navarro1996}. The behavior near the Galactic Center where $r \ll r_s$ is $\rho \propto r^{-\gamma}$. The dark matter annihilation flux as a function of energy $E_{\gamma}$ and galactic coordinates $l$ and $b$ can be written as \cite{Baltz2008,Gordon:2013vta}:
\begin{equation}
    \Phi (E_{\gamma}, b, l) = \Phi^{\rm PP} (E_{\gamma}) \times J(b,l)
\end{equation}
\noindent where $\Phi^{\rm PP}$ is a particle physics contribution:
\begin{equation}
    \Phi^{\rm PP} = \frac{1}{2} \frac{\langle \sigma v \rangle}{ 4 \pi m^2_{\rm DM}} \sum_f \frac{\dd N_f}{\dd E_{\gamma}} B_f
\end{equation}
\noindent where $\langle \sigma v \rangle$ is the mean self-annihilation cross-section multiplied by the relative velocity of the dark matter particles, $m_{\rm DM}$ is the WIMP mass, $\dd N_f / \dd E_{\gamma}$ is the spectrum produced by annihilation channel $f$ with branching ratio $B_f$. The factor $J(b,l)$ accounts for the dark matter profile along the line of sight at $l$ and $b$, where the flux produced at a point in space is proportional to the square of the density of dark matter particles:
\begin{equation}
    J(b,l) = \int \dd s \rho^2(r(s,b,l))
\end{equation}
\noindent Hooper and Goodenough \cite{Hooper:2010mq} found $7$--$10$ GeV WIMPs annihilating to the tau lepton pair $\tau^{+} \tau^{-}$ provided a good fit to the data. Abazajian and Kaplinghat \cite{Abazajian:2012pn} find annihilations of $10$ GeV to $1$ TeV WIMPS to $b \Bar{b}$ quarks and $10$--$30$ GeV WIMPs to $\tau^{+} \tau^{-}$. Gordon and Macias \cite{Gordon:2013vta} suggest WIMPs with mass $20$--$60$ GeV annihilating to a mixture of $\tau^{+} \tau^{-}$ and $b \Bar{b}$. In addition to finding the spectrum and spatial morphology of the GCE is consistent with self-annihilating WIMPs distributed according to an NFW profile, these studies also find annihilation cross-sections similar to the thermal relic value of $\langle \sigma v \rangle \approx 2.2 \times 10^{-26}$ cm$^3$ s$^{-1}$, the predicted value, given the current dark matter density, for WIMPs of mass $\gtrsim 10$ GeV frozen out of thermal equilibrium in the early universe as it cooled \cite{Steigman_2012}.

An alternative possibility is a population of point sources in the Galactic Center unresolved by Fermi-LAT, such as MSPs which have similar gamma-ray spectra to the GCE \cite{Abazajian:2010zy}. The possibility of a population of MSPs in the inner Galaxy was proposed as early as 2005 in order to explain diffuse gamma-ray emission observed by EGRET in the Galactic Center region \cite{Wang2005}. As the flux would be proportional to their density, they would instead have a density distribution $\rho(r) \propto r^{-2 \gamma}$. Brandt and Kocsis \cite{Brandt:2015ula} proposed that this profile could have arisen from the disruption of globular clusters. During this process, dynamical friction causes the orbits of globular clusters to decay and tidal forces disrupt the cluster as it approaches the Galactic Center. According to Gnedin et al.~\cite{Gnedin:2013cda} this would lead to a distribution of stars, including MSPs, with inner profile $\rho(r) \sim r^{-2.2}$.

Yuan and Zhang \cite{Yuan2014} used resolved gamma-ray MSP data from the second Fermi-LAT catalog of gamma-ray pulsars \cite{TheFermi-LAT:2013ssa} to show that a population of MSPs with an $r^{-2.4}$ profile and with a luminosity distribution similar to those in the disk could explain the GCE without any being resolved. A similar analysis was performed by Petrovic et al.~\cite{Petrovic2015}, who also suggested the possibility that there may be a secondary emission contribution to the GCE caused by inverse Compton scattering by relativistic leptons escaping MSPs. Hooper and Mohlabeng \cite{Hooper:2015jlu} later argued, fitting the spatial and luminosity distribution of MSPs in the Fermi-LAT third source catalog \cite{Acero:2015hja}, that between $15$ and $43$ Galactic Center MSPs would have been resolved. However, Ploeg et al.~\cite{Ploeg:2017vai} found, using different methods, that there existed a region of parameter space that was consistent with the observed MSP data and where a population of bulge MSPs with the same luminosity distribution could explain the GCE while having a high probability that none were resolved. Hooper and Mohlabeng \cite{Hooper:2015jlu} performed a binned fit to resolved MSP Galactic coordinates and flux, whereas Ploeg et al.~\cite{Ploeg:2017vai} used an unbinned fit and also included parallax distance estimates where available.

There is statistical evidence of unresolved point sources near the Galactic Center below the Fermi-LAT detection threshold that may explain at least a significant fraction of the GCE \cite{Lee:2015fea,Bartels:2015aea,Buschmann2020,Calore2021}. In addition, in recent years it has become increasingly apparent that the GCE may in fact be correlated with the distribution of stellar mass in the Galactic Center \citep{Macias_2018,Bartels2017, Macias19, Abazajian2020,Coleman19}. These studies perform fits using spatial templates modelling the contributions from sources distributed as the boxy bulge and nuclear bulge, as well as from NFW squared templates modelling dark matter annihilation. They do not find a significant NFW squared component, strongly disfavouring a scenario in which a spherically symmetric distribution of annihilating dark matter is the dominant source of the GCE. Bartels et al.~\cite{Bartels2017} find the best fitting boxy bulge model of Cao et al.~\cite{Cao:2013dwa} (Eq.~\ref{eq:intro_cao13}) provided a good fit to the data. Using the Freudenreich \cite{Freudenreich:1997bx} model (Eq.~\ref{eq:intro_f98}), Macias et al.~\cite{Macias19} argue for an admixture formation scenario where there is a mixture of primordial and dynamical formation. In the primordial case, the MSPs would simply follow the bar density $\rho_{\rm bar}$, whereas for dynamical formation their density would be proportional to the stellar encounter rate which in turn is proportional to the bar density squared, $\rho_{\rm bar}^2$. Adding a parameter $s$, they find using the density $\rho_{\rm F98,~bar}^s$, a best fit value of $s = 1.4$. They also show, by fitting a combination of the primordial and dynamical templates, the primordial channel contribution is $(52 \pm 23)\%$. These results are dependent on modelled Galactic diffuse gamma-ray emission; for example, Di Mauro \cite{DiMauro2021} recently argued, based on 11 years of Fermi-LAT data, that the spatial morphology of the GCE remains consistent with dark matter annihilating with a spherically symmetric NFW profile.

In summary, while WIMPs annihilating to gamma-rays cannot at present be ruled out, the evidence for sub-threshold point sources and a non-spherical spatial distribution suggests that the GCE may be produced by a population of unresolved astrophysical point sources. The spectrum of the GCE is also similar to Fermi-LAT detected MSPs, and it is these we consider in this thesis.

\section{Millisecond Pulsars}
\label{sec:intro_msps}

\begin{figure}
    \centering
    \includegraphics[width=0.99\linewidth]{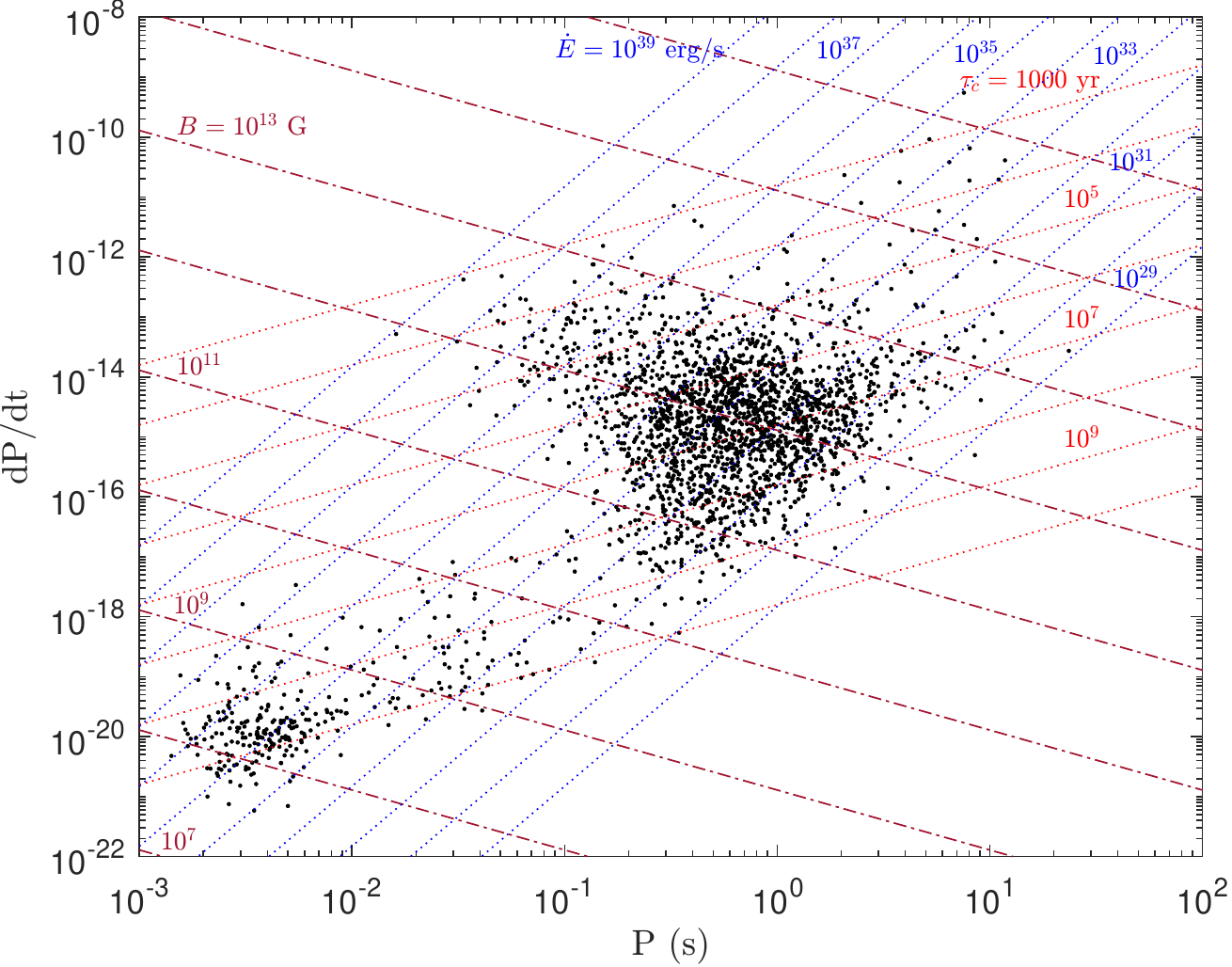}
    \caption[Periods and period derivatives of pulsars.]{Periods $P$ and measured period derivatives $\dot{P}$ of pulsars in ATNF catalog \cite{Manchester:2004bp}. Included are lines of constant magnetic field strength $B$, spin-down power $\dot{E}$ and characteristic age $\tau_c$.}
    \label{fig:intro_P_vs_Pdot}
\end{figure}

Pulsars are rapidly rotating neutron stars. Their radio and gamma-ray emission is driven by the acceleration of charged particles in strong magnetic fields. This emission is not isotropic, with the strongest emission along the magnetic axis leading to pulsations as the pulsar rotates. They can be divided into two groups: MSPs, and young pulsars. MSPs are pulsars with rotation periods $P$ on the order of milliseconds, they also tend to spin down at a slower rate, with $\dot{P} \lesssim 10^{-18}$. In Fig.~\ref{fig:intro_P_vs_Pdot}, the periods and rate of spindown of pulsars in the Australia Telescope National Facility (ATNF) pulsar catalog are shown, the MSPs are clearly seen in the bottom left corner separated by a region of relatively few pulsars from the larger population of young pulsars.
For MSPs, $\dot{P}$ is small enough that a correction should be applied, where the true period rate of change is \cite{TheFermi-LAT:2015kwa}:
\begin{equation}
    \dot{P}_{\rm int} = \dot{P} - \Delta \dot{P}
\end{equation}
\noindent where $\Delta \dot{P}$ accounts for the Shklovskii effect \cite{Shklovskii1970} and radial acceleration in the Galactic potential. This correction of order $\sim 10^{-21}$ will be important later in Chapter \ref{ch:msp_pop}. The surface magnetic field strength of a pulsar $B \propto \sqrt{P \dot{P}}$ is much smaller for MSPs ($\sim 10^8$--$10^9$ G) than young pulsars ($\sim 10^{11}$--$10^{13}$ G). Assuming pulsars spin down via magnetic dipole braking and that the birth period $P_I \ll P$, we can define the characteristic age as \cite{Ferrario2007}:
\begin{equation}
    \tau_c = P / 2 \dot{P}
\end{equation}
For MSPs these are $\sim 1$--$10$ Gyr and for young pulsars $\lesssim 100$ Myr. As a result of their far weaker magnetic fields, the MSPs spin down very slowly when compared to young pulsars. This is part of the reason we do not further consider young pulsars as a likely explanation for the GCE, a lack of recent star formation means they would have long ago spun down to the point where they no longer produce sufficient gamma radiation \cite{Ploeg:2017vai}. However, O'Leary et al.~\cite{OLeary:2015qpx} have argued there may be enough star formation in the inner $\lesssim 200$ pc for them to make a substantial contribution.

The ``recycling'' model of MSP formation involves the accretion of material from a binary companion onto an old pulsar formed in a core collapse supernova, transferring angular momentum and spinning up the pulsar to millisecond periods \cite{Bhattacharya1991,Lorimer2005,TaurisLangerKramer2012}. This process can occur when, assuming the binary system remains intact after the supernova, the companion evolves and overfills its Roche lobe, losing mass to the neutron star. Given the much stronger magnetic fields of young pulsars, this model of MSP formation also requires that there is some mechanism by which the magnetic field strengths decay to around the $10^8$--$10^9$ G range. For example, if the magnetic field is generated by currents in the neutron star crust, one possibility is that decay of the magnetic field may be accelerated by heating of the crust during the accretion process. When the crust is heated, its conductivity is decreased and as a result the rate of Ohmic dissipation of currents is increased \cite{Geppert1994,Konar1997}. During the accretion phase the system could potentially be seen as a low mass X-ray binary (LMXB) and Haggard et al.~\cite{Haggard_2017} have argued, assuming the ratio of total MSP luminosity in globular clusters to the number of LMXBs is similar, that $\sim 1000$ Galactic bulge LMXBs should have been detected, and therefore the detection of only $42$ by the INTEGRAL telescope is evidence against the MSP scenario of the GCE. A similar argument is made by Cholis et al.~\cite{Cholis:2014lta}; however, these environments are different, and there are a number of pathways that could lead to recycled MSPs \cite{Ivanova_2008}, the ratios of which may differ between globular clusters and the Galactic bulge. Bartels et al.~\cite{Bartels_UCXBs} suggest that a population of dim ultra-compact X-ray binaries, a hydrogen-deficient sub-class of LMXBs that form at a significantly higher rate than hydrogen-rich LMXBs \cite{van_Haaften_2015}, could be consistent with the detected numbers and could produce enough MSPs to explain the GCE.

An alternative channel for MSP formation is Accretion Induced Collapse (AIC), in which a white dwarf accretes enough material to exceed the Chandrasekhar mass limit of about $1.4 \msun$ and collapses to a neutron star \cite{Bhattacharya1991,Ferrario2007,Tauris2013}. Conservation of angular momentum could result in periods on the order of milliseconds, similarly magnetic flux conservation could produce $\sim 10^8$--$10^9$ G magnetic field strengths after the collapse of a white dwarf where generally $B \lesssim 10^3$ G \cite{Ferrario2007}. We therefore do not need any process by which magnetic fields decay to explain observed MSP magnetic field strengths. Ferrario and Wickramasinghe \cite{Ferrario2007} note that the distribution of magnetic field strengths for white dwarfs is bimodal, for the second group in the range $\sim 10^6$--$10^9$ G AIC would produce MSPs with very strong magnetic fields $\gsim 10^{10}$ G. However, they would spin down very rapidly. The kick associated with AIC would be much smaller than that occurring during core collapse of a massive star, this may explain the relatively low velocities of MSPs as well as the number in globular clusters, which have small escape velocities \cite{Tauris2013}.

Neutron stars formed in core collapse supernovae receive a natal kick due to asymmetries in the supernova explosion, and these kick velocities are on the order of several hundred km s$^{-1}$ \cite{Wongwathanarat2013,Bear2018}. In Hobbs et al.~\cite{Hobbs_2005}, where the proper motions of $233$ pulsars were studied, the young pulsars had mean transverse velocities of $246\pm22$ km s$^{-1}$, however, the MSPs had significantly lower mean transverse velocities of $87\pm13$ km s$^{-1}$. If MSPs are primarily recycled pulsars, then the lower velocities may be a result of the requirement that the binary system is not disrupted by a large natal kick. If it is disrupted, the neutron star may be observed as an isolated young pulsar with a high transverse velocity, but cannot later be spun up to millisecond periods by a binary companion. As part of the modelling of the Galactic MSP population in Chapter \ref{ch:msp_pop}, we fit a Maxwell distributed peculiar velocity distribution to the resolved MSP proper motions, where peculiar velocity is the velocity relative to circular motion at a particular location in the Galaxy. We use the result as an input to the work described in Chapter \ref{ch:nbody}, where we relate it to a natal kick velocity distribution.

The gamma-ray emission of MSPs is believed to be due to curvature radiation from charged particles accelerated in the electromagnetic field \cite{Kalapotharakos_2019,Petri2019}. Petri \cite{Petri2019} uses a simple model in which a spherically symmetric distribution of positrons and electrons are assumed to be in equilibrium between acceleration in the electromagnetic field and braking due to the emission of radiation. In this radiation reaction limit, where the speed is assumed to be the speed of light $c$, the velocity of a positron or electron at a point depends on the local electric field $\pmb{E}$ and magnetic field $\pmb{B}$ \cite{Gruzinov2013}:
\begin{equation}
\label{eq:radiation_reaction_velocity}
    \pmb{v}_{\pm} = \frac{\pmb{E} \cross \pmb{B} \pm \left(E_0 \pmb{E} / c + c B_0 \pmb{B} \right)}{E_0^2 / c^2 + B^2}
\end{equation}
\noindent where $\pmb{v}_{+}$ and $\pmb{v}_{-}$ are the velocity vectors for positrons and electrons respectively; and $E_0$ and $B_0$ are two electromagnetic invariants which, requiring $E_0 \geq 0$, satisfy:
\begin{equation}
    E^2 - c^2 B^2 = E_0^2 - c^2 B_0^2
\end{equation}
\begin{equation}
    \pmb{E} \cdot \pmb{B} = E_0 B_0
\end{equation}
Then the curvature radius $\rho_c$ can be found using:
\begin{equation}
    \abs{\pmb{a_{\pm}}} = \abs{\frac{\dd \pmb{v_{\pm}}}{\dd t}} = \frac{c^2}{\rho_c}
\end{equation}
\noindent and the Lorentz factor $\gamma$ is:
\begin{equation}
    \gamma^4 = \frac{6 \pi \epsilon_0}{e} E_0 \rho_c^2
\end{equation}
\noindent where $\epsilon_0$ is the vacuum permittivity and $e$ is the elementary charge. Petri \cite{Petri2019} then uses for the curvature radiation the following spectrum:
\begin{equation}
\label{eq:curvature_radiation_spectrum}
    \frac{\dd I}{\dd \omega} = \frac{\sqrt{3} e^2}{4 \pi \epsilon_0 c} \gamma \frac{\omega}{\omega_c} \int_{\omega/\omega_c}^{\infty} K_{5/3} (x) \dd x
\end{equation}
\noindent where $I$ is the intensity, $\omega$ is the angular frequency, $K_{5/3}$ is the modified Bessel function of the second kind and where the characteristic frequency is:
\begin{equation}
\label{eq:curvature_radiation_characteristic_frequency}
    \omega_c = \frac{3}{2} \gamma^3 \frac{c}{\rho_c}
\end{equation}
In this simple model, with photons approximately emitted in the lepton direction of motion, $\gamma \gsim 10^8$ producing pulsar spectra with peaks of a few GeV. If a particle trajectory is at a high pitch-angle relative to the asymptotic trajectory given by Eq.~\ref{eq:radiation_reaction_velocity}, the particle will emit synchrotron radiation where the radius of curvature used in Eqs.~\ref{eq:curvature_radiation_spectrum} and \ref{eq:curvature_radiation_characteristic_frequency} is instead the gyro-radius of the particle motion. Kalapotharakos et al.~\cite{Kalapotharakos_2019} show that synchrotron radiation and curvature radiation will lead to different relationships between the luminosity of a pulsar and its spectral energy cutoff, magnetic field strength and spin down power. They show using a least squares fit to Fermi-LAT detected pulsars that the curvature radiation relationship provides a better fit to the data, and we confirm this in Chapter \ref{ch:msp_pop}.

More realistic models of MSP emission focus on three regions, the polar caps, the slot gap and the outer gap \cite{Harding2007,Venter2009,Venter2012,Pierbattista2012,Petri2016}. The polar caps are a region where open magnetic field lines cross the surface of the pulsar. These field lines are those that cross the light cylinder at a radius $r_L = c P / 2 \pi$, beyond which particles cannot corotate with the neutron star with $v < c$. In the polar cap model, the emission is associated with particles accelerated along open magnetic field lines from near the surface \cite{Sturrock1971,Ruderman1975,Harding1978,Daugherty1982}. The slot gap is the region near the boundary of the closed field lines, extending from the surface of the pulsar to the light cylinder \cite{Arons1983,Muslimov2003,Muslimov2004}. The outer gap is the volume in the light cylinder between closed field line boundary and the null surface where $\pmb{\Omega} \cdot \pmb{B} = 0$, where $\pmb{\Omega}$ is the rotation axis \cite{Cheng1986,Chiang1992,Romani1995,Romani1996}. Kalapotharakos et al.~\cite{Kalapotharakos2018} use a three dimensional particle-in-cell method to model the distribution and trajectories of charged particles in the pulsar magnetosphere and their associated magnetic fields. Using the radiation reaction limit, they simulate sky maps and spectra of the gamma-ray emission, they also show there is a relationship between spin down power and spectral energy cutoff. 

\chapter{Modelling the Galactic Millisecond Pulsar Population}
\label{ch:msp_pop}

\graphicspath{{msp_population/Figs/}}

\section{Introduction}
\label{sec:intro}

In previous investigations, the spectrum of the MSPs in the bulge have been assumed to be the same as those in the disk. However, the bulge MSPs are expected to be on average  older than the disk MSPs due to their different star formation histories \cite{Crocker:2016zzt}.
Therefore, we would expect them to have on average lower luminosities. In addition to this, a correlation between luminosities and the spectral parameters is seen in the data \cite{TheFermi-LAT:2013ssa,Kalapotharakos_2019}
and so one would expect the bulge MSPs to be spectrally different from the disk ones. This motivates the  more detailed modelling of the MSP populations that is performed in the current chapter.

We extend the model of the Galactic population of gamma-ray MSPs of Ploeg et al.~\cite{Ploeg:2017vai}. 
In that work, only distance estimates from parallax measurements were included. Here we also use the dispersion measure estimates of the MSP distances and incorporate the corresponding uncertainties in the free electron densities.
Also, instead of empirically parameterising the MSP luminosity function, we start from empirical distributions describing MSP initial period, magnetic field strength, age, and gamma-ray spectra. We  assume there is a relationship between the luminosity of a pulsar and some of its other properties such as its period, period derivative or spectral energy cutoff. 
This supposition is motivated by work such as by Kalapotharakos et al.~\cite{Kalapotharakos_2019} in which, based on data on  resolved MSPs and young pulsars in Abdo et al.~\cite{TheFermi-LAT:2013ssa}, it was determined that  there is
a relationship between
MSP
luminosity, on the one hand, 
and MSP spectral energy cutoff $E_{\rm cut}$, magnetic field strength $B$, and the spin-down power $\dot{E}$,
on the other.
Specifically,
Kalapotharakos et al.~\cite{Kalapotharakos_2019}
found that gamma-ray emission from MSPs is via curvature radiation and scales like $L \propto E_{\rm cut}^{1.18 \pm 0.24} B^{0.17 \pm 0.05} \dot{E}^{0.41 \pm 0.08}$ where we use 68\% confidence intervals when quoting uncertainties unless otherwise specified.
Similarly, Gonthier et al.~\cite{Gonthier2018} assumed that the radio and gamma ray luminosities of MSPs
were dependent on period and period derivative 
and, under this assumption,
successfully
determined
the parameters of that relationship using a model of the distribution and properties of the Galactic population of MSPs, radio and gamma-ray detection thresholds, and a model of how the observed flux of an individual MSP depends on viewing angle and magnetic axis angle.

\section{Method}
\label{sec:method}
To fit our model of the Galactic MSP population we used MSPs with confirmed gamma ray pulsations according to the Public List of LAT-Detected Gamma-Ray Pulsars.\footnote{\label{footnote:pulsars}\url{https://confluence.slac.stanford.edu/display/GLAMCOG/Public+List+of+LAT-Detected+Gamma-Ray+Pulsars}} From that list, we obtained names and periods of pulsars. We then used the gamma-ray data for the corresponding sources in the Fermi Large Area Telescope fourth source catalog data release 2 \citep[4FGL-DR2:][]{Ballet:2020hze}. Additional data were obtained from the ATNF pulsar catalog \citep{Manchester:2004bp} if available. As in Bartels et al.~\cite{Bartels2018} we used pulsars with periods less than $30$ ms that were not associated with globular clusters. For the GCE, we use the boxy bulge and nuclear bulge spectra\footnote{Available from \url{https://github.com/chrisgordon1/Galactic_bulge_spectra}.} from Macias et al.~\cite{Macias19}. We included both the systematic and statistical errors of these spectra which were added in quadrature to get the total error. The systematic error accounts for variation in the GCE spectra caused by using different maps of Inverse Compton emission.

\subsection{Modeling the Galactic Millisecond Pulsar population}
\label{ssec:method_model_MSP_pop}
\subsubsection{Spatial distribution}
The spatial model of MSPs has three components: a disk distribution, a boxy bulge distribution, and a nuclear bulge distribution. The disk component models the population from which we expect the resolved MSPs to mainly come and has density:
\begin{equation}
\label{eq:rho_disk}
\rho_{\rm disk} (R, z) \propto \exp(-R^2/2\sigma_r^2) \exp(-\abs{z}/z_0)
\end{equation}
\noindent where $R^2 = x^2 + y^2$ is the radial coordinate in the Galactic disk and $z$ is the height above the Galactic Plane. We treat $\sigma_r$ and $z_0$ as free parameters to be fit to the data. The modeled GCE is produced by the boxy bulge and nuclear bulge components. The boxy bulge has density \citep{Freudenreich:1997bx,Macias19}:
\begin{equation}
\label{eq:rho_main_bulge}
\rho_{\rm boxy~bulge} (R_s) \propto \sech^2(R_s)
\times \begin{cases}
      1 & R \leq R_{\rm end} \\
      \exp(-(R - R_{\rm end})^2/h^2_{\rm end}) & R > R_{\rm end} \\
\end{cases}
\end{equation}
\noindent where $R_{\rm end} = 3.128$ kpc, $h_{\rm end} = 0.461$ kpc, and:
\begin{equation}
\label{eq:rho_main_bulge_r_perp}
R_\perp^{C_\perp} = \left(\frac{\abs{x'}}{1.696~\textrm{kpc}} \right)^{C_\perp} + \left(\frac{\abs{y'}}{0.6426~\textrm{kpc}} \right)^{C_\perp}
\end{equation}
\begin{equation}
\label{eq:rho_main_bulge_r_s}
R_s^{C_\parallel} = R_\perp^{C_\parallel} + \left(\frac{\abs{z'}}{0.4425~\textrm{kpc}} \right)^{C_\parallel}
\end{equation}
\noindent where $C_\parallel = 3.501$ and $C_\perp = 1.574$. The coordinates $x'$, $y'$ and $z'$ are Cartesian coordinates in the boxy bulge frame. Relative to the frame in which $x_\odot=-R_0$, $y_\odot=z_\odot=0$, this frame is rotated $13.79^{\circ}$ around the z-axis then $0.023^{\circ}$ around the new y-axis. We assume $R_0=8.3$ kpc as adopted from the YMW16 free electron density model \citep{Yao2017} which we use to convert between distance and dispersion measure.
The nuclear bulge MSP density is proportional to the sum of the mass densities of the NSC and NSD \citep{Bartels2017}:
\begin{equation}
\label{eq:rho_nuclear_bulge}
\rho_{\rm nuclear~bulge} (r,z) \propto \rho_{\rm NSC}(r) + \rho_{\rm NSD}(r,z)
\end{equation}
\noindent where $r^2 = x^2 + y^2 + z^2$ and where the NSC has density:
\begin{equation}
\label{eq:rho_nuclear_stellar_cluster}
\rho_{\rm NSC} (r) = 
\begin{cases} 
      \frac{\rho_{0 \rm ,NSC}}{1 + \left( \frac{r}{r_0} \right)^2} & r \leq 6~\textrm{pc} \\
      \frac{\rho_{1 \rm ,NSC}}{1 + \left( \frac{r}{r_0} \right)^3} & 6~\textrm{pc} < r \leq 200~\textrm{pc} \\
      0 & r > 200~\textrm{pc}
   \end{cases}
\end{equation}
\noindent where $r_0 = 0.22~\textrm{pc}$, $\rho_{0 \rm ,NSC} = 3.3 \times 10^6~\textrm{M}_\odot~\textrm{pc}^{-3}$ and $\rho_{1 \rm ,NSC}$ is set so that $\rho_{\rm NSC} (r)$ is continuous at $r = 6~\textrm{pc}$. The NSD has density:
\begin{equation}
\label{eq:rho_nuclear_stellar_density}
\rho_{\rm NSD} (r,z) = 
\begin{cases} 
      \rho_{0 \rm ,NSD}\left(\frac{r}{1~\textrm{pc}} \right)^{-0.1} e^{-\frac{\abs{z}}{45~\textrm{pc}}} & r < 120~\textrm{pc} \\
      \rho_{1 \rm ,NSD}\left(\frac{r}{1~\textrm{pc}} \right)^{-3.5} e^{-\frac{\abs{z}}{45~\textrm{pc}}} & 120~\textrm{pc} \leq r < 220~\textrm{pc} \\
      \rho_{2 \rm ,NSD}\left(\frac{r}{1~\textrm{pc}} \right)^{-10} e^{-\frac{\abs{z}}{45~\textrm{pc}}} & r \geq 220~\textrm{pc}
   \end{cases}
\end{equation}
\noindent where $\rho_{0 \rm ,NSD} = 301~\textrm{M}_\odot~\textrm{pc}^{-3}$ and $\rho_{1 \rm ,NSD}$ and $\rho_{2 \rm ,NSD}$ are set so that $\rho_{\rm NSD} (r,z)$ is continous at both $r = 120~\textrm{pc}$ and $r = 220~\textrm{pc}$.

\subsubsection{Age distribution}
The MSP age distribution is dependent on two distributions: the SFR, and the delay time distribution (DTD). The delay time is the time between star and MSP formation. We assume for the DTD a five bin distribution, linearly spaced between 0 Gyr and the age of the universe.

\begin{figure}
    \centering
    \includegraphics[width=0.99\linewidth]{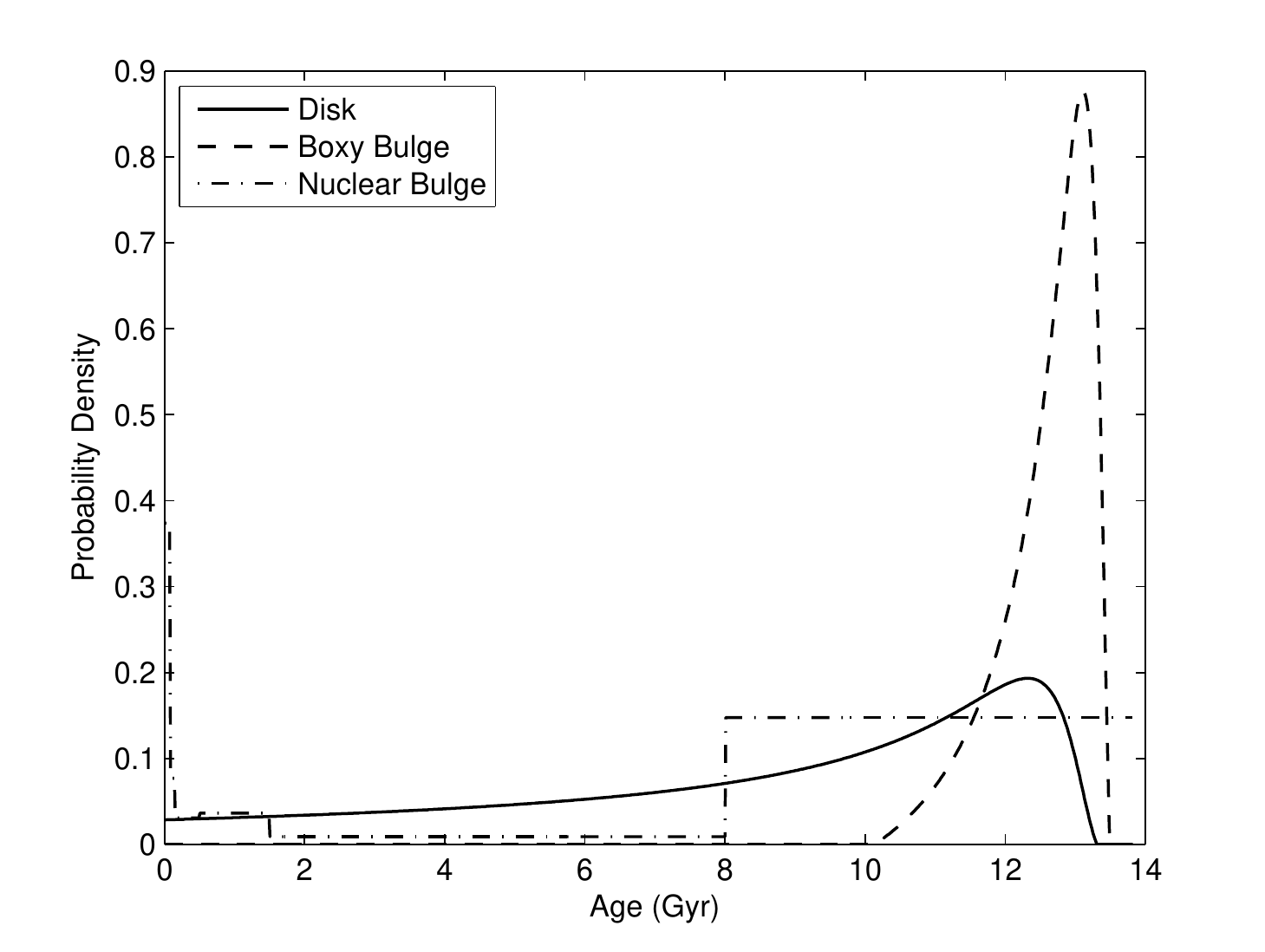}
    \caption[Star formation rates for the disk, boxy bulge, and nuclear bulge.]{Star formation rates for the disk, boxy bulge, and nuclear bulge. Here, this is given as a normalized probability density of stellar mass having formed at a particular age. 
    }
    \label{fig:sfr}
\end{figure}

For the disk and boxy bulge components of the MSP population, the SFR is \citep{Crocker:2016zzt}:
\begin{equation}
\label{eq:SFR}
\textrm{SFR}(z) = \max(10^{A z^2 + B z + C} - D, 0)
\end{equation}
\noindent where for the disk MSPs $A = -4.06 \times 10^{-2}$, $B = 0.331$, $C = 0.338$ and $D = 0.771$. For the boxy bulge $A = -2.62 \times 10^{-2}$, $B = 0.384$, $C = -8.42 \times 10^{-2}$ and $D = 3.254$. The relationship between cosmological time $t$ and redshift  $z$ is \citep{Weinberg:2008zzc}:
\begin{equation}
\label{eq:z_to_t}
t(z) = \frac{9.778~\textrm{Gyr}}{h} \int^{1/(1+z)}_0 \frac{\dd x}{x \sqrt{\Omega_\Lambda + \Omega_K x^{-2} + \Omega_M x^{-3} + \Omega_R x^{-4}}}
\end{equation}
\noindent where we have used $h = 0.67$, $\Omega_\Lambda = 0.68$, $\Omega_M = 1 - \Omega_\Lambda$ and $\Omega_K = \Omega_R = 0$ \citep{Planck2014}. For the nuclear bulge, we use the MIST star formation rate from Nogueras-Lara et al.~\cite{Nogueras-Lara2020}. The three star formation rates are shown in Fig.~\ref{fig:sfr}.

From the SFR and DTD, we can find the probability density function of MSP ages:
\begin{equation}
    p({\rm age}) = \frac{\int_{0}^{t_0 - {\rm age}} {\rm SFR}(z(\tau)) {\rm DTD}(t_0 - {\rm age} - \tau) \dd\tau}{\int_{0}^{t_0}\int_{0}^{t'} {\rm SFR}(z(\tau)) {\rm DTD}(t' - \tau) \dd\tau \dd t'}
\end{equation}
\noindent where $t_0$ is the age of the universe, and $z(t)$ is the inverse of Eq.\ \ref{eq:z_to_t}.

As an alternative age distribution, we try a uniform distribution where MSPs form at a constant rate over the last 10 Gyr, similar to that assumed by Gonthier et al.~\cite{Gonthier2018}.

\subsubsection{Angular velocity}
In our model, the angular transverse velocities of pulsars in the directions of Galactic longitude and latitude, $\mu_l$ and $\mu_b$ respectively, are determined by assuming a pulsar travels in a circular orbit around the center of the Galaxy using a  parametric form of the   potential \citep{Carlberg1987,Kuijken1989}
with a random normally distributed peculiar velocity in every direction of scale $\sigma_v$ (i.e., a Maxwell distribution). For the Sun, we assume circular motion in the same potential with a peculiar velocity of (11.1, 12.24, 7.25) km s$^{-1}$ where the velocity components are in the direction of the Galactic Center, the direction of rotation and in the direction perpendicular to the plane, respectively \citep{Schonrich2010}. The relationship between velocity ($v$) and angular velocity ($\mu$) at a distance $d$ is:
\begin{equation}
\label{eq:proper_motion_to_linear_velocity}
    v \approx 4.74 \left(\frac{d}{\textrm{kpc}}\right) \left(\frac{\mu}{\textrm{ mas yr}^{-1}}\right) \textrm{ km s}^{-1}
\end{equation}

\subsubsection{Galactic Center Excess}
To simulate the GCE, we need to assign each bulge MSP a spectrum that describes its photon number flux ($N$) at energy $E$:%
\begin{equation}
\label{eq:plsec_spectrum}
    \frac{\dd N}{\dd E} = K E^{-\Gamma} \exp(-\left(E/E_{\rm cut}\right)^{2/3})
\end{equation}
\noindent where the spectral parameters are $K$, $\Gamma$, and $E_{\rm cut}$. The proportionality constant $K$ is determined via:
\begin{equation}
    F = \int_{0.1\textrm{ GeV}}^{100\textrm{ GeV}} E \frac{\dd N}{\dd E} \dd E
\end{equation}
{\textcolor{black}{where $F$ is the energy flux}}.
\noindent The simulated boxy bulge and nuclear bulge GCE spectra are then the sum of all MSP spectra in each of those two bulge populations of MSPs. This spectrum is equivalent to the one used to fit the resolved MSPs in the 4FGL-DR2 
{\textcolor{black}{catalog}}
\citep{Ballet:2020hze} and it will also be the one we use for our resolved MSPs. 

\subsubsection{Millisecond pulsar parameters}
Force-free electrodynamic solutions have given the following expression for the spin-down luminosity
 \cite{Spitkovsky2006}
\begin{equation}
    L_{\mathrm{sd}} \sim \frac{\mu^{2} \Omega^{4}}{c^{3}}\left(1+\sin ^{2} \alpha\right)
    \label{eq:spindownlum}
\end{equation}
where $\mu$ is the magnetic dipole moment, $\Omega$ is the rotational angular velocity, $c$ is the speed of light, and $\alpha$ is the angle between the rotation and magnetic field axes. The magnetic field strength ($B$)
at the magnetic pole of the star  is related to $\mu$ by \cite{Spitkovsky2006}
\begin{equation}
    \mu=\frac{B R^3}{2}
\label{eq:mu}
\end{equation}
where $R$ is the radius of the neutron star and we use $R = 12~\textrm{km}$.
The rotational kinetic energy  of the neutron star is given by the standard formula for a rotating body
\begin{equation}
E=\frac{1}{2} I \Omega^{2}
\end{equation}
where 
 $I$ is the neutron star's moment of inertia and we use
 $I = 1.7 \times 10^{45}~\textrm{g cm}^2$.
Therefore, the spin-down power  satisfies
\begin{equation}
    \dot{E} = 4 \pi^2 I \dot{P}_{\rm int} / P^3
    \label{eq:Edot}
\end{equation}
where $\dot{P}_{\rm int}$ is the time derivative of the intrinsic  period, which may be different to the observed period derivative ($\dot{P}$),
and $P=2\pi/\Omega$ is the period.
Equating the spin-down luminosity (Eq.~\ref{eq:spindownlum}) to the spin down power (Eq.~\ref{eq:Edot}) and using Eq.~\ref{eq:mu} yields
the following expression for the magnetic field strength of an MSP
\begin{equation}
\label{eq:magnetic_field_strength}
B^2 = \frac{c^3 I P \dot{P}_{\rm int}}{\pi^2 R^6 (1 + \sin^2(\alpha))}.
\end{equation}
The angle $\alpha$ is chosen randomly from the probability density:
\begin{equation}
    p(\alpha) = \frac{1}{2} \sin(\alpha)
\end{equation}
\noindent which corresponds to a uniformly random magnetic field axis relative to the rotation axis.

The intrinsic period time derivative is related to the observed period time  derivative $\dot{P}$ by \citep{TheFermi-LAT:2013ssa}:
\begin{equation}
\label{eq:period_derivative_observed_corrections}
    \dot{P} = \dot{P}_{\rm int} + \dot{P}_{\rm Shklovskii} + \dot{P}_{\rm Galactic}
\end{equation}
\noindent where the 
contribution to the observed period derivative
from the 
Shklovskii effect is given by 
\begin{equation}
\label{eq:Shklovskii}
    \dot{P}_{\rm Shklovskii} = 2.43 \times 10^{-21} \left(\frac{\mu}{\textrm{mas yr}^{-1}}\right)^2 \left(\frac{d}{\textrm{kpc}}\right) \left(\frac{P}{\textrm{s}}\right)
\end{equation}
\noindent and the contribution due to the relative acceleration in the Galactic potential is:
\begin{equation}
\label{eq:period_derivative_Galactic_correction}
    \dot{P}_{\rm Galactic} = \frac{1}{c} \boldsymbol{n}_{10} \cdot (\boldsymbol{a}_p - \boldsymbol{a}_\odot) P
\end{equation}
\noindent where $\boldsymbol{n}_{10}$ is the unit vector from the Sun to the pulsar, and $\boldsymbol{a}_p$ and $\boldsymbol{a}_\odot$ are the accelerations, due to the Galactic potential  \cite{Carlberg1987,Kuijken1989},  of the pulsar and Sun 
respectively.
Note that $P=\left(1+v_{\mathrm{R}} / c\right) P_{\mathrm{int}}$, where $v_{\mathrm{R}}$ is the radial velocity of the pulsar and $P_{\mathrm{int}}$ is the intrinsic period. 
Given $v_{\mathrm{R}} \ll c$, we approximate $P_{\mathrm{int}}=P$. Making the common assumption that magnetic field strength remains constant over time, Eq.~\ref{eq:magnetic_field_strength} results in:
\begin{equation}
\label{eq:current_period}
P = \sqrt{P_I^2 + \frac{2 \pi^2 R^6}{I c^3} (1 + \sin^2(\alpha)) B^2 t}
\end{equation}
\noindent where $P_I$ is the initial period of the MSP at birth and $t$ is the age.

We %
{\textcolor{black}{consider}}
multiple relationships between pulsar parameters and the $0.1$-$100$ GeV luminosity ($L$).
Our most general form is 
that     used by Kalapotharakos et al.~\cite{Kalapotharakos_2019}:
    \begin{equation}
    \label{eq:Model1}
        L = \eta E_{\rm cut}^{a_{\gamma}} B^{b_{\gamma}} \dot{E}^{d_{\gamma}}
    \end{equation}
    where $\eta$ is a proportionality factor.
 We also consider the model used by Gonthier et al.~\cite{Gonthier2018}:
    \begin{equation}
    \label{eq:Model6}
        L = \eta P^{\alpha_{\gamma}} \dot{P}^{\beta_{\gamma}}\,.
    \end{equation}
The simplest form we consider is that the luminosity is an entirely independent parameter as used by  Ploeg et al.~\cite{Ploeg:2017vai}:
    \begin{equation}
    \label{eq:Model9}
       L = \eta\,. 
    \end{equation}
\noindent

The likelihood probability density distributions of  $B$, $E_{\rm cut}$, and $\eta$ are assumed to be  log-normal
{\textcolor{black}{as this functional form gives a good fit to their histogrammed data of the resolved Fermi-LAT MSPs:}}
\begin{equation}
\label{eq:lognormal_def}
p(\log_{10}(x)|x_{\rm med},\sigma_x) = \frac{1}{\sqrt{2 \pi} \sigma_x} \exp\left(-\frac{\left(\log_{10}(x) - \log_{10}(x_{\rm med}) \right)^2}{2 \sigma_x^2}\right)
\end{equation}
\noindent where $x_{\rm med}$ is the median of $x$ and $\sigma_x$ is the standard deviation of $\log_{10}(x)$. 
We also assume $P_I$ has this form but our results are not sensitive to {\textcolor{black}{to this assumption.}}
In particular, we found that the cut off power law model used by Gonthier et al.~\cite{Gonthier2018}
{\textcolor{black}{gives similar results}}. 
Note that, 
even though $E_{\rm cut}$ is obtained from a fit of an individual MSP's spectral data,
for our purposes it is treated as a directly measured datum rather than a parameter to be estimated. 

Fits to the MSP gamma-ray data \citep{TheFermi-LAT:2013ssa} have uncovered correlations between the spectral parameters and $\dot{E}$. To allow  for this to potentially be an intrinsic property of the MSPs, we parameterise the median of the $E_{\rm cut}$ likelihood as:
\begin{equation}
    \log_{10}(E_{\rm cut, med} / \text{MeV}) = a_{E_{\rm cut}} \log_{10}(\dot{E}/(10^{34.5} \textrm{ erg s}^{-1})) + b_{E_{\rm cut}}
    \label{eq:Ecut}
\end{equation}
\noindent where $a_{E_{\rm cut}}$ and $b_{E_{\rm cut}}$ are allowed to vary in our model fits.
We also model the likelihood of spectral index $\Gamma$ using a normal distribution with mean $\mu_{\Gamma}$ and standard deviation $\sigma_{\Gamma}$. We assume that
\begin{equation}
    \mu_{\Gamma} = a_{\Gamma} \log_{10}(\dot{E}/(10^{34.5} \textrm{ erg s}^{-1})) + b_{\Gamma}
    \label{eq:Gamma}
\end{equation}
\noindent where $a_{\Gamma}$ and $b_{\Gamma}$ are parameters. We also have a correlation coefficient, $r_{\Gamma, E_{\rm cut}}$ between $\Gamma$ and $\log_{10}(E_{\rm cut})$, so the likelihood is:
\begin{equation}
    p(\Gamma, \log_{10}(E_{\rm cut})|\thetab) = \frac{1}{2 \pi \sigma_{\Gamma} \sigma_{E_{\rm cut}} \sqrt{1 - r_{\Gamma, E_{\rm cut}}^2}} \exp\left(-\frac{z_{\Gamma, E_{\rm cut}}}{2 (1 - r_{\Gamma, E_{\rm cut}}^2)}\right)
\end{equation}
\noindent where:
\begin{equation}
    \begin{multlined}
    z_{\Gamma, E_{\rm cut}} = \frac{\left(\Gamma - \mu_{\Gamma} \right)^2}{\sigma_{\Gamma}^2} + \frac{\left(\log_{10}(E_{\rm cut}) - \log_{10}(E_{\rm cut, med}) \right)^2}{\sigma_{E_{\rm cut}}^2} \\ - \frac{2 r_{\Gamma, E_{\rm cut}} \left(\Gamma - \mu_{\Gamma} \right) \left(\log_{10}(E_{\rm cut}) - \log_{10}(E_{\rm cut, med}) \right)}{\sigma_{\Gamma} \sigma_{E_{\rm cut}}}\, .
    \end{multlined}
\end{equation}
\noindent
Note that here and in the rest of this chapter we will use $\thetab$ to indicate the relevant model parameters. In this case they are $\thetab=\{a_{E_{\rm cut }},b_{E_{\rm cut }},a_\Gamma, B_\Gamma, \sigma_{E_{\rm cut}},\sigma_\Gamma,r_{\Gamma,E_{\rm cut}}\}$. See Table~\ref{tab:prior_ranges} for the corresponding  priors that we use.

As alternatives, we try two likelihoods of $P_I$ where there is a dependence on $B$. In the first,  a bivariate normal distribution relating $\log_{10}(B)$  and $\log_{10}(P_I)$ with a correlation parameter $r_{B, P_I}$ is assumed.
In the second, we adopt the relationship used by Gonthier et al.~\cite{Gonthier2018}:
\begin{equation}
\label{eq:gonthier_initial_period_dist}
    P_I = 0.18 \times 10^{C_{P_I} + 3 \delta / 7} B_8^{6/7} \textrm{ms}
\end{equation}
\noindent where there is a lower bound of $1.3$ ms, $\delta$ is drawn from a uniform distribution between $0$ and $2$. Also, $B_8 = (B/10^8 \textrm{G})$.
We fit $C_{P_I}$  as model parameter while it is set to $0$ by Gonthier et al.~\cite{Gonthier2018}.

\subsubsection{Millisecond pulsar detection}
The flux of an MSP is related to the luminosity in the usual way:
\begin{equation}
\label{eq:flux}
F = \frac{L}{4 \pi d^2}
\end{equation}
\noindent and the MSP detection threshold flux $F_{\rm th}$ is drawn from a log-normal distribution so that the probability of a detection is \citep{Hooper:2015jlu,Ploeg:2017vai,Bartels2018}:
\begin{equation}
\label{eq:detection_probability}
p(F_{\rm th} \leq F \vert l, b,\thetab) = \frac{1}{2} \left(1 + \erf \left(\frac{\log_{10}(F) - \left(\log_{10}(\mu_{\rm th}(l,b)) + K_{\rm th}\right)}{\sqrt{2} \sigma_{\rm th}}\right)\right)
\end{equation}
\noindent where $K_{\rm th}$ and $\sigma_{\rm th}$ are free parameters, and a map in longitude ($l$) and latitude ($b$) of $\mu_{\rm th}(l,b)$ associated with the 4FGL-DR2 catalog can be found online.\footnote{\url{https://fermi.gsfc.nasa.gov/ssc/data/access/lat/10yr\_catalog/}}

\subsection{Fitting the model to data}
\label{ssec:method_fitting_data}
To fit the model parameters to the resolved MSP and GCE data, we use an adaptive Markov Chain Monte Carlo (MCMC) algorithm \citep{Haario01}. The Metropolis-Hastings algorithm \cite{Hastings1970,Gelman2013} is a method for generating a sequence of samples $\pmb{x}_0, ..., \pmb{x}_n$ that, for a sufficiently large number of samples, are distributed according to some probability distribution $p(\pmb{x})$. Starting from some initial point $\pmb{x}_0$, and for some proposal distribution $q(\pmb{y} \vert \pmb{x})$, we draw a sample $\pmb{x}_t$ as follows:
\begin{enumerate}
    \item Draw a point $\pmb{y}$ from the proposal distribution $q(\pmb{y} \vert \pmb{x}_{t-1})$
    \item Calculate the acceptance probability $r$:
    \begin{equation}
        \label{eq:metropolis_hastings_r}
        r = \frac{p(\pmb{y}) q(\pmb{x}_{t-1} \vert \pmb{y})}{p(\pmb{x}_{t-1}) q(\pmb{y} \vert \pmb{x}_{t-1})}
    \end{equation}
    \item Set $\pmb{x}_t$ to $\pmb{y}$ with probability $\min(1,r)$, otherwise $\pmb{x}_t = \pmb{x}_{t-1}$
\end{enumerate}
\noindent The proposal distribution would typically be symmetric (such as a Gaussian centered at $\pmb{x}_{t-1}$) so that $r$ simplifies to $p(\pmb{y})/p(\pmb{x}_{t-1})$. An inappropriate choice of proposal distribution could lead to a situation in which the number of samples required to converge to the target distribution $p(\pmb{x})$ could be unreasonably large. For example, a proposal distribution that generates too many small steps would take a long time to explore the parameter space. A proposal distribution that often produces steps where $r$ is very small will take a long time to make a successful move.

The adaptive Metropolis algorithm of Haario et al.~\cite{Haario01} is very similar to the Metropolis-Hastings algorithm except that the proposal distribution now depends on the history of the chain. We replace $q(\pmb{y} \vert \pmb{x}_{t-1})$ with $q(\pmb{y} \vert \pmb{x}_{t-1},...\pmb{x}_0)$. This proposal distribution is a Gaussian centered at $\pmb{x}_{t-1}$ with a covariance matrix $C_t$ that adapts so that the proposal distribution is of a scale similar to the target distribution. For the first $t_0$ samples of the chain we use some initial covariance matrix $C_0$, then use $C_t$ related to the sample covariance matrix of all past samples $\pmb{x}_0,...,\pmb{x}_{t-1}$:
\begin{equation}
    C_t = 
    \begin{cases}
        C_0 & t \leq t_0 \\
        s_d \textrm{cov}(\pmb{x}_0,...,\pmb{x}_{t-1}) + s_d \epsilon I & t > t_0 \\
    \end{cases}
\end{equation}
\noindent where for $d$ dimensions $s_d = 2.4^2 / d$ and the extra term $s_d \epsilon I$ with small $\epsilon$ prevents the covariance matrix from becoming singular.

The resolved MSPs have an unbinned Poisson likelihood \citep{Cash1979}:
\begin{equation}
\label{eq:likelihood_definition}
\mathcal{L}_{\rm res}
\propto \exp(-\lambda_{\rm res})  \prod_{i=1}^{N_{\rm res} }  \rho(\D{i})
\end{equation}
\noindent 
 where $N_{\rm res}$ is the number of resolved MSPs, $\lambda_{\rm res}$ is the expected number of resolved MSPs, 
and $\rho(\D{i})$ is the phase space density of the resolved MSPs which have 
the data 
\begin{equation}
    \D{i}=\{
l_i, b_i, d_i, P_i, \dot{P}_i, \mu_{l, i}, \mu_{b, i}, F_i, E_{{\rm cut}, i}, \Gamma_i
\} \,.
\end{equation}
That is, the resolved MSP
is located
at longitude $l_i$, latitude $b_i$,  distance $d_i$, period $P_i$, observed period derivative $\dot{P}_i$,  proper motion  in longitude  $\mu_{l, i}$, proper motion  in  latitude $\mu_{b, i}$, flux $F_i$, spectrum energy cut-off $E_{{\rm cut}, i}$, and  spectral index $\Gamma_i$.

The expected number of resolved MSPs can be obtained by integrating the phase space density over the  phase space volume as follows \cite{Cash1979}:
\begin{equation}
\label{eq:lambdarho}
    \lambda_{\rm res}=\int \rho(\D{i})\, \d{\D{i}}%
\end{equation}
This relation implies that 
\begin{equation}
\label{eq:rhop}
\rho(\D{i})=p(\D{i}\vert {\rm obs}, \thetab) \lambda_{\rm res}
\end{equation}
where ${\rm obs}$ indicates the MSP was observed, i.e.\ it was resolved. Also,
$p( \D{i}|{\rm obs},\thetab) $ is 
the probability density that a resolved MSP has data $\D{i}$
given that  the model parameter values are $\thetab$. It then follows from the above two equations that 
\begin{equation}
     \int p(\D{i}\vert {\rm obs}, \thetab) \, \d{\D{i}}=1
\end{equation}
as required.

There are many tens of thousands of unresolved MSPs in the Milky Way \cite{Ploeg:2017vai,Gonthier2018} and we have only resolved of order 100 in gamma rays. It follows that the probability of observing an individual MSP must be a very small. Therefore, from the law of rare events (see for example Section 1.1.1 of  \cite{Cameron1998}), 
the total number of resolved and unresolved MSPs ($N_{\rm tot}$) is well approximated by:
\begin{equation}
\label{eq:pobs1}
    N_{\rm tot} = \lambda_{\rm res} / p(\obs|\thetab)\, .
\end{equation}
We can find $p({\rm obs}|\thetab)$ by noting that a luminosity threshold distribution $p(L_{\rm th})$ is determined by the combination of the flux threshold and spatial distributions. Thus:
\begin{equation}
\label{eq:pobs}
    p({\rm obs}|\thetab) = \int \dd L_{\rm th} p(L_{\rm th}) p(L \geq L_{\rm th})
\end{equation}
\noindent where $p(L \geq L_{\rm th})$ is the probability that the luminosity is greater than or equal to the threshold.
Another useful relation that follows from standard probability theory is: 
\begin{equation}
\label{eq:PobsD}
    p( {\rm obs}, \D{i}\vert \thetab) = p(\D{i}\vert {\rm obs}, \thetab)p({\rm obs}|\thetab)\,.
\end{equation}
Combining the above equation with Eqs.~\ref{eq:pobs1} and \ref{eq:rhop} gives
 \begin{equation}
 \label{eq:rhoD}
   \rho(\D{i})=p({\rm obs}, \D{i}\vert \thetab) N_{\rm tot} \,.
\end{equation}

We have two distinct types of resolved MSPs: those with parallax measurements and those without. To accommodate this we have two separate probability density functions
\begin{equation}
p({\rm obs},\D{i}\vert \thetab)=  p({\rm obs},\D{i},{\rm parallax}_i\vert\thetab)+p({\rm obs},\D{i},\textrm{not parallax}_i|\thetab)
\end{equation}
The $\D{i}$ components are the same in both cases as we can estimate the distance for those MSPs that do not have parallax measurements by  their dispersion measures. 
The probability of a parallax measurement given distance $d$ is modelled as:
\begin{equation}
\label{eq:parallax_model}
    p(\textrm{parallax}_i \vert d_i,\thetab) = \min(1, C_{\rm parallax} \exp(-d_i / d_{\rm parallax})) \,.
\end{equation}
and $ p(\textrm{not parallax}_i \vert d_i,\thetab)=1- p(\textrm{parallax}_i \vert d_i,\thetab)$.

To take into account measurement uncertainty in our values of $\D{i}$ we marginalise over the true values $\hat{\pmb{D}}_i$
\begin{equation}
\label{eq:likelihood_integral_over_true_vals}
  p(\obs, \D{i}\vert \theta) = \int  p(\D{i}|\hat{\pmb{D}}_i)  p({\rm obs}, \hat{\pmb{D}}_i\vert\thetab)\, \d\hat{\pmb{D}}_i
\end{equation}
If MSP $i$ has the $j$th component of its data $\D{i}$ missing then we account for that by making $p(\D{i}|\hat{\pmb{D}}_i)$ uniform in the $j$th component.

Some of the data used are given in the form of $\pmb{y}_j$ where $\pmb{y}$ is not one of the directly modeled MSP parameters. In this case we perform a transformation to the observed parameter:
\begin{equation}
    \label{eq:uncertainty_integral_transformed}
    p(..., \pmb{y}_j, ...) = \int \dd \pmb{y} p(\pmb{y}_j \vert \pmb{y}) \lvert \pmb{J} \rvert p(..., \pmb{x}(\pmb{y}), ...)
\end{equation}
\noindent where $\lvert \pmb{J} \rvert$ is the Jacobian determinant of the transformation from $\pmb{y}$ to $\pmb{x}$, so $\pmb{J}_{i,j} = \partial x_i (\pmb{y}) / \partial y_j$. This transformation is needed for integrating over uncertainty in distance where we have either parallax ($\omega$), where $\omega = 1/d$, or dispersion measure (DM), where $\textrm{DM} = \int_0^d n_e(s) \dd s$ with $n_e(s)$ being the free electron density at distance $s$.

See Appendix~\ref{app:probability} for more details of the resolved MSP probability density function and Appendix~\ref{appendix:uncertainties} for more details about the measurement uncertainties.

We fit the boxy bulge and nuclear bulge GCE to spectra found by Macias et al.~\cite{Macias19}. We use a Gaussian likelihood for each bin in the GCE spectra:
\begin{equation}
\label{eq:likelihood_GCE}
    \mathcal{L}_{\rm GCE} \propto \prod_{i=1}^N \exp(-\left(\left(\frac{\dd N}{\dd E}\right)_{\textrm{sim,}i} - \left(\frac{\dd N}{\dd E}\right)_{\textrm{data,}i} \right)^2 / \left(2 \sigma^2_{\textrm{data,}i}\right))
\end{equation}
\noindent where $\left(\frac{\dd N}{\dd E}\right)_{\textrm{sim,}i}$ is the simulated GCE for bin $i$ and $\left(\frac{\dd N}{\dd E}\right)_{\textrm{data,}i}$ is the data with uncertainty $\sigma^2_{\textrm{data,}i}$. We fit only energy bins lower than $10$ GeV as the higher energy bins may contain significant secondary emission \cite{Macias19}.  The combined likelihood for our resolved MSPs and the GCE is then given by substituting Eqs.~\ref{eq:likelihood_GCE} and \ref{eq:likelihood_definition} into
\begin{equation}
    \mathcal{L}_{\rm total}=\mathcal{L}_{\rm res}\times \mathcal{L}_{\rm GCE} \,.
    \label{eq:Ltotal}
\end{equation}
This is then multiplied with the priors given in Table~\ref{tab:prior_ranges} to get the posterior which is then sampled using MCMC:
\begin{equation}
    p(\thetab \vert \D{0},...,\D{N_{\rm res}}) \propto p(\thetab) \mathcal{L}_{\rm total}
\end{equation}
\noindent where $p(\thetab)$ is the prior probability density function. To test our models we generate posterior predictive distributions \cite{Gelman2013}. See Appendix~\ref{appendix:sampling} for more details on the sampling methods we used.

\section{Results}
\label{sec:results}

In this section we present results for various 
assumed
luminosity functions, age distributions, and relationships between magnetic field strength and initial period. We rank these various models using the Watanabe Akaike Information Criterion (WAIC) \cite{Wantabe2010,Gelman2013} as described in Appendix \ref{appendix:WAIC}. The WAIC provides a measure of the expected predictive accuracy of a model and it takes into account the number of model parameters and their posterior uncertainty. 
Under suitable regularity conditions, in the limit of a large amount of data, the difference in the WAIC between two models ($\Delta$WAIC) tends towards minus two times the log of their likelihood ratio \cite{Gelman2013}. So when adding an extra parameter to a model, the ``number of sigma'' in favour of adding  that parameter is given approximately by $\sqrt{\Delta{\rm WAIC}}$   \cite{Wilks1938}. This provides a rough benchmark in evaluating the significance of $\Delta$WAIC values.

We can gain a more detailed view of a model's fit
by comparing its posterior predictive distributions to data using single dimensional binned plots showing medians and $68\%$ and $95\%$ intervals, as well as corner plots constructed using the software by Foreman-Mackey~\cite{corner}. These posterior predictive distributions provide an effective method of evaluating the goodness of fit \citep{Gelman2013,Gelman2013a}. 
In particular, major failures of the model correspond to extreme posterior predictive p-values. These are defined as the proportion of posterior simulations which are more extreme than the data or some statistic of the data \citep{Gelman2013}.
Our corner plots show two dimensional distributions of real data and simulated data with $68\%$, $95\%$ and $99.7\%$ contours. In producing simulated data we model both missing data and uncertainties by picking a random real MSP and removing simulated data that are not available for the real MSP. For data with uncertainties attached, we take the relative error for the real MSP and 
add Gaussian noise to the simulated MSP which has a standard deviation with the same relative error as the real MSP.
We have a distance dependent model of the probability of a parallax measurement being available. If a simulated MSP has a parallax measurement, with probability given by Eq.\ \ref{eq:parallax_model}, we select a random real MSP out of those with available parallax measurements and use its relative uncertainty to simulate a parallax error.

The WAIC allows us to compare the various models used while accounting for the varying number of parameters they involve. In Table \ref{tab:WAIC} we show the WAIC averaged over the eight chains for each of a set of models of the Galactic MSP population relative to that of the best model. We also show the sample standard deviation over the WAIC for the eight MCMC chains run for each model. 
This variation occurs because we used Monte Carlo integration to compute the integrals in Section~\ref{sec:method}. The random numbers used to compute these integrals (such as over the resolved MSP data uncertainty distributions) were generated once per MCMC chain so that a calculation of the likelihood for a given set of parameters will always return the same result. However, between chains the likelihood may shift slightly as the set of randomly generated numbers used will be different.
As a result of this variation in the likelihood, we did not generate a single WAIC for all eight chains combined. We found that the posterior distributions of the model parameters were generally indistinguishable despite these variations, and the variation in WAIC for each model is typically small compared to $\Delta \textrm{WAIC}$ between models.

In Table \ref{tab:mcmc_results_parameters} we report medians and $68\%$ confidence intervals for the parameters of a subset of these models. The first three parameters in this table
are related to the number of MSPs in each of the three spatial components of the model: The parameter $\lambda_{\rm res}$ is the expected number of resolved MSPs; $\log_{10}(N_{\rm disk} / N_{\rm bulge})$ and $\log_{10}(N_{\rm nb} / N_{\rm bb})$  are parameters defining, respectively, the ratio of $N_{\rm disk}$ to $N_{\rm bulge}$ and $N_{\rm nb}$ to $N_{\rm bb}$. 
Here $N_{\rm disk}$ is the number of disk MSPs, $N_{\rm bulge}=N_{\rm nb}+N_{\rm bb}$ is the total number of  bulge MSPs, $N_{\rm nb}$ is the number of nuclear bulge MSPs, and $N_{\rm bb}$ is the number of boxy bulge MSPs. 

\begin{figure}
    \centering
    \includegraphics[width=0.99\linewidth]{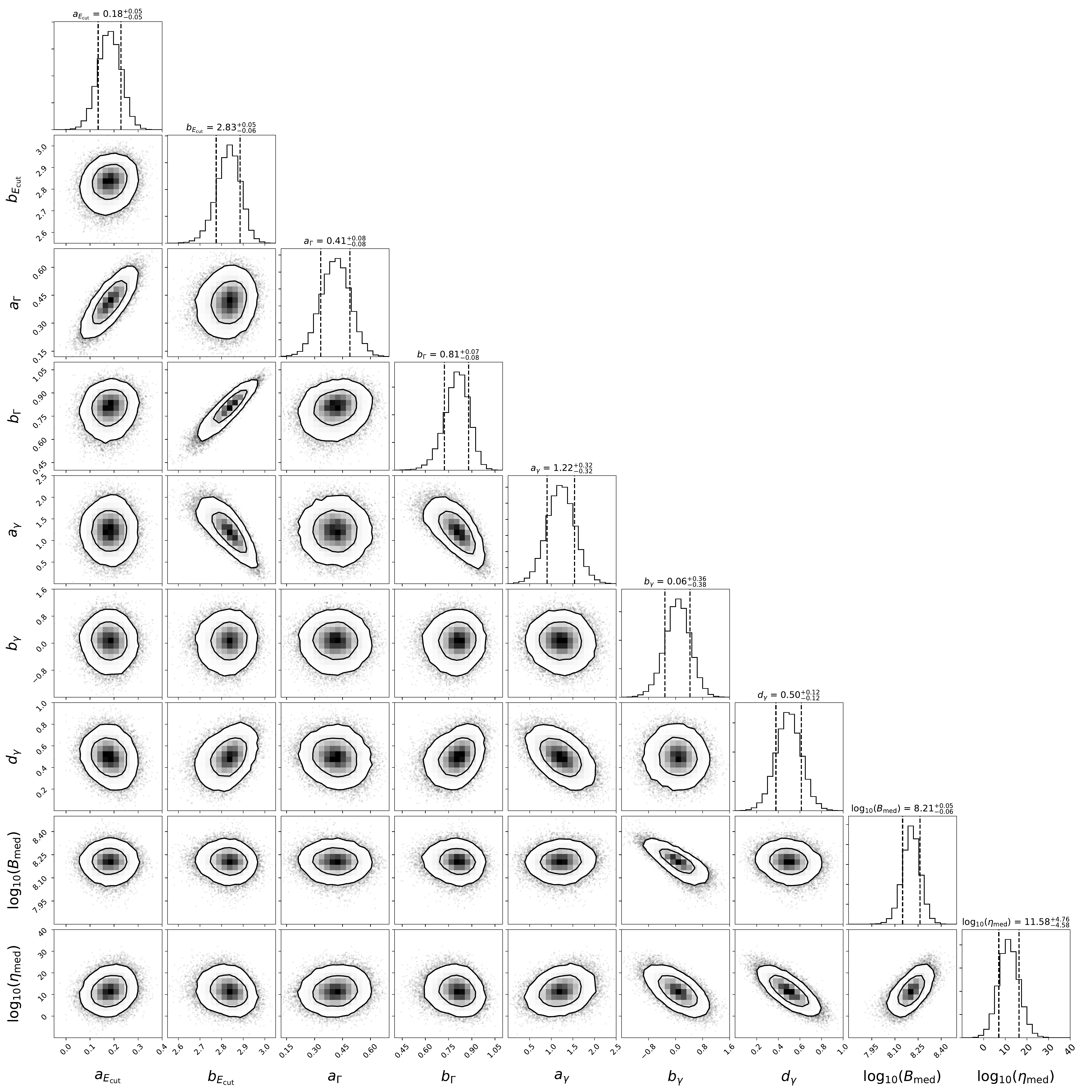}
    \caption[Corner plot showing a selection of parameters for {\textcolor{black}{Model A1}} ($L = \eta E_{\rm cut}^{a_{\gamma}} B^{b_{\gamma}} \dot{E}^{d_{\gamma}}$)  with $68\%$ and $95\%$ contours.]{Corner plot showing a selection of parameters for {\textcolor{black}{Model A1}} ($L = \eta E_{\rm cut}^{a_{\gamma}} B^{b_{\gamma}} \dot{E}^{d_{\gamma}}$)  with $68\%$ and $95\%$ contours. These parameters relate to the luminosity, $E_{\rm cut}$, $\Gamma$ and $B$ distributions. The $B_{\rm med}$ parameter has units of Gauss.}
    \label{fig:E_cut_B_E_dot_MCMC_Params_1}
\end{figure}

\begin{figure}
    \centering
    \includegraphics[width=0.99\linewidth]{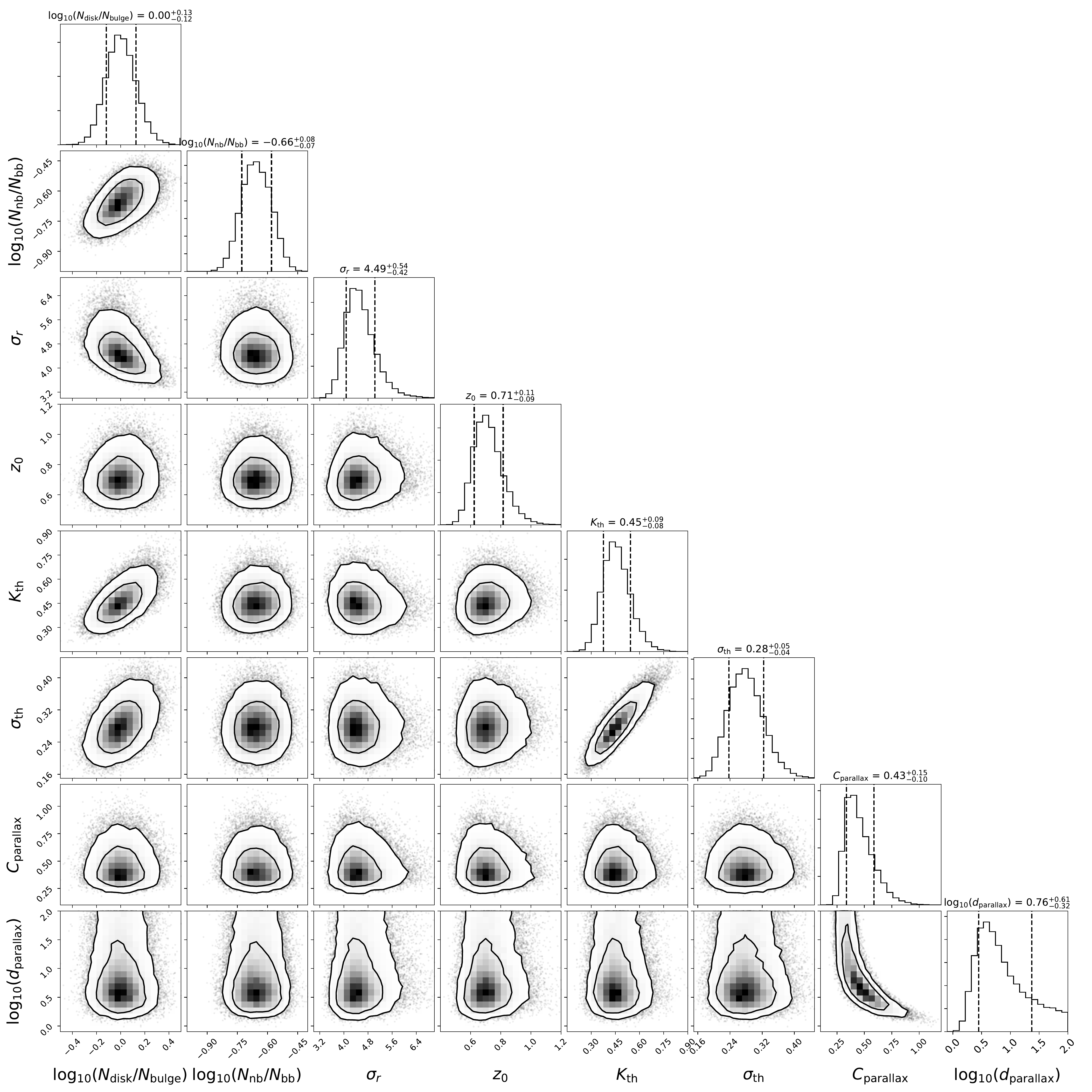}
    \caption[Corner plot showing a different selection of parameters for Model A1 with $68\%$ and $95\%$ contours.]{Corner plot showing a selection of parameters for Model A1 with $68\%$ and $95\%$ contours. These parameters relate to the number of MSPs in different components of the spatial distribution, the flux threshold, the model of the parallax measurement availability and the initial period distribution. 
    The parameters $\sigma_r$, $z_0$, and $d_{\rm parallax}$ have units of kpc.
    }
    \label{fig:E_cut_B_E_dot_MCMC_Params_2}
\end{figure}

\begin{figure}
    \centering
    \includegraphics[width=0.99\linewidth]{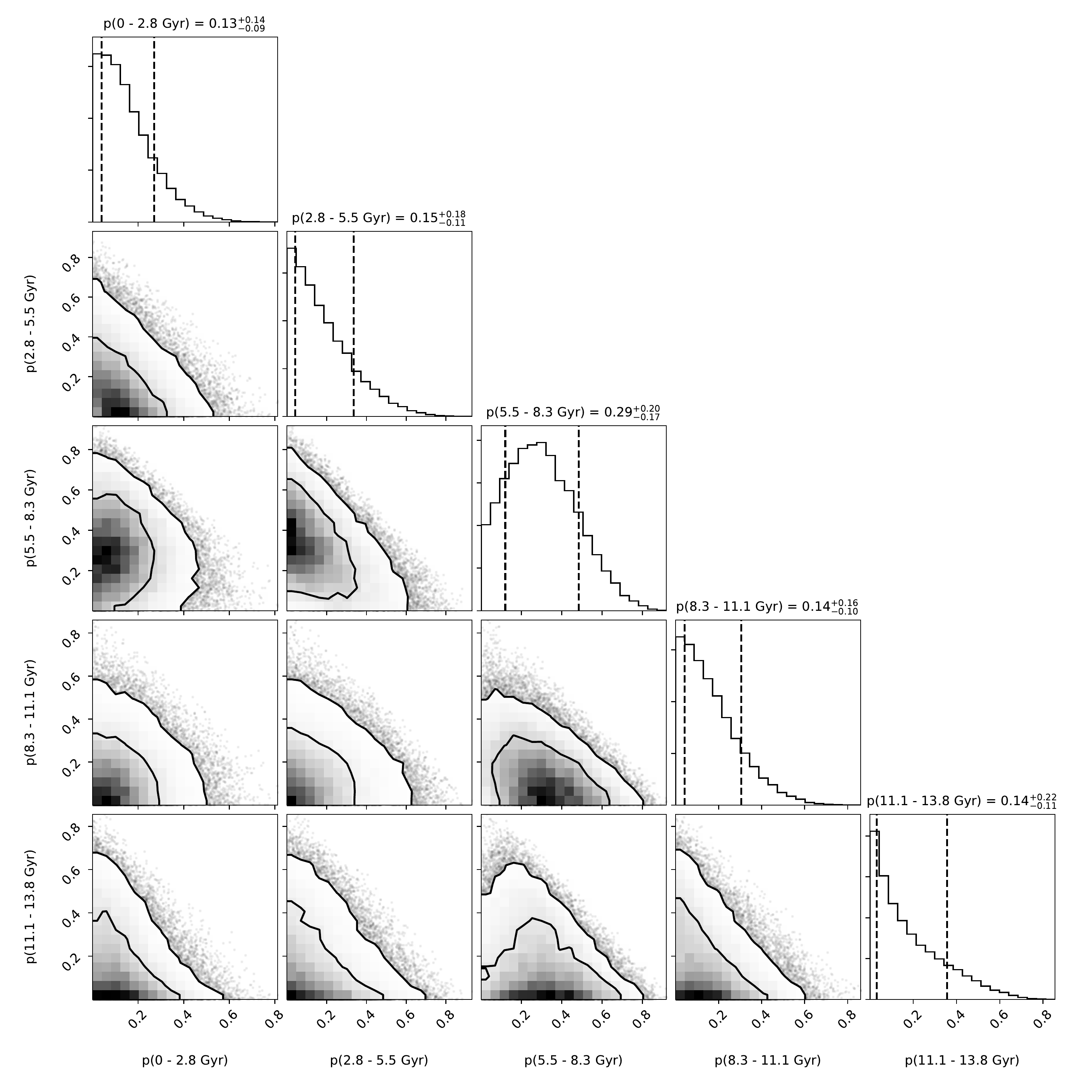}
    \caption[Model A1  probability of delay time within each DTD bin.]{Model A1  probability of delay time within each DTD bin.}
    \label{fig:E_cut_B_E_dot_MCMC_DTD_bin_Params}
\end{figure}

\begin{figure}
\vspace{-2cm}
    \centering
    \subfigure{\centering\includegraphics[width=0.329\linewidth]{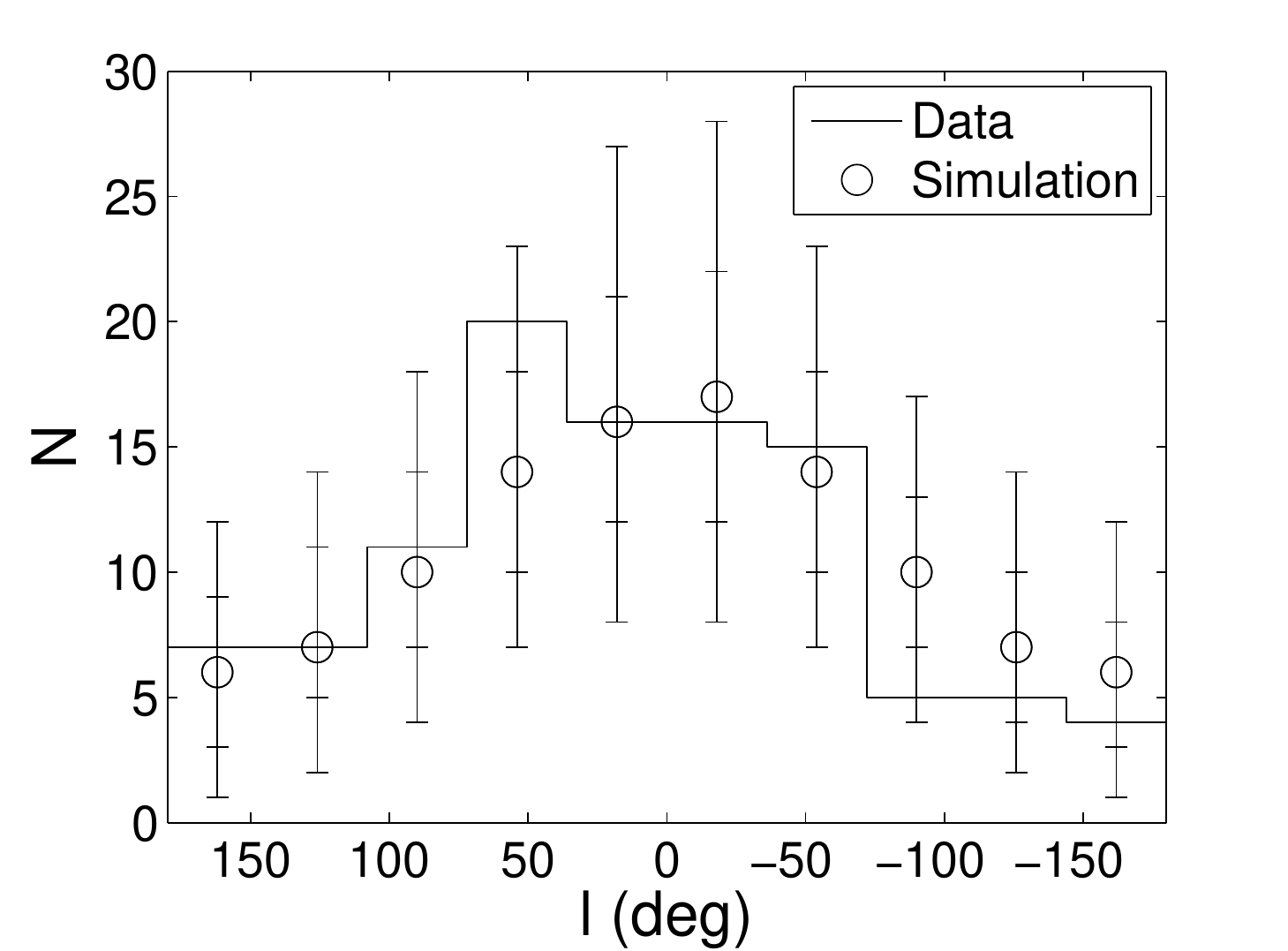}}
    \subfigure{\centering\includegraphics[width=0.329\linewidth]{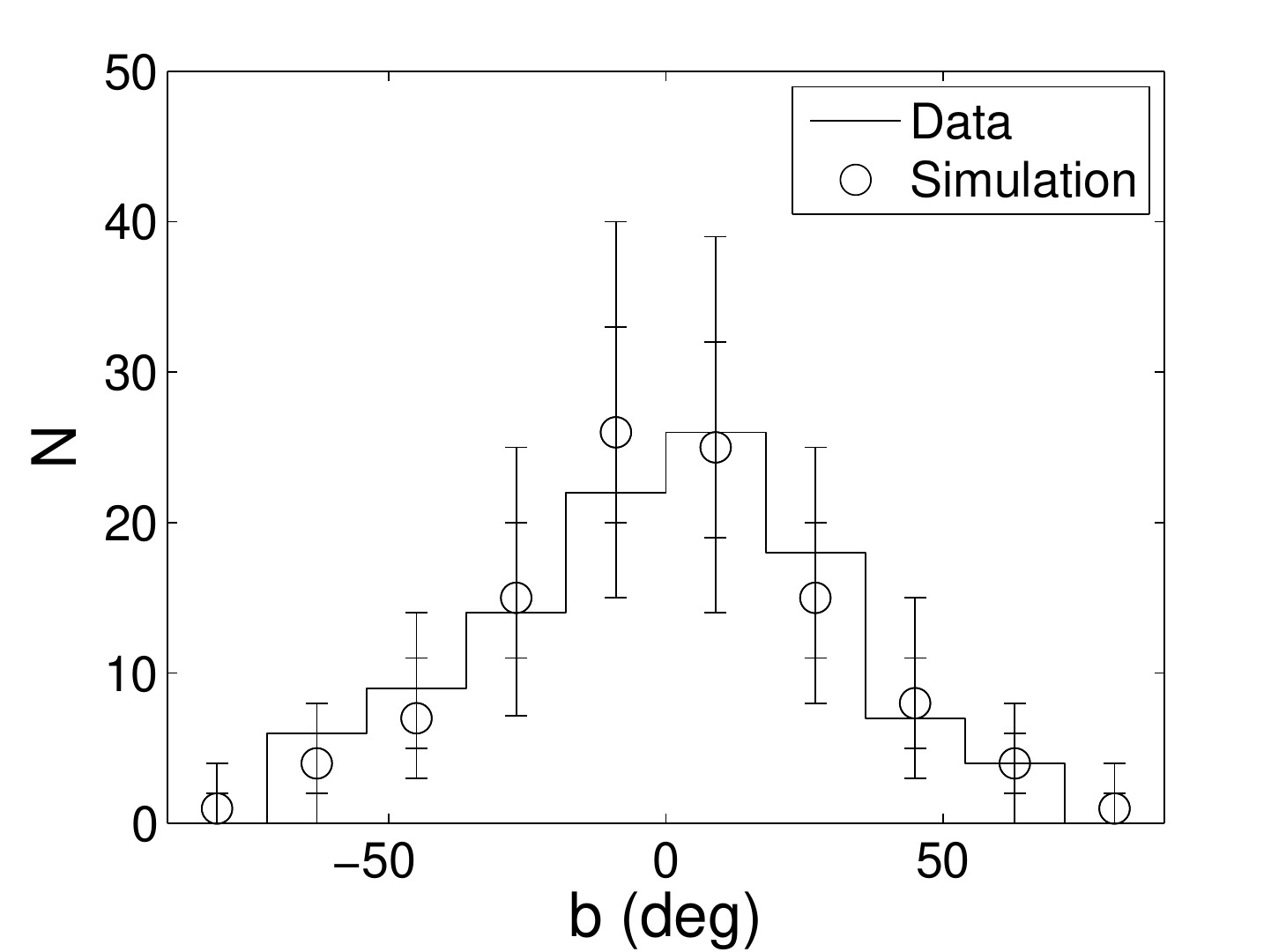}}
    \subfigure{\centering\includegraphics[width=0.329\linewidth]{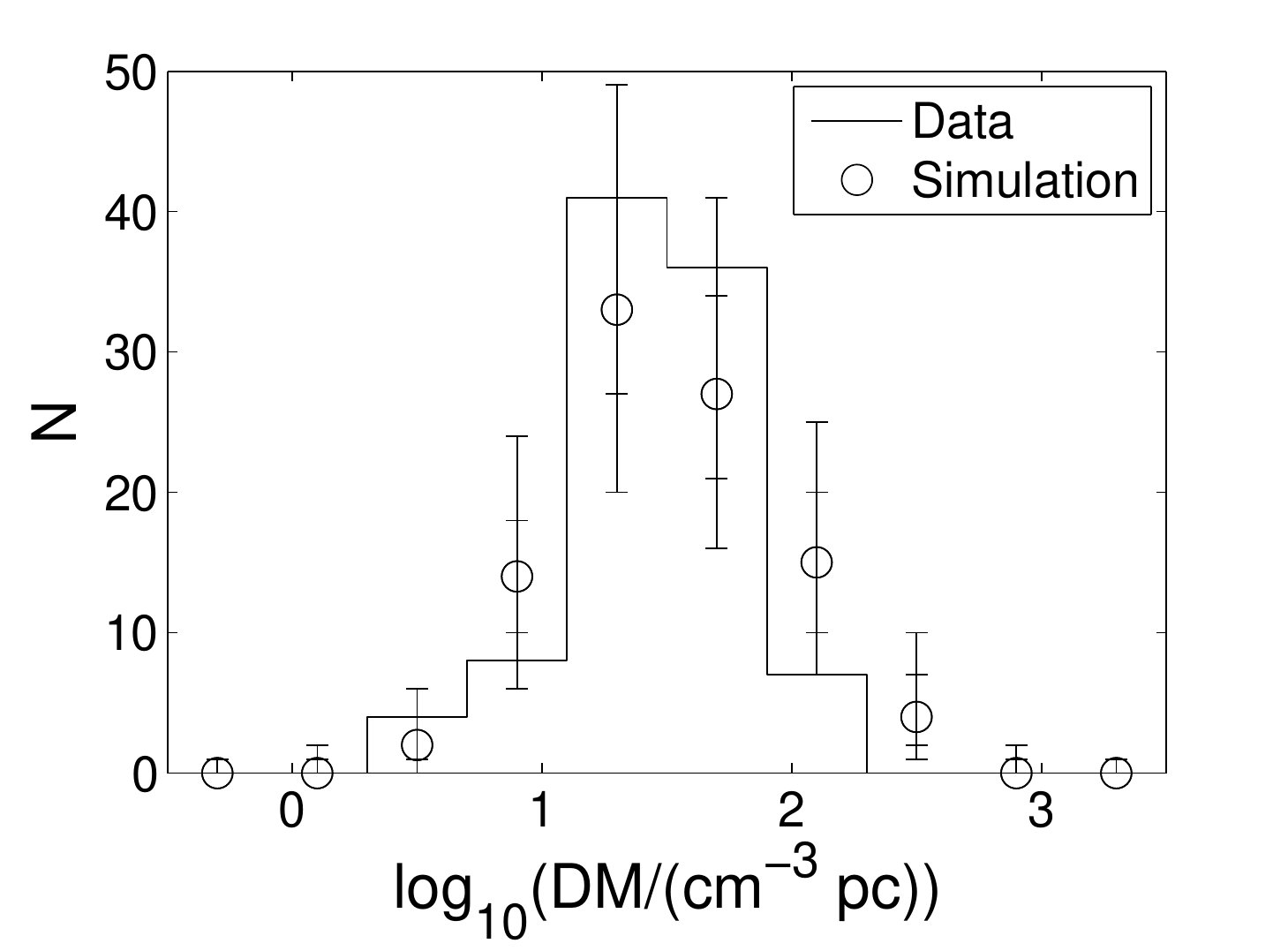}}
    \subfigure{\centering\includegraphics[width=0.329\linewidth]{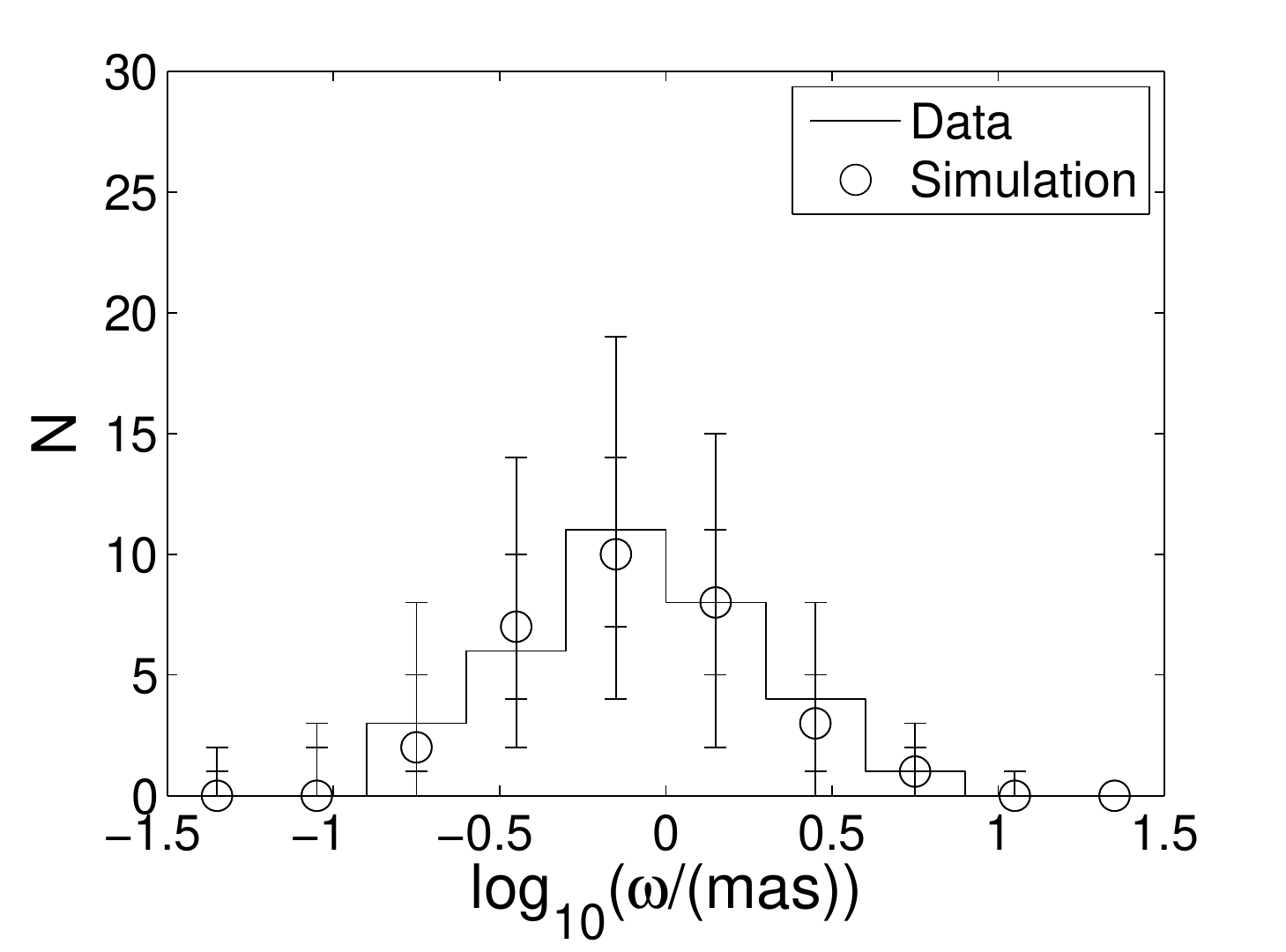}}
    \subfigure{\centering\includegraphics[width=0.329\linewidth]{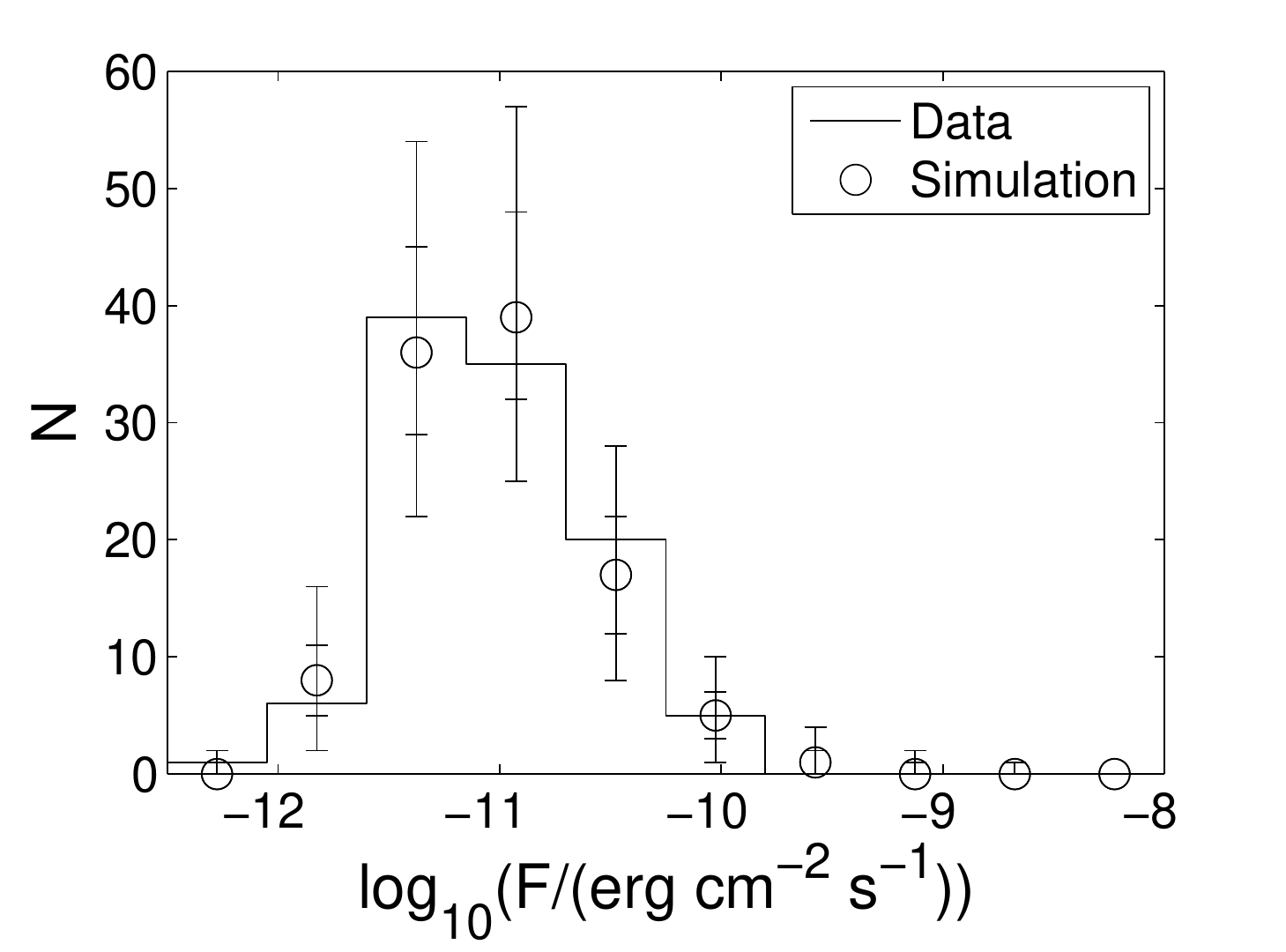}}
    \subfigure{\centering\includegraphics[width=0.329\linewidth]{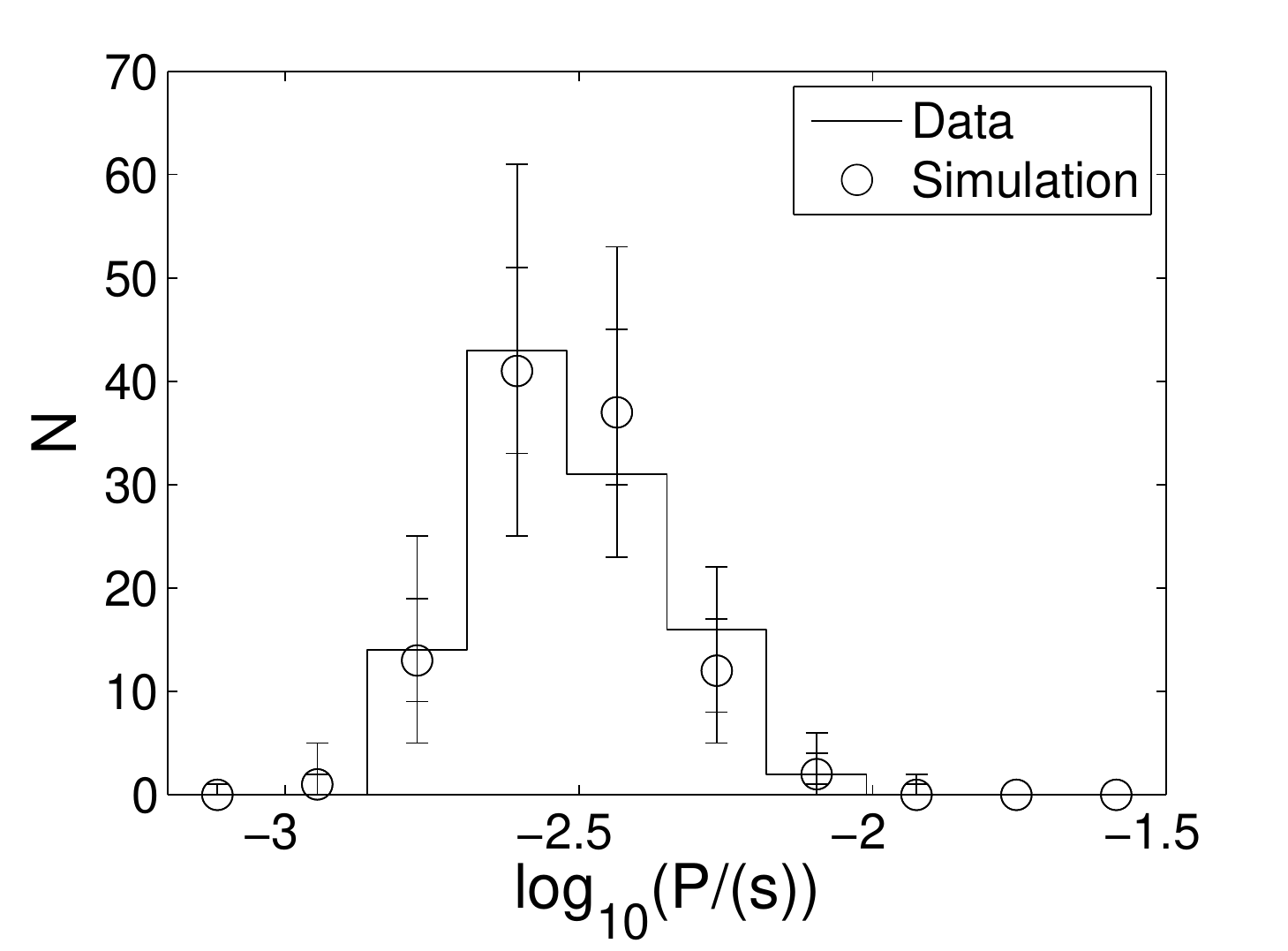}}
    \subfigure{\centering\includegraphics[width=0.329\linewidth]{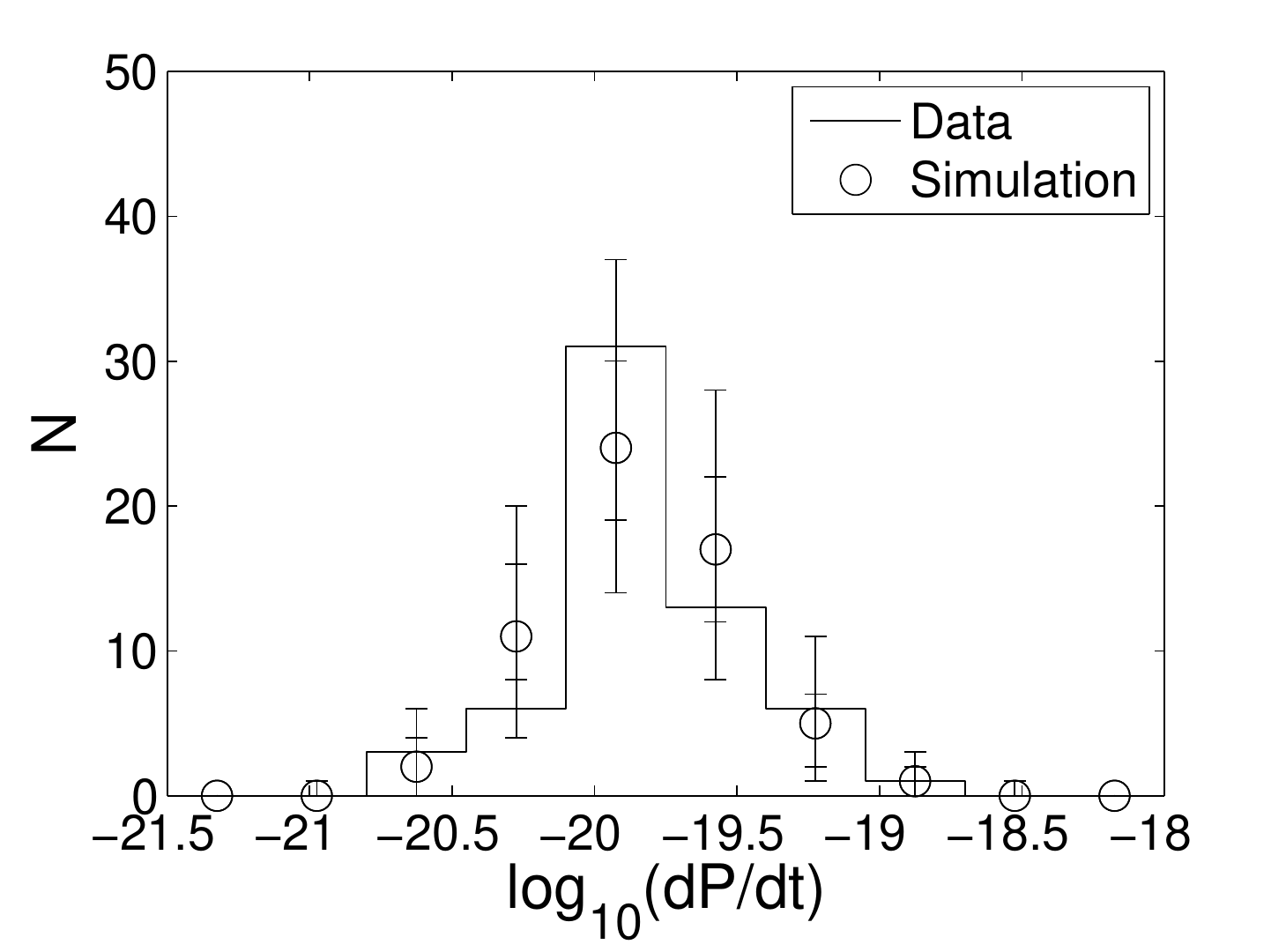}}
    \subfigure{\centering\includegraphics[width=0.329\linewidth]{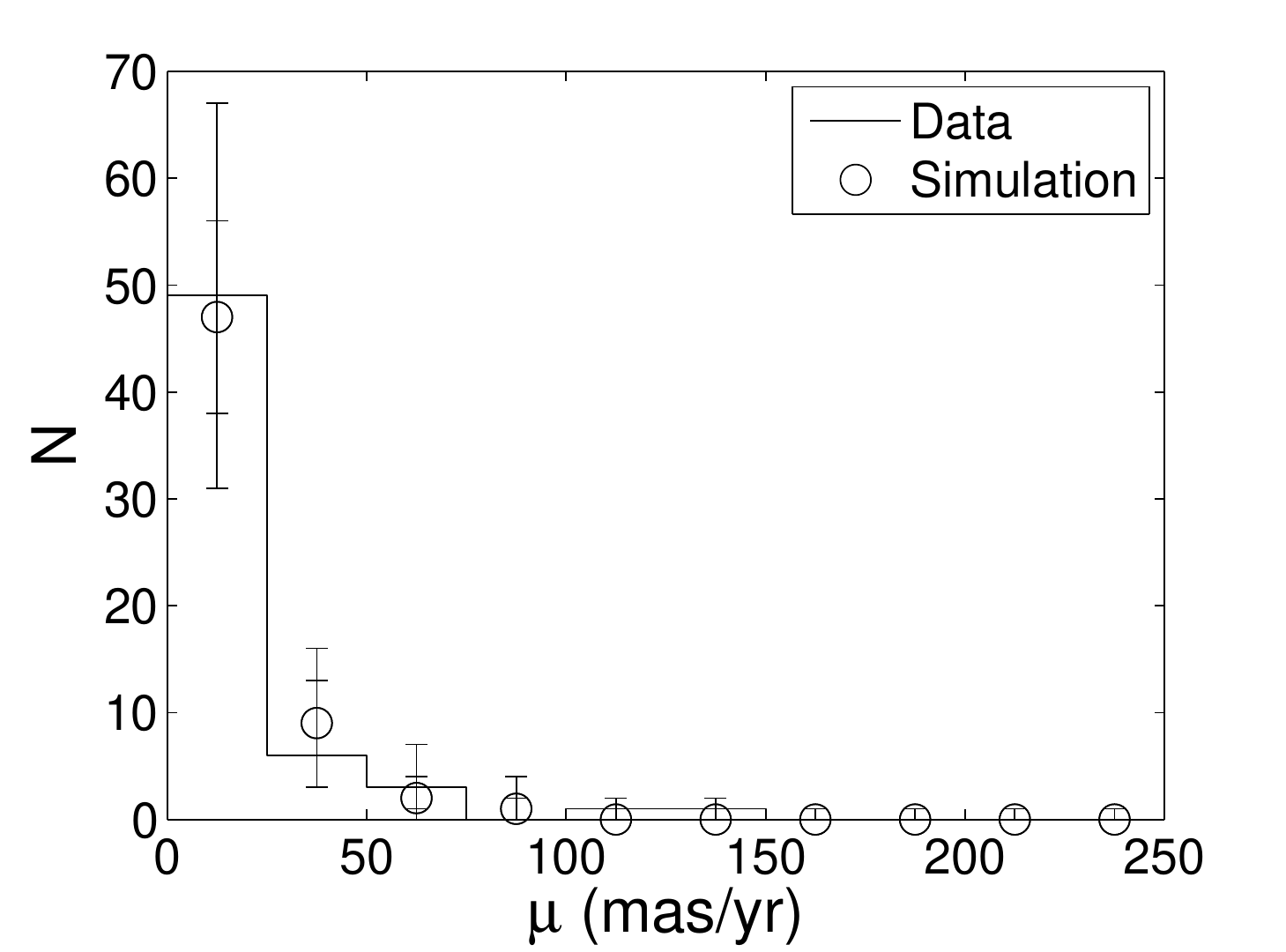}}
    \subfigure{\centering\includegraphics[width=0.329\linewidth]{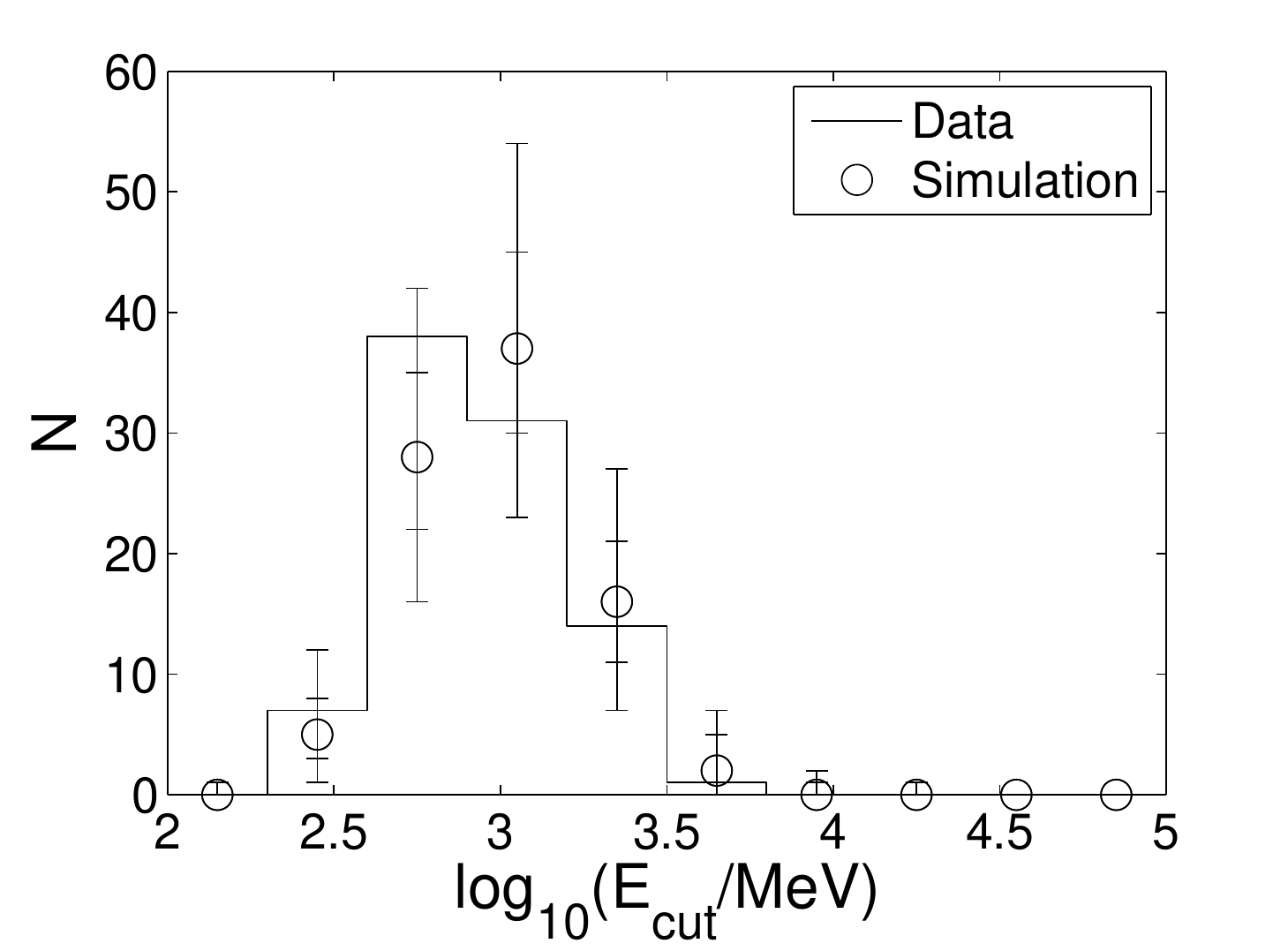}}
    \subfigure{\centering\includegraphics[width=0.329\linewidth]{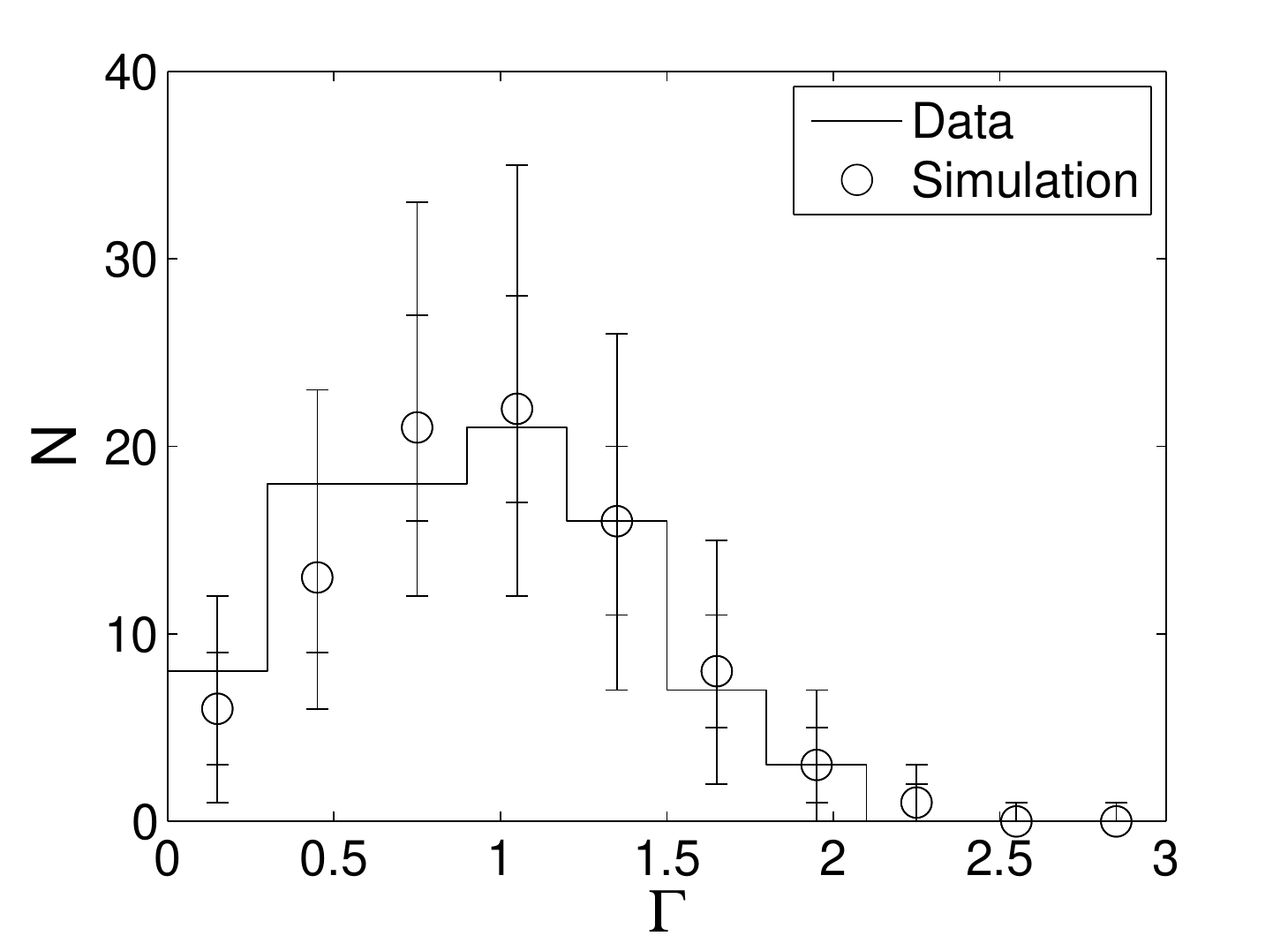}}
    \subfigure{\centering\includegraphics[width=0.329\linewidth]{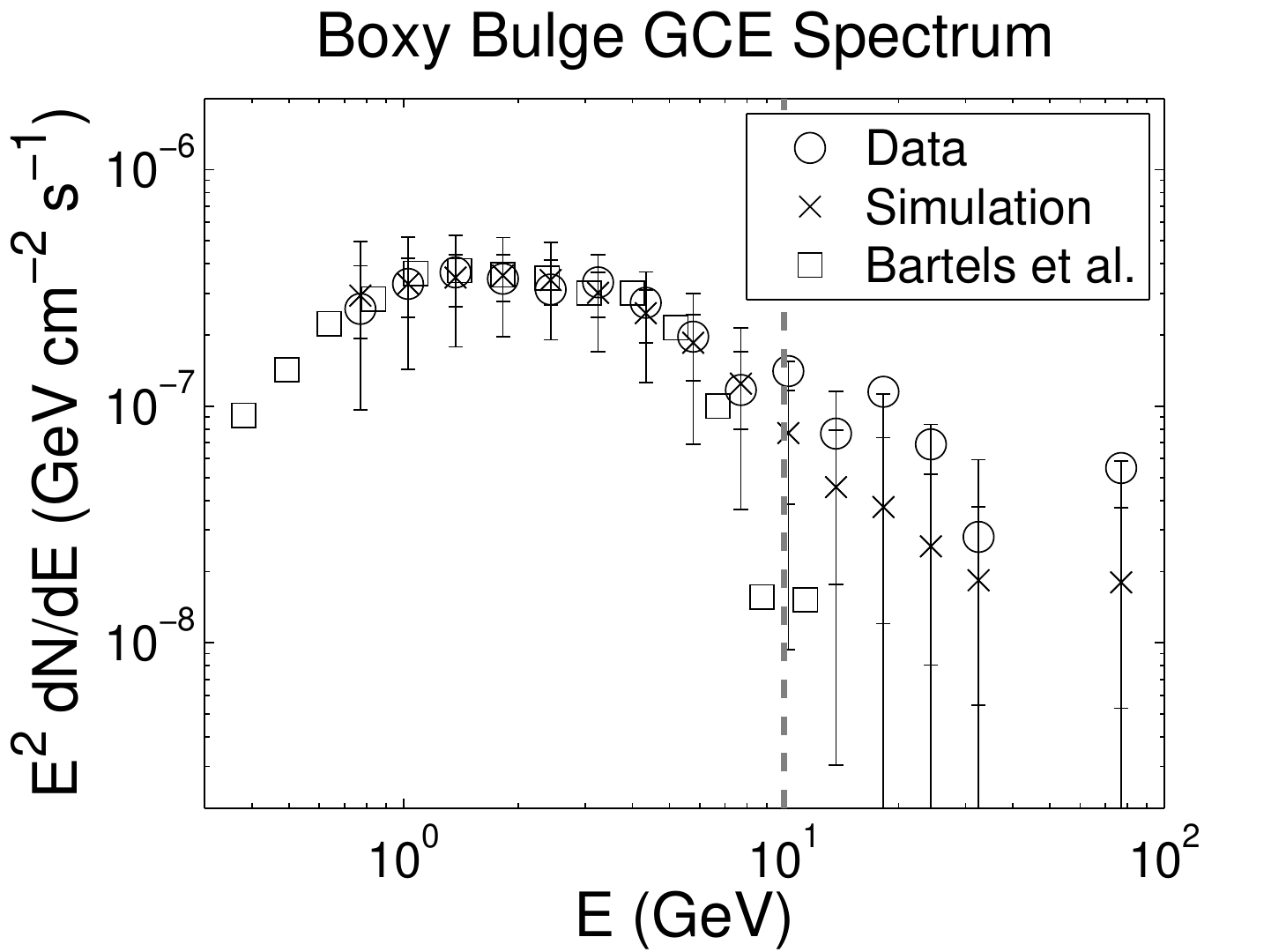}}
    \subfigure{\centering\includegraphics[width=0.329\linewidth]{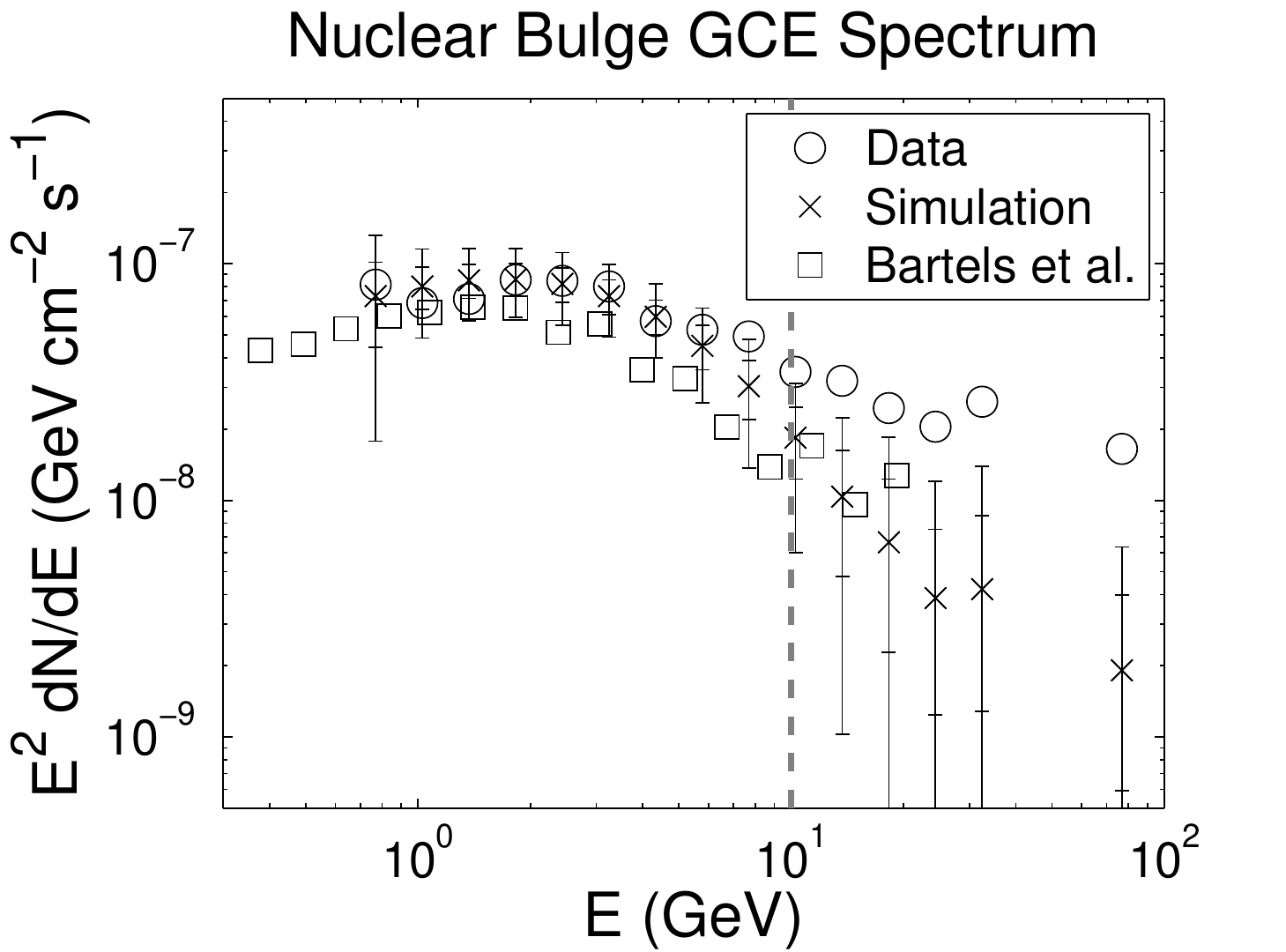}}
    \caption[Observed data compared to simulated observations for Model A1.]{Observed data compared to simulated observations for Model A1 of longitude $l$ (deg), latitude $b$ (deg), dispersion measure DM (cm$^{-3}$ pc), parallax $\omega$ (mas), flux $F$ (erg cm$^{-2}$ s$^{-1}$), period $P$ (s), period derivative $\dot{P}$, proper motion $\mu$ (mas yr$^{-1}$), spectral cutoff $E_{\rm cut}$ (MeV), spectral index $\Gamma$ and GCE spectra for the boxy and nuclear bulges. The GCE spectra are for the inner $40^{\circ} \times 40^{\circ}$ region and the vertical dashed line at $10$ GeV shows the maximum energy up to which we fitted. The simulated data are shown as medians, $68\%$ and $95\%$ intervals in each bin. The intervals on the simulated GCE data include the errors on the observed data. For the GCE plots, we also show boxy bulge and nuclear bulge spectra from Bartels et al.~\cite{Bartels2017} in addition to those we fitted to of Macias et al.~\cite{Macias19}. }
    \label{fig:E_cut_B_E_dot_posterior_predictive_plots}
\end{figure}

\begin{figure}
    \centering
    \subfigure{\centering\includegraphics[width=0.49\linewidth]{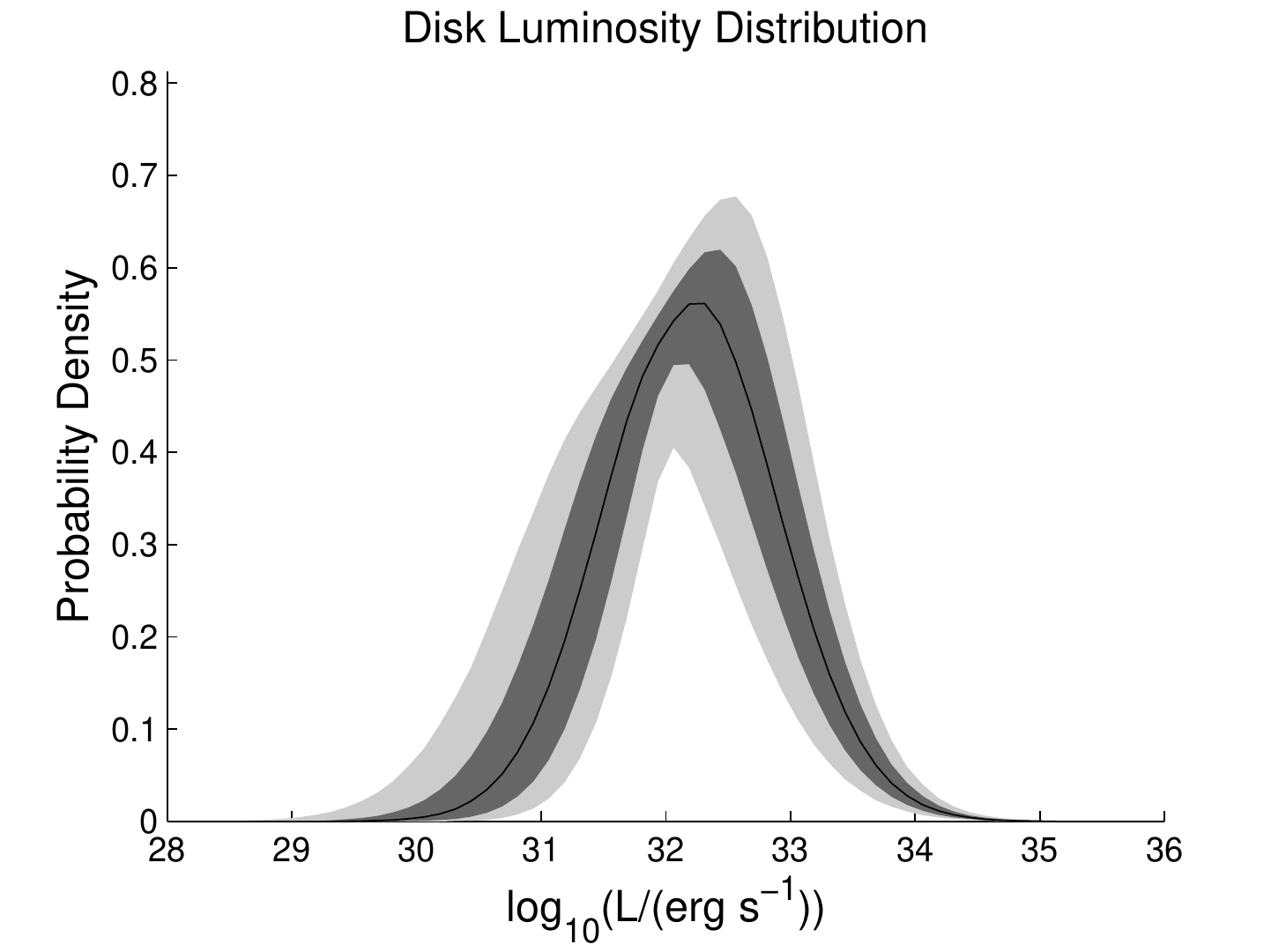}}
    \subfigure{\centering\includegraphics[width=0.49\linewidth]{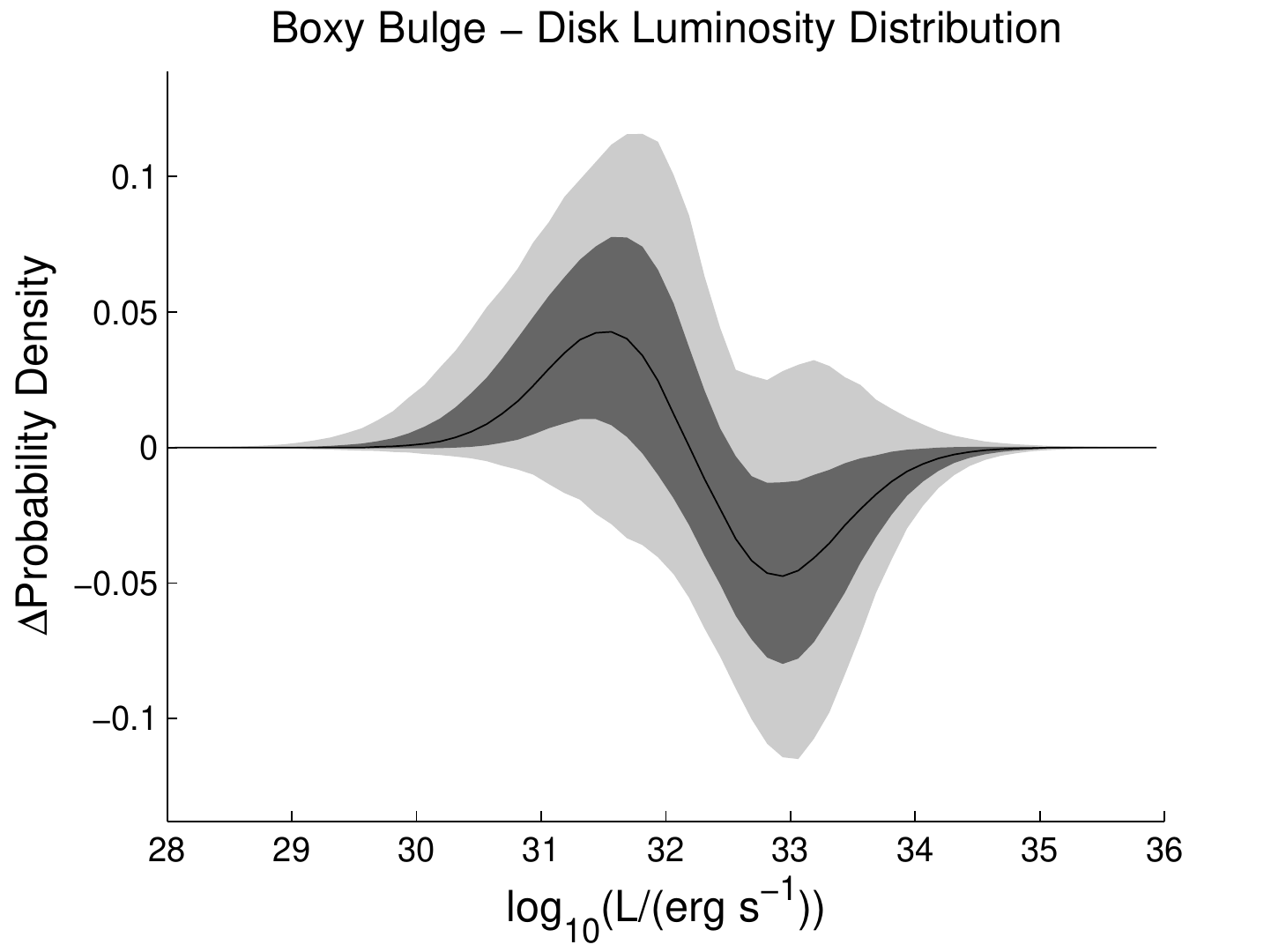}}
    \subfigure{\centering\includegraphics[width=0.49\linewidth]{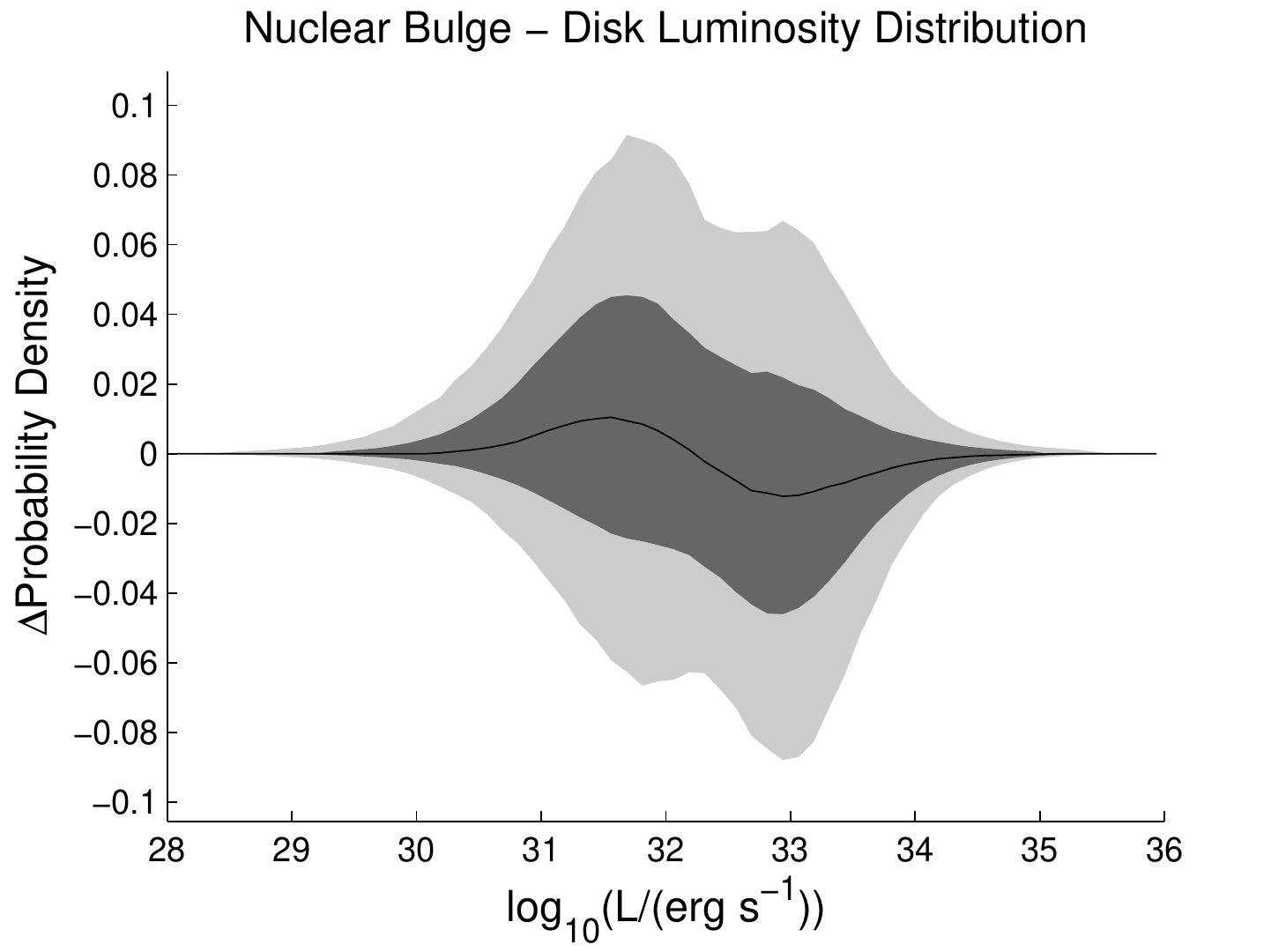}}
    \caption[MSP luminosity distributions for Model A1.]{MSP luminosity distribution in the disk and the change in probability density for the boxy bulge and nuclear bulge for Model A1. The black line shows the median probability density in each bin, dark grey the $68\%$ interval and light grey the $95\%$ interval. Note that the probability density is of $\log_{10}(L/({\rm erg\, s^{-1}}))$ rather than of $L/({\rm erg\, s^{-1}})$.
    }
    \label{fig:E_cut_B_E_dot_luminosity_distribution}
\end{figure}

\begin{figure}
    \centering
    \subfigure{\centering\includegraphics[width=0.49\linewidth]{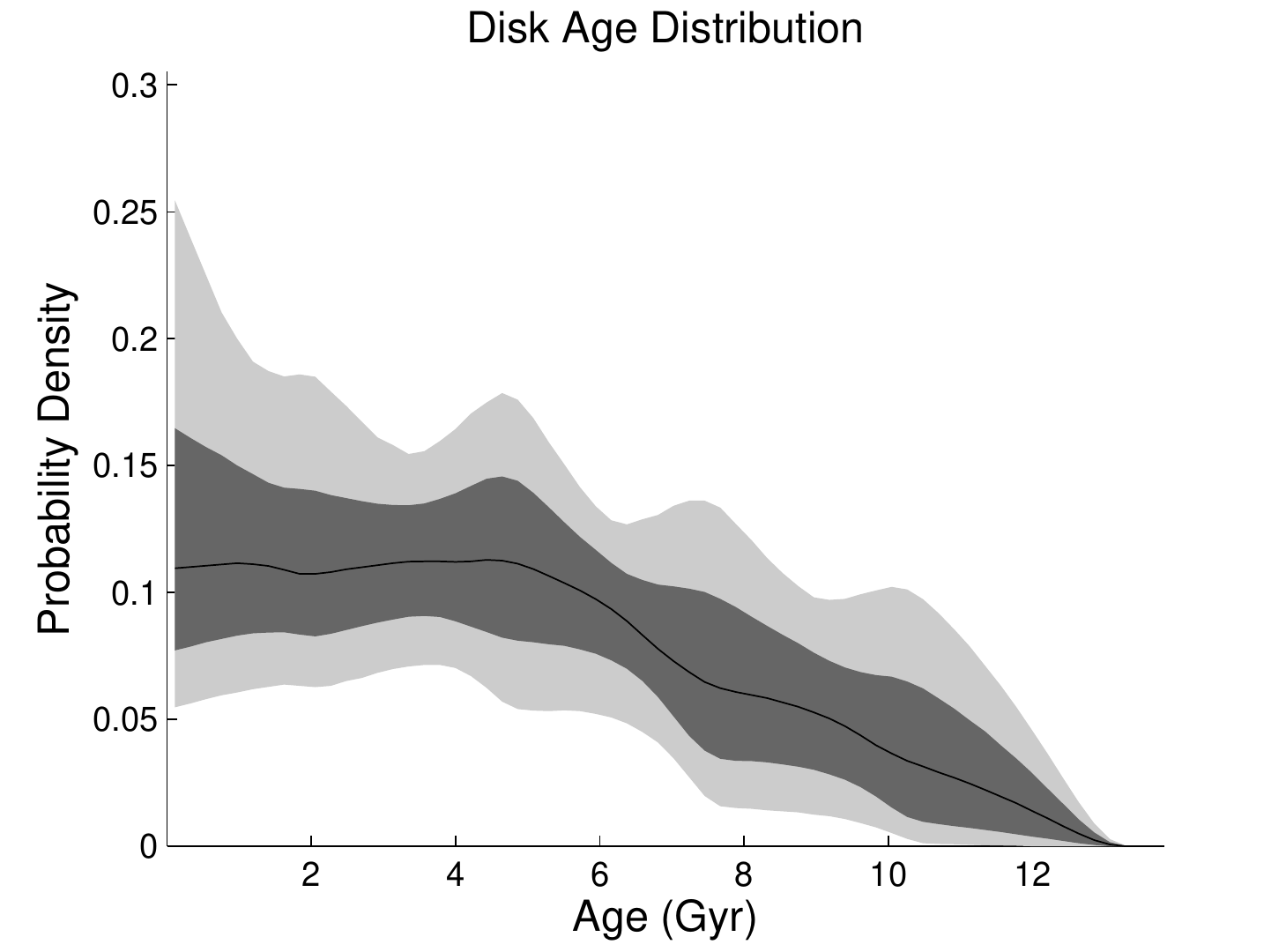}}
    \subfigure{\centering\includegraphics[width=0.49\linewidth]{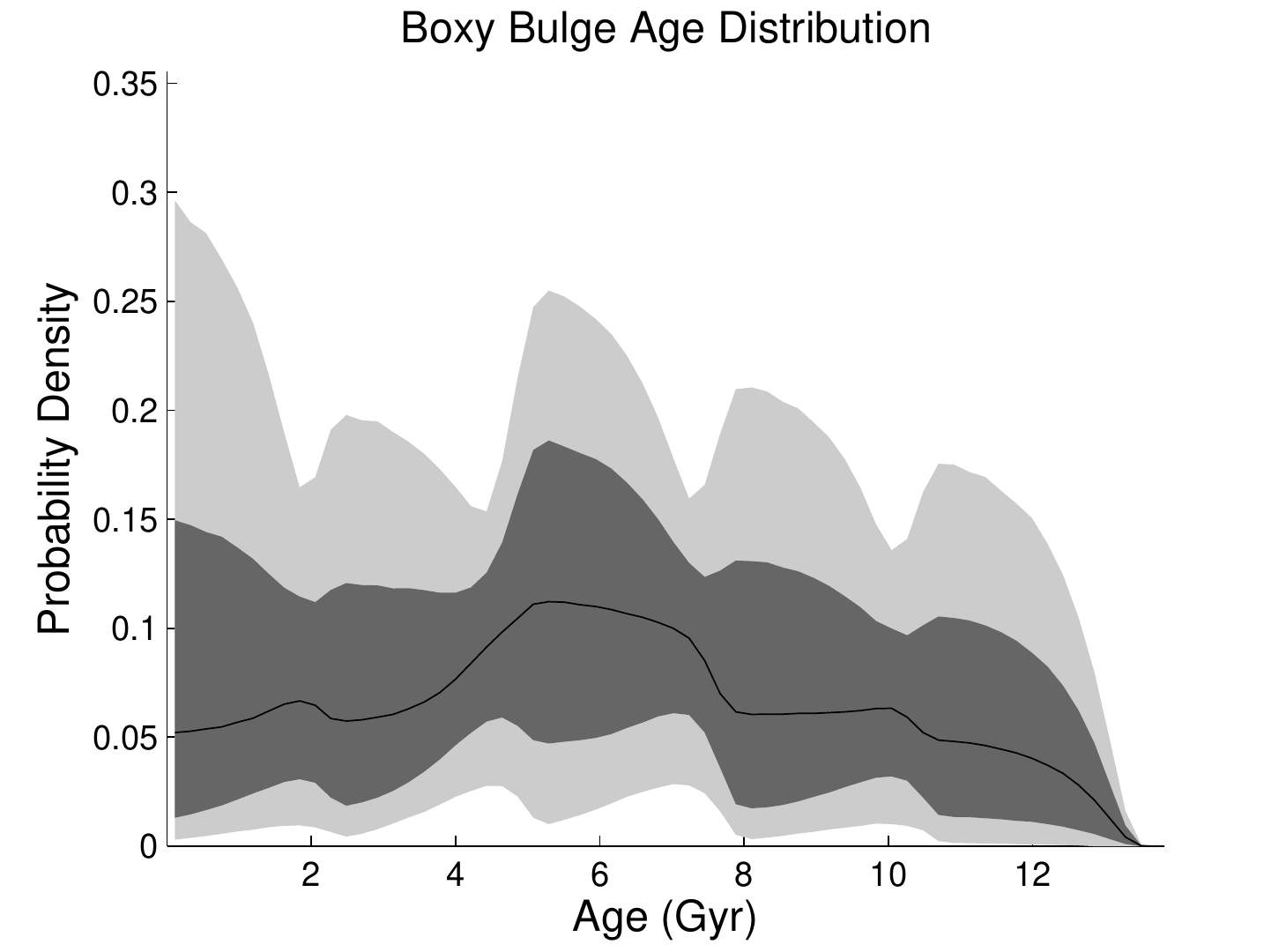}}
    \subfigure{\centering\includegraphics[width=0.49\linewidth]{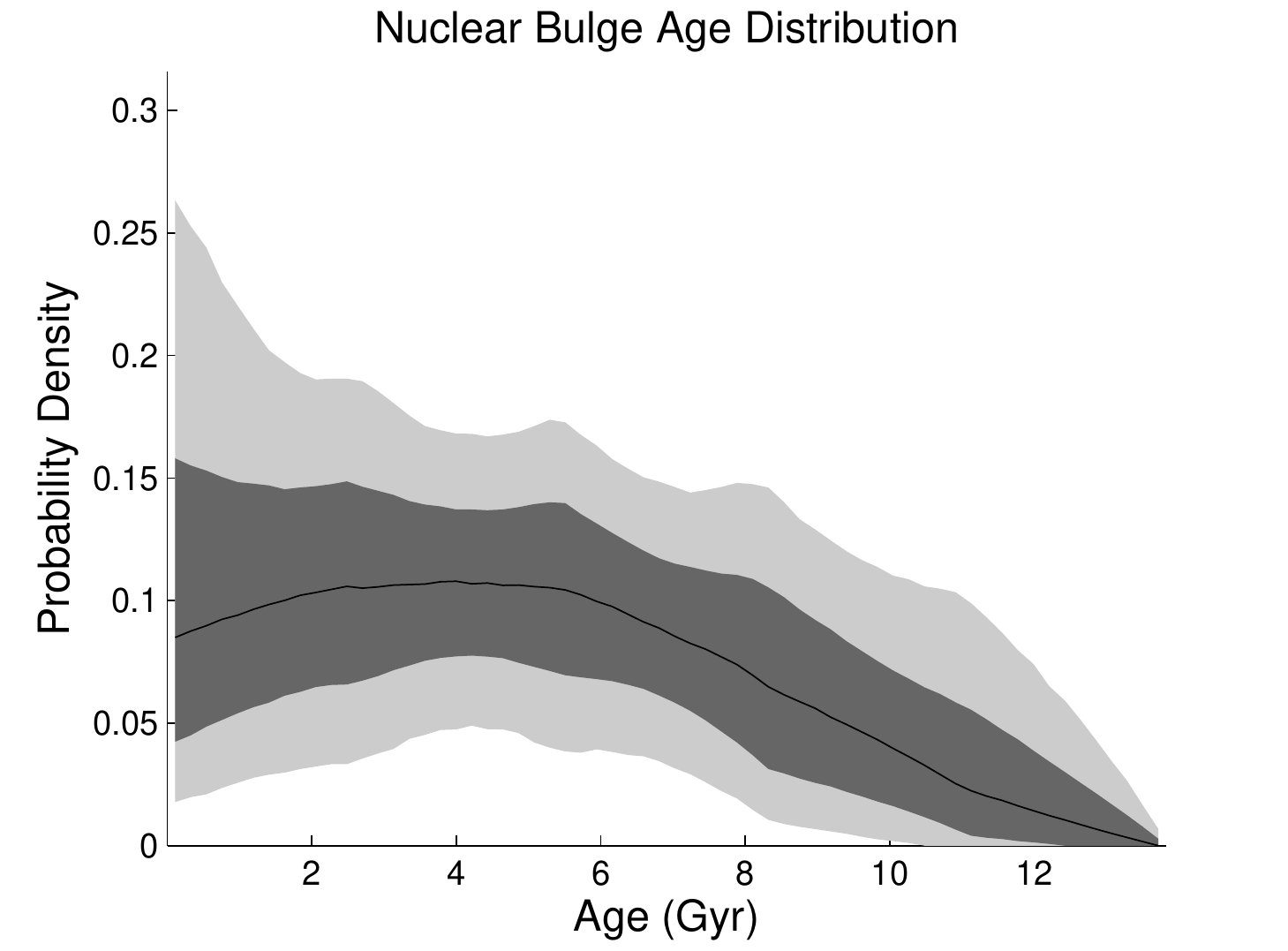}}
    \caption[MSP age distributions for Model A1.]{MSP age distribution in the disk, boxy bulge and nuclear bulge for Model A1. The black line shows the median probability density in each bin, dark grey the $68\%$ interval and light grey the $95\%$ interval.}
    \label{fig:E_cut_B_E_dot_age_distribution}
\end{figure}

\begin{figure}
    \centering
    \includegraphics[width=0.99\linewidth]{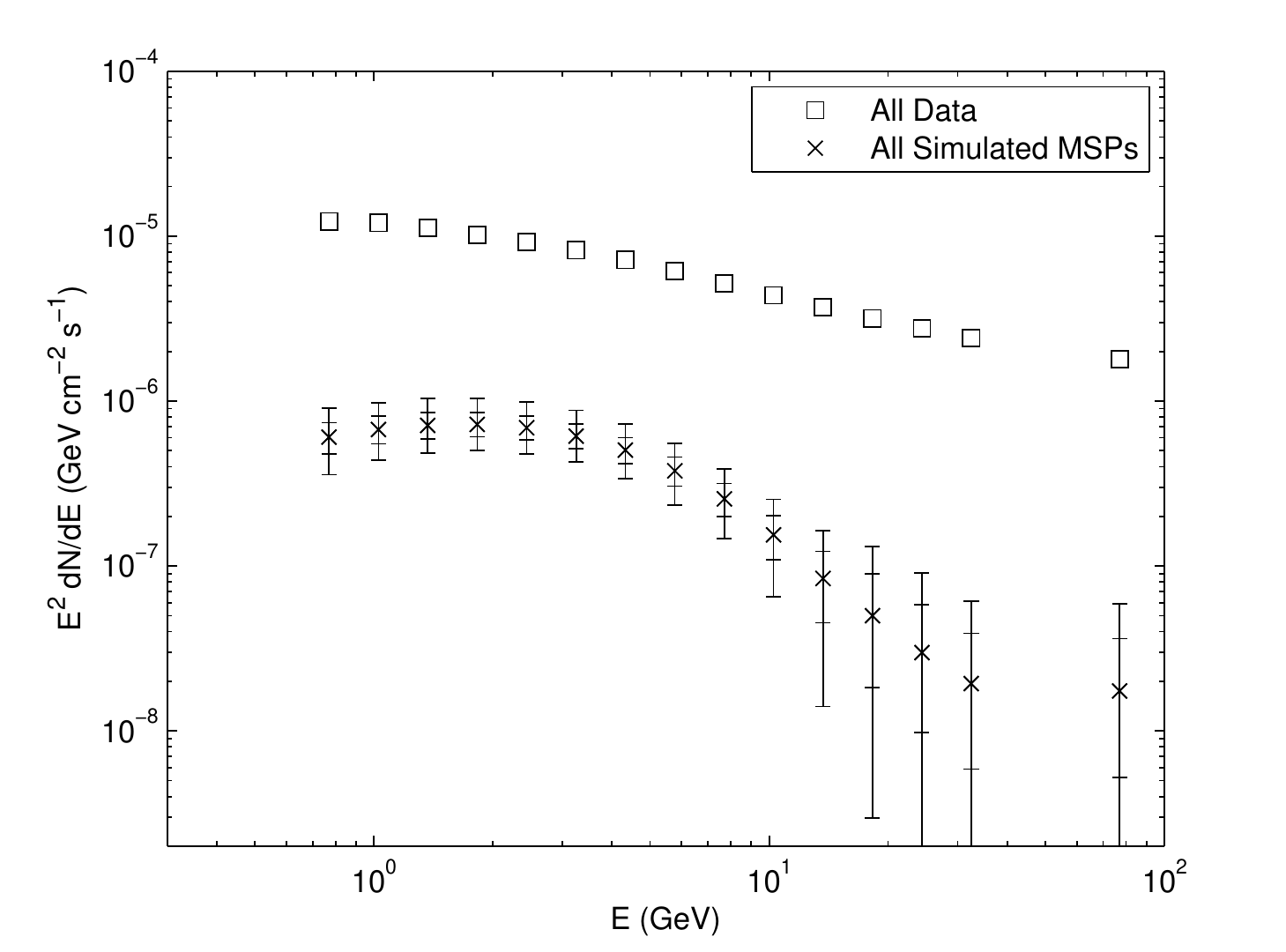}
    \caption[Total observed gamma ray emission compared to all simulated MSP emission from inner $40^{\circ} \times 40^{\circ}$ region for Model A1.]{ Total observed gamma ray emission compared to all simulated MSP emission from inner $40^{\circ} \times 40^{\circ}$ region for Model A1. Data are from Abazajian et al.~\cite{Abazajian2020}. }
    \label{fig:E_cut_B_E_dot_all_msp_emission}
\end{figure}

\begin{figure}
    \centering
    \subfigure{\centering\includegraphics[width=0.49\linewidth]{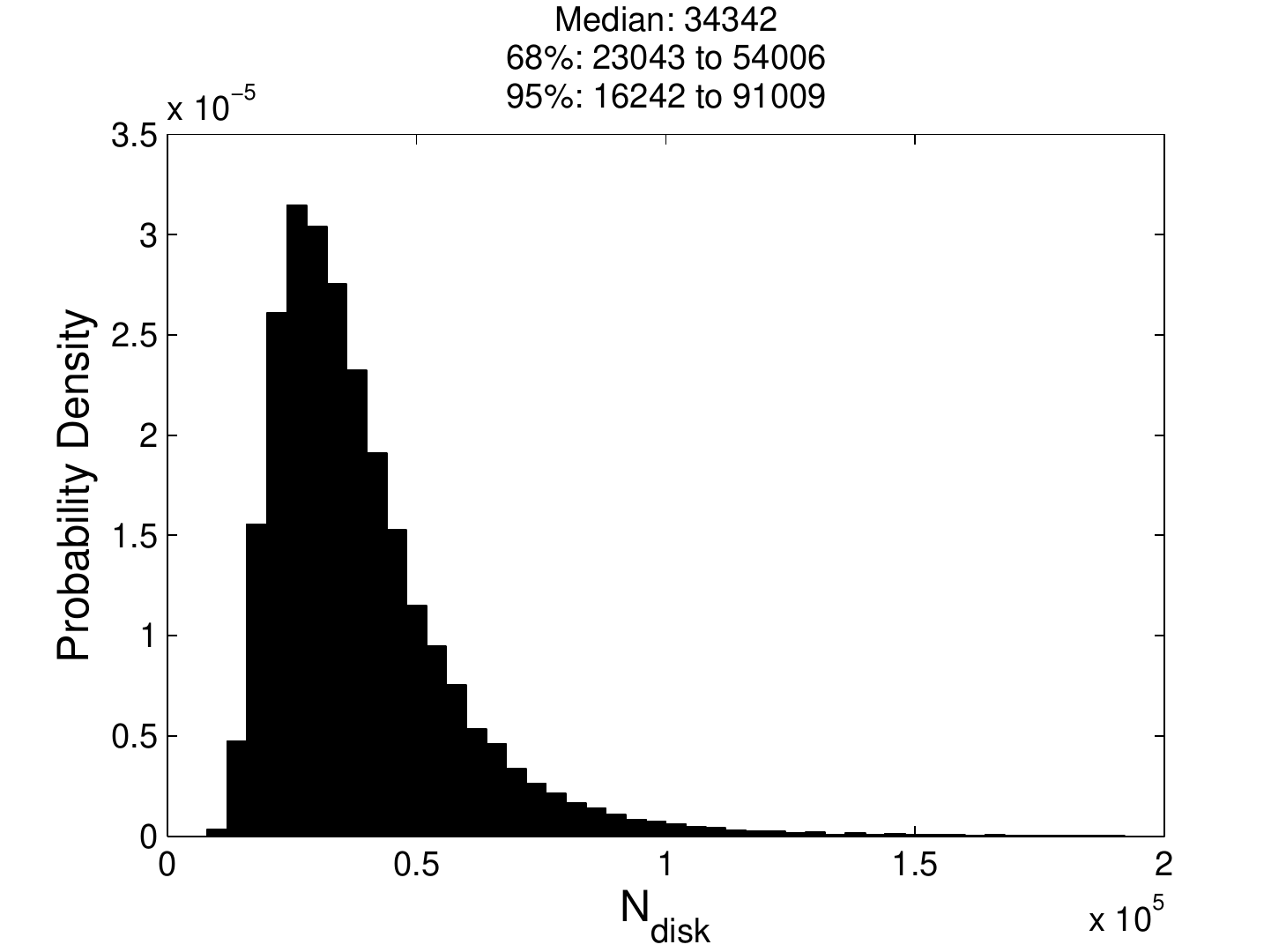}}
    \subfigure{\centering\includegraphics[width=0.49\linewidth]{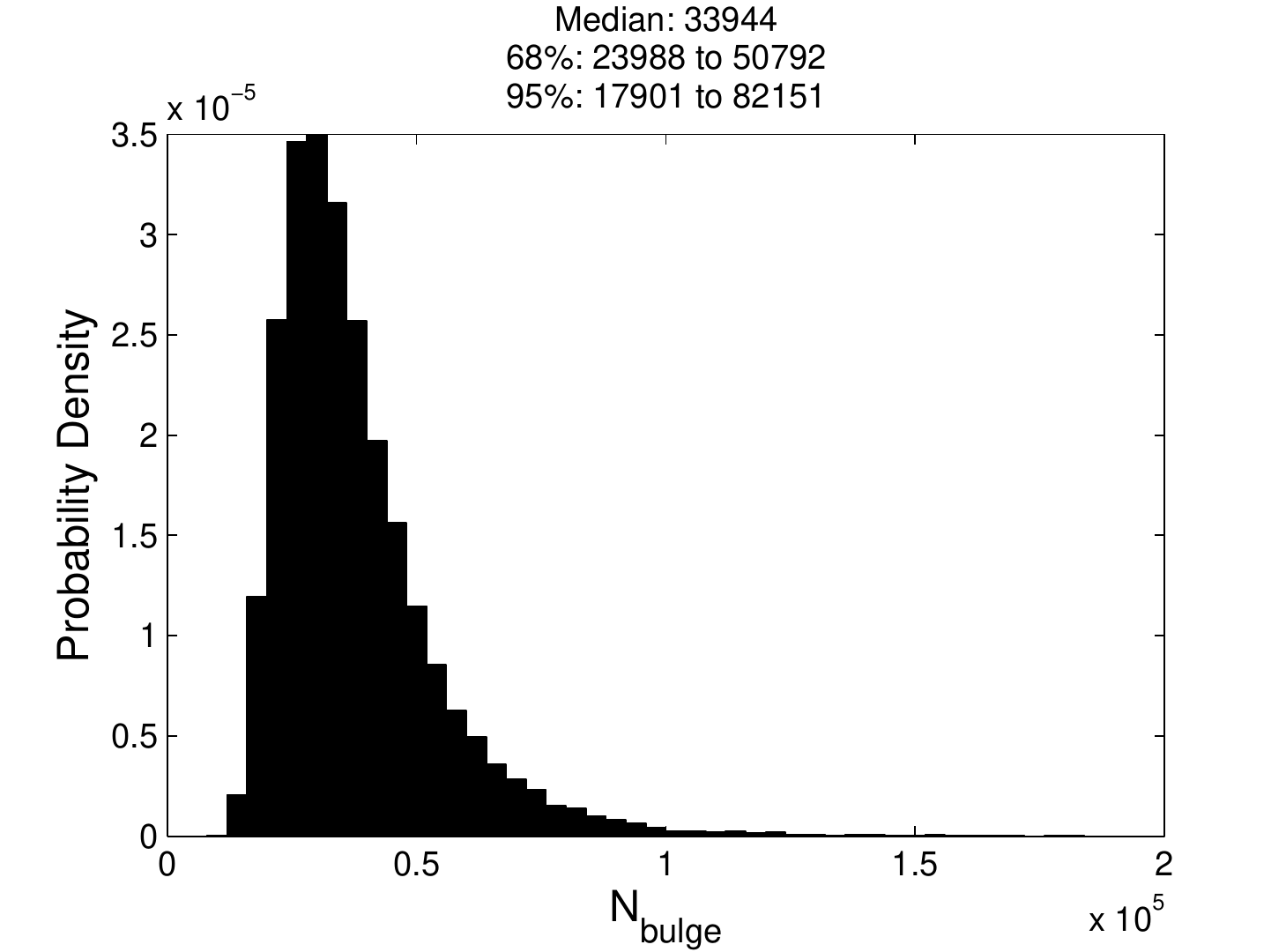}}
    \caption[Distribution of the number of disk and bulge MSPs for Model A1.]{Distribution of the number of disk and bulge MSPs for Model A1.}
    \label{fig:E_cut_B_E_dot_N_MSPs}
\end{figure}

\begin{figure}
    \centering
    \subfigure{\centering\includegraphics[width=0.49\linewidth]{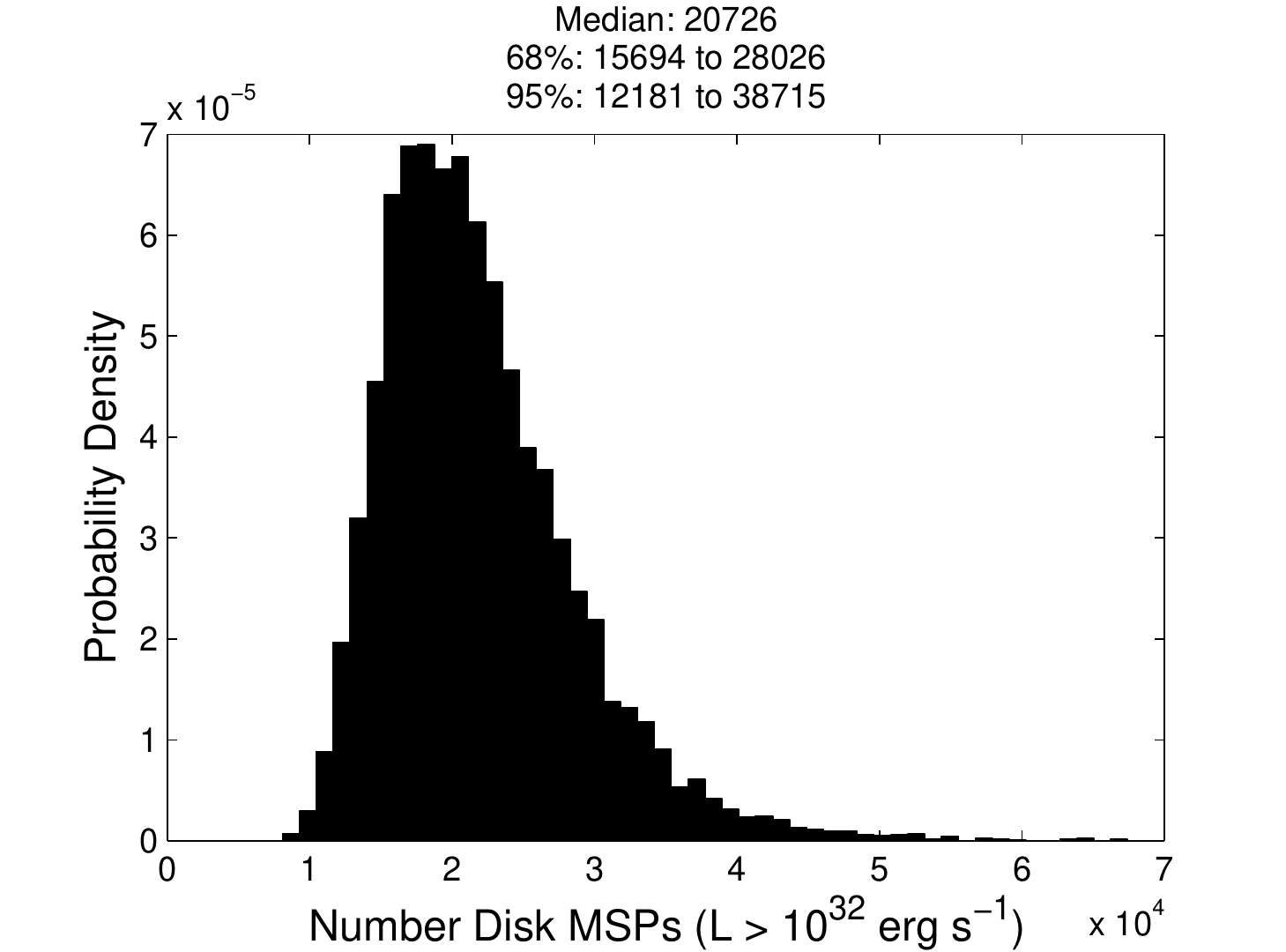}}
    \subfigure{\centering\includegraphics[width=0.49\linewidth]{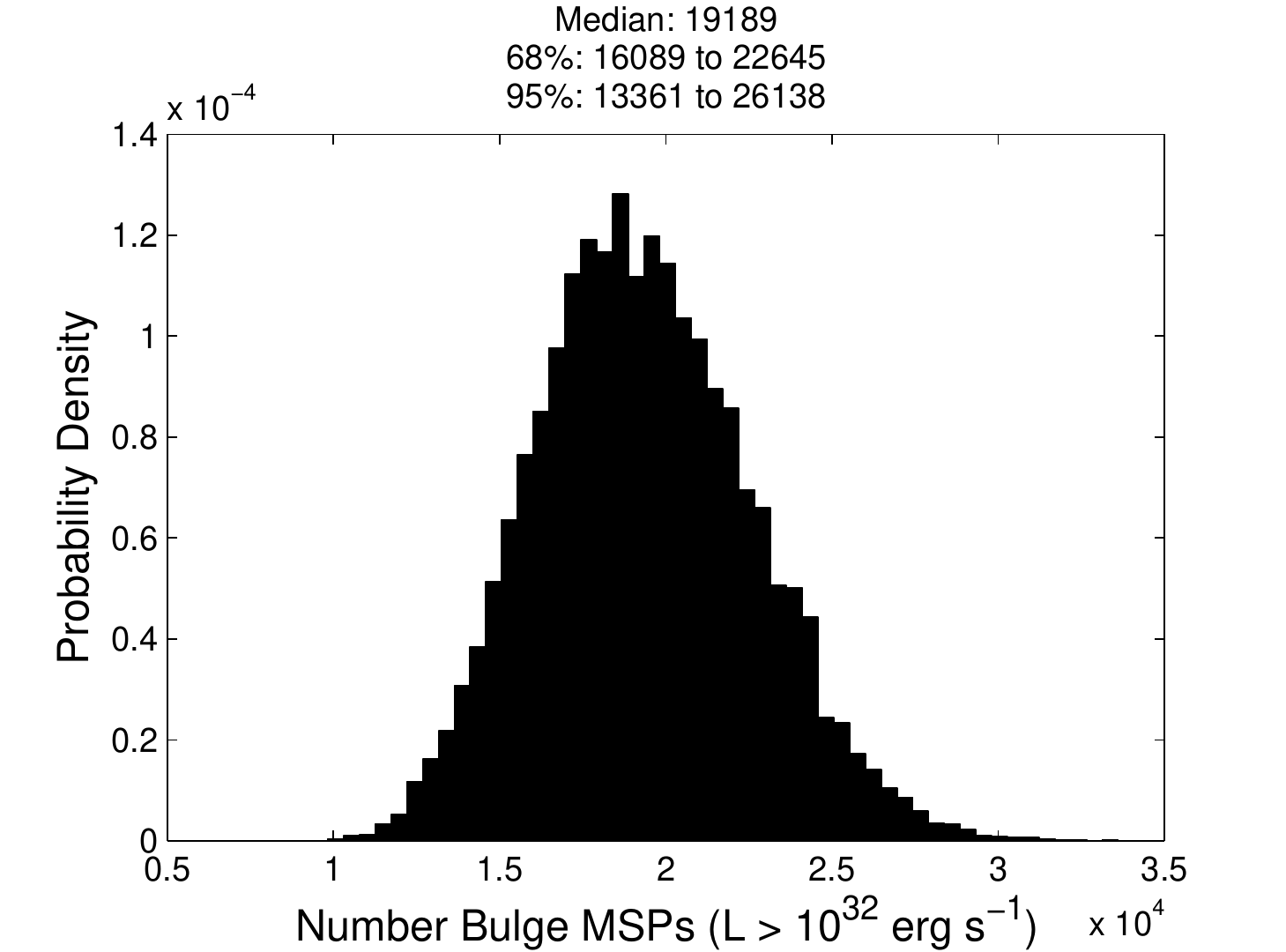}}
    \caption[Distribution of the number of disk and bulge MSPs with $L > 10^{32} \textrm{ erg s}^{-1}$ for 
    Model A1.]{Distribution of the number of disk and bulge MSPs with $L > 10^{32} \textrm{ erg s}^{-1}$ for 
    Model A1.}
    \label{fig:E_cut_B_E_dot_N_MSPs_greater_than_log_L_32}
\end{figure}

\begin{figure}
    \centering
    \subfigure{\centering\includegraphics[width=0.49\linewidth]{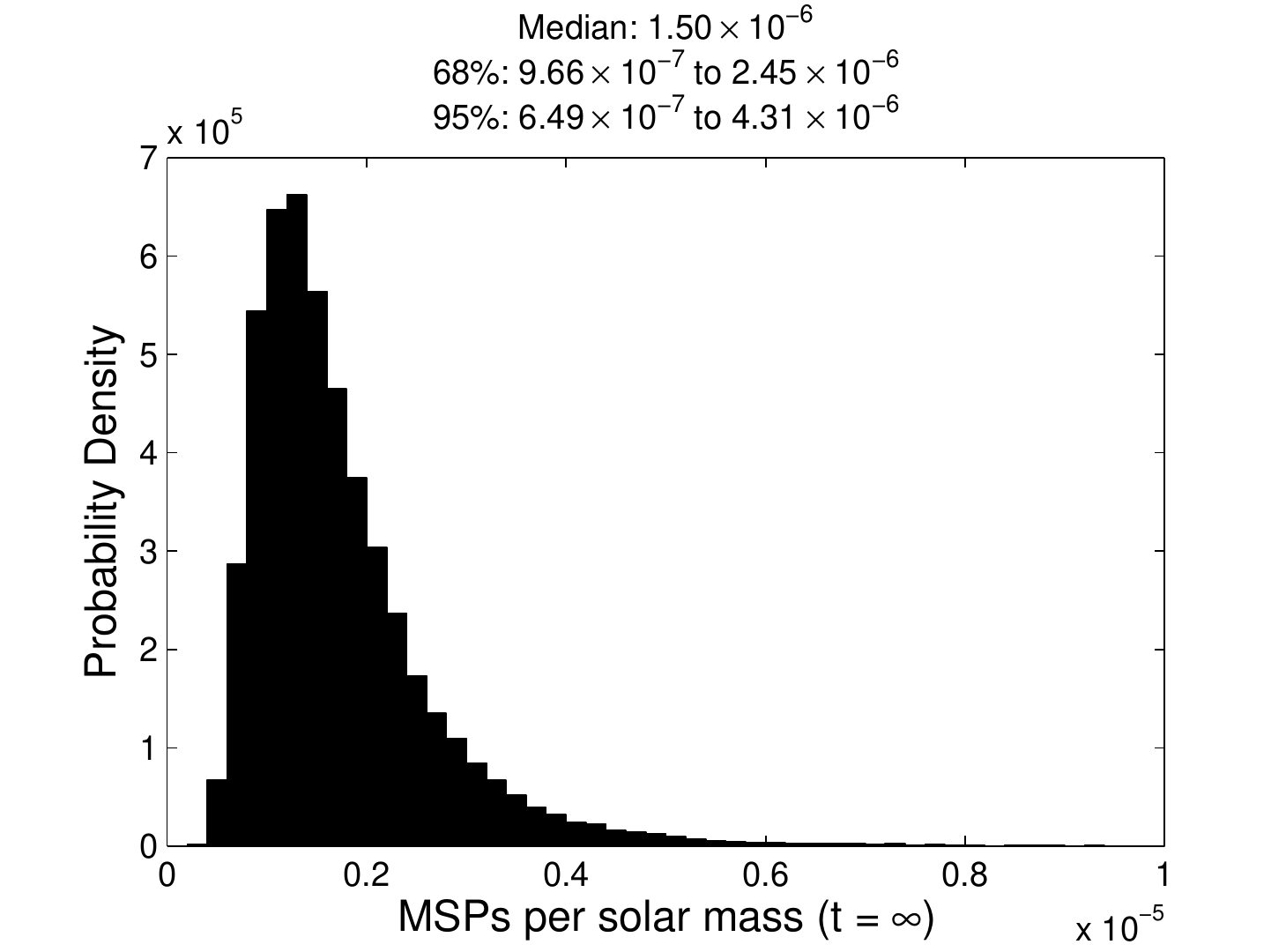}}
    \subfigure{\centering\includegraphics[width=0.49\linewidth]{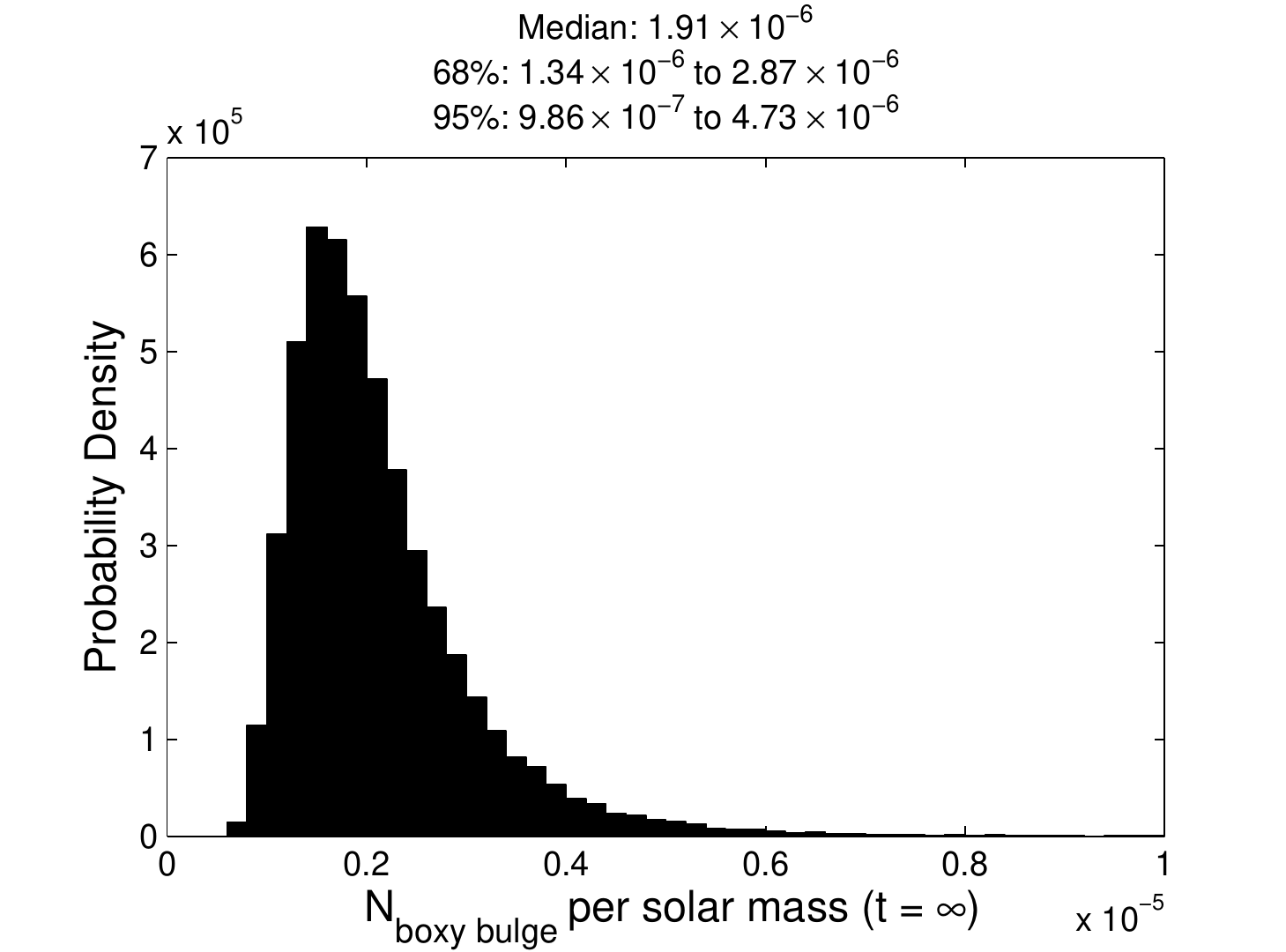}}
    \subfigure{\centering\includegraphics[width=0.49\linewidth]{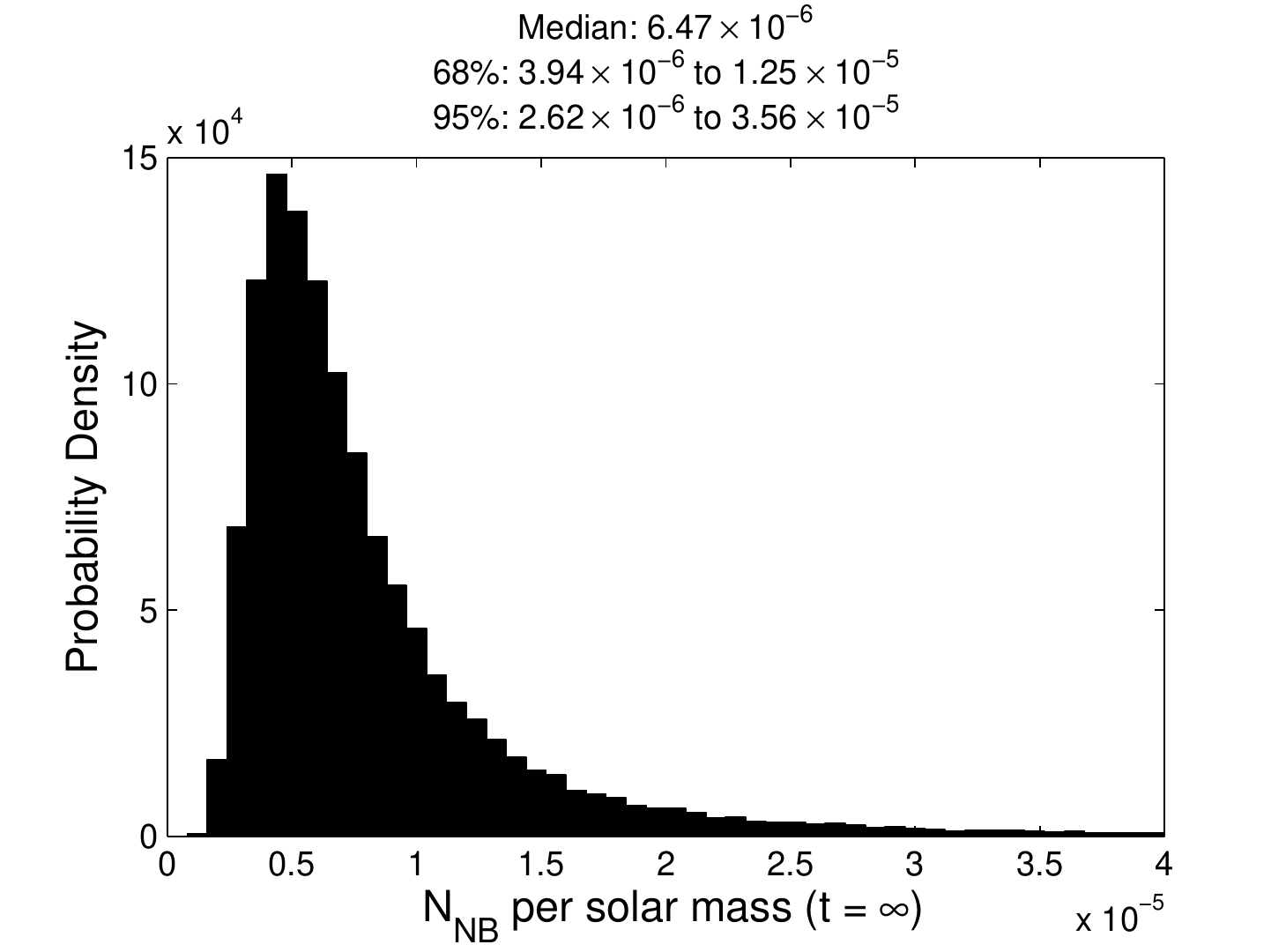}}
    \subfigure{\centering\includegraphics[width=0.49\linewidth]{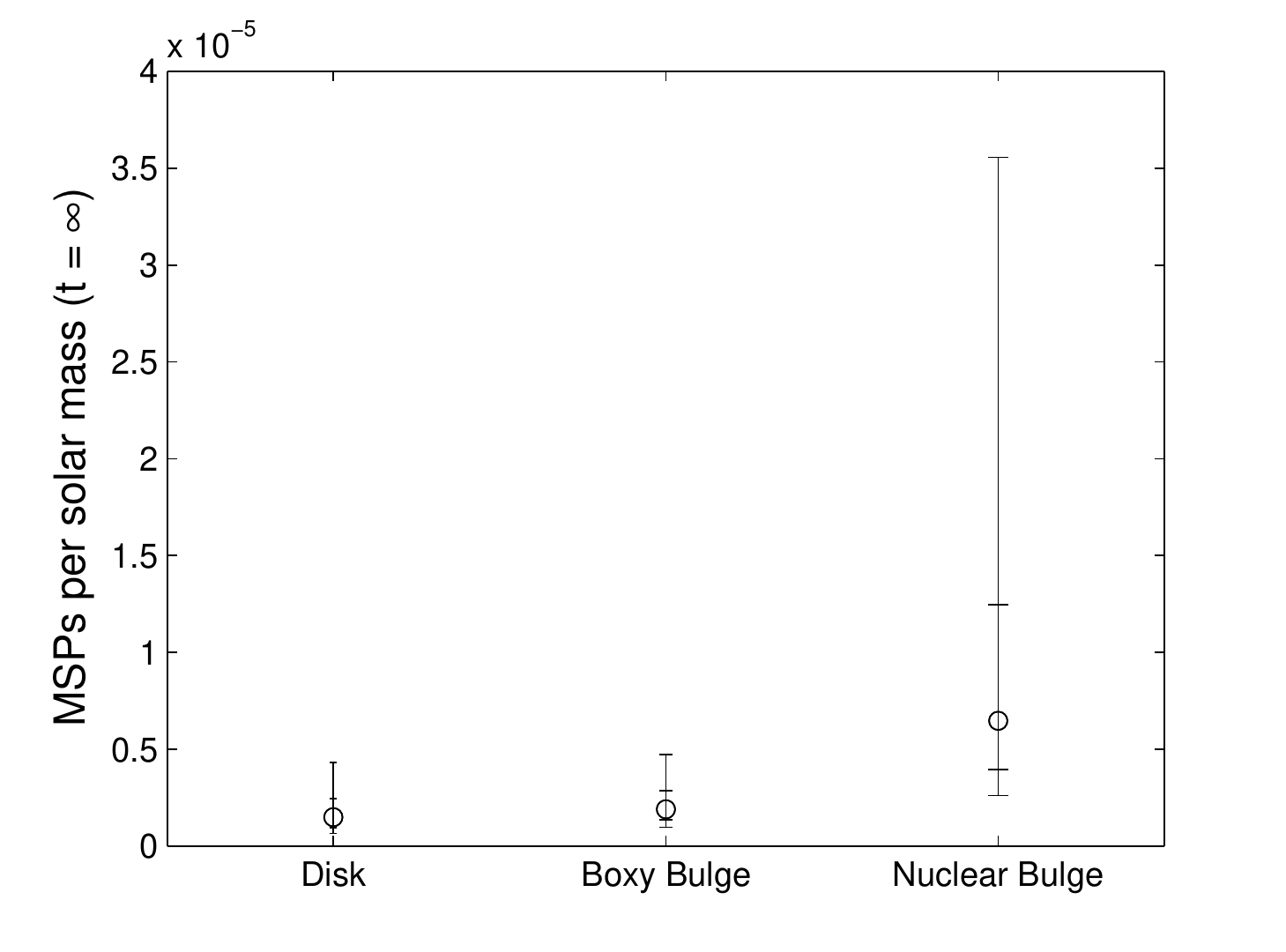}}
    \caption[Distribution of the number of disk, boxy bulge and nuclear bulge MSPs produced  per solar mass for Model A1.]{Distribution of the number of disk, boxy bulge and nuclear bulge MSPs produced  per solar mass at $t = \infty$ assuming no further star formation beyond today for Model A1. In the bottom right plot, the medians, $68\%$ and $95\%$ intervals are shown side by side. }
    \label{fig:E_cut_B_E_dot_N_MSPs_per_solar_mass}
\end{figure}

\begin{figure}
    \centering
    \subfigure{\centering\includegraphics[width=0.49\linewidth]{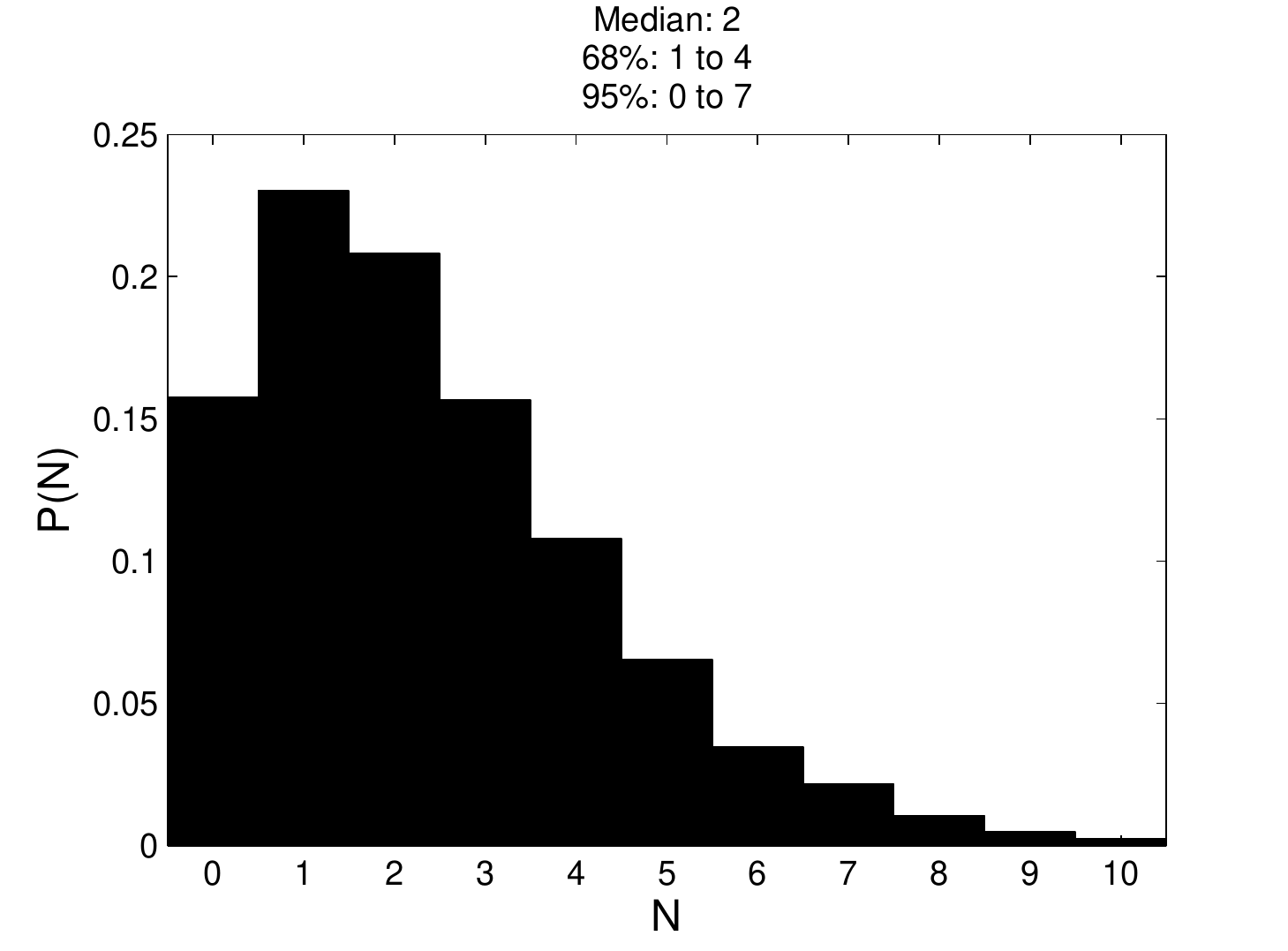}}
    \subfigure{\centering\includegraphics[width=0.49\linewidth]{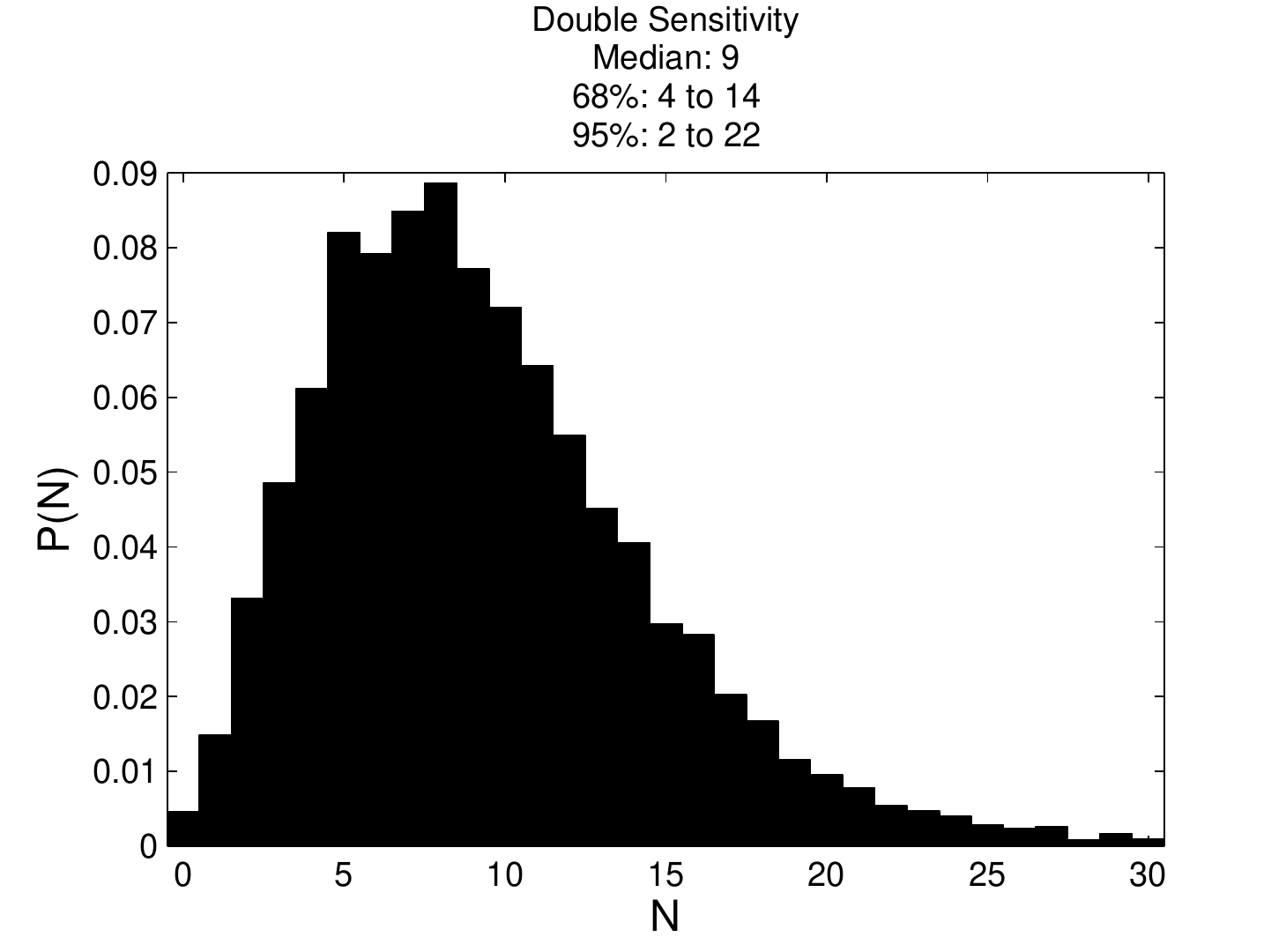}}
    \subfigure{\centering\includegraphics[width=0.49\linewidth]{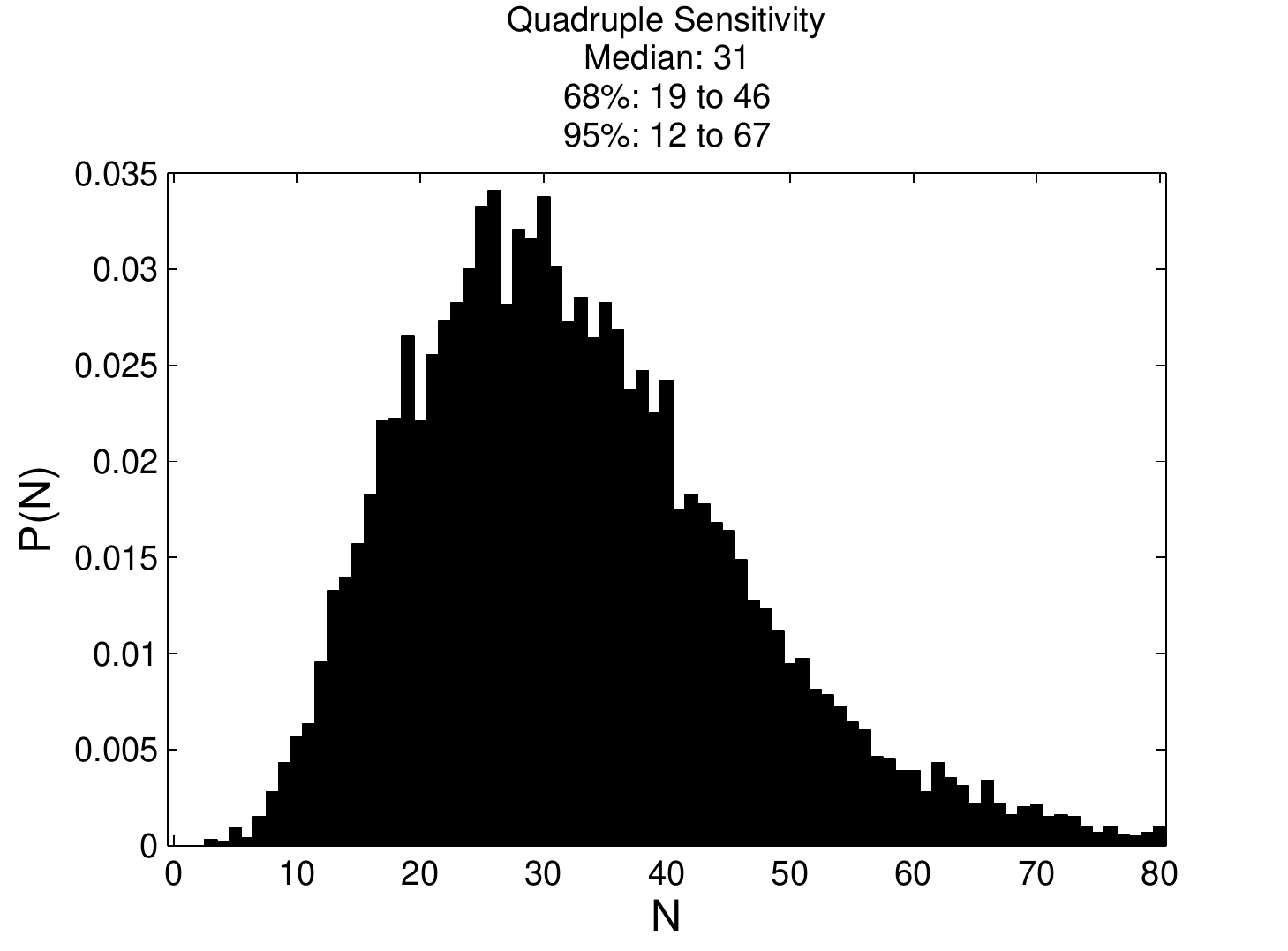}}
    \caption[Probability of observing $N$ bulge MSPs with the current sensitivity as well as for two and four times the current sensitivity for Model A1.]{Probability of observing $N$ bulge MSPs with the current sensitivity as well as for two and four times the current sensitivity for Model A1. The double and quadruple sensitivity distributions were produced by subtracting $0.3$ and $0.6$ respectively from $K_{\rm th}$.}
    \label{fig:E_cut_B_E_dot_N_observed_bulge_MSPs}
\end{figure}

\begin{figure}
    \centering
    \includegraphics[width=0.99\linewidth]{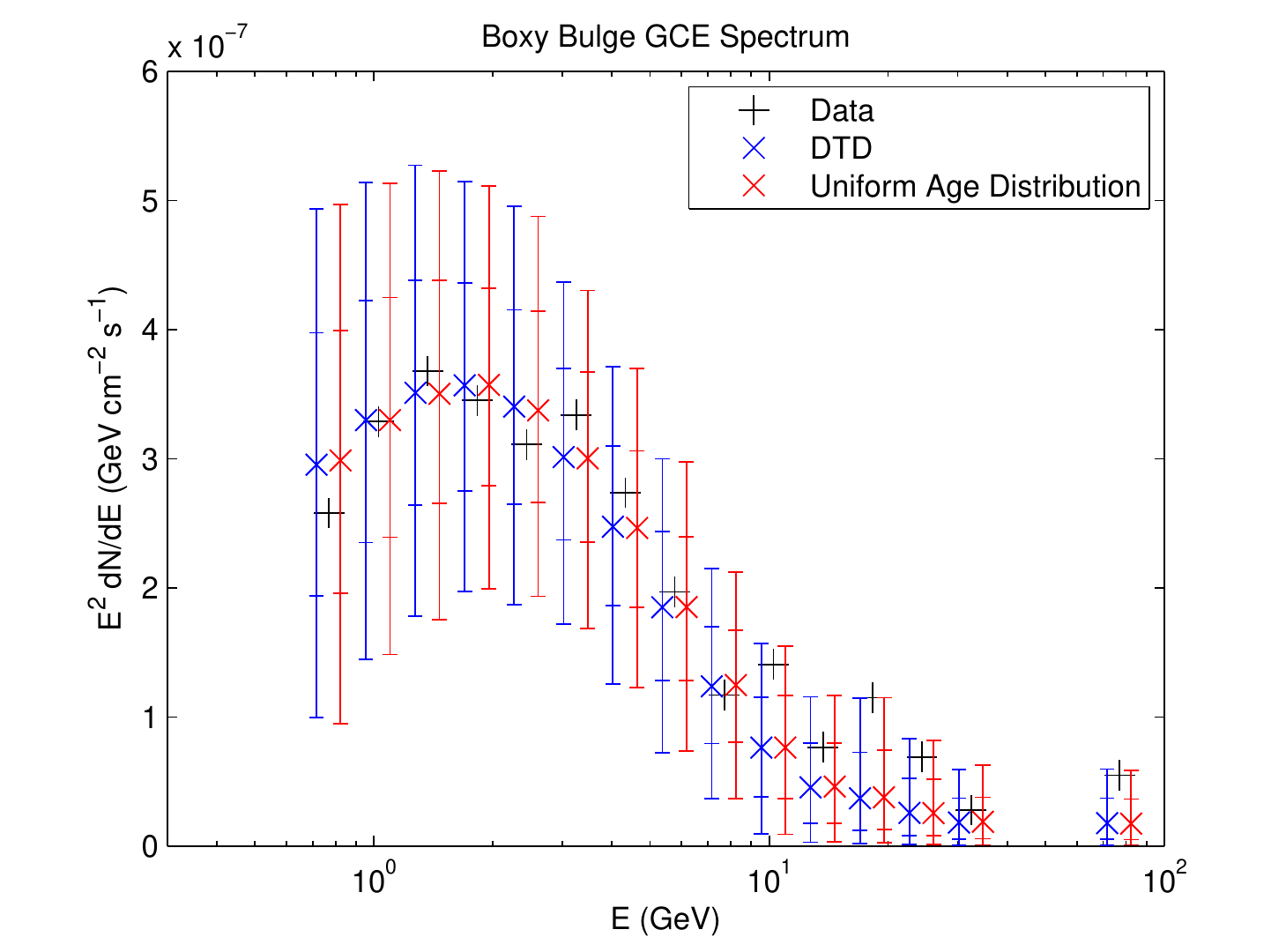}
    \caption[Comparison of posterior predictive boxy bulge GCE spectra for the $L = \eta E_{\rm cut}^{a_{\gamma}} B^{b_{\gamma}} \dot{E}^{d_{\gamma}}$ model in the case where we have a DTD 
    (Model A1) and the case where a uniform MSP age distribution was used (Model A2).]{ Comparison of posterior predictive boxy bulge GCE spectra for the $L = \eta E_{\rm cut}^{a_{\gamma}} B^{b_{\gamma}} \dot{E}^{d_{\gamma}}$ model in the case where we have a DTD 
    (Model A1) and the case where a uniform MSP age distribution was used (Model A2). The DTD case has been shifted slightly to the left and the uniform case to the right in order to make comparison easier. Only bins with $E<10$~GeV were fitted. }
    \label{fig:E_cut_B_E_dot_dtd_vs_uniform_boxy_bulge_spectra}
\end{figure}

In Figs.~\ref{fig:E_cut_B_E_dot_MCMC_Params_1}, \ref{fig:E_cut_B_E_dot_MCMC_Params_2}, and \ref{fig:E_cut_B_E_dot_MCMC_DTD_bin_Params}, we display corner plots showing some of these parameters for the best model, which was A1 which had  $L = \eta E_{\rm cut}^{a_{\gamma}} B^{b_{\gamma}} \dot{E}^{d_{\gamma}}$.
In Fig.~\ref{fig:E_cut_B_E_dot_posterior_predictive_plots} the resolved MSP and GCE data are compared to simulated data. 
{\textcolor{black}{We exclude data that led to an apparently negative $\dot{P}_{\rm int}$ from the binned $\dot{P}$ data.}} This exclusion affects all bins with data except the highest two. The luminosity distribution is shown in Fig.~\ref{fig:E_cut_B_E_dot_luminosity_distribution}, and  the MSP age distributions are shown in Fig.~\ref{fig:E_cut_B_E_dot_age_distribution}. In Fig.~\ref{fig:E_cut_B_E_dot_all_msp_emission} we compare the total gamma ray emission from MSPs in the region of interest to the observed total. In Fig.~\ref{fig:E_cut_B_E_dot_N_MSPs} we show the number of MSPs in the disk and in the bulge.
The number with luminosity greater than $10^{32} \textrm{ erg s}^{-1}$ is shown in Fig.~\ref{fig:E_cut_B_E_dot_N_MSPs_greater_than_log_L_32}, and in Fig.~\ref{fig:E_cut_B_E_dot_N_MSPs_per_solar_mass} we show the number of MSPs produced per solar mass at $t = \infty$ assuming no further star formation {\textcolor{black}{after today}}. The masses used were $(3.7 \pm 0.5) \times 10^{10} \msun$ for the disk, $(1.6 \pm 0.2) \times 10^{10} \msun$ for the boxy bulge, and $(1.4 \pm 0.6) \times 10^{9} \msun$ for the nuclear bulge \citep{Bland-Hawthorn2016, Crocker:2016zzt}. In Fig.~\ref{fig:E_cut_B_E_dot_N_observed_bulge_MSPs} we show the modeled probability of resolving $N$ bulge MSPs, as well as the number for double and quadruple the current sensitivity. In Fig.~\ref{fig:E_cut_B_E_dot_dtd_vs_uniform_boxy_bulge_spectra} we compare the posterior predictive distributions for the boxy bulge spectra in the cases where we fitted the DTD and where the MSPs were uniformly distributed in age. %

\begin{figure}
    \centering
    \includegraphics[width=0.99\linewidth]{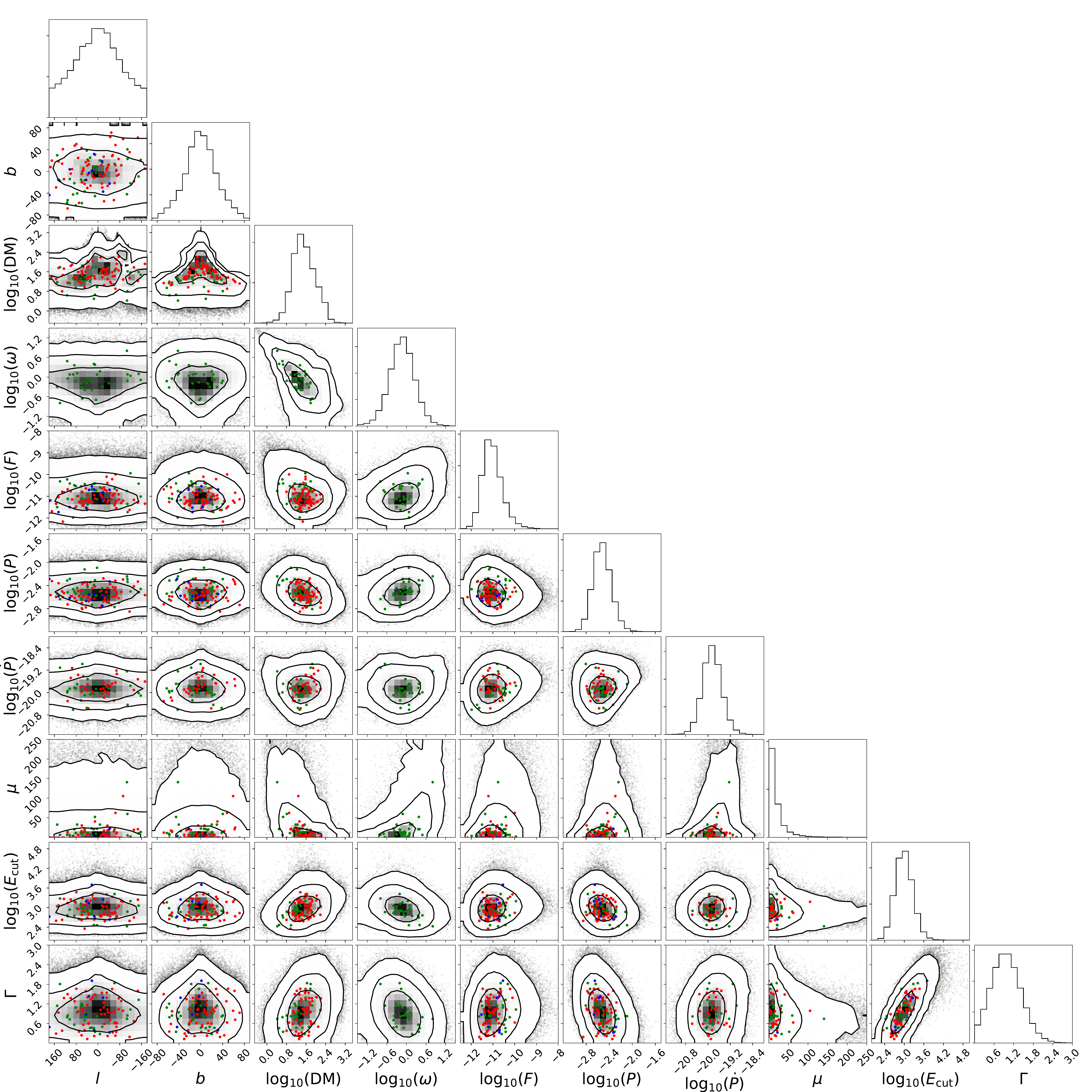}
    \caption[Model A1 ($L = \eta E_{\rm cut}^{a_{\gamma}} B^{b_{\gamma}} \dot{E}^{d_{\gamma}}$) posterior predictive corner plots.]{Model A1 ($L = \eta E_{\rm cut}^{a_{\gamma}} B^{b_{\gamma}} \dot{E}^{d_{\gamma}}$) posterior predictive corner plots showing longitude $l$ (deg), latitude $b$ (deg), dispersion measure DM (cm$^{-3}$ pc), parallax $\omega$ (mas), flux $F$ (erg cm$^{-2}$ s$^{-1}$), period $P$ (s), period derivative $\dot{P}$, proper motion $\mu$ (mas yr$^{-1}$), spectral cutoff $E_{\rm cut}$ (MeV) and spectral index $\Gamma$ simulated and real data. Red points are MSPs with dispersion measure distances, green is parallax, and blue is no distance measurement. The contours show regions containing $68\%$, $95\%$ and $99.7\%$ of the simulated data.}
    \label{fig:E_cut_B_E_dot_MSP_Params}
\end{figure}

\begin{figure}
    \centering
    \includegraphics[width=0.99\linewidth]{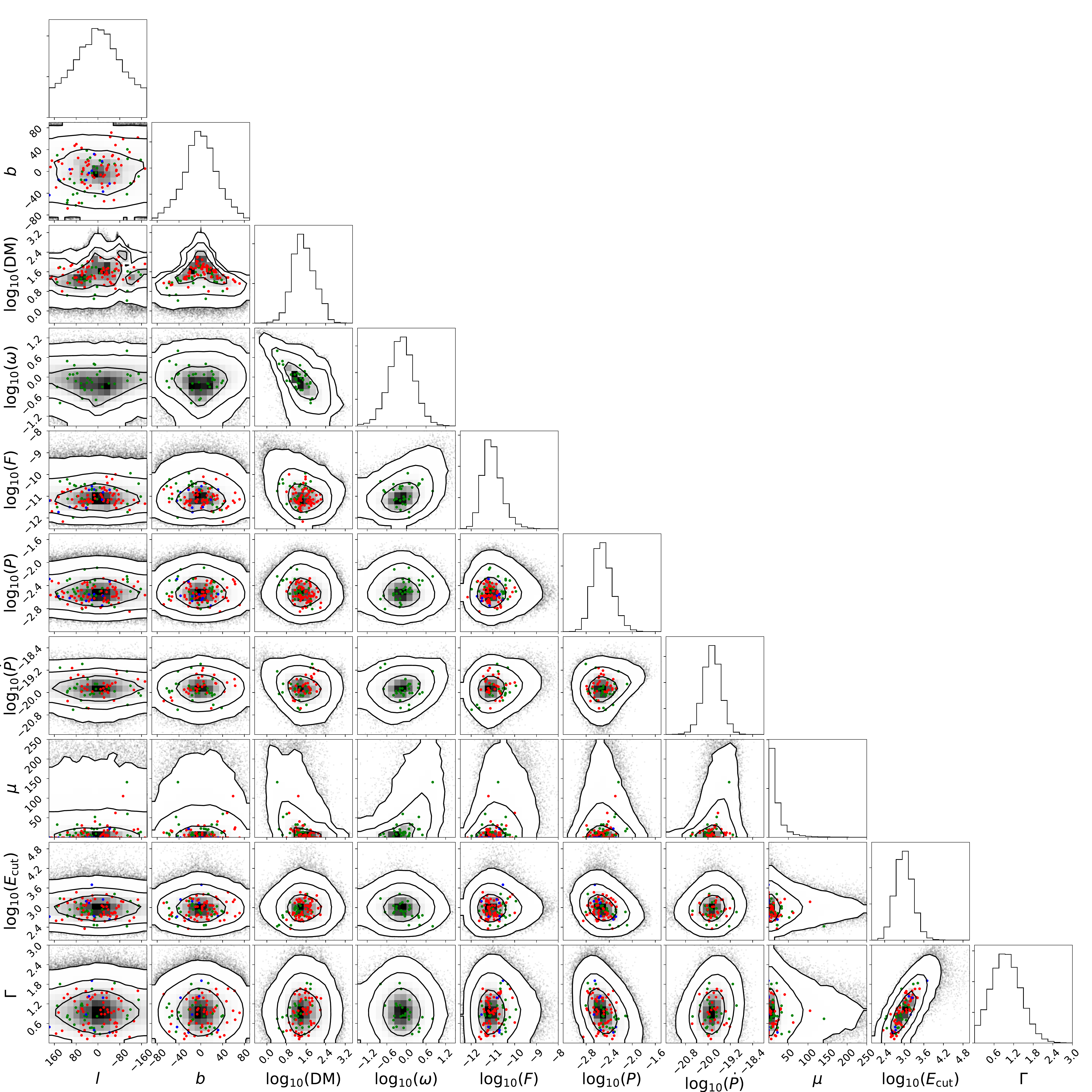}
    \caption[Model A9 ($L=\eta$) posterior predictive corner plots.]{
    The same as Fig.~\ref{fig:E_cut_B_E_dot_MSP_Params} except for 
    Model A9  which has $L=\eta$.}
    \label{fig:efficiency_only_MSP_Params}
\end{figure}

\begin{figure}
    \centering
    \includegraphics[width=0.99\linewidth]{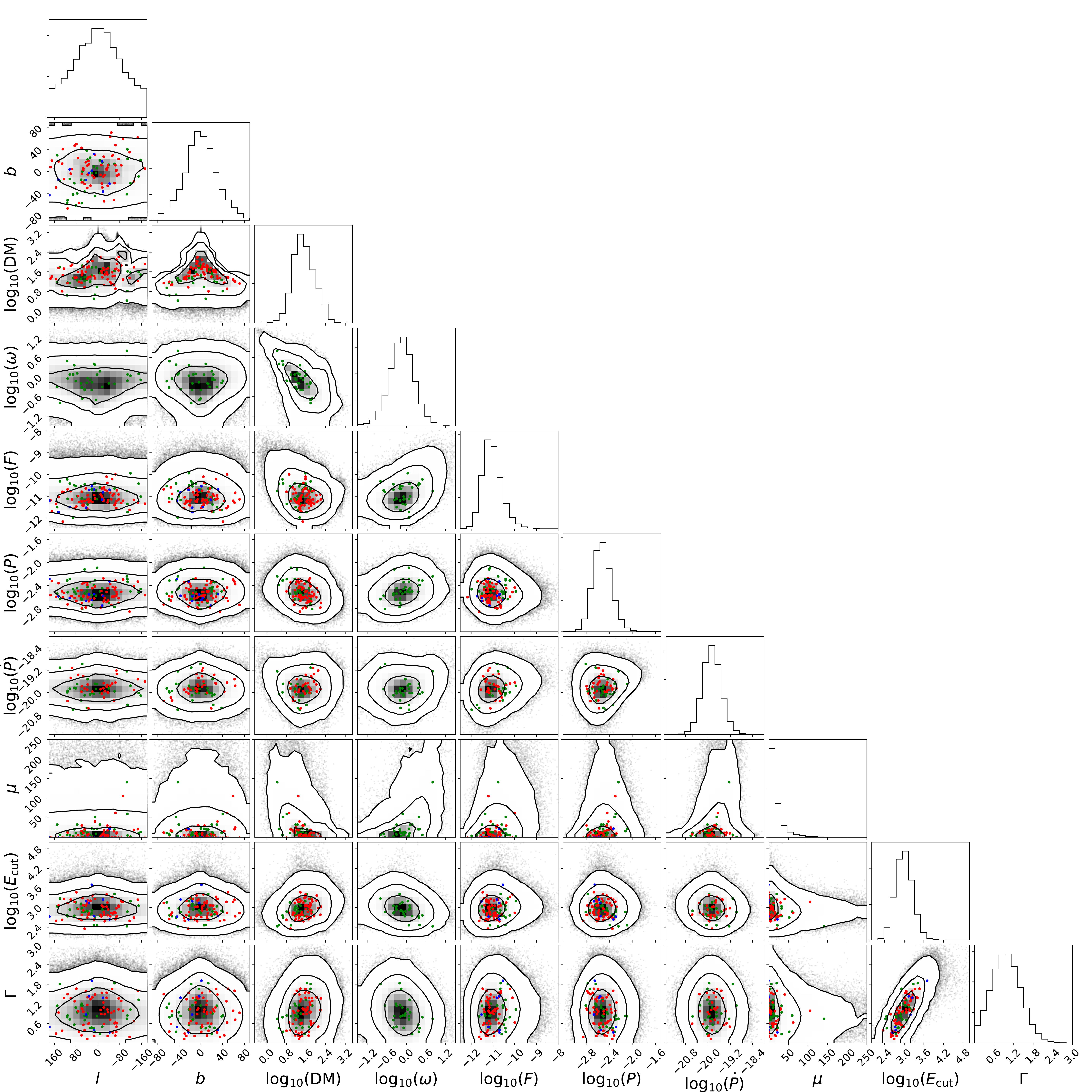}
    \caption[Model A7 ($L = \eta E_{\rm cut}^{a_{\gamma}} B^{b_{\gamma}} \dot{E}^{d_{\gamma}}$, $a_{E_{\rm cut}} = a_{\Gamma} = 0$) posterior predictive corner plots.]{The same as Fig.~\ref{fig:E_cut_B_E_dot_MSP_Params} except for 
    Model A7  which has the same luminosity function as Model A1 but has $E_{\rm cut}$ independent of $\dot{E}$ ($a_{E_{\rm cut}} = a_{\Gamma} = 0$).}
    \label{fig:E_cut_B_E_dot_NoSpectrumDependenceOnEdot_MSP_Params}
\end{figure}

Finally, we also display corner plots showing the distributions of simulated, resolved MSPs for different models of the MSP luminosity function: Fig.~\ref{fig:E_cut_B_E_dot_MSP_Params} shows the $L = \eta E_{\rm cut}^{a_{\gamma}} B^{b_{\gamma}} \dot{E}^{d_{\gamma}}$ case, Fig.~\ref{fig:efficiency_only_MSP_Params} shows the $L = \eta$ case, and Fig.~\ref{fig:E_cut_B_E_dot_NoSpectrumDependenceOnEdot_MSP_Params} shows the case for the $L = \eta E_{\rm cut}^{a_{\gamma}} B^{b_{\gamma}} \dot{E}^{d_{\gamma}}$ with $E_{\rm cut}$ and $\Gamma$ independent of $\dot{E}$.

\begin{table}
\centering
    \begin{tabular}{c|c|c}
         Parameter & Prior Minimum & Prior Maximum \\ \hline \hline
         
         $\lambda_{\rm res}$ & $0$ & $1000$ \\ \hline
         $\log_{10}(N_{\rm disk} / N_{\rm bulge})$ & $-100$ & $100$ \\ \hline
         $\log_{10}(N_{\rm nb} / N_{\rm bb})$ & $-100$ & $100$ \\ \hline
         $\sigma_r$ (kpc) & $0$ & $15$ \\ \hline
         $z_0$ (kpc) & $0$ & $1.5$ \\ \hline
         $K_{\rm th}$ & $-20$ & $20$ \\ \hline
         $\sigma_{\rm th}$ & $0$ & $5$ \\ \hline
         $C_{\rm parallax}$ & $0$ & $5$ \\ \hline
         $\log_{10}(d_{\rm parallax}/{\rm kpc})$ & $-1$ & $2$ \\ \hline
         $a_{E_{\rm cut}}$ & $-5$ & $5$ \\ \hline
         $b_{E_{\rm cut}}$ & $2$ & $5$ \\ \hline
         $\sigma_{E_{\rm cut}}$ & $0$ & $3$ \\ \hline
         $a_{\Gamma}$ & $-5$ & $5$ \\ \hline
         $b_{\Gamma}$ & $0$ & $5$ \\ \hline
         $\sigma_{\Gamma}$ & $0$ & $5$ \\ \hline
         $r_{\Gamma, E_{\rm cut}}$ & $-1$ & $1$ \\ \hline
         $a_{\gamma}$ & $-10$ & $10$ \\ \hline
         $b_{\gamma}$ & $-10$ & $10$ \\ \hline
         $d_{\gamma}$ & $-10$ & $10$ \\ \hline
         $\alpha_{\gamma}$ & $-10$ & $10$ \\ \hline
         $\beta_{\gamma}$ & $-10$ & $10$ \\ \hline
         $\textrm{DTD } p(0 \textrm{ - } 2.8 \textrm{ Gyr})$ & $0$ & $1$ \\ \hline
         $\textrm{DTD } p(2.8 \textrm{ - } 5.5 \textrm{ Gyr})$ & $0$ & $1$ \\ \hline
         $\textrm{DTD } p(5.5 \textrm{ - } 8.3 \textrm{ Gyr})$ & $0$ & $1$ \\ \hline
         $\textrm{DTD } p(8.3 \textrm{ - } 11.1 \textrm{ Gyr})$ & $0$ & $1$ \\ \hline
         $\textrm{DTD } p(11.1 \textrm{ - } 13.8 \textrm{ Gyr})$ & $0$ & $1$ \\ \hline
         $\log_{10}(P_{I, \textrm{ med}}/\textrm{s})$ & $-4$ & $10$ \\ \hline
         $\sigma_{P_{I}}$ & $0$ & $10$ \\ \hline
         $C_{P_I}$ & $-2$ & $5$ \\ \hline
         $\log_{10}(B_{\rm med}/\textrm{G})$ & $-1$ & $20$ \\ \hline
         $\sigma_{B}$ & $0$ & $10$ \\ \hline
         $r_{B, P_I}$ & $-1$ & $1$ \\ \hline
         $\log_{10}(\eta_{\rm med})$ & $-50$ & $50$ \\ \hline
         $\sigma_{\eta}$ & $0$ & $10$ \\ \hline
         $\sigma_{v}$ (km s$^{-1}$) & $0$ & $10000$ \\ \hline
    \end{tabular}
    \caption[Prior ranges for MSP population model parameters.]{Prior ranges for model parameters. The priors are uniform within these ranges except for the DTD bin probabilities which have  a Dirichlet prior \citep{Betancourt2013}. I.e.\ the prior for the DTD bin probabilities is uniform on a four dimensional hyperplane in the five dimensional DTD bin space. The hyperplane consist of all points for which the five DTD bin probabilities add up to one. }
    \label{tab:prior_ranges}
\end{table}

\begin{landscape}
\begin{table}
    \begin{tabular}{c|c|c|c|c}
         Model Label & Description & $\Delta \textrm{WAIC}$ & WAIC & $p_\text{WAIC}$ \\
          &  &  & Std. Dev. &  \\ \hline \hline
         
      A1&   $L = \eta E_{\rm cut}^{a_{\gamma}} B^{b_{\gamma}} \dot{E}^{d_{\gamma}}$ & $0$ & $0.8$ & $26.9$ \\ \hline
      A2&   $L = \eta E_{\rm cut}^{a_{\gamma}} B^{b_{\gamma}} \dot{E}^{d_{\gamma}}$, Uniform Age Distribution & $0.4$ & $1.0$ & $26.1$ \\ \hline
    A3&     $L = \eta E_{\rm cut}^{a_{\gamma}} B^{b_{\gamma}} \dot{E}^{d_{\gamma}}$, Covariance $B$ and $P_I$ & $1.5$ & $1.0$ & $27.5$ \\ \hline
        A4& $L = \eta E_{\rm cut}^{a_{\gamma}} B^{b_{\gamma}} \dot{E}^{d_{\gamma}}$, Eq.\ \ref{eq:gonthier_initial_period_dist} Initial Period Distribution & $7.3$ & $1.5$ & $25.9$ \\ \hline
        A5& $L = \eta \dot{E}^{\alpha_{\gamma}}$  & $9.5$ & $1.2$ & $24.4$ \\ \hline
        A6& $L = \eta P^{\alpha_{\gamma}} \dot{P}^{\beta_{\gamma}}$ & $11.0$ & $0.7$ & $25.3$ \\ \hline
        A7& $L = \eta E_{\rm cut}^{a_{\gamma}} B^{b_{\gamma}} \dot{E}^{d_{\gamma}}$, $a_{E_{\rm cut}} = a_{\Gamma} = 0$ & $19.4$ & $0.5$ & $25.1$ \\ \hline
        A8& $L = \eta E_{\rm cut}^{a_{\gamma}} B^{b_{\gamma}} \dot{E}^{d_{\gamma}}$, $a_{E_{\rm cut}} = a_{\Gamma} = 0$, Uniform Age Distribution & $21.1$ & $0.7$ & $24.0$ \\ \hline
    A9&     $L = \eta$ & $42.2$ & $0.7$ & $24.1$ \\ \hline
         \hline
      B1&   $L = \eta E_{\rm cut}^{a_{\gamma}} B^{b_{\gamma}} \dot{E}^{d_{\gamma}}$, Uniform Age Distribution, No GCE & $0$ & $1$ & $24$ \\ \hline
    B2&     $L = \eta E_{\rm cut}^{a_{\gamma}} B^{b_{\gamma}} \dot{E}^{d_{\gamma}}$, No GCE & $1.2$ & $0.8$ & $25.3$ \\ \hline
      B3&   $L = \eta E_{\rm cut}^{a_{\gamma}} B^{b_{\gamma}} \dot{E}^{d_{\gamma}}$, Covariance $B$ and $P_I$, No GCE & $2.3$ & $0.8$ & $26.1$ \\ \hline
    B4&     $L = \eta E_{\rm cut}^{a_{\gamma}} B^{b_{\gamma}} \dot{E}^{d_{\gamma}}$, Eq.\ \ref{eq:gonthier_initial_period_dist} Initial Period Distribution, No GCE & $9$ & $4$ & $24$ \\ \hline
      B5&  $L = \eta \dot{E}^{\alpha_{\gamma}}$, No GCE & $13$ & $1$ & $23$\\ \hline
    B6&     $L = \eta P^{\alpha_{\gamma}} \dot{P}^{\beta_{\gamma}}$, No GCE & $15.2$ & $0.9$ & $23.8$ \\ \hline
      B7&  $L = \eta E_{\rm cut}^{a_{\gamma}} B^{b_{\gamma}} \dot{E}^{d_{\gamma}}$, $a_{E_{\rm cut}} = a_{\Gamma} = 0$, No GCE & $21.3$ & $0.9$ & $23.6$ \\ \hline
    B8&     $L = \eta E_{\rm cut}^{a_{\gamma}} B^{b_{\gamma}} \dot{E}^{d_{\gamma}}$, $a_{E_{\rm cut}} = a_{\Gamma} = 0$, Uniform Age Distribution, No GCE & $22.9$ & $0.9$ & $22.5$ \\ \hline
      B9&  $L = \eta$, No GCE & $47$ & $2$ & $23$ \\ \hline
    \end{tabular}
    \caption[Average WAIC for each model.]{Average WAIC for each model relative to the model with best (lowest) WAIC. The average is taken over eight MCMC chains run for each model. We separate cases where we do not fit a GCE (B1-B9) from those where we do (A1-A9). The WAIC standard deviation column shows the sample standard deviation of WAIC for the eight MCMC chains run for each model. The $p_{\rm WAIC}$ column gives a measure of the effective number of parameters for the corresponding model \citep{Wantabe2010, Gelman2013}. }
    \label{tab:WAIC}
\end{table}
\end{landscape}

\begin{landscape}
\begin{longtable}{c|c|c|c|c|c|c}
    \hline
    Parameter & Model A1 & Model B2 & Model A5 & Model A6 & Model A7 & Model A9 \\
    \hline \hline
    \endfirsthead
    \hline
    Parameter & Model A1 & Model B2 & Model A5 & Model A6 & Model A7 & Model A9 \\
    \hline \hline
    \endhead
    
         $\lambda_{\rm res}$ & $108\substack{+11 \\ -10}$ & $107\substack{+11 \\ -10}$ & $108\substack{+10 \\ -10}$ & $108\substack{+10 \\ -10}$ & $108\substack{+11 \\ -10}$ & $108\substack{+11 \\ -10}$ \\ \hline
         $\log_{10}(N_{\rm disk} / N_{\rm bulge})$ & $0.00\substack{+0.13 \\ -0.12}$ & $-$ & $0.00\substack{+0.13 \\ -0.12}$ & $0.01\substack{+0.13 \\ -0.12}$ & $0.04\substack{+0.12 \\ -0.11}$ & $0.09\substack{+0.11 \\ -0.10}$ \\ \hline
         $\log_{10}(N_{\rm nb} / N_{\rm bb})$ & $-0.66\substack{+0.08 \\ -0.07}$ & $-$ & $-0.66\substack{+0.08 \\ -0.07}$ & $-0.66\substack{+0.07 \\ -0.07}$ & $-0.61\substack{+0.06 \\ -0.07}$ & $-0.58\substack{+0.04 \\ -0.04}$ \\ \hline
         $\sigma_r$ (kpc) & $4.5\substack{+0.5 \\ -0.4}$ & $4.5\substack{+0.5 \\ -0.4}$ & $4.4\substack{+0.5 \\ -0.4}$ & $4.5\substack{+0.5 \\ -0.4}$ & $4.4\substack{+0.5 \\ -0.4}$ & $4.5\substack{+0.6 \\ -0.4}$ \\ \hline
         $z_0$ (kpc) & $0.71\substack{+0.11 \\ -0.09}$ & $0.70\substack{+0.10 \\ -0.09}$ & $0.71\substack{+0.11 \\ -0.09}$ & $0.72\substack{+0.11 \\ -0.09}$ & $0.71\substack{+0.11 \\ -0.09}$ & $0.72\substack{+0.11 \\ -0.09}$ \\ \hline
         $K_{\rm th}$ & $0.45\substack{+0.09 \\ -0.08}$ & $0.43\substack{+0.09 \\ -0.08}$ & $0.45\substack{+0.09 \\ -0.08}$ & $0.45\substack{+0.09 \\ -0.08}$ & $0.46\substack{+0.09 \\ -0.08}$ & $0.46\substack{+0.09 \\ -0.08}$ \\ \hline
         $\sigma_{\rm th}$ & $0.28\substack{+0.05 \\ -0.04}$ & $0.27\substack{+0.05 \\ -0.04}$ & $0.28\substack{+0.05 \\ -0.04}$ & $0.28\substack{+0.05 \\ -0.04}$ & $0.28\substack{+0.05 \\ -0.04}$ & $0.28\substack{+0.05 \\ -0.04}$ \\ \hline
         $C_{\rm parallax}$ & $0.43\substack{+0.15 \\ -0.10}$ & $0.45\substack{+0.16 \\ -0.11}$ & $0.43\substack{+0.15 \\ -0.10}$ & $0.42\substack{+0.15 \\ -0.09}$ & $0.43\substack{+0.15 \\ -0.10}$ & $0.40\substack{+0.13 \\ -0.08}$ \\ \hline
         $\log_{10}(d_{\rm parallax}/{\rm kpc})$ & $0.8\substack{+0.6 \\ -0.3}$ & $0.7\substack{+0.6 \\ -0.3}$ & $0.8\substack{+0.6 \\ -0.3}$ & $0.8\substack{+0.6 \\ -0.3}$ & $0.8\substack{+0.6 \\ -0.3}$ & $0.9\substack{+0.6 \\ -0.4}$ \\ \hline
         $a_{E_{\rm cut}}$ & $0.18\substack{+0.05 \\ -0.05}$ & $0.21\substack{+0.05 \\ -0.05}$ & $0.19\substack{+0.05 \\ -0.05}$ & $0.19\substack{+0.05 \\ -0.05}$ & $-$ & $0.17\substack{+0.05 \\ -0.05}$ \\ \hline
         $b_{E_{\rm cut}}$ & $2.83\substack{+0.05 \\ -0.06}$ & $2.79\substack{+0.06 \\ -0.07}$ & $2.99\substack{+0.03 \\ -0.02}$ & $2.99\substack{+0.03 \\ -0.02}$ & $2.79\substack{+0.06 \\ -0.06}$ & $2.99\substack{+0.03 \\ -0.03}$ \\ \hline
         $\sigma_{E_{\rm cut}}$ & $0.23\substack{+0.02 \\ -0.02}$ & $0.23\substack{+0.02 \\ -0.02}$ & $0.22\substack{+0.02 \\ -0.02}$ & $0.22\substack{+0.02 \\ -0.02}$ & $0.25\substack{+0.02 \\ -0.02}$ & $0.23\substack{+0.02 \\ -0.02}$ \\ \hline
         $a_{\Gamma}$ & $0.41\substack{+0.08 \\ -0.08}$ & $0.43\substack{+0.08 \\ -0.08}$ & $0.43\substack{+0.07 \\ -0.08}$ & $0.43\substack{+0.07 \\ -0.08}$ & $-$ & $0.39\substack{+0.08 \\ -0.08}$ \\ \hline
         $b_{\Gamma}$ & $0.81\substack{+0.07 \\ -0.08}$ & $0.77\substack{+0.08 \\ -0.10}$ & $1.00\substack{+0.04 \\ -0.04}$ & $1.00\substack{+0.04 \\ -0.04}$ & $0.70\substack{+0.09 \\ -0.10}$ & $0.99\substack{+0.04 \\ -0.04}$ \\ \hline
         $\sigma_{\Gamma}$ & $0.36\substack{+0.04 \\ -0.03}$ & $0.36\substack{+0.04 \\ -0.03}$ & $0.34\substack{+0.03 \\ -0.03}$ & $0.34\substack{+0.03 \\ -0.03}$ & $0.42\substack{+0.04 \\ -0.03}$ & $0.35\substack{+0.03 \\ -0.03}$ \\ \hline
         $r_{\Gamma, E_{\rm cut}}$ & $0.75\substack{+0.05 \\ -0.06}$ & $0.77\substack{+0.05 \\ -0.06}$ & $0.73\substack{+0.06 \\ -0.07}$ & $0.73\substack{+0.06 \\ -0.07}$ & $0.79\substack{+0.04 \\ -0.05}$ & $0.75\substack{+0.05 \\ -0.06}$ \\ \hline
         $a_{\gamma}$ & $1.2\substack{+0.3 \\ -0.3}$ & $1.4\substack{+0.3 \\ -0.3}$ & $-$ & $-$ & $1.3\substack{+0.3 \\ -0.3}$ & $-$ \\ \hline
         $b_{\gamma}$ & $0.1\substack{+0.4 \\ -0.4}$ & $0.1\substack{+0.4 \\ -0.4}$ & $-$ & $-$ & $0.1\substack{+0.4 \\ -0.4}$ & $-$ \\ \hline
         $d_{\gamma}$ & $0.50\substack{+0.12 \\ -0.12}$ & $0.48\substack{+0.12 \\ -0.12}$ & $-$ & $-$ & $0.56\substack{+0.11 \\ -0.10}$ & $-$ \\ \hline
         $\alpha_{\gamma}$ & $-$ & $-$ & $0.74\substack{+0.11 \\ -0.10}$ & $-2.2\substack{+0.4 \\ -0.4}$ & $-$ & $-$ \\ \hline
         $\beta_{\gamma}$ & $-$ & $-$ & $-$ & $0.79\substack{+0.20 \\ -0.20}$ & $-$ & $-$ \\ \hline
         $\textrm{DTD } p(0 \textrm{ - } 2.8 \textrm{ Gyr})$ & $0.13\substack{+0.14 \\ -0.09}$ & $0.13\substack{+0.15 \\ -0.09}$ & $0.13\substack{+0.14 \\ -0.09}$ & $0.13\substack{+0.15 \\ -0.09}$ & $0.10\substack{+0.11 \\ -0.07}$ & $0.02\substack{+0.02 \\ -0.01}$ \\ \hline
         $\textrm{DTD } p(2.8 \textrm{ - } 5.5 \textrm{ Gyr})$ & $0.15\substack{+0.18 \\ -0.11}$ & $0.15\substack{+0.19 \\ -0.11}$ & $0.15\substack{+0.18 \\ -0.11}$ & $0.15\substack{+0.19 \\ -0.11}$ & $0.12\substack{+0.16 \\ -0.09}$ & $0.02\substack{+0.03 \\ -0.02}$ \\ \hline
         $\textrm{DTD } p(5.5 \textrm{ - } 8.3 \textrm{ Gyr})$ & $0.29\substack{+0.20 \\ -0.17}$ & $0.26\substack{+0.18 \\ -0.16}$ & $0.29\substack{+0.20 \\ -0.17}$ & $0.29\substack{+0.20 \\ -0.17}$ & $0.25\substack{+0.18 \\ -0.14}$ & $0.09\substack{+0.06 \\ -0.05}$ \\ \hline
         $\textrm{DTD } p(8.3 \textrm{ - } 11.1 \textrm{ Gyr})$ & $0.14\substack{+0.16 \\ -0.10}$ & $0.11\substack{+0.14 \\ -0.08}$ & $0.15\substack{+0.17 \\ -0.10}$ & $0.14\substack{+0.16 \\ -0.10}$ & $0.14\substack{+0.17 \\ -0.10}$ & $0.06\substack{+0.11 \\ -0.05}$ \\ \hline
         $\textrm{DTD } p(11.1 \textrm{ - } 13.8 \textrm{ Gyr})$ & $0.14\substack{+0.22 \\ -0.11}$ & $0.22\substack{+0.20 \\ -0.15}$ & $0.13\substack{+0.21 \\ -0.10}$ & $0.14\substack{+0.21 \\ -0.11}$ & $0.27\substack{+0.24 \\ -0.19}$ & $0.78\substack{+0.08 \\ -0.14}$ \\ \hline
         $\log_{10}(P_{i, \textrm{ med}}/\textrm{s})$ & $-2.61\substack{+0.05 \\ -0.04}$ & $-2.60\substack{+0.05 \\ -0.04}$ & $-2.61\substack{+0.05 \\ -0.04}$ & $-2.61\substack{+0.05 \\ -0.04}$ & $-2.63\substack{+0.04 \\ -0.03}$ & $-2.69\substack{+0.03 \\ -0.03}$ \\ \hline
         $\sigma_{P_{i}}$ & $0.13\substack{+0.02 \\ -0.02}$ & $0.13\substack{+0.02 \\ -0.02}$ & $0.13\substack{+0.02 \\ -0.02}$ & $0.12\substack{+0.02 \\ -0.02}$ & $0.12\substack{+0.02 \\ -0.02}$ & $0.13\substack{+0.02 \\ -0.02}$ \\ \hline
         $\log_{10}(B_{\rm med}/\textrm{G})$ & $8.21\substack{+0.05 \\ -0.06}$ & $8.21\substack{+0.06 \\ -0.06}$ & $8.21\substack{+0.03 \\ -0.04}$ & $8.20\substack{+0.06 \\ -0.06}$ & $8.21\substack{+0.05 \\ -0.05}$ & $8.25\substack{+0.02 \\ -0.02}$ \\ \hline
         $\sigma_{B}$ & $0.21\substack{+0.03 \\ -0.02}$ & $0.21\substack{+0.03 \\ -0.02}$ & $0.22\substack{+0.03 \\ -0.03}$ & $0.22\substack{+0.03 \\ -0.02}$ & $0.21\substack{+0.03 \\ -0.03}$ & $0.19\substack{+0.02 \\ -0.02}$ \\ \hline
         $\log_{10}(\eta_{\rm med})$ & $12\substack{+5 \\ -5}$ & $12\substack{+5 \\ -5}$ & $7\substack{+4 \\ -4}$ & $-4\substack{+4 \\ -4}$ & $9\substack{+4 \\ -4}$ & $32.17\substack{+0.23 \\ -0.31}$ \\ \hline
         $\sigma_{\eta}$ & $0.52\substack{+0.06 \\ -0.05}$ & $0.53\substack{+0.06 \\ -0.05}$ & $0.58\substack{+0.06 \\ -0.05}$ & $0.58\substack{+0.07 \\ -0.06}$ & $0.51\substack{+0.06 \\ -0.05}$ & $0.72\substack{+0.08 \\ -0.06}$ \\ \hline
         $\sigma_{v}$ (km s$^{-1}$) & $77\substack{+6 \\ -6}$ & $76\substack{+6 \\ -5}$ & $77\substack{+6 \\ -6}$ & $77\substack{+6 \\ -6}$ & $77\substack{+6 \\ -6}$ & $78\substack{+6 \\ -6}$ \\ \hline

    \caption[Medians and $68\%$ intervals for model parameters of a selection of MSP population models.]{ Medians and $68\%$ confidence intervals for a selection of different models of the Galactic MSP population. See Table~\ref{tab:WAIC} for a description of the model labels.    }
    \label{tab:mcmc_results_parameters}
\end{longtable}
\end{landscape}

\section{Discussion}
\label{sec:discussion}

\begin{figure}
    \centering
    \subfigure{\centering\includegraphics[width=0.49\linewidth]{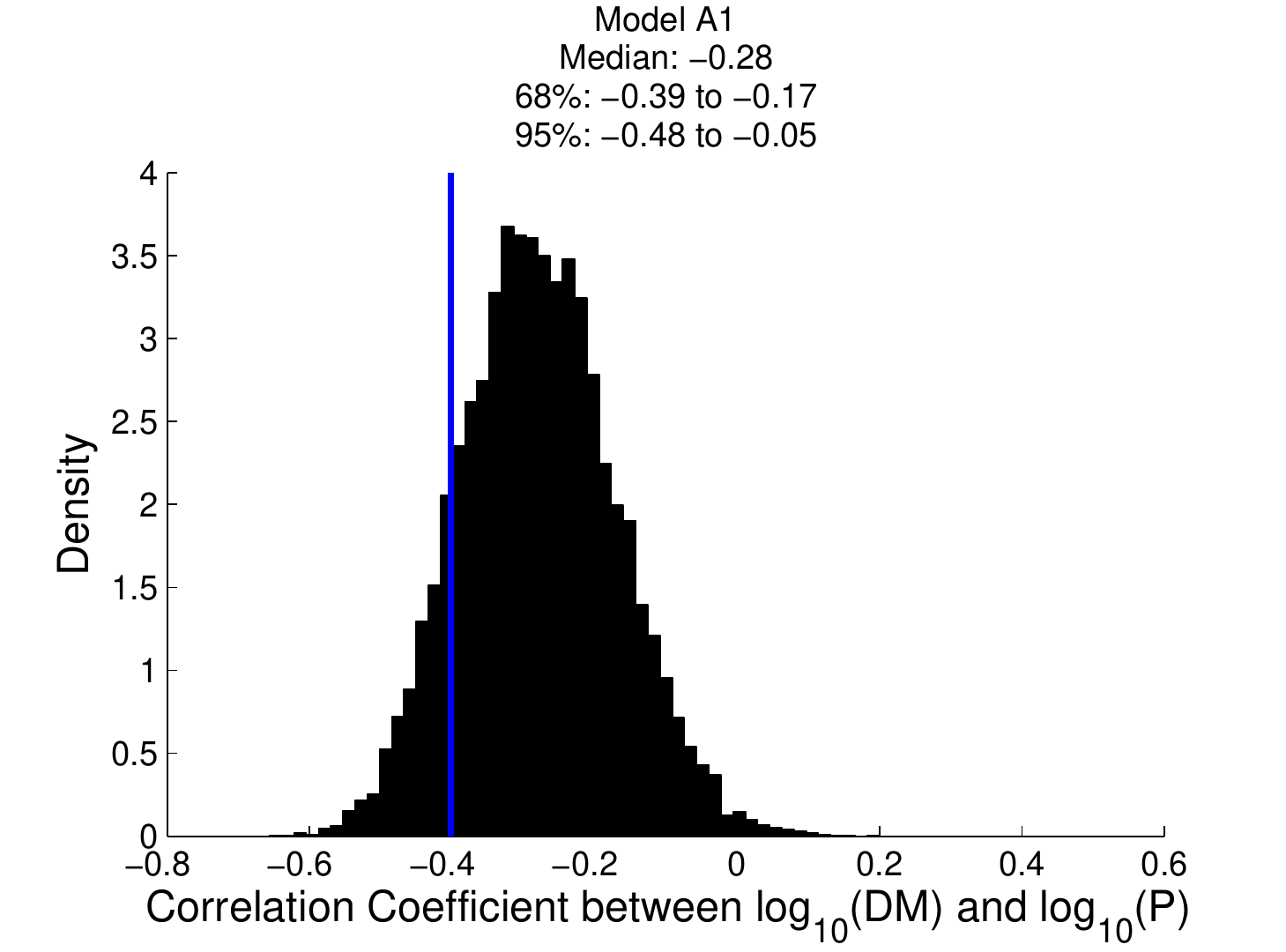}}
    \subfigure{\centering\includegraphics[width=0.49\linewidth]{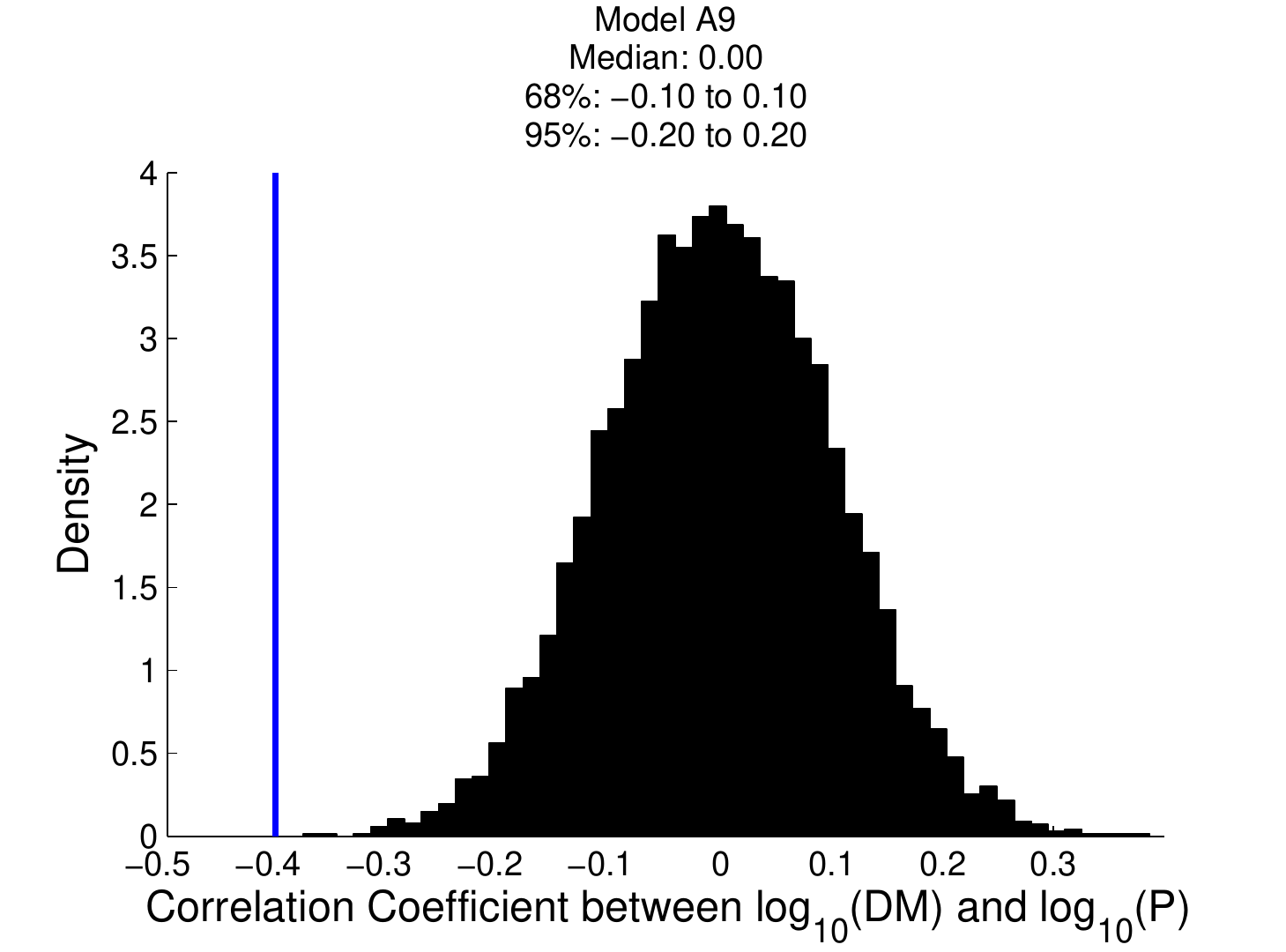}}
    \caption[Comparison of correlation coefficients between $\log_{10}(P)$ and $\log_{10}({\rm DM})$ for resolved MSP simulated and real data for  Model A1 ($L = \eta E_{\rm cut}^{a_{\gamma}} B^{b_{\gamma}} \dot{E}^{d_{\gamma}}$) and Model A9 ($L=\eta$).]{Comparison of correlation coefficients between $\log_{10}(P)$ and $\log_{10}({\rm DM})$ for resolved MSP simulated and real data for  Model A1 
    ($L = \eta E_{\rm cut}^{a_{\gamma}} B^{b_{\gamma}} \dot{E}^{d_{\gamma}}$) 
    on the left and Model A9 ($L=\eta$) on the right. The blue line shows the correlation coefficient for the resolved MSP data at -0.40. The posterior predictive p-values are $0.13$ for A1, and $0$ for A9. }
    \label{fig:E_cut_B_E_dot_vs_efficiency_only_log_DM_log_P_correlation_coefficients}
\end{figure}

Some of our luminosity functions are nested. We can go from Model A1 with 
$L = \eta E_{\rm cut}^{a_{\gamma}} B^{b_{\gamma}} \dot{E}^{d_{\gamma}}$ to Model A9 with $L=\eta$ by setting $a_{\gamma}=b_{\gamma}=d_{\gamma}=0$. As can be seen from the Model A1 fit in Fig.~\ref{fig:E_cut_B_E_dot_MCMC_Params_1} and Table \ref{tab:mcmc_results_parameters}, the data prefer the $a_\gamma=1.2\pm 0.3$ and $d_\gamma=0.5\pm0.1$ parameters to be larger than zero at high significance.  This is consistent with the results of Table~\ref{tab:WAIC} where Model~A9 has a $\Delta{\rm WAIC}=42.2$ relative to
Model~A1. Comparing the posterior predictive corner plots 
 for Model~A1 in 
 Fig.~\ref{fig:E_cut_B_E_dot_MSP_Params} and Model~A9 in Fig.~\ref{fig:efficiency_only_MSP_Params} we can see that Model~A9  does not capture some of the correlations in the data, 
 particularly those between distance (shown indirectly in the form of dispersion measure) and the spectral parameter
 $E_{\rm cut}$ 
 as well as between distance and period. A comparison of correlation coefficients between the real data and simulated data is shown in Fig.~\ref{fig:E_cut_B_E_dot_vs_efficiency_only_log_DM_log_P_correlation_coefficients}. 
 More detail can be seen by comparing Tables~\ref{tab:model_a1_pvals} and \ref{tab:model_a9_pvals} which list the posterior predictive p-values for the correlations between observables for Models A1 and A9 respectively.
Following  the recommendation given in Section 6.3 of ref.~\cite{Gelman2013}, we consider posterior predictive p-values below 1\% and above 99\% to be of concern.
 As can be seen, the posterior predictive p-values for Model A1 are all comfortably within the 1\% to 99\% interval. In contrast to this, seven correlation coefficients for Model A9 are outside this interval.
 The correlation coefficient 
  between $\log_{10}(P)$ and $\log_{10}({\rm DM})$ and also the one between $\log_{10}(E_{\rm cut})$ and $\log_{10}({\rm DM})$ are both particularly discrepant with the data.
These logarithmic correlations are  associated with the flux threshold which implies that, the 
more distant the MSP, the larger the intrinsic luminosity it is required to have in order to have a significant probability of being resolved. 
The point distinguishing Model A1 from Model A9 in  regards to the relationship between distance and $E_{\rm cut}$ arises because Model A1 has a significant positive $a_{\gamma}=1.2\pm 0.3$ in Eq.~\ref{eq:Model1}, while in Model A9 we have set $a_{\gamma}=0$.
{\textcolor{black}{Regarding the logarithmic correlation between distance  and $P$, it follows from
the presence of $P$ in the denominator of Eq.~\ref{eq:Edot} that 
 the significantly positive $d_{\gamma}=0.5\pm0.1$ in Eq.~\ref{eq:Model1} leads to the negative correlation with the distance measures in Model A1, while, again, in Model A9 we have $d_\gamma=0$ by construction.}}

\begin{table}
\centering
    \begin{tabular}{|c||c|c|c|c|c|c|c|}
         \cline{1-2}
        $\log_{10}(\omega)$ & $0.417$ \\ \cline{1-3}
        $\log_{10}(F)$ & $0.627$ & $0.159$ \\ \cline{1-4}
        $\log_{10}(P)$  & $0.126$ & $0.948$ & $0.752$ \\ \cline{1-5}
        $\log_{10}(\dot{P})$ & $0.911$ & $0.245$ & $0.451$ & $0.159$ \\ \cline{1-6}
        $\mu$ & $0.034$   & $0.398$ & $0.294$ & $0.938$ & $0.338$ \\ \cline{1-7}
        $\log_{10}(E_{\rm cut})$ & $0.789$ & $0.039$ & $0.437$ & $0.527$ & $0.382$ & $0.477$ \\ \hline
        $\Gamma$   & $0.550$ & $0.163$ & $0.773$ & $0.583$ & $0.352$ & $0.707$ & $0.612$ \\ \hline \hline
       & $\log_{10}(\textrm{DM})$ & $\log_{10}(\omega)$ & $\log_{10}(F)$ & $\log_{10}(P)$ & $\log_{10}(\dot{P})$ & $\mu$ & $\log_{10}(E_{\rm cut})$ \\ \hline
    \end{tabular}
    \caption[Correlation coefficient posterior predictive p-values between observables for model A1.]{ Posterior predictive p-values for the correlation coefficients between observables for  Model A1 which has $L = \eta E_{\rm cut}^{a_{\gamma}} B^{b_{\gamma}} \dot{E}^{d_{\gamma}}$. }
    \label{tab:model_a1_pvals}
\end{table}

\begin{table}
\centering
\begin{tabular}{|c||c|c|c|c|c|ll} 
\cline{1-2}
 $\log_{10}(\omega)$       & $0.426$                   & \multicolumn{1}{l}{} & \multicolumn{1}{l}{} & \multicolumn{1}{l}{} & \multicolumn{1}{l}{}  &                               &                                                 \\ 
\cline{1-3}
 $\log_{10}(F)$            & $0.615$                   & $0.158$              & \multicolumn{1}{l}{} & \multicolumn{1}{l}{} & \multicolumn{1}{l}{}  &                               &                                                 \\ 
\cline{1-4}
 $\log_{10}(P)$            & $0.138$                   & $0.941$              & $0.743$              & \multicolumn{1}{l}{} & \multicolumn{1}{l}{}  &                               &                                                 \\ 
\cline{1-5}
 $\log_{10}(\dot{P})$      & $0.891$                   & $0.265$              & $0.431$              & $0.145$              & \multicolumn{1}{l}{}  &                               &                                                 \\ 
\cline{1-6}
 $\mu$                     & $0.430$                   & $0.395$              & $0.282$              & $0.932$              & $0.366$               &                               &                                                 \\ 
\cline{1-7}
 $\log_{10}(E_{\rm cut})$  & $\textcolor{red}{0.992}$                   & $\textcolor{red}{0.007}$              & $0.697$              & $0.609$              & $0.216$               & \multicolumn{1}{c|}{$0.175$ } &                                                 \\ 
\hline
 $\Gamma$                  & $0.852$                   & $0.069$              & $0.879$              & $0.667$              & $0.223$               & \multicolumn{1}{c|}{$0.515$ } & \multicolumn{1}{c|}{$0.625$ }                   \\ 
\hhline{|=::=======|}
                           & $\log_{10}(\textrm{DM})$  & $\log_{10}(\omega)$  & $\log_{10}(F)$       & $\log_{10}(P)$       & $\log_{10}(\dot{P})$  & \multicolumn{1}{c|}{$\mu$ }   & \multicolumn{1}{c|}{$\log_{10}(E_{\rm cut})$ }  \\
\hline
\end{tabular}
\caption[Correlation coefficient posterior predictive p-values between observables for model A6.]{ Posterior predictive p-values for the correlation coefficients between observables for  Model A6 which has $L = \eta P^{\alpha_{\gamma}} \dot{P}^{\beta_{\gamma}}$. Values  greater than 0.99 and less than 0.01 are marked in red.  }
\label{tab:model_a6_pvals}
\end{table}

\begin{table}
\centering
\begin{tabular}{|c||c|c|c|c|c|ll}
\cline{1-2}
 $\log_{10}(\omega)$       & $0.384$                   & \multicolumn{1}{l}{} & \multicolumn{1}{l}{} & \multicolumn{1}{l}{}          & \multicolumn{1}{l}{}  &                               &                                                 \\ 
\cline{1-3}
 $\log_{10}(F)$            & $0.719$                   & $0.120$              & \multicolumn{1}{l}{} & \multicolumn{1}{l}{}          & \multicolumn{1}{l}{}  &                               &                                                 \\ 
\cline{1-4}
 $\log_{10}(P)$            & $0.057$                   & $0.965$              & $0.646$              & \multicolumn{1}{l}{}          & \multicolumn{1}{l}{}  &                               &                                                 \\ 
\cline{1-5}
 $\log_{10}(\dot{P})$      & $0.938$                   & $0.225$              & $0.502$              & $0.157$                       & \multicolumn{1}{l}{}  &                               &                                                 \\ 
\cline{1-6}
 $\mu$                     & $0.412$                   & $0.415$              & $0.244$              & $0.963$                       & $0.316$               &                               &                                                 \\ 
\cline{1-7}
 $\log_{10}(E_{\rm cut})$  & $0.942$                   & $0.017$              & $0.583$              & $\textcolor{red}{0.004}$                       & $0.819$               & \multicolumn{1}{c|}{$0.339$ } &                                                 \\ 
\hline
 $\Gamma$                  & $0.889$                   & $0.059$              & $0.906$              & $\bf \textcolor{red}{0.000}$  & $0.894$               & \multicolumn{1}{c|}{$0.509$ } & \multicolumn{1}{c|}{$0.661$ }                   \\ 
\hhline{|=::=======|}
                           & $\log_{10}(\textrm{DM})$  & $\log_{10}(\omega)$  & $\log_{10}(F)$       & $\log_{10}(P)$                & $\log_{10}(\dot{P})$  & \multicolumn{1}{c|}{$\mu$ }   & \multicolumn{1}{c|}{$\log_{10}(E_{\rm cut})$ }  \\
\hline
\end{tabular}
\caption[Correlation coefficient posterior predictive p-values between observables for model A7.]{
Same as Table~\ref{tab:model_a6_pvals} except for
  Model A7 which has $L = \eta E_{\rm cut}^{a_{\gamma}} B^{b_{\gamma}} \dot{E}^{d_{\gamma}}$, $a_{E_{\rm cut}} = a_{\Gamma} = 0$.}
\label{tab:model_a7_pvals}
\end{table}

\begin{table}
\centering
\begin{tabular}{|c||c|c|c|c|c|ll}
\cline{1-2}
 $\log_{10}(\omega)$       & $0.422$                         & \multicolumn{1}{l}{} & \multicolumn{1}{l}{} & \multicolumn{1}{l}{} & \multicolumn{1}{l}{}  &                               &                                                 \\ 
\cline{1-3}
 $\log_{10}(F)$            & $0.670$                         & $0.141$              & \multicolumn{1}{l}{} & \multicolumn{1}{l}{} & \multicolumn{1}{l}{}  &                               &                                                 \\ 
\cline{1-4}
 $\log_{10}(P)$            & $\bf \textcolor{red}{0.000}$    & $\textcolor{red}{0.998}$              & $0.289$              & \multicolumn{1}{l}{} & \multicolumn{1}{l}{}  &                               &                                                 \\ 
\cline{1-5}
 $\log_{10}(\dot{P})$      & $\textcolor{red}{0.997}$                         & $0.104$              & $0.697$              & $0.140$              & \multicolumn{1}{l}{}  &                               &                                                 \\ 
\cline{1-6}
 $\mu$                     & $0.424$                         & $0.396$              & $0.275$              & $\textcolor{red}{0.998}$              & $0.155$               &                               &                                                 \\ 
\cline{1-7}
 $\log_{10}(E_{\rm cut})$  & ${\bf \textcolor{red}{1.000}}$  & $\textcolor{red}{0.002}$              & $0.869$              & $0.541$              & $0.316$               & \multicolumn{1}{c|}{$0.053$ } &                                                 \\ 
\hline
 $\Gamma$                  & $\textcolor{red}{0.998}$                         & $0.012$              & $0.978$              & $0.621$              & $0.354$               & \multicolumn{1}{c|}{$0.192$ } & \multicolumn{1}{c|}{$0.631$ }                   \\ 
\hhline{|=::=======|}
                           & $\log_{10}(\textrm{DM})$        & $\log_{10}(\omega)$  & $\log_{10}(F)$       & $\log_{10}(P)$       & $\log_{10}(\dot{P})$  & \multicolumn{1}{c|}{$\mu$ }   & \multicolumn{1}{c|}{$\log_{10}(E_{\rm cut})$ }  \\
\hline
\end{tabular}
\caption[Correlation coefficient posterior predictive p-values between observables for model A9.]{Same as Table~\ref{tab:model_a6_pvals} except for  Model A9 which has $L=\eta$.}
\label{tab:model_a9_pvals}
\end{table}

Model A6 is related to model A1 as follows from Eqs.~\ref{eq:Edot} and \ref{eq:magnetic_field_strength} which
demonstrate
that we can go from the luminosity function in Model~A1 to the luminosity function in Model~A6 by setting $b_{\gamma }=\frac{\alpha _{\gamma }}{2}+\frac{3 \beta _{\gamma }}{2}$ and $d_{\gamma }=\frac{\beta _{\gamma }}{4}-\frac{\alpha _{\gamma }}{4}$. However, Model~A6 does not have a dependence on $E_{\rm cut}$ and therefore assumes $a_\gamma=0$. As can be seen in Table~\ref{tab:mcmc_results_parameters}, $a_\gamma=1.2\pm0.3$ is significantly positive for Model~A1 and so  Model~A1 is preferred over Model~A6.
This is consistent with Model~A6 having a $\Delta{\rm WAIC=}11.0$ relative to Model A1. As can be seen in Table~\ref{tab:model_a6_pvals} the problem with Model A6 is that it does not predict the observed logarithmic correlation between distance and $E_{\rm cut}$.

 Model A7 is the same as Model A1 except that we set $a_{E_{\rm cut}} = a_{\Gamma} = 0$. 
 We see from Table~\ref{tab:mcmc_results_parameters} that this choice for Model A7 is disfavoured as  we found that Model A1 had a significantly positive  $a_\Gamma=0.41\pm0.08$
 for Eq.~\ref{eq:Gamma}. This preference for Model A1 over A7 is confirmed in Table~\ref{tab:WAIC} where it can be seen that Model A7 has a $\Delta{\rm WAIC}=19.4$ relative to Model A1.
We can see how Model A7  produced a worse fit to the data by comparing Figs.~\ref{fig:E_cut_B_E_dot_MSP_Params} and \ref{fig:E_cut_B_E_dot_NoSpectrumDependenceOnEdot_MSP_Params}: in the latter case where $a_{E_{\rm cut}} = a_{\Gamma} = 0$, the relationship between spectral index and period 
has clearly disappeared. 
This is confirmed in Table~\ref{tab:model_a7_pvals} where two of the correlation coefficients are of concern and, in particular, the correlation coefficient between $\Gamma$ and $\log_{\rm 10}(P)$ has a posterior predictive p-value of 0.000.

Our best fitting model of the Galactic MSP population was Model A1 for which $L = \eta E_{\rm cut}^{a_{\gamma}} B^{b_{\gamma}} \dot{E}^{d_{\gamma}}$. From inspection of Figs.~\ref{fig:E_cut_B_E_dot_posterior_predictive_plots} and \ref{fig:E_cut_B_E_dot_MSP_Params} it is evident that this model generally provided a good fit to the resolved MSP data. This is also confirmed in Table~\ref{tab:model_a1_pvals} where all the posterior predictive p--values are within the 1\% to 99\% range.
We find $a_{\gamma} = 1.2\substack{+0.3 \\ -0.3}$, $b_{\gamma} = 0.1\substack{+0.4 \\ -0.4}$ and $d_{\gamma} = 0.5\substack{+0.1 \\ -0.1}$. These results are consistent with those obtained by 
Kalapotharakos et al.~\cite{Kalapotharakos_2019}. These authors performed a least squares fit to both the MSPs and young pulsars in the Second Fermi Pulsar Catalog \citep{TheFermi-LAT:2013ssa} and find $a_{\gamma} = 1.12 \pm 0.24$, $b_{\gamma} = 0.17 \pm 0.05$ and $d_{\gamma} = 0.41 \pm 0.08$. Kalapotharakos et al.~\cite{Kalapotharakos_2019} point out that their results are consistent with predicted values of $a_{\gamma} = 4/3$, $b_{\gamma} = 1/6$ and $d_{\gamma} = 5/12$ in the case that curvature radiation is the source of gamma-ray emission.
This stands in contrast to the case of synchrotron radiation for which $a_{\gamma} = 1$, $b_{\gamma} = 0$ and $d_{\gamma} = 1$ is expected. Our posterior distributions for $a_{\gamma}$ and $b_{\gamma}$ are consistent with both cases to within $2 \sigma$, but our $d_{\gamma}$ is inconsistent with synchrotron radiation. One major difference between the work of Kalapotharakos et al.~\cite{Kalapotharakos_2019} and ours is that we use many more MSPs, but no young pulsars. Overall we have a similar total number of pulsars. However, because of the Shklovskii effect, 
the intrinsic period derivative $\dot{P}_{\rm int}$ is poorly determined relative to the case of young pulsars.
Also, $B$ and $\dot{E}$ depend on $\dot{P}_{\rm int}$ through Eqs.~\ref{eq:magnetic_field_strength} and \ref{eq:Edot} respectively. It follows that some MSPs with smaller values of $\dot{P}$
may have poorly-determined period derivatives relative to the case presented by young pulsars.

\begin{figure}
    \centering
    \subfigure{\centering\includegraphics[width=0.49\linewidth]{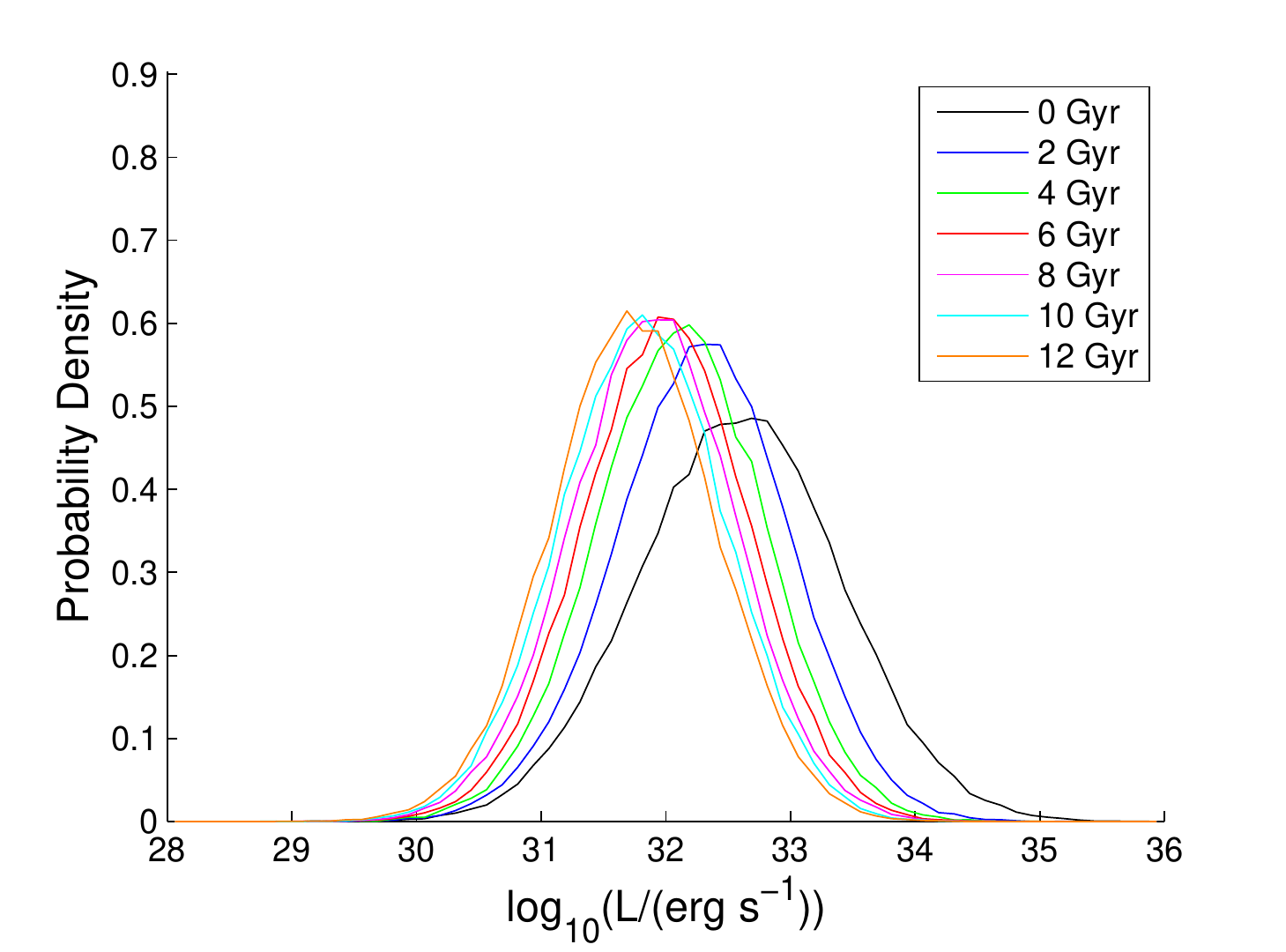}}
    \subfigure{\centering\includegraphics[width=0.49\linewidth]{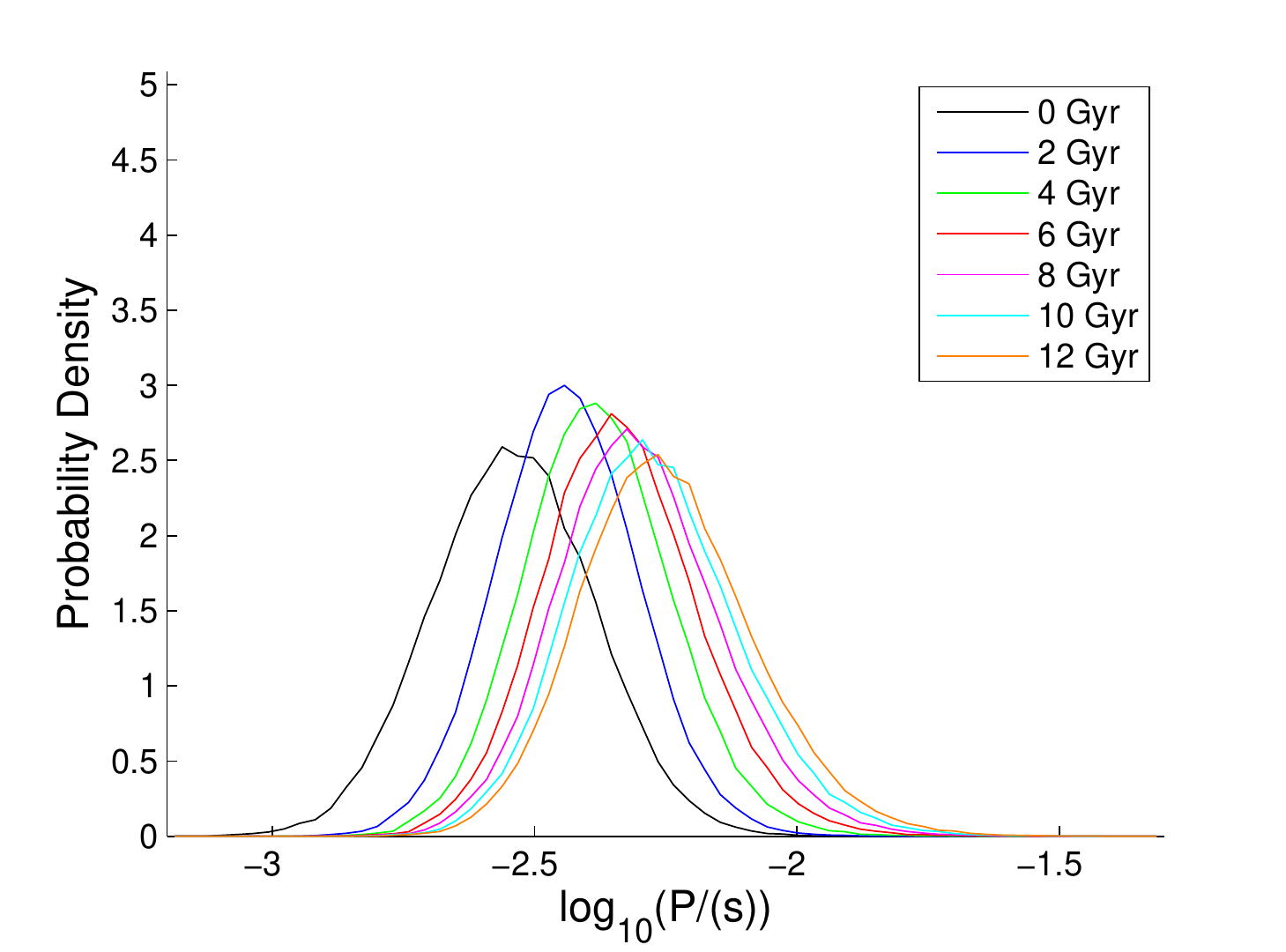}}
    \caption[Evolution of the luminosity and period distributions with time for Model A1.]{Evolution of the luminosity and period distributions with time for Model A1 which has $L = \eta E_{\rm cut}^{a_{\gamma}} B^{b_{\gamma}} \dot{E}^{d_{\gamma}}$. The legend gives the age of the MSPs in the corresponding distribution. These figures were made by selecting randomly from the highest likelihood parameter sets in our eight Markov chains for this model.
    \label{fig:E_cut_B_E_dot_luminosity_period_evolution}}
\end{figure}

In Model A1, which has
$L = \eta E_{\rm cut}^{a_{\gamma}} B^{b_{\gamma}} \dot{E}^{d_{\gamma}}$,
relatively young members of the underlying population of MSPs are more likely to be resolved as they are brighter. This follows as $\dot{E}$, and therefore luminosity, decreases with age. As the magnetic field strength is assumed constant for each MSP over time, it can be seen from Eqs.~\ref{eq:magnetic_field_strength} and \ref{eq:Edot} that $\dot{E} \propto P^{-4}$. We show in Fig.~\ref{fig:E_cut_B_E_dot_luminosity_period_evolution} an example of how the $\log_{10}(L)$ and $\log_{10}(P)$ probability density functions evolve with age. 

As can be seen from Fig.~\ref{fig:E_cut_B_E_dot_MCMC_DTD_bin_Params},  %
the DTD peaks in the central $5.5$ to $8.3$ Gyr bin producing an age distribution that tends to plateau starting around $5$ Gyr ago as shown in Fig.~\ref{fig:E_cut_B_E_dot_age_distribution}.
It can be seen in Table \ref{tab:WAIC} that
there is no significant difference in the WAIC for the DTD versus the uniform age distribution case.
As can be seen in Fig.~\ref{fig:E_cut_B_E_dot_age_distribution}, the fitted DTD produces an age distribution that is similar to the uniform case. For the uniform age distribution, the probability density is $0.1$ for ages less than $10$ Gyr, and $0$ elsewhere. 

The models that were significantly worse when the GCE was included remained significantly worse when it was not. The model for which the luminosity obeys $L = \eta E_{\rm cut}^{a_{\gamma}} B^{b_{\gamma}} \dot{E}^{d_{\gamma}}$ remained the best model whether the GCE was included or not, i.e.,\ Models A1 and A2 were the best models when the GCE was included and Models B1 and B2 were the best models when the GCE was not included. As can be seen in Table \ref{tab:mcmc_results_parameters} the posterior distributions of the parameters are similar for the GCE and No GCE cases. All the fitted parameters of Model B2 are consistent with those of Model A1, which is the same except the former lacks a bulge MSP population and GCE.

As can be seen from Fig.~\ref{fig:E_cut_B_E_dot_dtd_vs_uniform_boxy_bulge_spectra}, there is a small difference in the GCE spectrum for the Model A1 and A2. This is due to the spectral dependence on $\dot{E}$ in Eqs.~\ref{eq:Ecut} and \ref{eq:Gamma} and also the different star formation rates for the bulge and disk which are illustrated in Fig.~\ref{fig:sfr}. However, as can be seen in Fig.~\ref{fig:E_cut_B_E_dot_dtd_vs_uniform_boxy_bulge_spectra}, the differences in the predicted spectrum are negligible in comparison to the model prediction uncertainties. 
As can be seen in the top right panel of Fig.~\ref{fig:E_cut_B_E_dot_luminosity_distribution} the boxy bulge MSPs do have a median luminosity function that is less bright than the disk MSPs. However, the 95\% interval band encompasses zero which is the no difference case.
This shows that with the current levels of uncertainty the differences in the properties of the bulge and disk MSPs is not significant.

\begin{figure}
    \centering
    \includegraphics[width=0.99\linewidth]{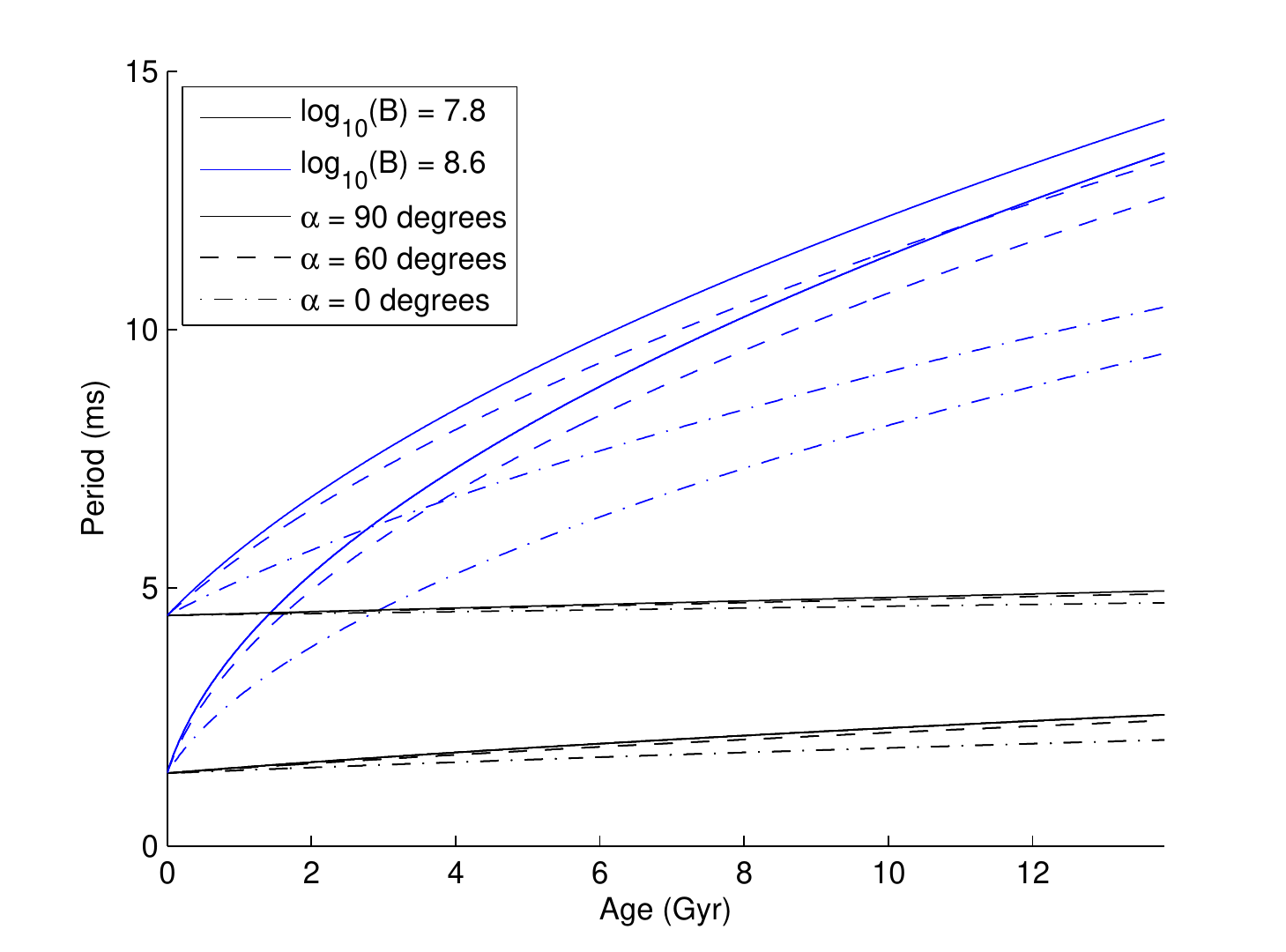}
    \caption[Evolution of period with time for various magnetic field strengths $B$, magnetic field axis angles $\alpha$ and initial periods $P_I$.]{Evolution of period with time for various magnetic field strengths $B$, magnetic field axis angles $\alpha$ and initial periods $P_I$. The two $P_I$ values used were $10^{-2.85}$ s and $10^{-2.35}$ s.}
    \label{fig:period_evolution}
\end{figure}

In Fig.~\ref{fig:period_evolution} we show the evolution of period with time (Eq.\ \ref{eq:current_period}) for MSPs with different properties. This indicates where the constraints on our models of the initial period and age distribution come from. For a resolved MSP with a weak magnetic field, the current period will be near the initial period even if the MSP is old, so these MSPs should be approximately distributed like the initial period distribution. This is a consequence of the fact that an MSP cannot be older than the universe. On the other hand, MSPs with strong magnetic fields will quickly, within a couple of Gyr, move out of the initial period distribution. This (relatively) rapid evolution of period means the distributions of $P_I$ and $\alpha$ produce (for a given $P$ and $B$) a range of possible ages for each MSP.

In Fig.~\ref{fig:E_cut_B_E_dot_N_MSPs} it is shown that, at 68\% confidence level, for
Model A1
between around 23000 and 54000 MSPs
are needed in the bulge to produce the observed GCE. 
This is consistent with Gonthier et al.~\cite{Gonthier2018} in which they find, though with a different bulge density model and GCE spectrum, 34,200 MSPs are needed with 11,500 in the region of interest 
associated with
the Gordon et al.~\cite{Gordon:2013vta} GCE spectrum. Calore et al.~\cite{Calore2016}, assuming the ratio of gamma-ray flux to number of radio-bright MSPs in globular clusters is the same as that of the bulge, estimate $\left(2.7 \pm 0.2 \right) \times 10^3$ radio-bright Galactic bulge MSPs, defined as those with a flux density at $1.4$ GHz of $\geq 10$ $\mu$Jy, and a total number of $\left(9.2 \pm 3.1 \right) \times 10^3$ assuming their adopted radio luminosity function.
In Ploeg et al.~\cite{Ploeg:2017vai} it was found that around $(4.0\pm 0.9)\times10^4$ MSPs with $L > 10^{32} \textrm{ erg s}^{-1}$ were needed to produce the GCE; here, as shown in Fig.~\ref{fig:E_cut_B_E_dot_N_MSPs_greater_than_log_L_32}, the required number is between about 13000 and 26000 at 95\% confidence interval. 
Note the difference between these two estimates is not statistically significant; the slight discrepancy is likely due to the more accurate bulge geometric model, luminosity function, and GCE spectrum used in the current study.

A number of studies have fitted luminosity functions to resolved MSPs. For our fit of an independent log-normal luminosity distribution, Model A9 where $L = \eta$, we found a median of $\log_{10}(L_{\rm med}) = 32.17\substack{+0.23 \\ -0.31}$ and a standard deviation in $\log_{10}(L)$ of $\sigma_L = 0.72\substack{+0.08 \\ -0.06}$. Similarly, fitting log-normal luminosity distributions to the data, Bartels et al.~\cite{Bartels2018} found $\log_{10}(L_{\rm med}) = 32.44\substack{+0.19 \\ -0.22}$ and $\sigma_L = 0.68\substack{+0.07 \\ -0.06}$, and Hooper and Mohlabeng \cite{Hooper:2015jlu} found best fit parameters of $\log_{10}(L_{\rm med}) = 32.67$ and $\sigma_L = 0.61$. For Model A6, where $L \propto P^{\alpha_{\gamma}} \dot{P}^{\beta_{\gamma}}$, we found $\alpha_{\gamma} = -2.2\substack{+0.4 \\ -0.4}$ and $\beta_{\gamma} = 0.79\substack{+0.20 \\ -0.20}$, consistent with Gonthier et al.~\cite{Gonthier2018} who found $(\alpha_{\gamma},\beta_{\gamma})$ of $(-2.3\pm0.3,0.8\pm0.2)$, $(-2.1\pm0.3,0.7\pm0.2)$ and $(-2.5\pm0.3,0.9\pm0.2)$ for three different models of gamma-ray emission geometry.

\begin{figure}
    \centering
    \subfigure{\centering\includegraphics[width=0.49\linewidth]{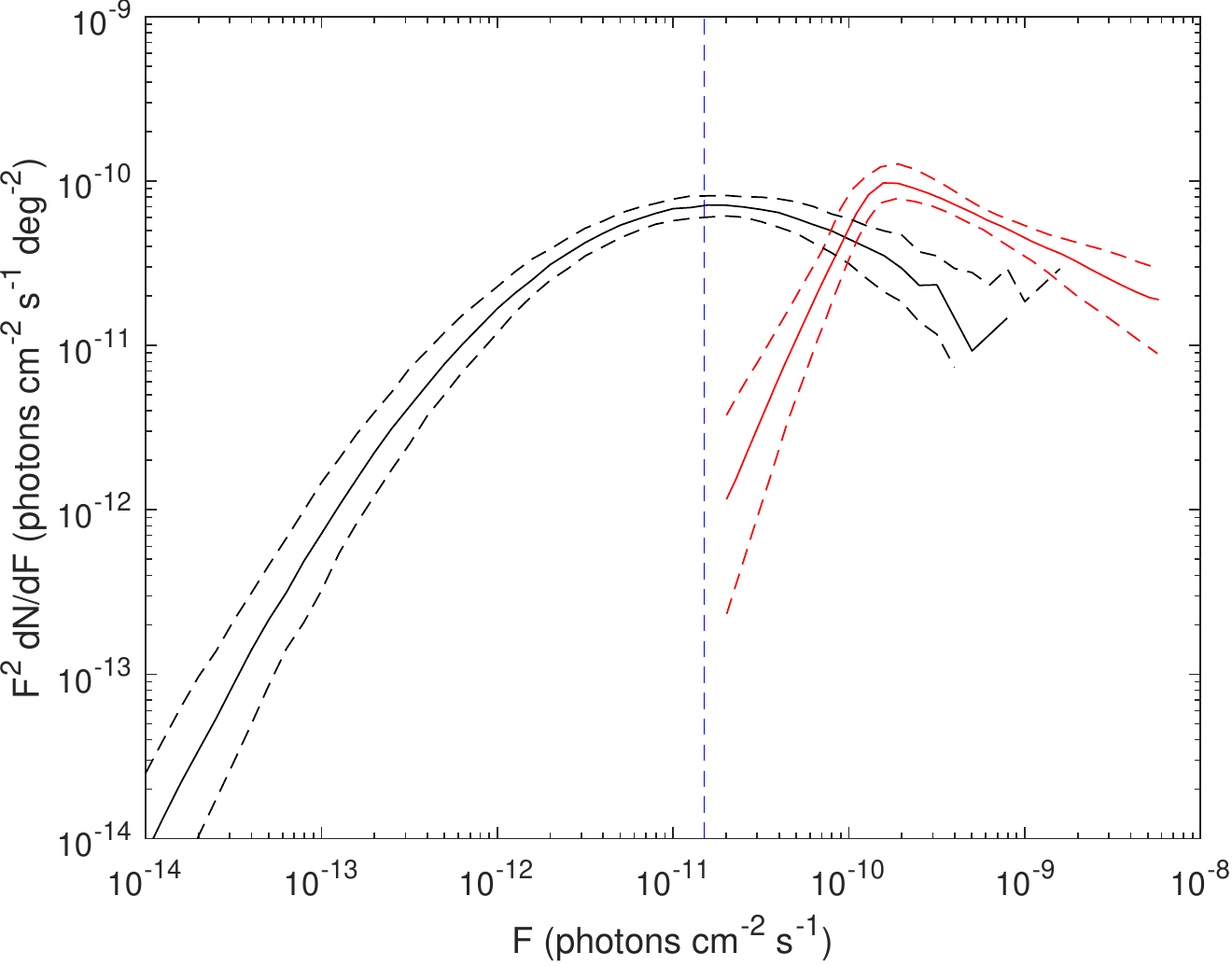}}
    \subfigure{\centering\includegraphics[width=0.49\linewidth]{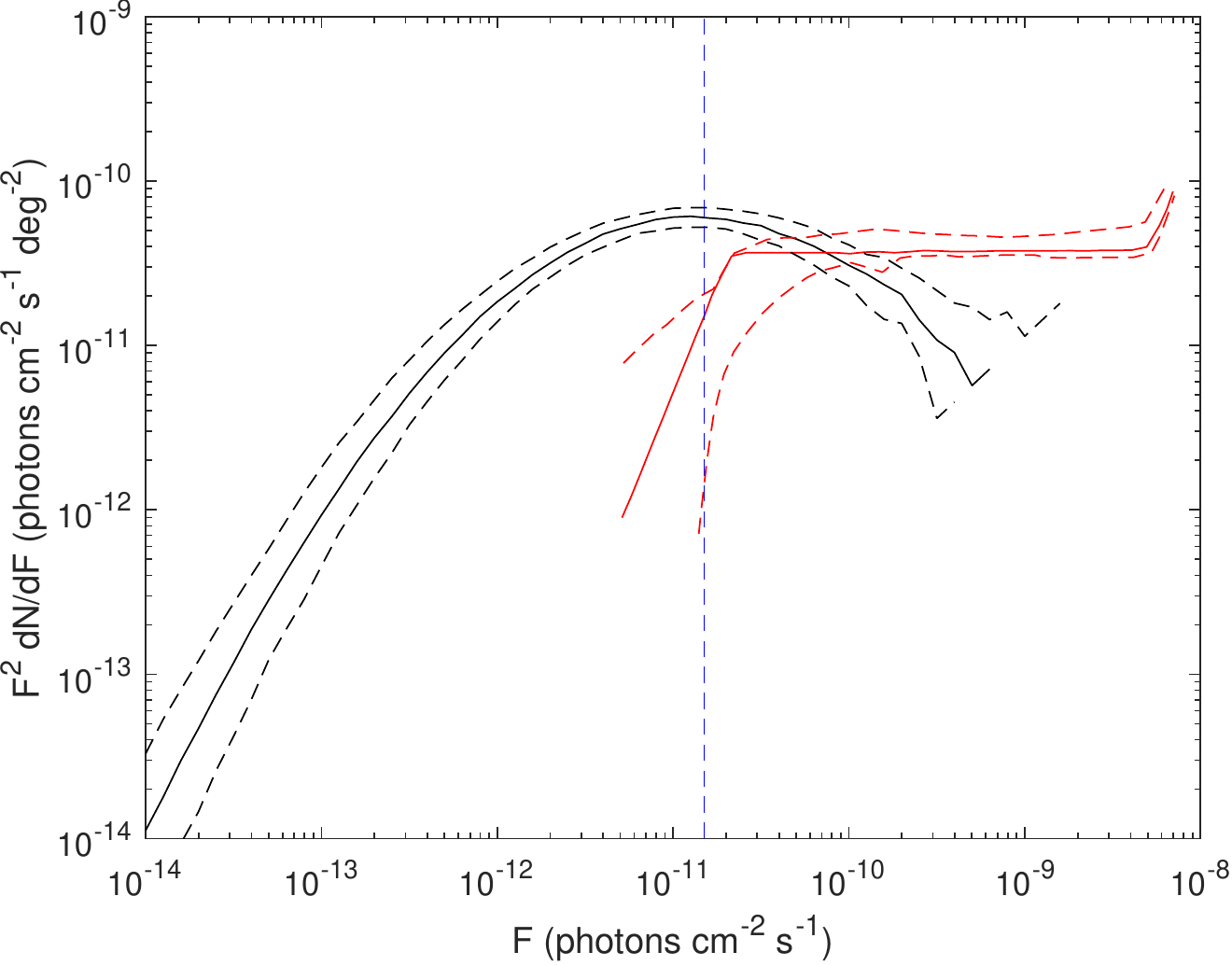}}
    \caption[Source count distribution for Model A1 MSP population.]{ Source count distribution for Model A1 compared to that of two studies applying non-Poissonian template fits to Fermi-LAT data. The reconstructed flux distributions are shown in red, in black is the modelled MSPs, the red and black dashed lines show $68\%$ intervals. The figure on the left is the photon flux distribution compared to Lee et al.~\cite{Lee:2015fea} for photons between $1.9$--$11.9$ GeV, for MSPs within $10^\circ$ of the Galacic Center and $\abs{b}>2^\circ$. The figure on the right compares to Calore et al.~\cite{Calore2021} for photons $2$--$5$ GeV from MSPs in the region $\abs{l}<10^\circ$ and $0.5^\circ < \abs{b} < 10^\circ$. The vertical blue dashed line shows the approximate flux of a source for which $\sim 1$ photon would have been detected according to Chang et al.~\cite{Chang_2020}.
    \label{fig:bulge_msp_source_counts}}
\end{figure}

In Fig.~\ref{fig:bulge_msp_source_counts} we compare, for Model A1, the photon flux distributions of MSPs to the sub-threshold distributions inferred from non-Poissonian template fits to Fermi-LAT photon count data from the inner Galaxy region \cite{Lee:2015fea,Calore2021}. Although there remains significant systematic uncertainty, these methods can be used to reconstruct the source count distribution down to around the level at which $\sim 1$ photon would have been detected, where sources would not be distinguishable from diffuse emission \cite{Chang_2020}. In Fig.~\ref{fig:bulge_msp_source_counts} the MSP flux distribution is generally similar to or below the sub-threshold point source distribution of the more recent work of Calore et al.~\cite{Calore2021}.

As can be seen in Fig.~\ref{fig:E_cut_B_E_dot_MCMC_Params_2}, the posterior distribution for parameter $\log_{10}(d_{\rm parallax}/{\rm kpc})$ is limited by the upper limit on the prior. In the limit as $d_{\rm parallax} \to \infty$, the probability, given by Eq.~\ref{eq:parallax_model}, of an MSP having a parallax measurement converges to $C_{\rm parallax}$ for all distances $d$; therefore, we imposed the prior upper limit at $d_{\rm parallax} = 100$ kpc to avoid this degeneracy. This upper limit will affect the $\log_{10}(d_{\rm parallax}/{\rm kpc})$ and $C_{\rm parallax}$ $68\%$ confidence intervals shown in Table \ref{tab:mcmc_results_parameters} and Fig.~\ref{fig:E_cut_B_E_dot_MCMC_Params_2}, however, they are not correlated with any other model parameters.

In order to check the sensitivity of the results to the particular GCE spectrum used we also fitted Model A1 to alternative spectra. Using Bartels et al.~\cite{Bartels2017}, we found a small number of changes where the median of a parameter was outside the $68\%$ ranges shown in Table \ref{tab:mcmc_results_parameters}. The ratio of nuclear bulge and boxy bulge MSPs is lower with $\log_{10}(N_{\rm nb} / N_{\rm bb}) = -0.90\substack{+0.06 \\ -0.06}$. The relationship between $\Gamma$ and $\dot{E}$ is steeper with $a_{\Gamma} = 0.52\substack{+0.07 \\ -0.08}$. The fifth DTD bin is constrained strongly to be near $0$ with $\textrm{DTD } p(11.1 \textrm{ - } 13.8 \textrm{ Gyr}) = 0.01\substack{+0.03 \\ -0.01}$. These changes are likely to be caused by the steep slope on both the low and high energy ends of the boxy bulge spectrum. For the spectrum in Calore et al.~\cite{Calore:2014xka}, we had no nuclear bulge, but otherwise the results were consistent with those using the spectra of Macias et al.~\cite{Macias19}.

It can be seen in Fig.~\ref{fig:E_cut_B_E_dot_N_observed_bulge_MSPs} that we find a probability of $0.16$ that no bulge MSPs would have been resolved at present, with a median of $2$ resolved. With a doubling and quadrupling of detection sensitivity, respectively, a median of $10$ and $35$ would be expected to be resolved. The detection probability model in Eq.~\ref{eq:detection_probability} models the flux threshold for detection of an MSP as a log-normal distribution. For the median, it uses the Fermi-LAT point source sensitivity at the Galactic coordinates of an MSP $\mu_{\rm th}(l,b)$ and multiplies it by a factor of $10^{K_{\rm th}}$, where we find for Model A1, $K_{\rm th} = 0.45\substack{+0.09 \\ -0.08}$. The standard deviation of $\log_{10}(F_{\rm th})$ around the median $F_{\rm th}$ of $10^{K_{\rm th}} \mu_{\rm th}(l,b)$ is $\sigma_{\rm th}$, which is $0.28\substack{+0.05 \\ -0.04}$ for Model A1. As we have used MSPs which are not only detected as a point source, but have confirmed gamma-ray pulsations, it is not a surprise that the MSP detection threshold is higher than the point source detection threshold. However, for most of the MSPs, gamma-ray pulsations have been discovered by searching for gamma-ray pulsations from a known radio MSP; therefore, the radio detection sensitivity, which is not explicitly accounted for, may be a significant contributor to the fitted values of $K_{\rm th}$ and $\sigma_{\rm th}$. A radio pulse is smeared by dispersion associated with the integrated density of free electrons and by scattering caused by small scale variation in that density, and this may be particularly severe in the region of the Galactic Center \cite{Macquart2015,Eatough2015,Calore2016,Rajwade2017}. As this would lead to lower sensitivity to radio pulsations, our model may in fact be overestimating the probability of resolving inner Galaxy MSPs. A number of young pulsars have been discovered in the Galactic Center region \cite{Johnston2006,Deneva2009}, but the smearing of an MSP pulse will be larger relative to its period. However, the discovery of a magnetar with a period of $3.76$ s near Sagittarius A*, the black hole at the center of the Milky Way, has suggested that the scattering of radio pulses from that region may be small enough that MSPs close to Sagittarius A* could be detected at radio frequencies $\gtrsim 10$ GHz \cite{Spitler2014,Bower2014}.

In Table \ref{tab:possible_resolved_bulge_msps} we provide a list of resolved MSPs with a significant probability (greater than $5\%$) of being bulge MSPs according to Model A1. These probabilities are worked out in Eq.~\ref{eq:bulgelikelihood} by evaluating the contribution of the bulge to the corresponding MSP's likelihood. 
The probability of resolved MSP $i$ being a bulge MSP can be thought of as a draw from a single trial binomial distribution with probability $p(\textrm{Bulge MSP}_i)$ and therefore also having an expectation value of $p(\textrm{Bulge MSP}_i)$. Using the linearity of expectation, this implies that the expected number of resolved bulge MSPs in the current data is $\sum_{i} p(\textrm{Bulge MSP}_i)=1.1$ where the sum is over all resolved MSPs including those listed in Table~\ref{tab:possible_resolved_bulge_msps}. This is at the lower end of, but consistent with, the range seen in Fig.~\ref{fig:E_cut_B_E_dot_N_observed_bulge_MSPs} produced based on the fitted model parameters.

Note that there is also some systematic uncertainty in the distance to the MSPs which is hard to quantify. For example, 
  in Table 4 of ref.~\cite{Camilo15} they have a distance of 3.4 kpc for J1747--4036 while we have a distance of $7.3\substack{+0.7 \\ -0.7}$~kpc. Also in Table 2.2 of ref.~\cite{Sanpaarsa2016} they have a distance of 3.1 kpc for J1855--1436 while we have a distance of $5.3\substack{+0.6 \\ -0.6}$~kpc. These difference may be due to a change in model of the Galactic free electron density. According to the ATNF online database\footnote{\url{https://www.atnf.csiro.au/research/pulsar/psrcat/}} J1747--4036 and J1855--1436  have distances of 7.15 kpc and 5.13 kpc respectively
  which are compatible with our values. Also, in Table 2 of ref.~\cite{Ng2020} they have a dispersion measure distance of 1.8 kpc for J1811--2405 while we have a parallax derived distance of $5\substack{+11 \\ -2}$~kpc. 

\begin{figure}
    \centering
    \subfigure{\centering\includegraphics[width=0.49\linewidth]{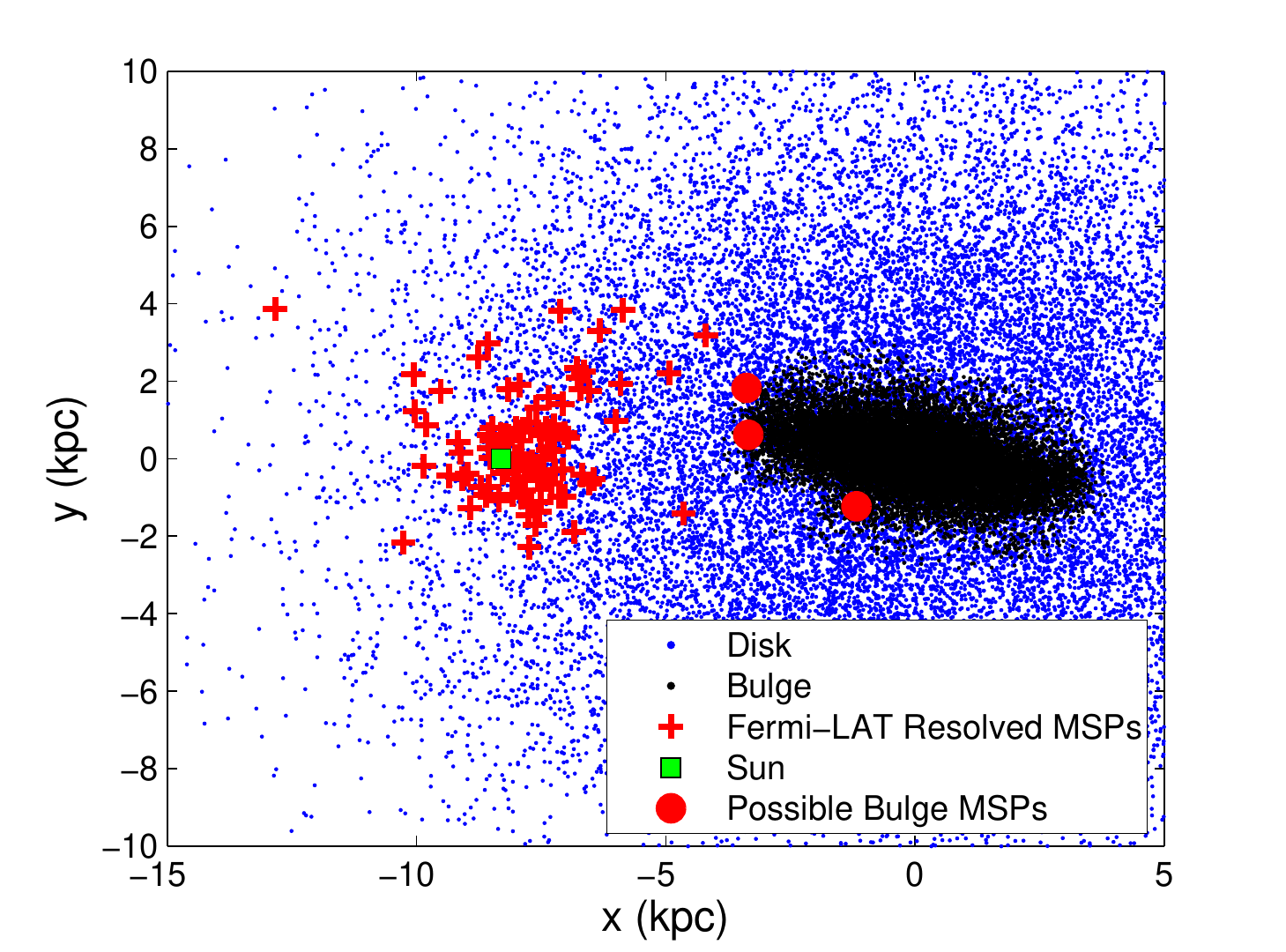}}
    \subfigure{\centering\includegraphics[width=0.49\linewidth]{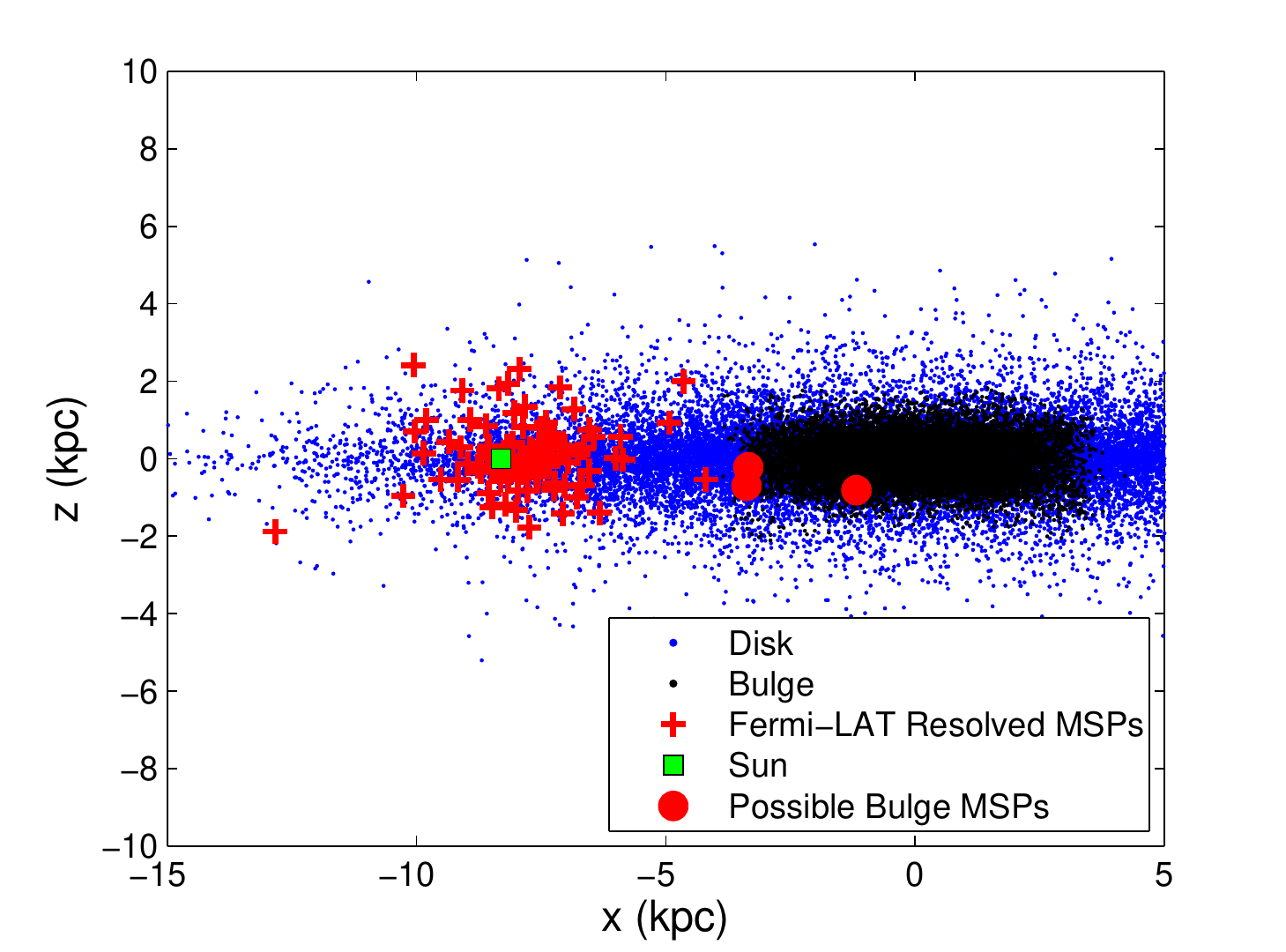}}
    \caption[Simulated spatial distribution of disk and bulge MSPs for Model A1.]{Simulated distribution of disk and bulge MSPs for Model A1 ($L = \eta E_{\rm cut}^{a_{\gamma}} B^{b_{\gamma}} \dot{E}^{d_{\gamma}}$). Also shown are those resolved MSPs with available distance estimates. These figures were made by selecting randomly from the highest likelihood parameter sets in our eight Markov chains for this model. Red circles are the MSPs in Table \ref{tab:possible_resolved_bulge_msps}. }
    \label{fig:sample_msp_pop}
\end{figure}

In Fig.~\ref{fig:sample_msp_pop} the locations of resolved MSPs are shown along with a simulated distribution of disk and bulge MSPs. The elongated nature of the bulge geometry does not play a big role in the probability of having a resolved bulge MSP. This can be seen by changing the boxy bulge geometry in Eq.~\ref{eq:rho_main_bulge} to a spherically symmetric geometry with $\rho_{\rm boxy~bulge}\propto r^{-2.4}$ up to $r=3.1$~kpc and $\rho_{\rm boxy~bulge}=0$ for larger radii \cite{Ploeg:2017vai}.
We then find that $2\pm 2$ MSPs are expected to be resolved for Model A1. Also, for this spherically symmetric bulge case, the MSPs in Table~\ref{tab:possible_resolved_bulge_msps} have probabilities of 0.5, 0.2, and $4\times 10^{-4}$ respectively. So only PSR J1855-1436, with its high $l=20.4^\circ$, is significantly affected by the bulge geometry.

As shown in Fig.~\ref{fig:E_cut_B_E_dot_N_MSPs_per_solar_mass}, the disk, nuclear bulge  and boxy bulge have a consistent MSP to stellar mass ratio. This is a good confirmation of our assumption that %
{\textcolor{black}{the population of individual, resolved MSPs belonging mostly to the disk population, on the one hand, and the apparently diffuse $\gamma$-ray emission from the GCE, on the other, can be self-consistently explained as arising from}}
MSPs drawn from the same underlying luminosity function given by Eq.~\ref{eq:Model1}.

\begin{landscape}
\begin{table}
\centering
    \begin{tabular}{c|c|c|c|c|c|c|c}
         Name & Bulge & $l$ & $b$ & $d$ & $L$ & $d$ & $L$ \\
          & Probability & (deg) & (deg) & (kpc) & ($\times 10^{34}$ erg s$^{-1}$) & percentile & percentile \\ \hline \hline
         
         PSR J1747-4036 & 0.4 & $-9.8$ & $-6.4$ & $7.3\substack{+0.7 \\ -0.7}$ & $7.7\substack{+1.8 \\ -1.5}$ & $100$ & $99$ \\ \hline
         PSR J1811-2405 & 0.5 & $7.1$ & $-2.5$ & $5\substack{+11 \\ -2}$ & $5\substack{+47 \\ -4}$ & $96$ & $97$ \\ \hline
         PSR J1855-1436 & 0.1 & $20.4$ & $-7.6$ & $5.3\substack{+0.6 \\ -0.6}$ & $1.8\substack{+0.5 \\ -0.4}$ & $98$ & $91$ \\ \hline
    \end{tabular}
    \caption[Details of MSPs with probability greater than $5\%$ of coming from bulge population for Model A1.]{ Details of MSPs with significant probability (greater than $5\%$) of coming from bulge population for Model A1. The names are from the Public List of LAT-Detected Gamma-Ray Pulsars.\cref{footnote:pulsars}
Distances are medians and $68\%$ intervals found by sampling the parallax uncertainty distribution if available or otherwise using the dispersion measure and randomly sampling the parameters of the YMW16 electron density model. Luminosities are medians and $68\%$ intervals found by additionally sampling from the flux uncertainty distribution. Percentiles are the fraction of resolved MSPs with smaller or equal luminosity/distance. This excludes MSPs without available distance estimates. }
    \label{tab:possible_resolved_bulge_msps}
\end{table}
\end{landscape}

\section{Conclusion}
\label{sec:conclusion}
We  compared a wide variety of luminosity function models for the Fermi-LAT gamma-ray MSP data. We found a convincing preference for Model A1 for which
$L = \eta E_{\rm cut}^{a_{\gamma}} B^{b_{\gamma}} \dot{E}^{d_{\gamma}}$
with a significantly positive $a_\gamma=1.2\pm0.3$ and $d_\gamma=0.5\pm 0.1$. Thus we confirm the result obtained by
Kalapotharakos et al.~\cite{Kalapotharakos_2019} that MSP gamma-ray  emission is consistent with curvature radiation and inconsistent with synchrotron radiation.
By comparing with other models, we showed that the main source of the positive 
$a_\gamma$ result was the need to account for a significant logarithmic correlation in the data between the $E_{\rm cut}$ and distance in the form of the dispersion measure. We also showed that the main source of the positive $d_\gamma$
was due to the significant logarithmic correlation in the data between the period and the distance in the form of the dispersion measure.

Additionally, we found that it was %
warranted
to include a relationship between the spectral parameters and $\dot{E}$. In particular we found that a linear relationship between the mean of the spectral index $\mu_\Gamma$ and the $\log_{10}(\dot{E})$, as specified in Eq.~\ref{eq:Gamma}, had a significantly positive slope of $a_\Gamma=0.4\pm0.08$. We identified the source of this positive slope to be the need to explain the significant correlation between $\Gamma$ and $\log_{\rm 10}(P)$ seen in the data.

We non-parametrically estimated the delay time distribution of the MSPs but found the current data do not strongly constrain it.
Our results obtained using a DTD prescription are also not significantly different from those obtained assuming a uniform age distribution for the MSPs.

Our results demonstrate that the population of MSPs that can explain the gamma-ray signal from the resolved MSPs in the Galactic disk and the unresolved MSPs in the boxy bulge and nuclear bulge
{\textcolor{black}{can consistently be described as arising from a common evolutionary trajectory for some subset of astrophysical sources common to all these different environments. 
We do not require that there is anything systematically different about the inner Galaxy MSPs to explain the GCE. We also found that the current data are not accurate enough to be sensitive to the small differences between the bulge and disk MSPs. 
}}

We estimated that there are 
between about 20 and 50 thousand
 MSPs in the boxy bulge at 68\% confidence interval.
We identified three candidate resolved MSPs (J1747-4036, J1811-2405, J1855-1436) that have significant probabilities (0.4, 0.5 and 0.1 respectively) of being members of the boxy bulge population. We estimate that this number would increase to 9 and 31 resolved boxy bulge MSPs were the sensitivity to gamma-ray MSPs doubled or quadrupled, respectively.

\chapter{The Effect of Birth Kicks on the Distribution of Millisecond Pulsars}
\label{ch:nbody}

\graphicspath{{nbody/Figs/}}

\section{Introduction}

In the ``recycling'' model of MSP formation a neutron star is spun up to millisecond periods through the transfer of mass from a binary companion. This requires that the binary system survives the kick produced by any asymmetry in the core collapse supernova explosion \cite{Bhattacharya1991}.
However, an alternative to the recycling channel is 
accretion induced collapse of white dwarfs into neutron stars which may produce more than half of all observed MSPs \citep{Ferrario2007,Hurley2010,Ruiter2019}. In this case the system does not receive a significant natal kick \citep{Fryer1999,Kitaura2006}. 
This would imply that the MSPs have much smaller peculiar velocities in comparison to the recycling model \citep{Lyne1994, Wongwathanarat2013, Bear2018}.

In Chapter \ref{ch:msp_pop}, we modelled Fermi-LAT detected MSPs as having a Maxwell distributed peculiar velocity with the scale parameter $\sigma_v$ found to be $77 \pm 6$ km s$^{-1}$ where we quote error bars at the 68\% confidence interval throughout this chapter. This velocity applies for disk MSPs and we assume it will not be significantly different for bulge MSPs. Although the star formation histories are very different in the bulge and disk, as can be seen from Fig.~\ref{fig:E_cut_B_E_dot_luminosity_distribution} the probability distribution of luminosities in the bulge and disk only differ by a few percent. 
Also, as can be seen from Fig.~\ref{fig:E_cut_B_E_dot_N_MSPs_per_solar_mass}, the bulge and disk have a ratio of number of MSPs formed per solar mass which is within one error bar of each other.
 Based on this, we assume that the bulge and disk have the same mix of MSP formation channels and thus the same probability distribution of natal kick velocities.

If the source of the GCE is a population of unresolved MSPs, then the spatial distribution may be smoothed to some degree relative to the stellar mass in the bulge. Eckner et al.~\cite{Eckner2018} used the  virial theorem to estimate the ``smoothing length" of MSPs as $700$ -- $900$ pc for kicks $\lesssim 70$ km s$^{-1}$. However, they assumed a spherically symmetric spatial distribution for the MSPs. 

In this chapter we use N-body simulations to estimate what are the effects of MSP kicks for a boxy bulge distribution. In Section~\ref{method} we explain our method. Our results are given in Section~\ref{results}, and our conclusions in Section~\ref{conclusion}.

\section{Method}
\label{method}
For this work we use the code of Bedorf et al.~\cite{Bedorf2012} to run $N$-body simulations in order to model the Milky Way.\footnote{Available at: https://github.com/treecode/Bonsai} We use parameters corresponding to models MWa, MWb and MWc0.8 as denoted by Fujii et al.~\cite{Fujii2019} as they were the best fitting models to Galaxy observations that Fujii et al.\ found. Comparing to bulge kinematics, bar length, and pattern speed observations they found $\chi^2=5.3,8.0, 12.6$ for MWa, MWb, MWc0.8 respectively.
For each model we generated a total of $30$ million disk, bulge and dark matter halo particles. These initial populations are generated using the methods of Kuijken and Dubinski \cite{Kuijken1995}, Widrow and Dubinski \cite{Widrow2005}, and Widrow et al.~\cite{Widrow2008}.\footnote{We used the implementation at: https://github.com/treecode/Galactics.parallel}
As in Fujii et al.~\cite{Fujii2019}, we use time-steps of $\sim0.6$ Myr, an opening angle of $0.4$ radians and ran the simulation for $10$ Gyr. However, we use a softening length of $30$ pc. Also, our dark-matter halo particles have a mass $8$ times larger than the disk and bulge particles. 
Taking into account the masses of the various components, this implies that,
out of the 30 million particles,  of order $10$ million represent  stellar mass and the remainder represent dark matter. 

In order to model the density of MSPs, we additionally include massless (so they do not affect the simulation) disk and bulge particles that are given a normally distributed perturbation to each component of their velocity vector with mean zero and standard deviation $\sigma_k$. The kick velocity magnitude is therefore Maxwell distributed. The probability density function of a Maxwell distribution can be written as:
\begin{equation}
    p(x) = \sqrt{\frac{2}{\pi}} \frac{x^2 \exp\left(-x^2/2\sigma^2\right)}{\sigma^3}
    \label{eq:Maxwell}
\end{equation}
\noindent where $x$ is the magnitude of a three dimensional vector with components sampled from the normal distribution $\mathcal{N}(0,\sigma^2)$. For each model we try a case where the kicks occurred at the beginning of the $N$-body simulations and a case where the kicks occur randomly with a uniform rate over the course of the $10$ Gyr.

The first step is to estimate the kick velocity scale required to produce a peculiar velocity distribution consistent with Chapter \ref{ch:msp_pop} where for the best model $\sigma_v=77 \pm 6$ km s$^{-1}$. We do this by running each model with $41$ populations of $10^5$ kicked particles with $\sigma_k$ between $70$ and $110$ km s$^{-1}$. We separate the velocity of each particle into two components:
\begin{equation}
    \boldsymbol{v} = \boldsymbol{v}_c + \boldsymbol{v}_p
\end{equation}
\noindent where $\boldsymbol{v}_c$ is the velocity of a particle on a circular orbit around the center of the galaxy and $\boldsymbol{v}_p$ is the peculiar velocity. The magnitude of $\boldsymbol{v}_c$ for a particle with coordinates $x$, $y$, and $z$ can be evaluated using the centripetal force: 
\begin{equation}
    \lVert \boldsymbol{v}_c \rVert = \sqrt{\lVert\boldsymbol{a}_c(x,y,0)\rVert R}
\end{equation}
\noindent where $\boldsymbol{a}_c(x,y,z)$ is the acceleration toward the center of the galaxy and $R^2=x^2+y^2$. For $R$ outside the bar region, and for small peculiar velocity, we are therefore assuming that particles are rotating with the disk, with $\boldsymbol{v}_c$ the rotation velocity of the disk at $R$ \cite{Verbunt2017}.

We use a maximum likelihood estimate of the final $\sigma_v$ for each initial $\sigma_k$. For a set of $N$ particles with peculiar velocities $v_1$, ..., $v_N$, the log-likelihood is obtained by assuming velocities have a Maxwell distribution:
\begin{equation}
    \log(L) = \frac{N}{2} \log(\frac{2}{\pi}) - 3 N \log(\sigma_v) + \sum_{i=1}^N 2 \log(v_i) - \frac{v_i^2}{2 \sigma_v^2}
\end{equation}
\noindent and therefore
\begin{equation}
    \frac{\dd\log(L)}{\dd\sigma_v} = -\frac{3 N}{\sigma_v} + \sum_{i=1}^N \frac{v_i^2}{\sigma_v^3} \, .
\end{equation}
\noindent Then solving for $\sigma_v$ where $\dd\log(L)/\dd\sigma_v = 0$, we find the maximum likelihood estimate for $\sigma_v$ is:
\begin{equation}
    \hat{\sigma_v} = \sqrt{\frac{\sum_{i=1}^N v_i^2}{3 N}}
\end{equation}
This is done for particles where $4 \textrm{ kpc} \leq R \leq 12 \textrm{ kpc}$ and $\lvert z \rvert \leq 2 \textrm{ kpc}$, ensuring we are estimating the peculiar velocity distribution scale parameter for particles in the disk region from which gamma-ray MSPs are most likely to be resolved and where $\boldsymbol{v}_c$ approximately represents disk rotation.

Once we have a best fitting $\sigma_k$ for each model, we rerun each $N$-body simulation using $2\times10^6$ kicked particles with that $\sigma_k$. We also have $2\times10^6$ massless particles which are not kicked with the same initial positions and velocities. We then use MCMC to fit a parametric model to both final particle distributions. This model consists of four components: a spherically symmetric bulge, bar, a long bar and a disk. The spherically symmetric bulge component uses the Hernquist model \cite{Hernquist1990}:
\begin{equation}
    \rho_{\rm Hernquist} (r) \propto \frac{1}{\left(r/a_b \right) \left(1 + r/a_b\right)^3}
\end{equation}
\noindent where $r^2 = x^2 + y^2 + z^2$ and $a_b$ is a free parameter. The initial conditions used here of Fujii et al.~\cite{Fujii2019} include a component distributed according the Hernquist model. The bar model is distributed as:
\begin{equation}
\rho_{\rm bar} (R_s) \propto K_0(R_s)
\times \begin{cases}
      1 & R \leq R_{\rm end} \\
      \exp(-(R - R_{\rm end})^2/h^2_{\rm end}) & R > R_{\rm end} \\
\end{cases}
\label{eq:bar}
\end{equation}
\noindent where $K_0$ is the modified Bessel function of the second kind and where:
\begin{equation}
R_\perp^{C_\perp} = \left(\frac{\abs{x}}{x_b} \right)^{C_\perp} + \left(\frac{\abs{y}}{y_b} \right)^{C_\perp}
\end{equation}
\begin{equation}
R_s^{C_\parallel} = R_\perp^{C_\parallel} + \left(\frac{\abs{z}}{z_b} \right)^{C_\parallel}
\end{equation}
\noindent where the free parameters are $C_\parallel$, $C_\perp$, $x_b$, $y_b$, $z_b$ and $R_{\rm end}$ with $h_{\rm end}$ fixed at $\sqrt{\frac{1}{2}}$ kpc. 
The effective radius is $R_s$; the scale lengths are $x_b$, $y_b$, and $z_b$; and $C_\perp$ and $C_{\parallel}$ are the face-on
and edge-on shape parameters.
The bar shape is elliptical in the corresponding direction when $C_\perp,C_{\parallel} = 2$, diamond-shaped when $C_\perp,C_{\parallel} < 2$, and boxy when $C_\perp,C_{\parallel} > 2$.
The Gaussian function with
scale length $h_{\rm end}$ in Eq. (\ref{eq:bar}) 
 truncates the bar at radius $R_{\rm end}$.
The modified Bessel function was also used in Cao et al.~\cite{Cao:2013dwa} to model the distribution of red clump giants, but with no cutoff and with $C_\parallel = 4$ and $C_\perp = 2$. For the long bar we use \cite{Wegg2015}:
\begin{equation}
\label{eq:long_bar}
    \rho_{\rm long~bar} (x, y, z) \propto \!\begin{multlined}[t] 
        \exp\left( -\left(\left(\frac{\abs{x}}{x_{\rm lb}} \right)^{C_{\perp,\rm lb}} + \left(\frac{\abs{y}}{y_{\rm lb}} \right)^{C_{\perp,\rm lb}}\right)^{1/C_{\perp,\rm lb}} \right) \exp\left(-\frac{\abs{z}}{z_{\rm lb}} \right) \\ \times \textrm{Cut}\left(\frac{R - R_{\rm out}}{\sigma_{\rm out}} \right) \textrm{Cut}\left(\frac{R_{\rm in} - R}{\sigma_{\rm in}} \right)
    \end{multlined}
\end{equation}
\noindent where $C_{\perp,\rm lb}$, $x_{\rm lb}$, $y_{\rm lb}$, $z_{\rm lb}$, $R_{\rm out}$ and $R_{\rm in}$ are free parameters, $\sigma_{\rm out} = \sigma_{\rm in} = \sqrt{\frac{1}{2}}$ kpc and:
\begin{equation}
    \textrm{Cut}\left(x\right) = \begin{cases}
      \exp(-x^2) & x > 0 \\
      1 & x \leq 0 \\
\end{cases}
\end{equation}
\noindent Finally, we have a disk with a central hole:
\begin{equation}
    \rho_{\rm disk} (x, y, z) \propto \exp(-R^2/2\sigma_r^2) \exp(-\abs{z}/z_0) H(x,y)
\end{equation}
\noindent where
$\sigma_r$ and $z_0$ are free parameters and for the hole we use the form adopted by Freudenreich \cite{Freudenreich:1997bx}:
\begin{equation}
    H(x,y) = 1 - \exp\left(-\left(R_H/O_R\right)^{O_N} \right)
\end{equation}
with:
\begin{equation}
    R_H^2 = (x)^2 + (\epsilon y)^2
\end{equation}
\noindent where $\epsilon$, $O_R$ and $O_N$ are also free parameters.

The total number of particles in our simulations are fixed. So we do not have to include the number of particles as part of our likelihood. Therefore the probability of having an $N$-body particle at position ${x,y,z}$ will be proportional to the density of our model ($\rho$) at ${x,y,z}$.
We have for each component of the model a parameter giving the probability a particle is from that component. We treat the probability of an $N$-body particle being from a component of the density distribution as parameters. These parameters, $P({\rm Disk})$, $P({\rm Bar})$, $P({\rm Hernquist})$ and $P({\rm Long~Bar})$, have a Dirichlet prior \cite{Betancourt2013}. This prior constrains
$$
 P({\rm Disk})  + P({\rm Bar}) + P({\rm Hernquist})  + P({\rm Long~Bar})=1
$$
and is uniformly distributed over any values of these parameter satisfying that condition.
The likelihood is then:
\begin{equation}
    \log(L) = \sum_i^N \log(\rho(x_i, y_i, z_i))
\end{equation}
\noindent where $x_i$, $y_i$ and $z_i$ are the coordinates of a particle, $N$ is the number of particles, and $\rho$ is the density of the model:
\begin{equation}
    \rho(x, y, z) =  \!\begin{multlined}[t] 
    P({\rm Disk}) \rho_{\rm disk}(x, y, z) + P({\rm Bar}) \rho_{\rm bar}(x, y, z) \\ + P({\rm Hernquist}) \rho_{\rm Hernquist}(x, y, z) + P({\rm Long~Bar}) \rho_{\rm long~bar}(x, y, z)
    \end{multlined}
\end{equation}
\sloppy All scale parameters are given a prior so they are uniform in $\log(\theta)$ where $\theta\in \{a_b, x_b, y_b, z_b, x_{\rm lb}, y_{\rm lb}, z_{\rm lb},\sigma_r,z_0, O_R\}$ and this implies $p(\theta) \propto 1/\theta$.
In calculating the likelihood, we do not include particles for which $R>12$ kpc or $\abs{z} > 3$. It can be seen in Fujii et al.~\cite{Fujii2019} that the scale height of the disk may start to decline between $10\lesssim R \lesssim 15$~kpc. We also don't want the fit to be affected by particles that may have been kicked well out of the galaxy. 
The likelihood ($L$) is insensitive to being multiplied by a constant but that constant has to be the same for all parameters of our combined model.
To accommodate this we normalize each density component such that
$$
\int_{R \leq 12~{\rm kpc},\abs{z} \leq 3~{\rm kpc}} \rho_i(x,y,z)\, {\rm d}x\,{\rm d}y\,{\rm d}z=1
$$
where $i\in$~\{disk, bar, Hernquist, long bar\}.
This integral is estimated with importance sampling.
We use a set of random numbers which are transformed into the points at which we evaluate the density models in order to estimate the normalization constant. In order to stabilize the estimation of the likelihood function these numbers are always the same every time we perform the importance sampling within a particular chain.

After running the $N$-body simulations, we shift the coordinates of the particles so that the center of mass is at the origin, then rotate so the bar is along the $x$-axis. The bar angle is estimated using the method described in Fujii et al.~\cite{Fujii2019}. However, we add four parameters that we expect to be near zero to allow a further shift in the center and clockwise rotation of the model. These are $\alpha$, $x_{\rm center}$, $y_{\rm center}$ and $z_{\rm center}$, with the latter three parameters in parsecs, so:
\begin{equation}
  \begin{aligned}
    x_{\rm data} &= \cos(\alpha) x + \sin(\alpha) y + x_{\rm center} / 1000 \\
    y_{\rm data} &= -\sin(\alpha) x + \cos(\alpha) y + y_{\rm center} / 1000 \\
    z_{\rm data} &= z + z_{\rm center} / 1000
  \end{aligned}
\end{equation}
\noindent where $x_{\rm data}$, $y_{\rm data}$ and $z_{\rm data}$ are coordinates in the coordinate system of the $N$-body simulation. In estimating the peculiar velocity distribution scale parameter, $\hat{\sigma_v}$, above, we assumed $x \approx x_{\rm data}$, $y \approx y_{\rm data}$ and $z \approx z_{\rm data}$.

In Chapter \ref{ch:msp_pop} we used for MCMC the adaptive Metropolis algorithm of Haario et al.~\cite{Haario01}, here, however, we found it was necessary to replace this with an alternative algorithm to ensure rapid convergence to the peak likelihood region of the parameter space. The MCMC algorithm used in this chapter is similar to that of Foreman-Mackay et al.~\cite{ForemanMackey2013} with a mixture of the Differential Evolution \cite{TerBraak2006} and snooker updates \cite{TerBraak2008}. Instead of performing a single random walk through the parameter space where proposed moves are accepted with the probability $r$ of Eq.~\ref{eq:metropolis_hastings_r}, we use an ensemble of $K$ ``walkers'' where a proposed update for a walker $j$ depends on the distribution of the other walkers.

A single step of the stretch move update suggested in Foreman-Mackay et al.~\cite{ForemanMackey2013} involves updating the $K$ walkers sequentially. Let $\pmb{x}_j$ be the state of walker $j$, an update for walker $\pmb{x}_j$ is performed as follows:
\begin{enumerate}
    \item Draw $k$ from $1,2,...,K$ where $k \neq j$
    \item Draw $z$ from the probability density function with parameter $a$ (Foreman-Mackay et al.~\cite{ForemanMackey2013} suggest $a=2$):
    \begin{equation}
        g(z) \propto \begin{cases}
            \frac{1}{\sqrt{z}} & z \in \left[\frac{1}{a}, a \right] \\
            0 & {\rm otherwise} \\
        \end{cases}
    \end{equation}
    \item Calculate proposal $\pmb{y} = \pmb{x}_k + z \left(\pmb{x}_j - \pmb{x}_k \right)$
    \item Calculate acceptance probability $r$:
    \begin{equation}
        r = z^{d - 1} \frac{p(\pmb{y})}{p(\pmb{x}_j)}
    \end{equation}
    where $d$ is the number of dimensions
    \item Set $\pmb{x}_j = \pmb{y}$ with probability $\min(1,r)$
\end{enumerate}
We found better results using a mixture of two alternative updates: $80\%$ the Differential Evolution update of ter Braak \cite{TerBraak2006} and $20\%$ the snooker update of ter Braak and Vrugt \cite{TerBraak2008}.

To update walker $\pmb{x}_j$ using the Differential Evolution update:
\begin{enumerate}
    \item Draw $k$ and $l$ from $1,2,...,K$ where $k \neq j$, $l \neq j$ and $l \neq k$
    \item Propose $\pmb{y} = \pmb{x}_j + \gamma \left(\pmb{x}_k - \pmb{x}_l \right) + \pmb{e}$ where $\gamma$ is a parameter and where $\pmb{e}$ is drawn from a small $d$ dimensional symmetric probability distribution
    \item Calculate acceptance probability $r$:
    \begin{equation}
        r = \frac{p(\pmb{y})}{p(\pmb{x}_j)}
    \end{equation}
    \item Set $\pmb{x}_j = \pmb{y}$ with probability $\min(1,r)$
\end{enumerate}
We drew $\pmb{e}$ from a $d$ dimensional Gaussian with standard deviation $10^{-5}$ in each dimension. In case the likelihood distribution had multiple modes, we used $\gamma = 1$ with probability $0.1$ as suggested by ter Braak \cite{TerBraak2006}, otherwise we used the default value of $\gamma = 2.38 / \sqrt{2 d}$.

Using the snooker update, we update $\pmb{x}_j$ as follows:
\begin{enumerate}
    \item Draw $k$, $l$ and $m$ from $1,2,...,K$ with no index repeated or equal to $j$
    \item Calculate the orthogonal projections of $\pmb{x}_l$ and $\pmb{x}_m$ onto the line $\pmb{x}_j - \pmb{x}_k$, $\mathrm{proj}_{\pmb{x}_j - \pmb{x}_k}(\pmb{x}_l)$ and $\mathrm{proj}_{\pmb{x}_j - \pmb{x}_k}(\pmb{x}_m)$, where:
    \begin{equation}
        \mathrm{proj}_{\pmb{u}}(\pmb{v}) = \frac{\pmb{v} \cdot \pmb{u}}{\pmb{u} \cdot \pmb{u}} \pmb{u}
    \end{equation}
    \item Propose $\pmb{y} = \pmb{x}_j + \gamma_s \left(\mathrm{proj}_{\pmb{x}_j - \pmb{x}_k}(\pmb{x}_l) - \mathrm{proj}_{\pmb{x}_j - \pmb{x}_k}(\pmb{x}_m) \right)$ where $\gamma_s$ is a parameter
    \item Calculate acceptance probability $r$:
    \begin{equation}
        r = \frac{p(\pmb{y}) \abs{\pmb{y} - \pmb{x}_k}^{d-1}}{p(\pmb{x}_j) \abs{\pmb{x}_j - \pmb{x}_k}^{d-1}}
    \end{equation}
    \item Set $\pmb{x}_j = \pmb{y}$ with probability $\min(1,r)$
\end{enumerate}
We use $\gamma_s = 2.38 / \sqrt{2}$ as suggested by ter Braak and Vrugt \cite{TerBraak2008}.

We also used a simple annealing method in which we divide the log-likelihood by a temperature $T$ which is gradually reduced to $1$. The posterior probability density for a parameter set $\theta$ at MCMC iteration $t$ is:
\begin{equation}
    p(\theta \vert \textrm{$N$-body data}) \propto p(\theta) L^{1/T_t}
\end{equation}
\noindent where $p(\theta)$ is the prior, $L$ is the likelihood, and $T_t$ is the temperature. We used a linearly decreasing $\log(T)$ from $\log(1000)$ to $\log(1)$ during the first half of each Markov chain, which we discard. This method allows the algorithm to explore a broad region in the parameter space, while slowly converging to the desired posterior distribution where $T = 1$. This appeared to help the Markov chains avoid getting stuck in local likelihood maxima.

\section{Results}
\label{results}

\begin{figure}
    \centering
    \includegraphics[width=0.8\linewidth]{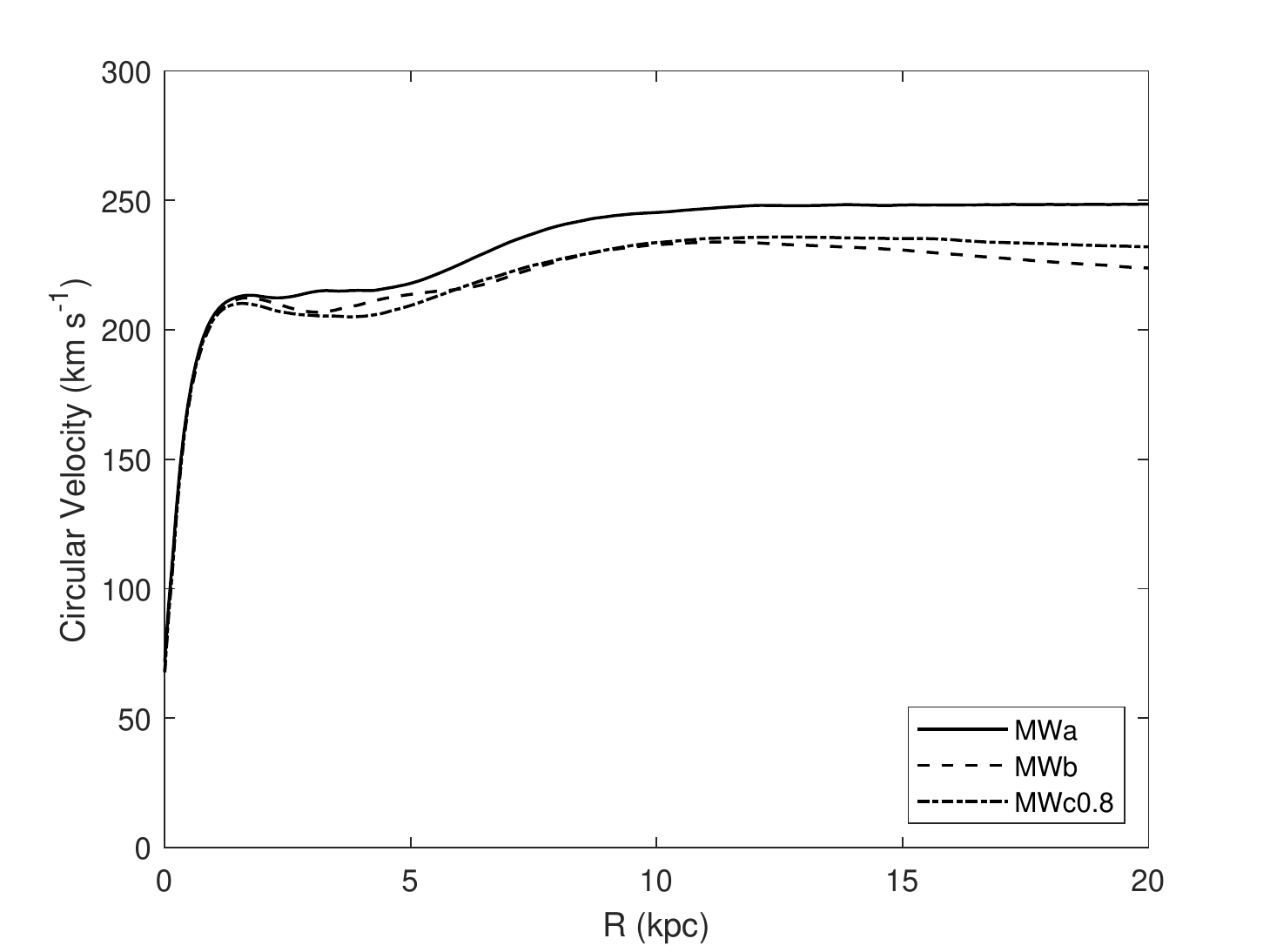}
    \caption[Simulated galaxy rotation curves.]{The rotation curves for the three models MWa, MWb and MWc0.8. The local circular velocity of the Sun is $238 \pm 15$ km s$^{-1}$ \cite{Bland-Hawthorn2016}. The distance between the Sun and the Galactic Center is approximately $8$ kpc. }
    \label{fig:rotation_curves}
\end{figure}

In Table~\ref{tab:kick_velocity} we present the kick velocities $\sigma_k$ that produce peculiar velocity distributions close to $\sigma_v = 77 \pm 6$ km s$^{-1}$ as estimated in Chapter \ref{ch:msp_pop}. We display the rotation curves at $t = 10$ Gyr for the three models in Fig.~\ref{fig:rotation_curves}. The central values in Table~\ref{tab:kick_velocity} were used to run $N$-body simulations with a larger number of particles to which we fitted a parametric model. The fitted parameters are shown in Tables~\ref{tab:MWa_params}, \ref{tab:MWb_params} and \ref{tab:MWc0.8_params}. There are three potential sources of uncertainty in the model parameters: the posterior density function, variation in the likelihood between chains as a result of the importance sampling method used to estimate the normalization constant for each model component, and the possibility that Markov chains may get stuck in different local likelihood maxima. We found that for most parameters the posterior distributions overlapped significantly or, in many cases, were indistinguishable. For a few parameters we had outlier chains; these can be seen in the tables as parameters with large, highly asymmetric uncertainties. For the kicked distributions, the disk model hole parameters were very uncertain with different Markov chains settling on a wide range of different values. As this hole tended to be smaller and/or less sharp, we simply removed it by setting $H(x,y) = 1$. We did find, including prior to removing the $\rho_{\rm disk}$ hole, the long bar model often acts like a second disk component when fitting to the kicked particles, this disk-like long bar was still allowed a hole through the parameter $R_{\rm in}$.

We ran the $N$-body simulations for the two kick rate scenarios separately, but we combine the two sets of Markov chains for the ``No Kick" columns in the parameter tables. Typically, the posterior distributions were very similar for Markov chains generated using these two sets of data, slightly expanding our $68\%$ intervals. However, we found a very significant difference was the location of the center. For this reason, we only report the change in parameters $\alpha$, $x_{\rm center}$, $y_{\rm center}$ and $z_{\rm center}$ for the fits to the kicked distributions by subtracting off the median of the corresponding Markov chains. For $\alpha$ the fitted $68\%$ intervals for no kick were  $-2^\circ$ to $-1^\circ$ for all models, $\abs{x_{\rm center}}$ and $\abs{y_{\rm center}}$ were always $\lesssim 40$ pc, and $\abs{z_{\rm center}}$ was $< 2$ pc.

\begin{figure}
    \centering
    \includegraphics[width=0.99\linewidth]{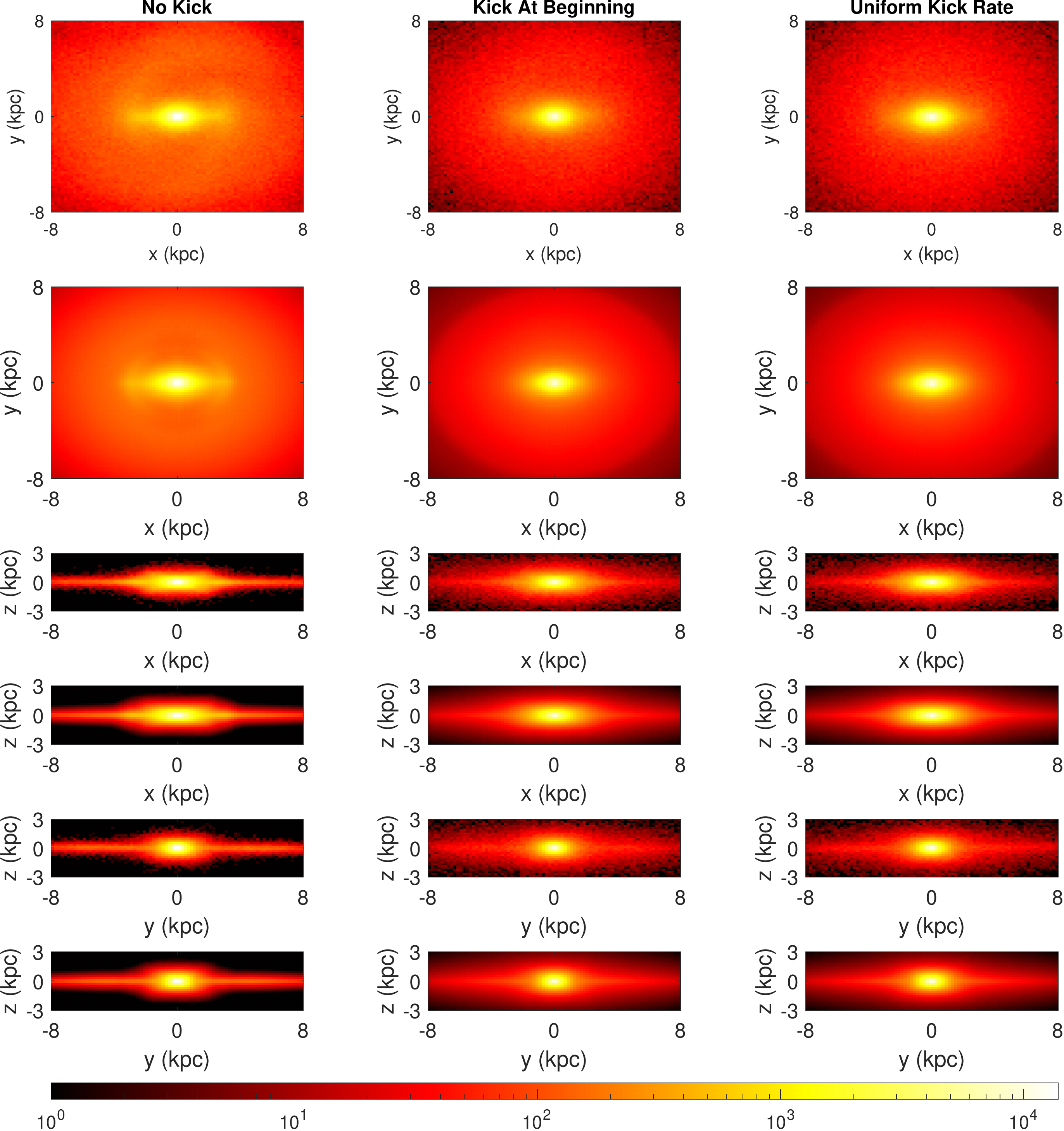}
    \caption[MWa model density maps.]{Density map of particles with no kick, a kick at the beginning and a uniform kick rate for the model MWa. Every second row shows the fitted model. }
    \label{fig:MWa_density_data_fit}
\end{figure}

\begin{figure}
    \centering
    \includegraphics[width=0.99\linewidth]{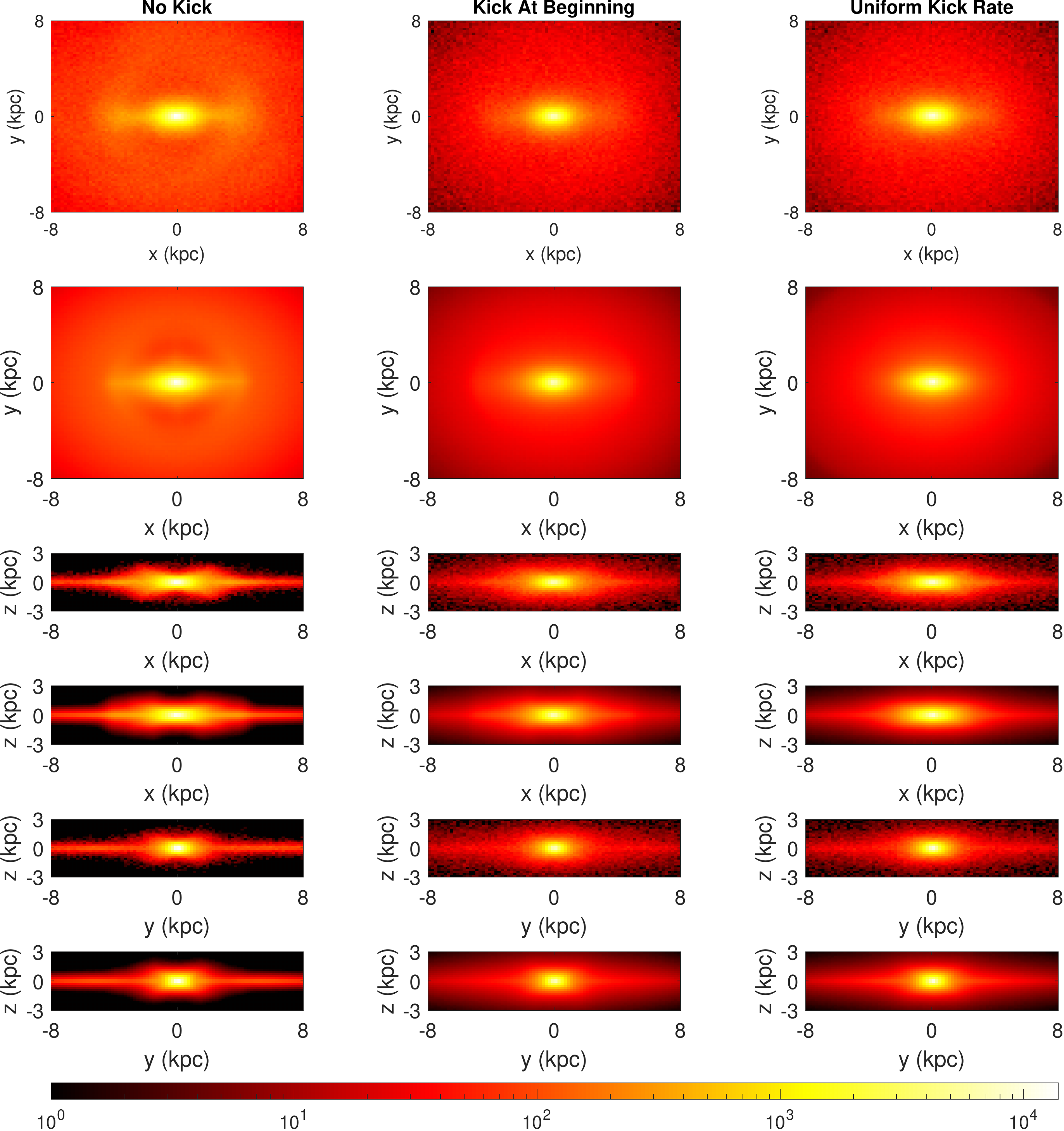}
    \caption[MWb model density maps.]{Density map of particles with no kick, a kick at the beginning and a uniform kick rate for the model MWb. Every second row shows the fitted model. }
    \label{fig:MWb_density_data_fit}
\end{figure}

\begin{figure}
    \centering
    \includegraphics[width=0.99\linewidth]{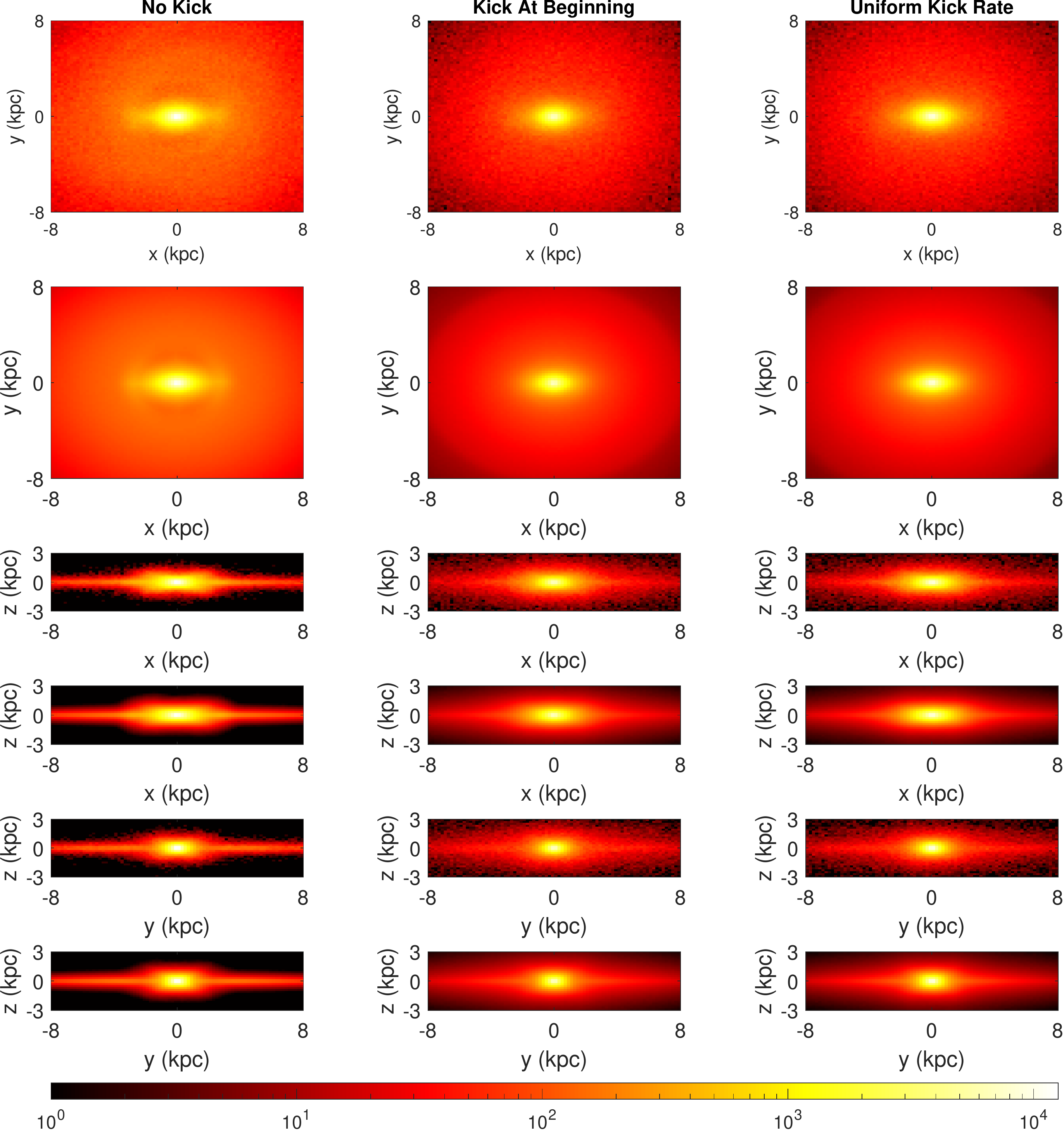}
    \caption[MWc0.8 model density maps.]{Density map of particles with no kick, a kick at the beginning and a uniform kick rate for the model MWc0.8. Every second row shows the fitted model. }
    \label{fig:MWc0.8_density_data_fit}
\end{figure}

\begin{figure}
    \centering
    \includegraphics[width=0.99\linewidth]{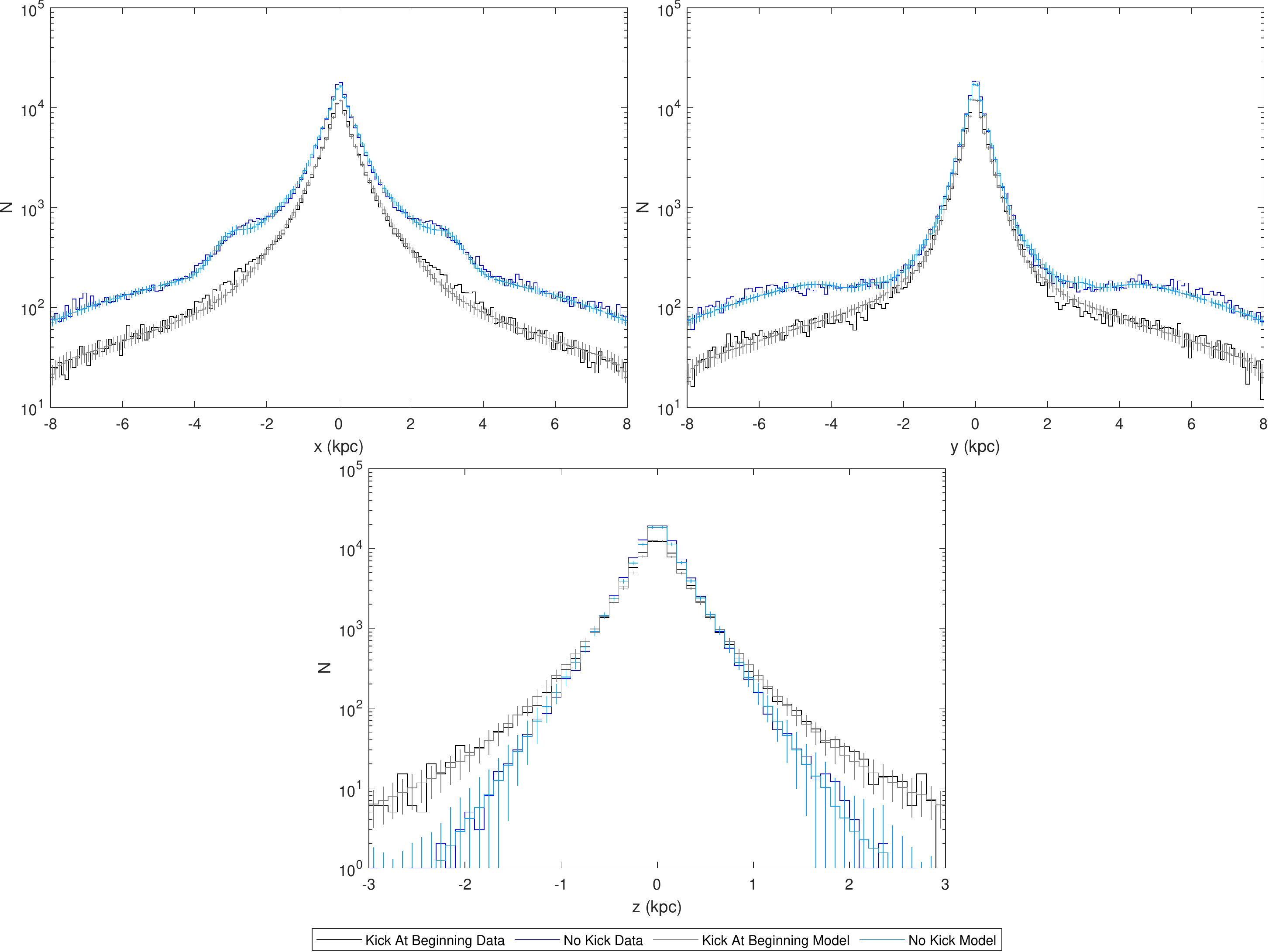}
    \caption[MWa $x$, $y$ and $z$ profiles with kicks at beginning.]{MWa profile along $x$, $y$ and $z$ axes with kicks occuring at the beginning. We show both $N$-body simulation data and data simulated using the fitted model. For the fitted model we show the mean number of particles in each bin and the standard deviation. }
    \label{fig:MWa_1d_profiles_kick_at_beginning}
\end{figure}

We show the density maps for MWa, MWb and MWc0.8 in Figs.~\ref{fig:MWa_density_data_fit}, \ref{fig:MWb_density_data_fit} and \ref{fig:MWc0.8_density_data_fit} respectively. These figures include both the $N$-body particle data as well as the fitted models. These density maps were produced by binning particles (either the $N$-body particles or particles drawn from the fitted model) within $0.25$ kpc of the $x$-$y$, $x$-$z$ and $y$-$z$ planes. The density maps for the fitted models were generated by taking the average in each bin for simulated data generated using 500 random parameter sets from our Markov chains in each case. In Fig.~\ref{fig:MWa_1d_profiles_kick_at_beginning}
we show the $N$-body simulation data and simulated particle distributions along the $x$, $y$ and $z$ axes for the MWa case with the kicks occurring at the beginning. The corresponding uniform kick rate case is displayed in Fig.~\ref{fig:MWa_1d_profiles_uniform_kick_rate}. The MWb and MWc0.8 cases are shown in Appendix \ref{appendix:extra_figures_nbody} as Figs.~\ref{fig:MWb_1d_profiles_kick_at_beginning}, \ref{fig:MWb_1d_profiles_uniform_kick_rate}, \ref{fig:MWc0.8_1d_profiles_kick_at_beginning} and \ref{fig:MWc0.8_1d_profiles_uniform_kick_rate}. In these figures, we bin all particles within $0.25$ kpc in the two perpendicular axes.
For MWa, we display in Fig.~\ref{fig:MWa_data_los} the integrated flux along lines of sight in the central $50^\circ \times 50^\circ$ of the galaxy. The Sun is placed at a distance of $7.9$ kpc, at an angle relative to the bar of $20^\circ$ and at a height of $15$ pc \cite{Coleman19}. These figures were generated by binning particles in Galactic latitude and longitude with weights of $1/d^2$ where $d$ is the distance of a particle from the Sun. We excluded particles within $1$ kpc of the Sun to reduce noise. The corresponding figures for MWb and MWc0.8 are shown in Appendix \ref{appendix:extra_figures_nbody} as Figs.~\ref{fig:MWb_data_los} and \ref{fig:MWc0.8_data_los}.

In Table~\ref{tab:alternative_bar} we show the change in $-2 \log(L)$ when replacing $\rho_{\rm bar} (R_s)$ with a range of different models from Freudenreich \cite{Freudenreich:1997bx} and Cao et al.~\cite{Cao:2013dwa}. As there is variation in the likelihood between chains, we also show the standard deviation in $-2 \log(L)$. Our choice of $\rho_{\rm bar} (R_s) \propto K_0(R_s)$ is clearly preferred over the others. The worst form, where $\rho_{\rm bar} (R_s) \propto \exp(R_s^{-n})$, was entirely removed with $P({\rm Bar}) = 0$ and the long bar component taking over the fit in the central region.

In order to estimate the kick effects on the Galactic bulge we used a linear fit to our simulation results of the form
\begin{equation}
\theta_{{\rm kicked},i}=\alpha_i \theta_i +\beta_i
\label{eq:simulation_parameters}
\end{equation}
where $\theta_{{\rm kicked},i}$ are the bulge parameters, $x_b$, $y_b$, $z_b$, $C_\perp$, and $C_\parallel$ for the kicked distribution  and 
$\theta_i$ are the corresponding parameters in the non-kicked case. The $\alpha_i$ and $\beta_i$ were found by performing a least squared fit for the values given in Tables~\ref{tab:MWa_params}, \ref{tab:MWb_params}, and \ref{tab:MWc0.8_params}. The results are shown in Table~\ref{tab:simulation_parameters} and Fig.~\ref{fig:simulation_parameters}. A prediction for the Milky Way bulge parameters found in ref.~\cite{Cao:2013dwa} are shown in Table~\ref{tab:milky_way_prediction}. The predicted line of sight contours for the kicked and unkicked Milky Way bulge are shown in Fig.~\ref{fig:los_prediction}.

\begin{table}
\begin{center}
    \begin{tabular}{r||c|c}
          & Kick At Beginning & Uniform Kick Rate \\ \hline \hline
         
         MWa & $93\pm10$ & $85\pm9$ \\ \hline
         MWb & $97\pm10$ & $84\pm10$ \\ \hline
         MWc0.8 & $95\pm10$ & $84\pm9$ \\ \hline
    \end{tabular}
    \caption[Kick velocity Maxwell distribution parameters that produce a peculiar velocity distribution consistent with resolved gamma-ray MSPs.]{ Kick velocity Maxwell distribution (Eq. (\ref{eq:Maxwell})) parameters ($\sigma_k$ in km/s)  that  produce a peculiar velocity distribution where $\sigma_v = 77\pm6$~km/s. }
    \label{tab:kick_velocity}
    \end{center}
\end{table}

\begin{table}
\begin{center}
    \begin{tabular}{r||c|c|c}
         Parameter & No Kick & Kick At Beginning & Uniform Kick Rate \\ \hline \hline
         
         $P({\rm Disk})$ & $0.499\substack{+0.006 \\ -0.005}$ & $0.418\substack{+0.006 \\ -0.006}$ & $0.365\substack{+0.006 \\ -0.004}$ \\ \hline
         $P({\rm Bar})$ & $0.354\substack{+0.004 \\ -0.005}$ & $0.3220\substack{+0.0019 \\ -0.0021}$ & $0.364\substack{+0.005 \\ -0.007}$ \\ \hline
         $P({\rm Hernquist})$ & $0.0145\substack{+0.0013 \\ -0.0018}$ & $0.059\substack{+0.003 \\ -0.002}$ & $0.0647\substack{+0.0022 \\ -0.0024}$ \\ \hline
         $P({\rm Long~Bar})$ & $0.1323\substack{+0.0025 \\ -0.0025}$ & $0.201\substack{+0.007 \\ -0.007}$ & $0.206\substack{+0.007 \\ -0.010}$ \\ \hline
         $\sigma_r$ (kpc) & $4.92\substack{+0.04 \\ -0.03}$ & $5.66\substack{+0.05 \\ -0.04}$ & $6.07\substack{+0.05 \\ -0.08}$ \\ \hline
         $z_0$ (kpc) & $0.2238\substack{+0.0008 \\ -0.0027}$ & $1.203\substack{+0.023 \\ -0.013}$ & $1.222\substack{+0.018 \\ -0.004}$ \\ \hline
         $O_R$ (kpc) & $2.82\substack{+0.08 \\ -0.11}$ & $-$ & $-$ \\ \hline
         $O_N$ & $4.3\substack{+0.3 \\ -0.2}$ & $-$ & $-$ \\ \hline
         $\epsilon$ & $0.780\substack{+0.020 \\ -0.013}$ & $-$ & $-$ \\ \hline
         $a_b$ (kpc) & $0.205\substack{+0.009 \\ -0.010}$ & $0.509\substack{+0.019 \\ -0.010}$ & $0.555\substack{+0.022 \\ -0.021}$ \\ \hline
         $C_\perp$ & $1.84\substack{+0.02 \\ -0.05}$ & $1.831\substack{+0.013 \\ -0.015}$ & $1.821\substack{+0.017 \\ -0.015}$ \\ \hline
         $C_\parallel$ & $3.08\substack{+0.11 \\ -0.11}$ & $2.49\substack{+0.03 \\ -0.03}$ & $2.542\substack{+0.019 \\ -0.019}$ \\ \hline
         $x_b$ (kpc) & $0.557\substack{+0.008 \\ -0.008}$ & $0.685\substack{+0.005 \\ -0.005}$ & $0.673\substack{+0.007 \\ -0.011}$ \\ \hline
         $y_b$ (kpc) & $0.367\substack{+0.006 \\ -0.003}$ & $0.455\substack{+0.003 \\ -0.003}$ & $0.445\substack{+0.006 \\ -0.011}$ \\ \hline
         $z_b$ (kpc) & $0.2557\substack{+0.0014 \\ -0.0010}$ & $0.3328\substack{+0.0014 \\ -0.0015}$ & $0.3155\substack{+0.0026 \\ -0.0029}$ \\ \hline
         $R_{\rm end}$ (kpc) & $2.00\substack{+0.05 \\ -0.06}$ & $4.6\substack{+0.4 \\ -0.3}$ & $4.9\substack{+0.6 \\ -0.6}$ \\ \hline
         $\Delta \alpha$ ($\deg$) & $-$ & $-0.34\substack{+0.15 \\ -0.14}$ & $-0.07\substack{+0.13 \\ -0.13}$ \\ \hline
         $\Delta x_{\rm center}$ (pc) & $-$ & $0.8\substack{+1.0 \\ -1.1}$ & $-0.7\substack{+1.0 \\ -1.0}$ \\ \hline
         $\Delta y_{\rm center}$ (pc) & $-$ & $-1.8\substack{+0.9 \\ -0.9}$ & $-0.3\substack{+0.7 \\ -0.7}$ \\ \hline
         $\Delta z_{\rm center}$ (pc) & $-$ & $-0.9\substack{+0.6 \\ -0.5}$ & $2.7\substack{+0.4 \\ -0.5}$ \\ \hline
         $x_{\rm lb}$ (kpc) & $5\substack{+5 \\ -1}$ & $2.78\substack{+0.03 \\ -0.03}$ & $2.46\substack{+0.04 \\ -0.03}$ \\ \hline
         $y_{\rm lb}$ (kpc) & $1.43\substack{+0.16 \\ -0.07}$ & $2.78\substack{+0.04 \\ -0.04}$ & $2.49\substack{+0.05 \\ -0.04}$ \\ \hline
         $z_{\rm lb}$ (kpc) & $0.315\substack{+0.029 \\ -0.010}$ & $0.541\substack{+0.006 \\ -0.011}$ & $0.495\substack{+0.010 \\ -0.014}$ \\ \hline
         $C_{\perp,\rm lb}$ & $0.88\substack{+0.04 \\ -0.11}$ & $1.764\substack{+0.026 \\ -0.025}$ & $1.983\substack{+0.028 \\ -0.026}$ \\ \hline
         $R_{\rm out}$ (kpc) & $2.86\substack{+0.04 \\ -0.05}$ & $7.46\substack{+0.06 \\ -0.04}$ & $8.37\substack{+0.06 \\ -0.06}$ \\ \hline
         $R_{\rm in}$ (kpc) & $1.676\substack{+0.018 \\ -0.016}$ & $2.09\substack{+0.03 \\ -0.04}$ & $2.03\substack{+0.07 \\ -0.12}$ \\ \hline
    \end{tabular}
    \caption[Model MWa fitted parameters.]{ Fitted parameters for model MWa. We have used the median of the MCMC chains for the central value and also included 68\% confidence intervals.  }
    \label{tab:MWa_params}
    \end{center}
\end{table}

\begin{table}
\begin{center}
    \begin{tabular}{r||c|c|c}
         Parameter & No Kick & Kick At Beginning & Uniform Kick Rate \\ \hline \hline
         
         $P({\rm Disk})$ & $0.506\substack{+0.010 \\ -0.005}$ & $0.6144\substack{+0.0024 \\ -0.0026}$ & $0.36\substack{+0.03 \\ -0.04}$ \\ \hline
         $P({\rm Bar})$ & $0.322\substack{+0.005 \\ -0.003}$ & $0.244\substack{+0.007 \\ -0.003}$ & $0.317\substack{+0.006 \\ -0.003}$ \\ \hline
         $P({\rm Hernquist})$ & $0.0140\substack{+0.0012 \\ -0.0009}$ & $0.0470\substack{+0.0022 \\ -0.0021}$ & $0.0819\substack{+0.0017 \\ -0.0016}$ \\ \hline
         $P({\rm Long~Bar})$ & $0.154\substack{+0.012 \\ -0.010}$ & $0.094\substack{+0.004 \\ -0.006}$ & $0.24\substack{+0.04 \\ -0.04}$ \\ \hline
         $\sigma_r$ (kpc) & $5.64\substack{+0.06 \\ -0.07}$ & $5.36\substack{+0.03 \\ -0.03}$ & $6.9\substack{+0.7 \\ -0.3}$ \\ \hline
         $z_0$ (kpc) & $0.2249\substack{+0.0003 \\ -0.0003}$ & $1.116\substack{+0.004 \\ -0.014}$ & $1.30\substack{+0.05 \\ -0.06}$ \\ \hline
         $O_R$ (kpc) & $3.19\substack{+0.05 \\ -0.09}$ & $-$ & $-$ \\ \hline
         $O_N$ & $4.25\substack{+0.09 \\ -0.17}$ & $-$ & $-$ \\ \hline
         $\epsilon$ & $0.762\substack{+0.019 \\ -0.019}$ & $-$ & $-$ \\ \hline
         $a_b$ (kpc) & $0.202\substack{+0.004 \\ -0.005}$ & $0.585\substack{+0.023 \\ -0.025}$ & $0.614\substack{+0.014 \\ -0.011}$ \\ \hline
         $C_\perp$ & $2.04\substack{+0.08 \\ -0.10}$ & $2.02\substack{+0.03 \\ -0.03}$ & $1.873\substack{+0.016 \\ -0.017}$ \\ \hline
         $C_\parallel$ & $3.85\substack{+0.14 \\ -0.08}$ & $3.06\substack{+0.05 \\ -0.05}$ & $2.90\substack{+0.04 \\ -0.04}$ \\ \hline
         $x_b$ (kpc) & $0.548\substack{+0.029 \\ -0.009}$ & $0.587\substack{+0.026 \\ -0.009}$ & $0.710\substack{+0.010 \\ -0.006}$ \\ \hline
         $y_b$ (kpc) & $0.339\substack{+0.006 \\ -0.005}$ & $0.406\substack{+0.013 \\ -0.005}$ & $0.411\substack{+0.010 \\ -0.005}$ \\ \hline
         $z_b$ (kpc) & $0.243\substack{+0.004 \\ -0.002}$ & $0.306\substack{+0.007 \\ -0.003}$ & $0.303\substack{+0.004 \\ -0.002}$ \\ \hline
         $R_{\rm end}$ (kpc) & $2.53\substack{+0.10 \\ -0.15}$ & $3.37\substack{+0.10 \\ -0.08}$ & $5.2\substack{+0.6 \\ -0.4}$ \\ \hline
         $\Delta \alpha$ ($\deg$) & $-$ & $-0.04\substack{+0.13 \\ -0.13}$ & $0.30\substack{+0.11 \\ -0.12}$ \\ \hline
         $\Delta x_{\rm center}$ (pc) & $-$ & $4.4\substack{+1.2 \\ -1.2}$ & $2.5\substack{+1.2 \\ -1.3}$ \\ \hline
         $\Delta y_{\rm center}$ (pc) & $-$ & $1.8\substack{+0.9 \\ -0.8}$ & $-1.2\substack{+0.9 \\ -0.8}$ \\ \hline
         $\Delta z_{\rm center}$ (pc) & $-$ & $-0.6\substack{+0.7 \\ -0.8}$ & $1.4\substack{+0.5 \\ -0.5}$ \\ \hline
         $x_{\rm lb}$ (kpc) & $2.7\substack{+0.8 \\ -0.3}$ & $2.1\substack{+0.3 \\ -0.1}$ & $2.78\substack{+0.04 \\ -0.02}$ \\ \hline
         $y_{\rm lb}$ (kpc) & $1.14\substack{+0.17 \\ -0.05}$ & $0.93\substack{+0.08 \\ -0.03}$ & $2.65\substack{+0.07 \\ -0.04}$ \\ \hline
         $z_{\rm lb}$ (kpc) & $0.414\substack{+0.006 \\ -0.016}$ & $0.656\substack{+0.011 \\ -0.020}$ & $0.56\substack{+0.04 \\ -0.03}$ \\ \hline
         $C_{\perp,\rm lb}$ & $0.97\substack{+0.03 \\ -0.11}$ & $1.61\substack{+0.04 \\ -0.04}$ & $1.86\substack{+0.03 \\ -0.03}$ \\ \hline
         $R_{\rm out}$ (kpc) & $3.66\substack{+0.22 \\ -0.11}$ & $4.67\substack{+0.05 \\ -0.05}$ & $9.32\substack{+0.07 \\ -0.25}$ \\ \hline
         $R_{\rm in}$ (kpc) & $1.69\substack{+0.04 \\ -0.03}$ & $1.78\substack{+0.04 \\ -0.03}$ & $1.71\substack{+0.10 \\ -0.06}$ \\ \hline
    \end{tabular}
    \caption[Model MWb fitted parameters.]{ Fitted parameters for model MWb. }
    \label{tab:MWb_params}
    \end{center}
\end{table}

\begin{table}
\begin{center}
    \begin{tabular}{r||c|c|c}
         Parameter & No Kick & Kick At Beginning & Uniform Kick Rate \\ \hline \hline
         
         $P({\rm Disk})$ & $0.555\substack{+0.003 \\ -0.009}$ & $0.467\substack{+0.012 \\ -0.010}$ & $0.357\substack{+0.013 \\ -0.003}$ \\ \hline
         $P({\rm Bar})$ & $0.318\substack{+0.004 \\ -0.005}$ & $0.262\substack{+0.003 \\ -0.005}$ & $0.305\substack{+0.003 \\ -0.002}$ \\ \hline
         $P({\rm Hernquist})$ & $0.0099\substack{+0.0006 \\ -0.0005}$ & $0.0603\substack{+0.0021 \\ -0.0025}$ & $0.0619\substack{+0.0019 \\ -0.0020}$ \\ \hline
         $P({\rm Long~Bar})$ & $0.116\substack{+0.014 \\ -0.004}$ & $0.210\substack{+0.014 \\ -0.010}$ & $0.277\substack{+0.005 \\ -0.017}$ \\ \hline
         $\sigma_r$ (kpc) & $5.49\substack{+0.03 \\ -0.04}$ & $6.11\substack{+0.07 \\ -0.09}$ & $6.99\substack{+0.06 \\ -0.19}$ \\ \hline
         $z_0$ (kpc) & $0.2246\substack{+0.0007 \\ -0.0022}$ & $1.266\substack{+0.016 \\ -0.028}$ & $1.262\substack{+0.026 \\ -0.022}$ \\ \hline
         $O_R$ (kpc) & $2.5\substack{+0.4 \\ -0.16}$ & $-$ & $-$ \\ \hline
         $O_N$ & $4.9\substack{+2.6 \\ -1.1}$ & $-$ & $-$ \\ \hline
         $\epsilon$ & $0.82\substack{+0.12 \\ -0.12}$ & $-$ & $-$ \\ \hline
         $a_b$ (kpc) & $0.220\substack{+0.006 \\ -0.006}$ & $0.72\substack{+0.01 \\ -0.04}$ & $0.74\substack{+0.03 \\ -0.03}$ \\ \hline
         $C_\perp$ & $1.88\substack{+0.03 \\ -0.03}$ & $1.861\substack{+0.020 \\ -0.020}$ & $1.909\substack{+0.024 \\ -0.022}$ \\ \hline
         $C_\parallel$ & $3.37\substack{+0.05 \\ -0.05}$ & $2.69\substack{+0.03 \\ -0.04}$ & $2.69\substack{+0.03 \\ -0.03}$ \\ \hline
         $x_b$ (kpc) & $0.551\substack{+0.012 \\ -0.014}$ & $0.657\substack{+0.006 \\ -0.007}$ & $0.620\substack{+0.006 \\ -0.005}$ \\ \hline
         $y_b$ (kpc) & $0.342\substack{+0.007 \\ -0.008}$ & $0.434\substack{+0.005 \\ -0.007}$ & $0.378\substack{+0.005 \\ -0.003}$ \\ \hline
         $z_b$ (kpc) & $0.2395\substack{+0.0015 \\ -0.0018}$ & $0.3217\substack{+0.0023 \\ -0.0023}$ & $0.2845\substack{+0.0025 \\ -0.0018}$ \\ \hline
         $R_{\rm end}$ (kpc) & $2.03\substack{+0.07 \\ -0.10}$ & $4.81\substack{+0.29 \\ -0.25}$ & $5.5\substack{+2.1 \\ -0.9}$ \\ \hline
         $\Delta \alpha$ ($\deg$) & $-$ & $-0.79\substack{+0.18 \\ -0.18}$ & $0.32\substack{+0.13 \\ -0.13}$ \\ \hline
         $\Delta x_{\rm center}$ (pc) & $-$ & $1.4\substack{+1.4 \\ -1.3}$ & $1.0\substack{+1.2 \\ -1.1}$ \\ \hline
         $\Delta y_{\rm center}$ (pc) & $-$ & $-0.2\substack{+0.9 \\ -0.9}$ & $0.5\substack{+0.8 \\ -0.8}$ \\ \hline
         $\Delta z_{\rm center}$ (pc) & $-$ & $2.4\substack{+0.7 \\ -0.6}$ & $-0.5\substack{+0.5 \\ -0.5}$ \\ \hline
         $x_{\rm lb}$ (kpc) & $3\substack{+7 \\ -1}$ & $2.533\substack{+0.023 \\ -0.022}$ & $2.533\substack{+0.020 \\ -0.021}$ \\ \hline
         $y_{\rm lb}$ (kpc) & $1.5\substack{+0.4 \\ -0.2}$ & $2.584\substack{+0.027 \\ -0.025}$ & $2.417\substack{+0.021 \\ -0.022}$ \\ \hline
         $z_{\rm lb}$ (kpc) & $0.335\substack{+0.006 \\ -0.005}$ & $0.564\substack{+0.019 \\ -0.006}$ & $0.530\substack{+0.010 \\ -0.011}$ \\ \hline
         $C_{\perp,\rm lb}$ & $0.95\substack{+0.07 \\ -0.11}$ & $1.964\substack{+0.029 \\ -0.029}$ & $1.954\substack{+0.022 \\ -0.022}$ \\ \hline
         $R_{\rm out}$ (kpc) & $3.0\substack{+0.1 \\ -0.4}$ & $7.88\substack{+0.07 \\ -0.09}$ & $9.13\substack{+0.13 \\ -0.08}$ \\ \hline
         $R_{\rm in}$ (kpc) & $1.61\substack{+0.03 \\ -0.06}$ & $1.91\substack{+0.05 \\ -0.07}$ & $1.60\substack{+0.05 \\ -0.03}$ \\ \hline
    \end{tabular}
    \caption[Model MWc0.8 fitted parameters.]{ Fitted parameters for model MWc0.8. }
    \label{tab:MWc0.8_params}
    \end{center}
\end{table}

\begin{figure}
    \centering
    \includegraphics[width=0.99\linewidth]{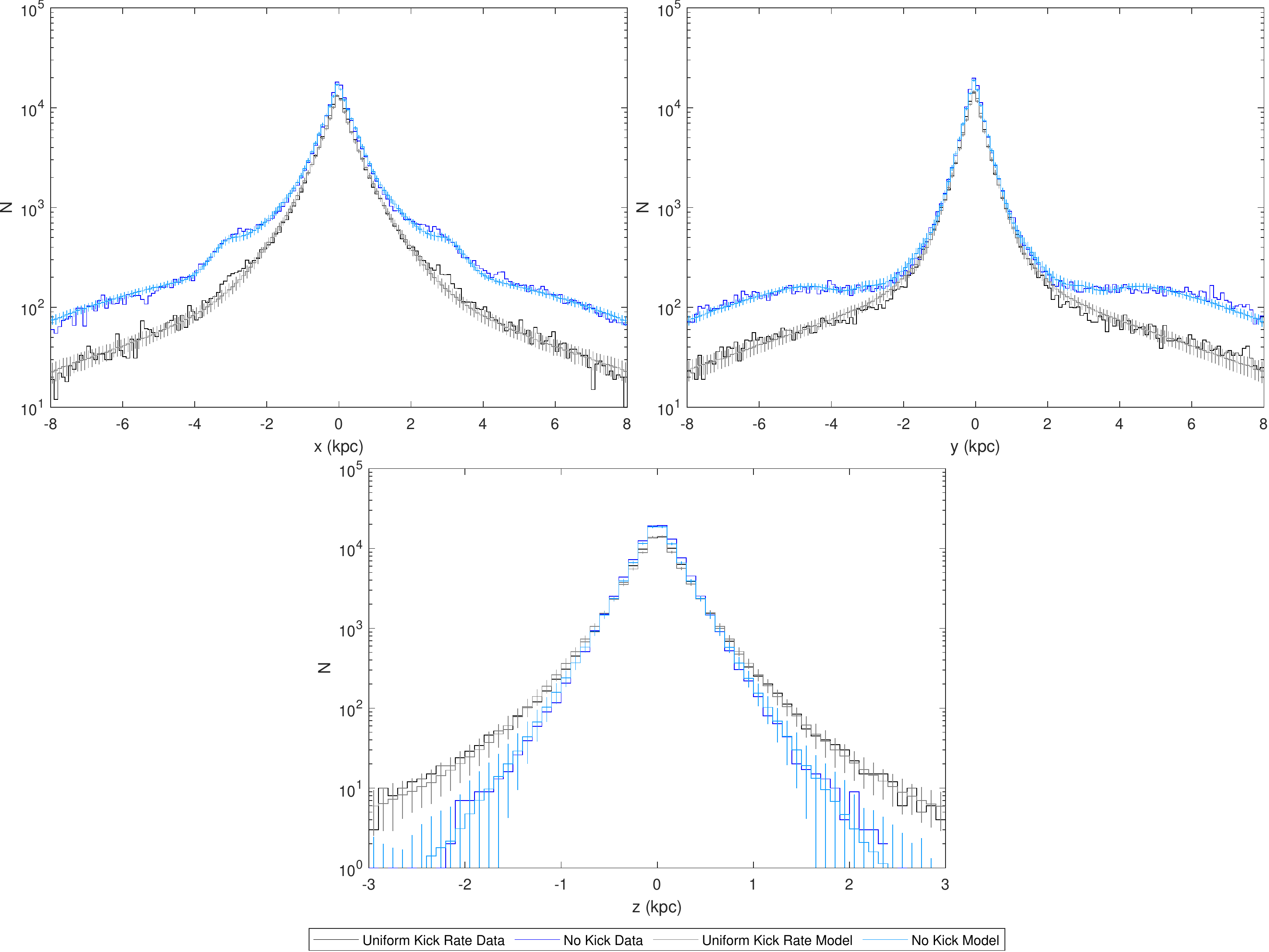}
    \caption[MWa $x$, $y$ and $z$ profiles with a uniform kick rate.]{MWa profile along $x$, $y$ and $z$ axes with a uniform kick rate. We show both $N$-body simulation data and data simulated using the fitted model. For the fitted model we show the mean number of particles in each bin and the standard deviation. }
    \label{fig:MWa_1d_profiles_uniform_kick_rate}
\end{figure}

\begin{figure}
    \centering
    \includegraphics[width=0.99\linewidth]{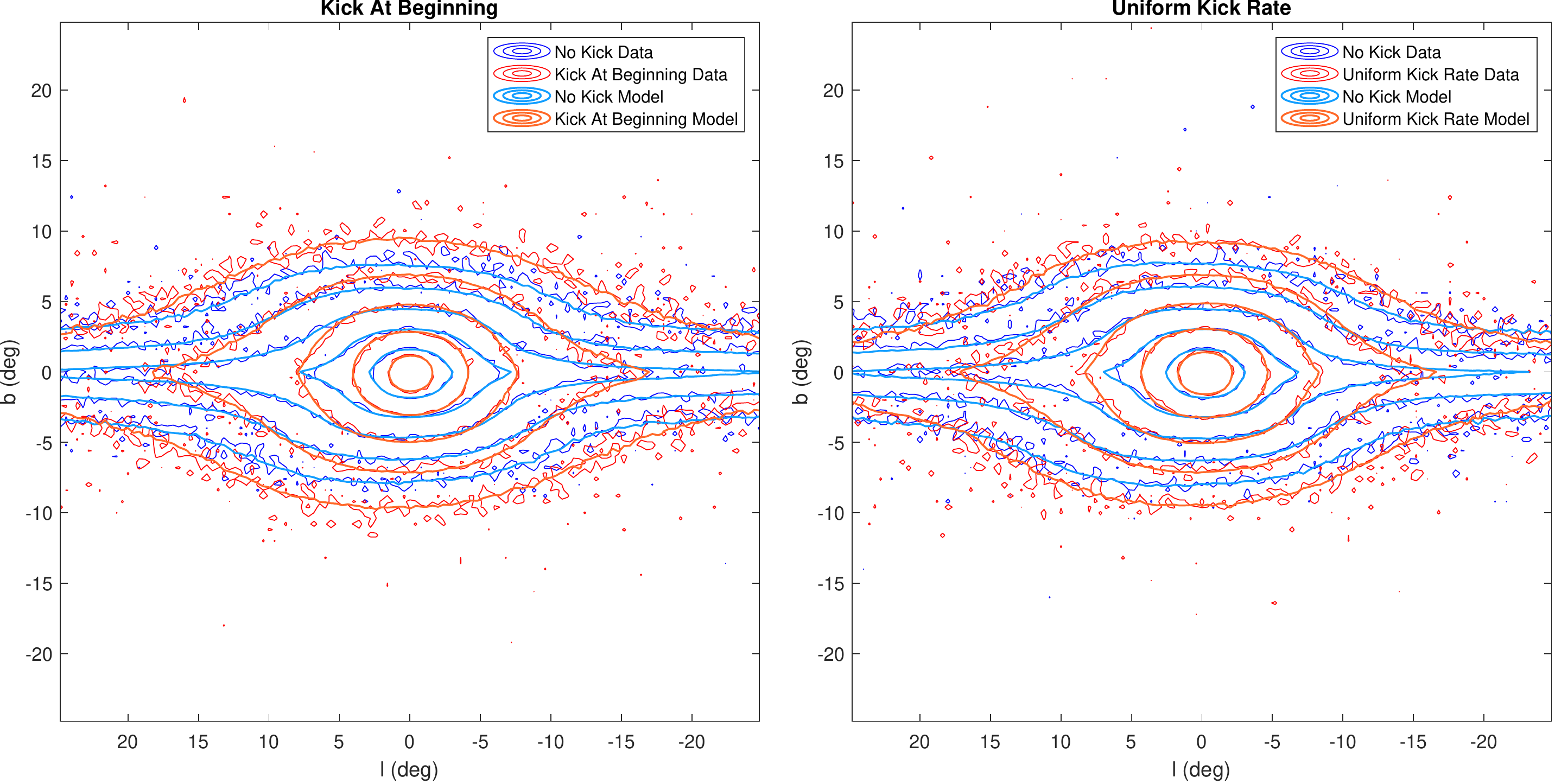}
    \caption[MWa flux distribution.]{MWa flux distribution in Galactic coordinates. The contours for each distribution are at $1$, $2$, $4$, $8$ and $16$ times the mean in this region. The Sun is placed at a distance of $7.9$ kpc, at an angle relative to the bar of $20^\circ$ and at a height of $15$ pc. }
    \label{fig:MWa_data_los}
\end{figure}

\begin{table}
\begin{center}
    \begin{tabular}{c||c|c}
          $\rho_{\rm bar} (R_s)$ & Mean $-2(\Delta \log(L))$ & Standard Deviation \\ \hline \hline
         
          $K_0(R_s)$ & $0$ & $739$ \\ \hline
          $\exp(-R_s)$ & $3581$ & $1606$ \\ \hline
          $\sech^2(R_s)$ & $12795$ & $1985$ \\ \hline
          $\exp(-0.5 R_s^2)$ & $19369$ & $391$ \\ \hline
          $(1 + R_s^n)^{-1}$ & $20403$ & $953$ \\ \hline
          $\exp(R_s^{-n})$ & $57114$ & $3152$ \\ \hline
    \end{tabular}
    \caption[Change in $-2 \log(L)$ using different bar models.]{ Change in mean $-2 \log(L)$ using different bar models $\rho_{\rm bar} (R_s)$. The mean was taken over all the samples in the MCMC chains.}
    \label{tab:alternative_bar}
    \end{center}
\end{table}

\begin{table}
\begin{center}
    \begin{tabular}{r||c|c}
         Parameter & $\alpha_i$ & $\beta_i$ \\ \hline \hline
         
         $x_b, y_b, z_b$ & 1.13$\pm$0.05 &  0.03$\pm$0.02 \\ \hline
         $C_\perp,C_{\parallel}$& 0.56$\pm$0.02 &0.81$\pm$0.05\\
          \hline
    \end{tabular}
\caption[Least square fit values for simulated bar parameters.]{ Least square fit values with 68\% confidence intervals for Eq.~\ref{eq:simulation_parameters} fitted to the points shown in Figure~\ref{fig:simulation_parameters}. }
    \label{tab:simulation_parameters}
    \end{center}
\end{table}

\begin{table}
\begin{center}
    \begin{tabular}{r||c|c|c|c|c}
         Parameter & $x_b$ (kpc)   & $y_b$ (kpc)     & $z_b$ (kpc)   & $C_\perp$            & $C_{\parallel}$ \\ \hline \hline
         Not kicked & 0.67         &   0.29          &  0.27         &    2                 & 4        \\ \hline
           Kicked  & 0.79$\pm$0.02 & 0.36$\pm$0.01 & 0.35 $\pm$ 0.01 &     1.93$\pm$0.02 &  3.05$\pm$0.03       \\ 
          \hline
    \end{tabular}
\caption[Prediction of Milky Way bar kicked spatial distribution.]{ Predictions with 68\% confidence intervals for the kicked spatial distribution for the Milky Way bulge model found in ref.~\cite{Cao:2013dwa} using Eq.~\ref{eq:simulation_parameters} and the parameter values given in Table~\ref{tab:simulation_parameters}. }
    \label{tab:milky_way_prediction}
    \end{center}
\end{table}

\begin{figure}
    \centering
    \includegraphics[width=0.8\linewidth]{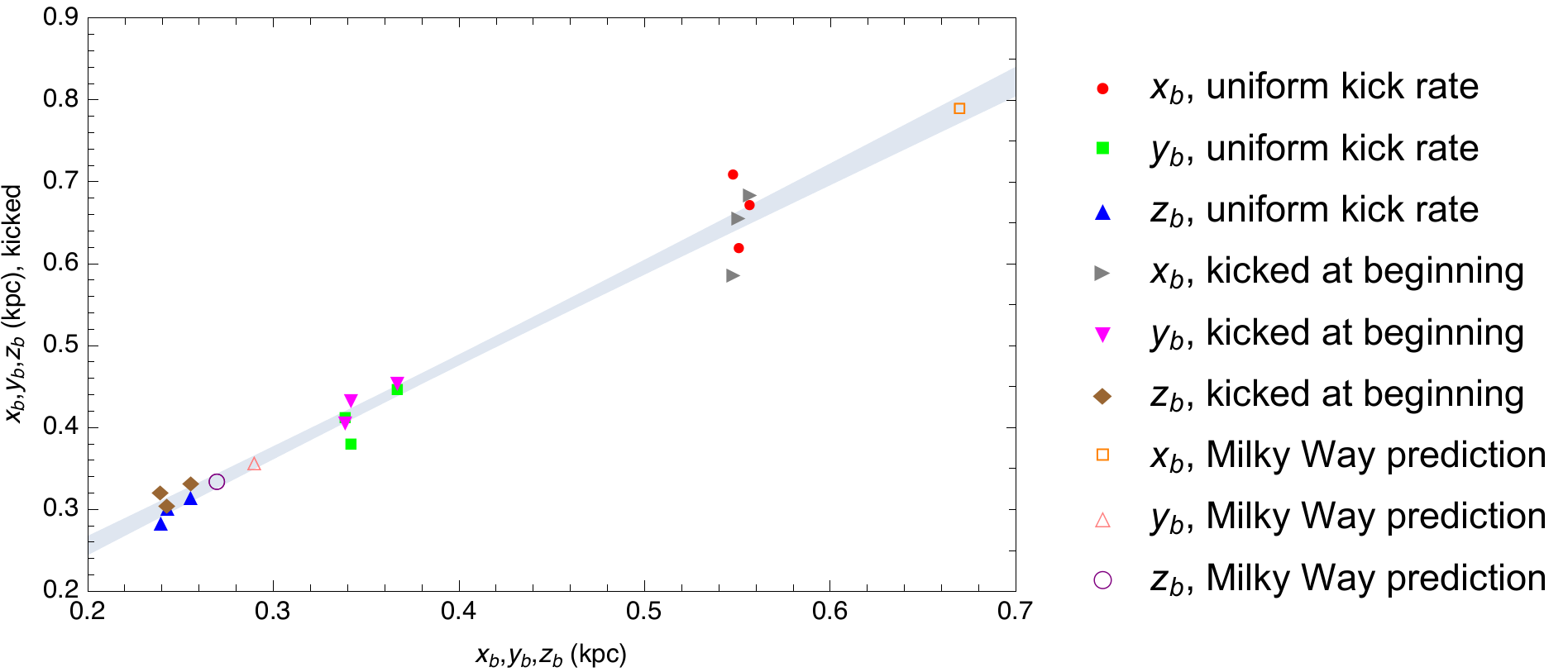}
    \includegraphics[width=0.8\linewidth]{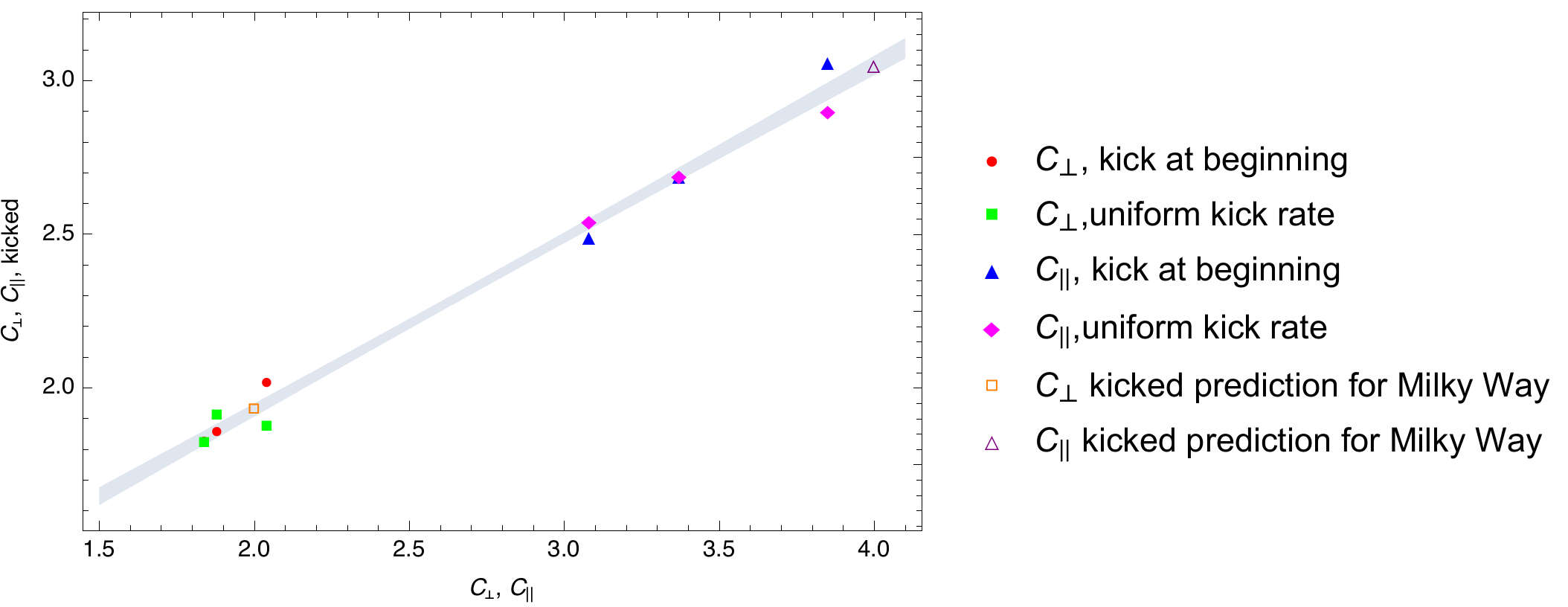}
    \caption[Simulated bar parameters with least square fit.]{Simulation parameters with 68\% confidence interval bands for straight line model fits.
    The closed symbol values are obtained from values given in Tables~\ref{tab:MWa_params}, \ref{tab:MWb_params}, and \ref{tab:MWc0.8_params}.
    The predictions for the ref.~\cite{Cao:2013dwa} model of the Milky Way Galaxy are given as open symbols. 
    \label{fig:simulation_parameters}}
\end{figure}

\begin{figure}
    \centering
    \includegraphics[width=0.8\linewidth]{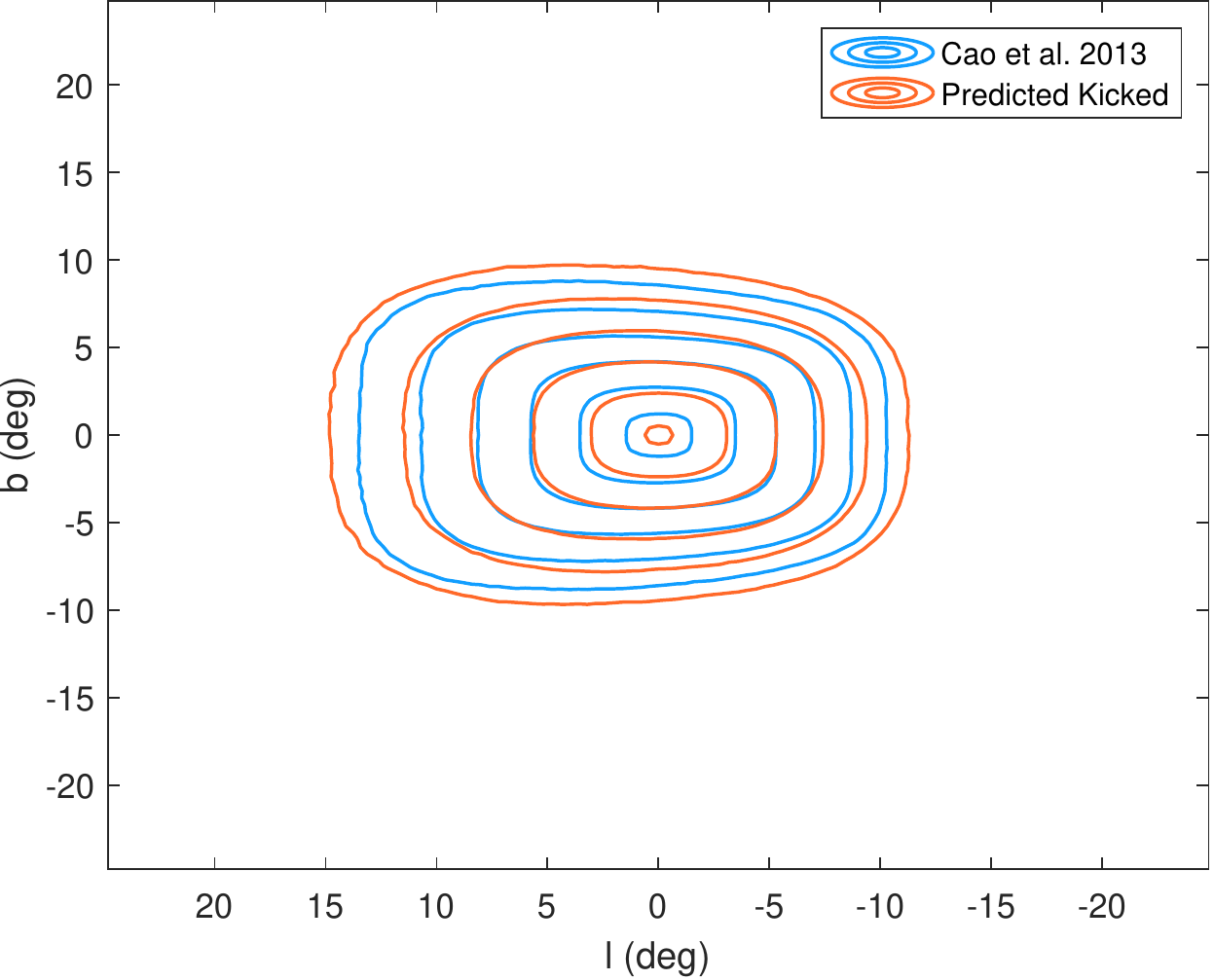}
    \caption[Line of sight contours for model of Galactic Center with predicted kicked version.]{Line of sight contours for model of the Milky Way given in ref.~\cite{Cao:2013dwa}. Both the model and its kicked version, obtained from the parameters in Table~\ref{tab:milky_way_prediction}, are shown. As in ref.~\cite{Cao:2013dwa} the Galactic Centre is taken to be 8.13 kpc away and the angle of the bulge to be 29$^\circ$. Contours are from 1 to 32 times the mean of the corresponding case given in steps of factors of 2.
    \label{fig:los_prediction}}
\end{figure}

\section{Discussion and Conclusions}
\label{conclusion}

Our goal was to investigate the effect of neutron star birth kicks on the distribution of MSPs in the Galactic Center.
We began by running $N$-body simulations with small populations of particles kicked with a range of scales in order to estimate the required Maxwellian kick to produce a peculiar velocity distribution similar to that of resolved gamma-ray MSPs. We then reran the simulations with a larger number of particles at the required kick velocity scale and used MCMC to fit the data with a model.

We used three models intended to approximate the Milky Way, these were the MWa, MWb and MWc0.8 models of Fujii et al.~\cite{Fujii2019}.
Our results were consistent with theirs as can be seen, for example, by comparing our Fig.~\ref{fig:rotation_curves}  to the top left hand panels of their Figs.~1, 2, and 3.
In Cao et al.~\cite{Cao:2013dwa} the bar scale lengths for a modified Bessel function of the second kind model fitted to red clump giant data are $0.67$, $0.29$ and $0.27$ for the $x$, $y$ and $z$ axes respectively, with the parameters $C_\parallel$ and $C_\perp$ fixed at $4$ and $2$. Our fits to $N$-body models without kicks find $(x_b, y_b, z_b)$ of $(0.56, 0.37, 0.26)$ for MWa, $(0.55, 0.34, 0.24)$ for MWb and $(0.55, 0.34, 0.24)$ for MWc0.8. We also have $C_\parallel$ between $3$ and $4$, producing a boxy structure in $x$-$z$ and $y$-$z$, this is visible in Fig.~\ref{fig:MWa_density_data_fit} for the MWa case and Figs.~\ref{fig:MWb_density_data_fit} and \ref{fig:MWc0.8_density_data_fit} for the MWb and
MWc0.8 cases. The other shape parameter $C_\perp$ was relatively close to $2$ in all cases, resulting in a more elliptical shape in $x$-$y$.
We found it was necessary to extend the bar structure using the long bar component given in Eq.~\ref{eq:long_bar}. Without it the bar scale parameters would be larger, but the bar would not be long enough to explain  the structure seen for $2\mbox{ kpc} \lesssim \abs{x} \lesssim 4$ kpc in Fig.~\ref{fig:MWa_1d_profiles_kick_at_beginning}
in the MWa, kicked at the beginning case, and Figs.~\ref{fig:MWa_1d_profiles_uniform_kick_rate}, \ref{fig:MWb_1d_profiles_kick_at_beginning}, \ref{fig:MWb_1d_profiles_uniform_kick_rate}, \ref{fig:MWc0.8_1d_profiles_kick_at_beginning} and \ref{fig:MWc0.8_1d_profiles_uniform_kick_rate} in the uniform kick rate, MWb and MWc0.8 cases.
We find the disk scale height to be $0.22$~kpc for the MWa, MWb, and MWc0.8 unkicked cases. This is at the lower end of the range of $220$ to $450$ pc given in Bland-Hawthorn and Gerhard \cite{Bland-Hawthorn2016}. There is also a small spherically symmetric Hernquist component $\sim1\%$ of the particles in the region of interest for the unkicked cases.

\begin{figure}
    \centering
    \includegraphics[width=0.99\linewidth]{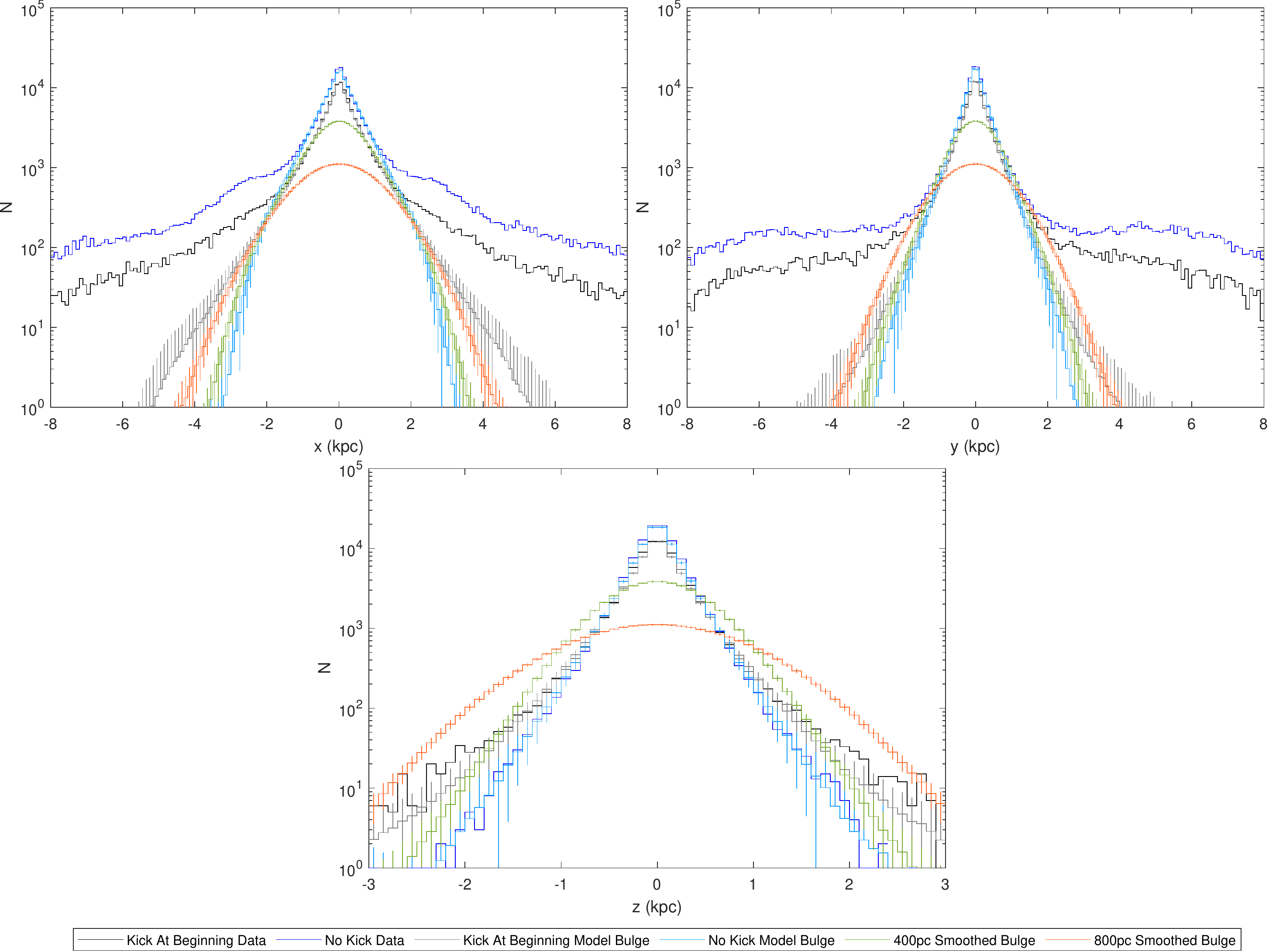}
    \caption[MWa $x$, $y$ and $z$ $400$ pc and $800$ pc smoothed fitted bulge profiles with kicks at beginning.]{MWa profile along $x$, $y$ and $z$ axes with kicks occurring at the beginning. Here we show the fitted bulge components as well as the no kick bulge smoothed with $400$ pc and $800$ pc Gaussians. We show both $N$-body simulation data and data simulated using the fitted model. For the fitted model we show the mean number of particles in each bin and the standard deviation. }
    \label{fig:MWa_bulge_smoothed_1d_profiles_kick_at_beginning}
\end{figure}

\begin{figure}
    \centering
    \includegraphics[width=0.99\linewidth]{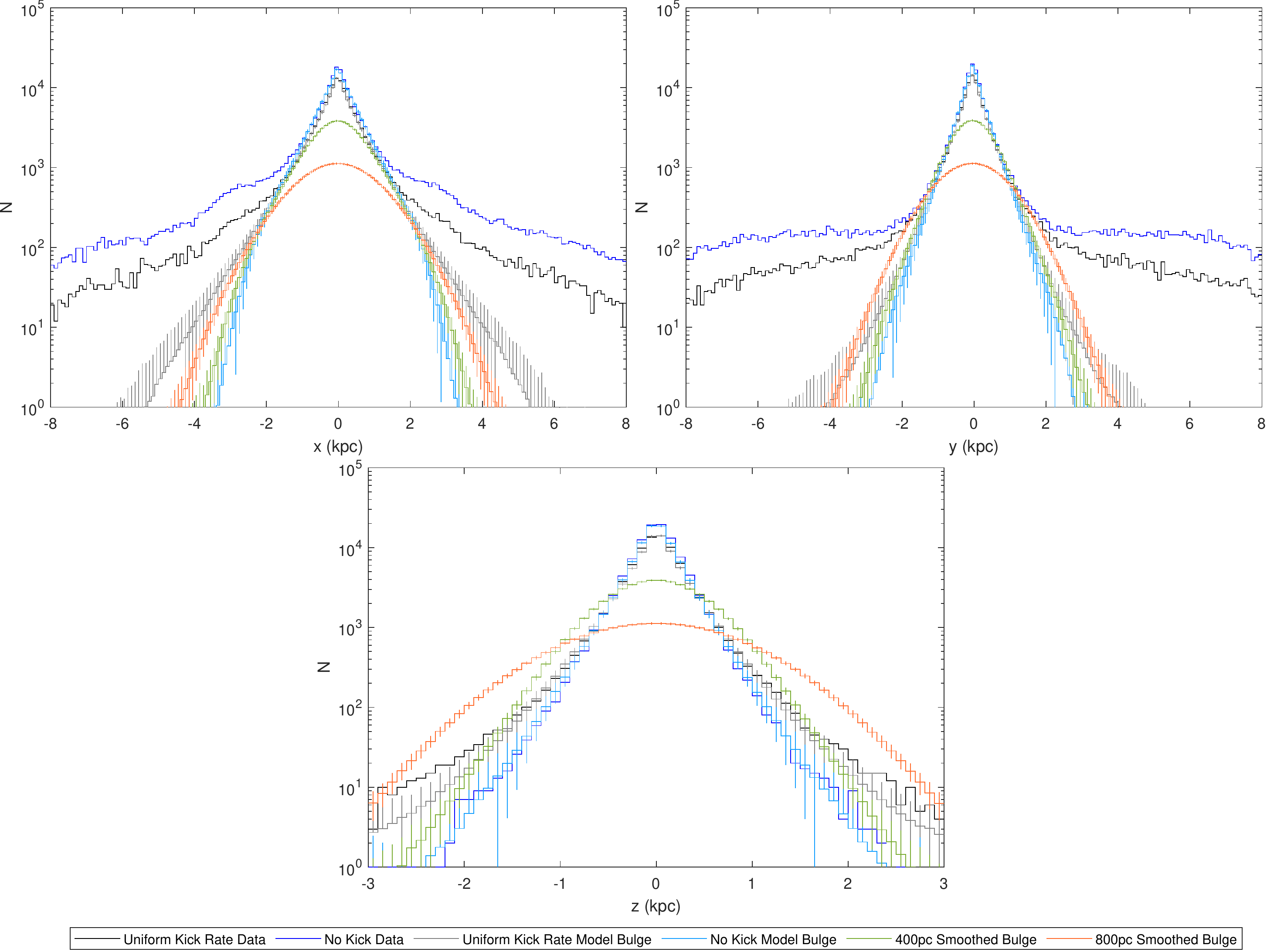}
    \caption[MWa $x$, $y$ and $z$ $400$ pc and $800$ pc smoothed fitted bulge profiles with a uniform kick rate.]{MWa profile along $x$, $y$ and $z$ axes with a uniform kick rate. Here we show the fitted bulge components as well as the no kick bulge smoothed with $400$ pc and $800$ pc Gaussians. We show both $N$-body simulation data and data simulated using the fitted model. For the fitted model we show the mean number of particles in each bin and the standard deviation. }
    \label{fig:MWa_bulge_smoothed_1d_profiles_uniform_kick_rate}
\end{figure}

\begin{figure}
    \centering
    \includegraphics[width=0.99\linewidth]{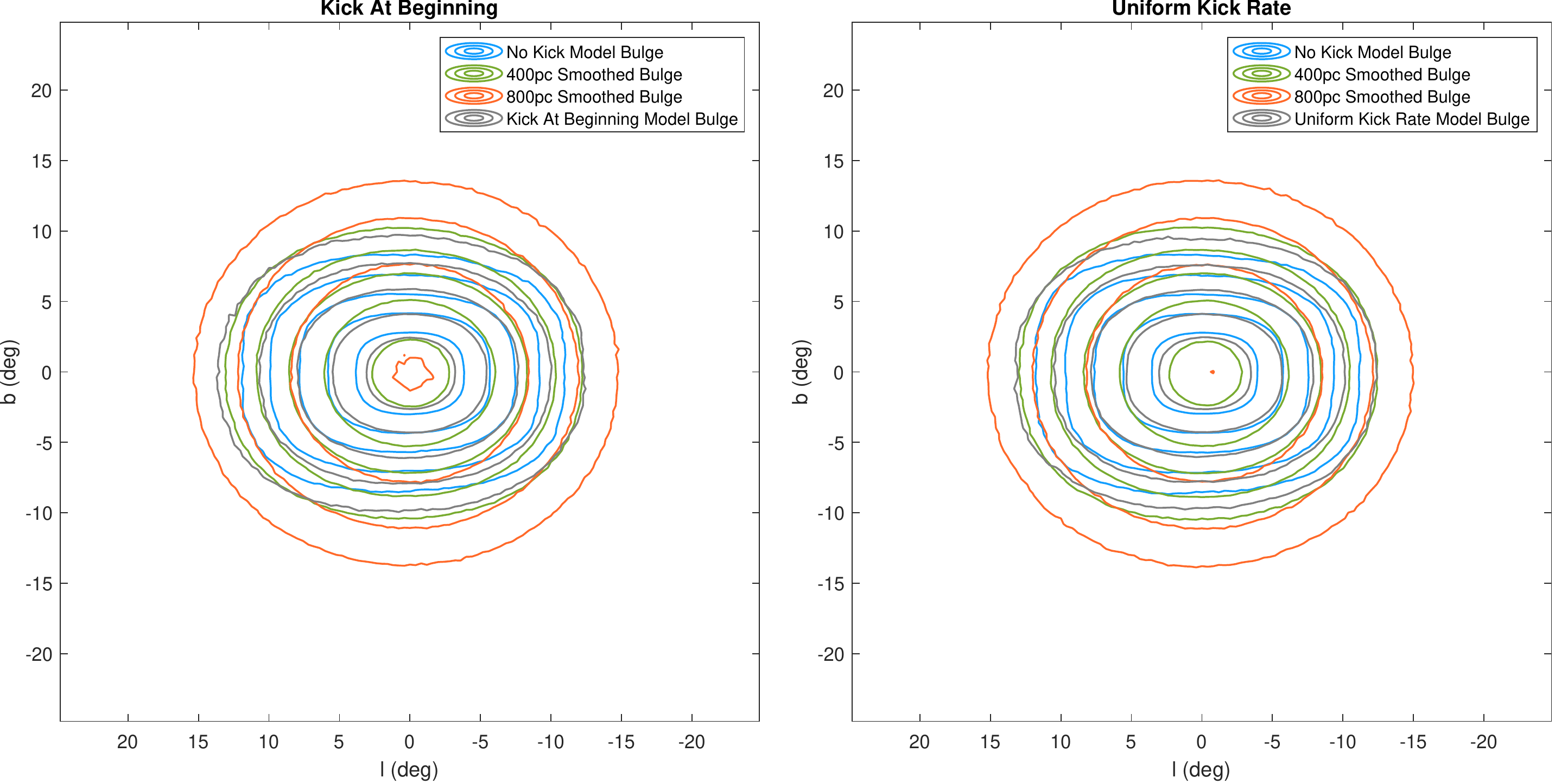}
    \caption[MWa $400$ pc and $800$ pc smoothed bulge flux distribution]{MWa bulge flux distribution in Galactic coordinates. We also show the no kick bulge smoothed with $400$ pc and $800$ pc Gaussians. The contours for each distribution are at $1$, $2$, $4$, $8$ and $16$ times the mean in this region. The Sun is placed at a distance of $7.9$ kpc, at an angle relative to the bar of $20^\circ$ and at a height of $15$ pc. }
    \label{fig:MWa_bulge_smoothed_los}
\end{figure}

\begin{figure}
    \centering
    \includegraphics[width=0.99\linewidth]{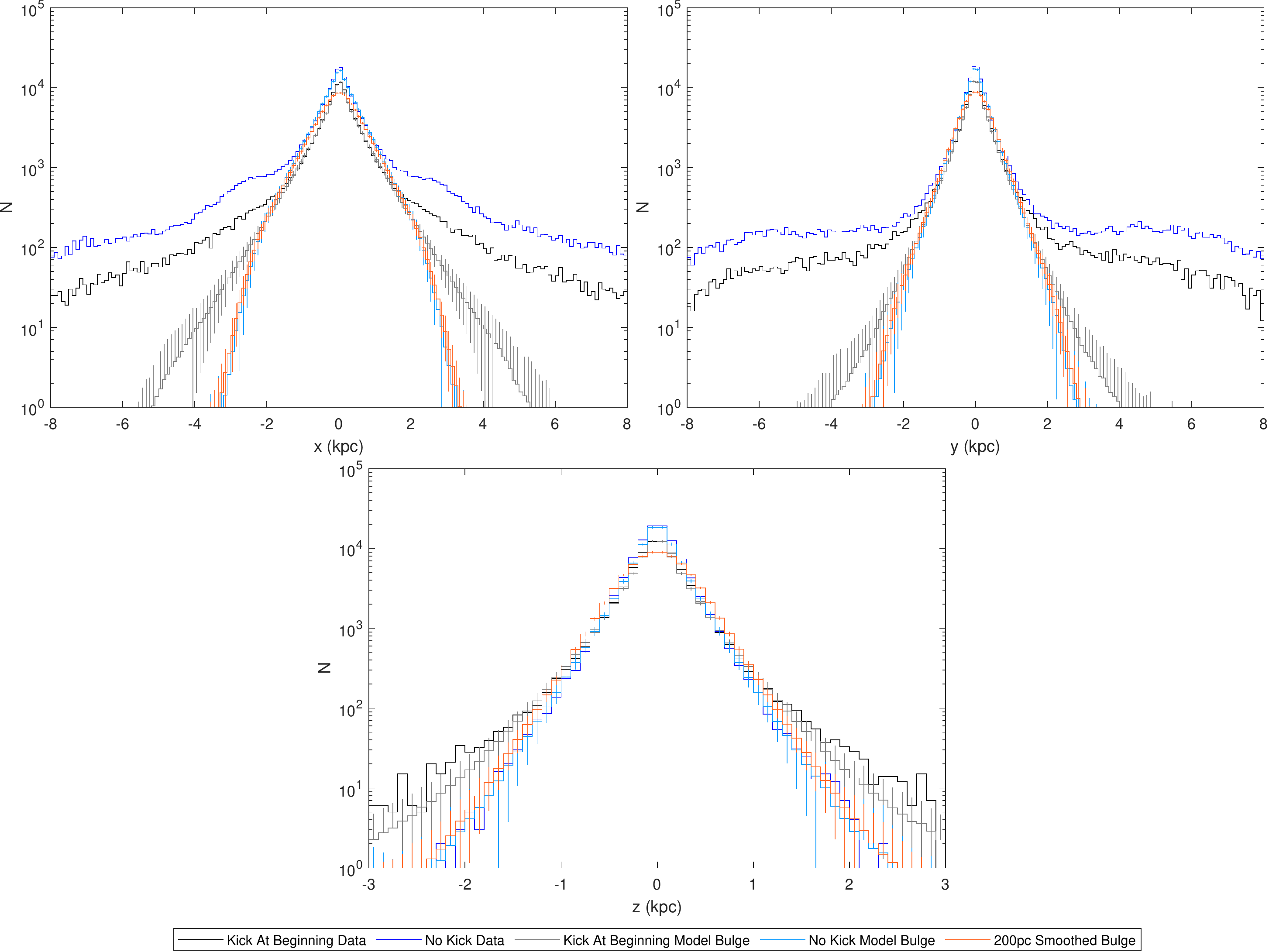}
    \caption[MWa $x$, $y$ and $z$ $200$ pc smoothed fitted bulge profiles with kicks at beginning.]{MWa profile along $x$, $y$ and $z$ axes with kicks occuring at the beginning. Here we show the fitted bulge components as well as the no kick bulge smoothed with a $200$ pc Gaussian. We show both $N$-body simulation data and data simulated using the fitted model. For the fitted model we show the mean number of particles in each bin and the standard deviation. }
    \label{fig:MWa_bulge_smoothed_200pc_1d_profiles_kick_at_beginning}
\end{figure}

\begin{figure}
    \centering
    \includegraphics[width=0.99\linewidth]{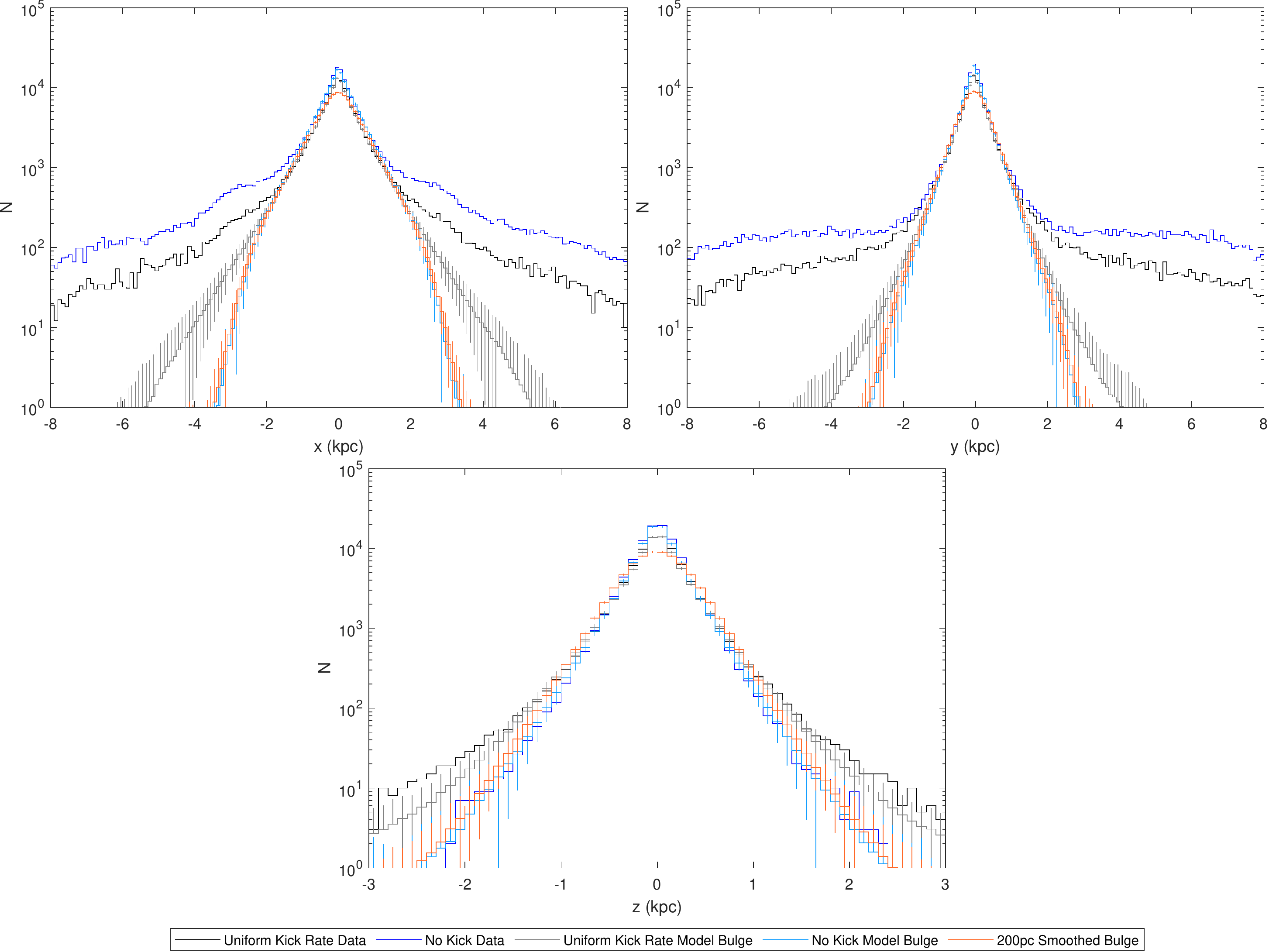}
    \caption[MWa $x$, $y$ and $z$ $200$ pc smoothed fitted bulge profiles with a uniform kick rate.]{MWa profile along $x$, $y$ and $z$ axes with a uniform kick rate. Here we show the fitted bulge components as well as the no kick bulge smoothed with a $200$ pc Gaussian. We show both $N$-body simulation data and data simulated using the fitted model. For the fitted model we show the mean number of particles in each bin and the standard deviation. }
    \label{fig:MWa_bulge_smoothed_200pc_1d_profiles_uniform_kick_rate}
\end{figure}

\begin{figure}
    \centering
    \includegraphics[width=0.99\linewidth]{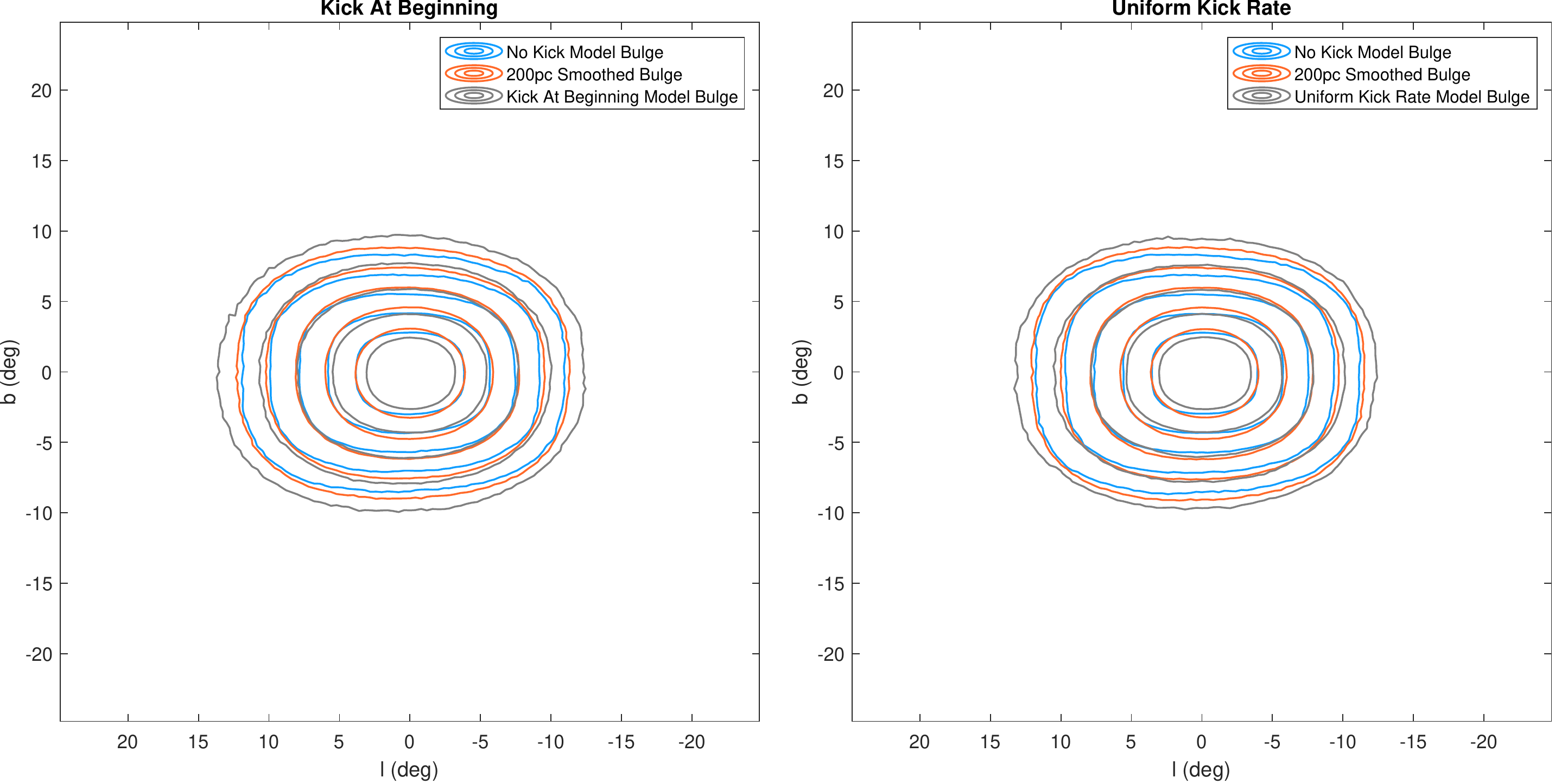}
    \caption[MWa $200$ pc smoothed bulge flux distribution]{MWa bulge flux distribution in Galactic coordinates. We also show the no kick bulge smoothed with a $200$ pc Gaussian. The contours for each distribution are at $1$, $2$, $4$, $8$ and $16$ times the mean in this region. The Sun is placed at a distance of $7.9$ kpc, at an angle relative to the bar of $20^\circ$ and at a height of $15$ pc. }
    \label{fig:MWa_bulge_smoothed_200pc_los}
\end{figure}

\begin{figure}
    \centering
    \includegraphics[width=0.99\linewidth]{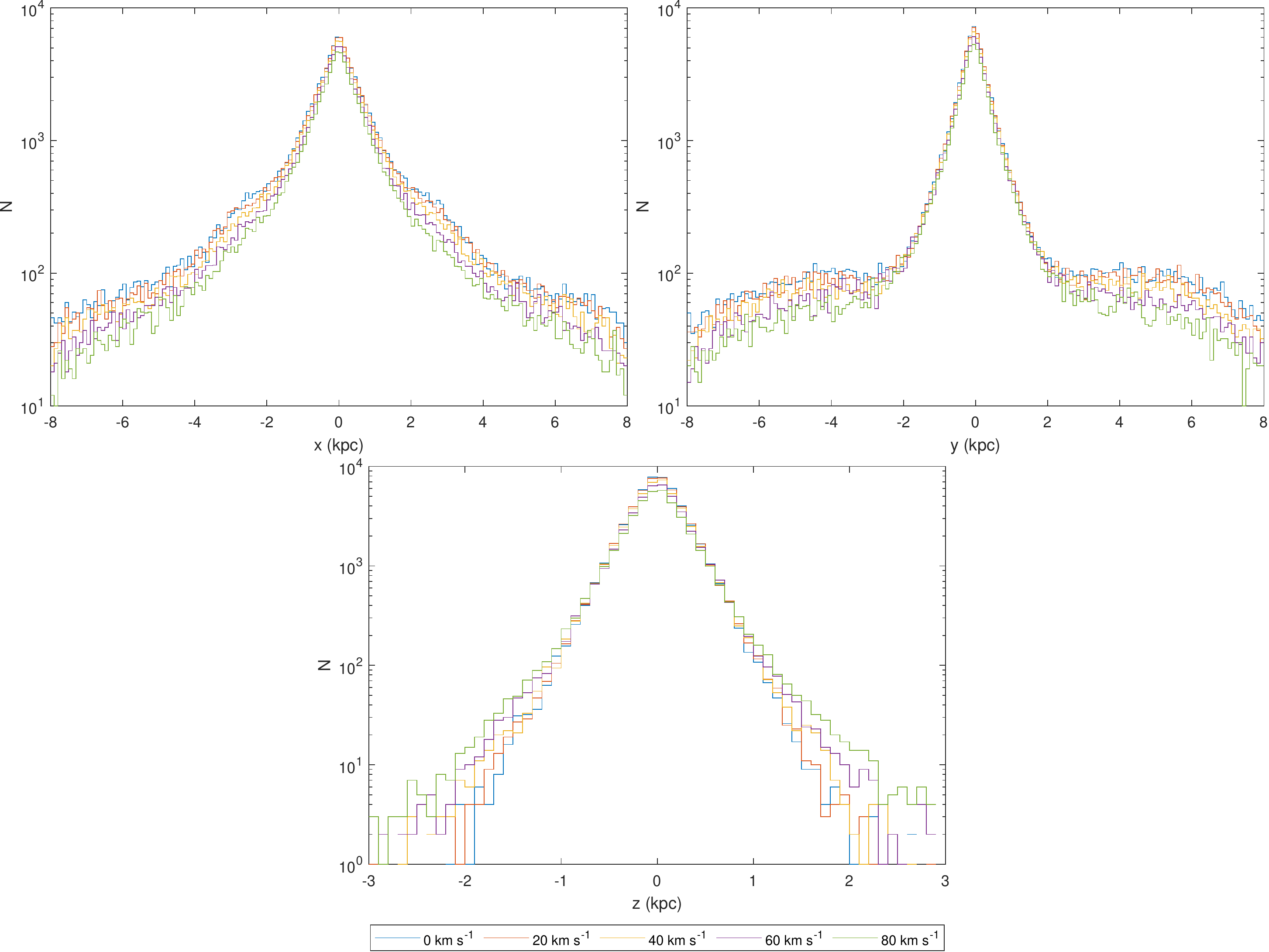}
    \caption[MWa $x$, $y$ and $z$ profiles with kicks at beginning for various kick scales. ]{MWa profile along $x$, $y$ and $z$ axes with kicks occuring at the beginning for kicks between $0$ km s$^{-1}$ and $80$ km s$^{-1}$ in $20$ km s$^{-1}$ increments. }
    \label{fig:MWa_less_80km_per_s_kick_1d_profiles_kick_at_beginning}
\end{figure}

\begin{figure}
    \centering
    \includegraphics[width=0.99\linewidth]{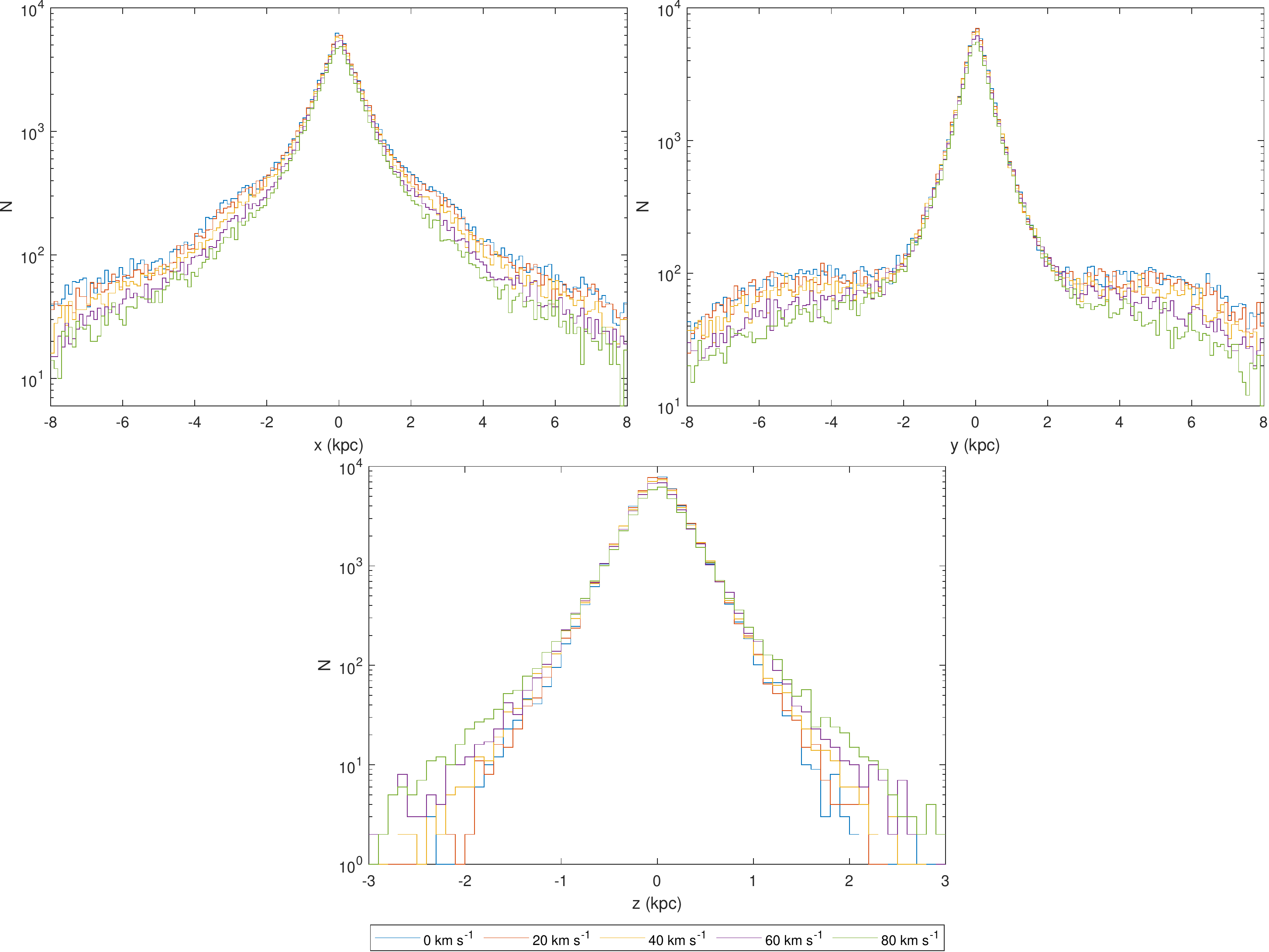}
    \caption[MWa $x$, $y$ and $z$ profiles with a uniform kick rate for various kick scales. ]{MWa profile along $x$, $y$ and $z$ axes with a uniform kick rate for kicks between $0$ km s$^{-1}$ and $80$ km s$^{-1}$ in $20$ km s$^{-1}$ increments. }
    \label{fig:MWa_less_80km_per_s_kick_1d_profiles_uniform_kick_rate}
\end{figure}

Eckner et al.~\cite{Eckner2018} argued using the virial theorem that kicks $\langle v^2 \rangle \lesssim (70 {\rm km~s^{-1}})^2$ would lead to a ``smoothing" of the distribution of $700$--$900$ pc.
We show the effect of a $400$ pc and an $800$ pc Gaussian smoothing on the fitted bulge (bar plus Hernquist bulge) distribution in Figs.~\ref{fig:MWa_bulge_smoothed_1d_profiles_kick_at_beginning} and \ref{fig:MWa_bulge_smoothed_los} for the kicked at the beginning case.
The uniform kick rate case is shown in Fig.~\ref{fig:MWa_bulge_smoothed_1d_profiles_uniform_kick_rate}.
Those figures also show the bulge component of the models with and without kicks for comparison. It is clear that a Gaussian smoothing kernel will remove the peak that survives in the $N$-body simulations of kicked distributions and, particularly for the $800$ pc case, will produce an apparently spherically symmetric bulge. 
From the peculiar velocity data we inferred kicks that are larger than assumed by Eckner et al.~\cite{Eckner2018} with $\sigma_k$ at around $80$--$100$ km s$^{-1}$ ($\langle v^2 \rangle = 3 \sigma_k^2$ for a Maxwell distribution where the angular brackets signify the mean value) so the smoothing effect of the Gaussian would be even more severe.
We show an even smaller Gaussian smoothing of $200$ pc in Figs.~\ref{fig:MWa_bulge_smoothed_200pc_1d_profiles_kick_at_beginning} and \ref{fig:MWa_bulge_smoothed_200pc_los}
for the kicked at beginning case and in Fig.~\ref{fig:MWa_bulge_smoothed_200pc_1d_profiles_uniform_kick_rate} for the uniform kick rate case. We also show in Fig.~\ref{fig:MWa_less_80km_per_s_kick_1d_profiles_kick_at_beginning}  the profile for particles with smaller kick scales between $0$ km s$^{-1}$ and $80$ km s$^{-1}$ for the kicked at the beginning case. The corresponding uniform kick rate case is shown in Fig.~\ref{fig:MWa_less_80km_per_s_kick_1d_profiles_uniform_kick_rate}. In these two figures, each kick scale has only $4\times10^5$ particles; therefore, to reduce noise, the bins in each of the other two dimensions are twice as big as in previous single dimensional plots, and particles within $0.5$ kpc (previously $0.25$ kpc) of the axis are included. The profiles along the $z$ axis in particular show that there is a reduction in the slope as the kick velocities increase, along the other two axes the general increase in scaleheight is seen as a reduction in density.
These results demonstrates that Gaussian smoothing is not a good way of modelling a kicked version of a boxy bulge template.

In every case, the bar fitted to the kicked data is both broader, with larger scale parameters $x_b$, $y_b$ and $z_b$, and less boxy, with smaller $C_\parallel$. For example, for model MWa $(x_b, y_b, z_b)$ increases from $(0.56, 0.37, 0.26)$ to $(0.69, 0.46, 0.33)$ and $(0.67, 0.45, 0.32)$ for the kick at beginning case and the uniform kick rate case respectively, while $C_\parallel$ declines from $3.08$ to $2.49$ and $2.54$. The spherically symmetric Hernquist bulge increases from $\sim1\%$ of the particles to $6\%$ for MWa. Like the bar, it becomes broader with $a_b$ increasing from around $0.2$ kpc to $0.51$ kpc and $0.56$ kpc. For the other two models similar changes occur, $P({\rm Long~Bar})$ and $a_b$ both increase significantly. In MSP model A1 of Chapter \ref{ch:msp_pop}, the disk parameters were $\sigma_r = 4.5\substack{+0.5 \\ -0.4}$ kpc and $z_0 = 0.71\substack{+0.11 \\ -0.09}$. In the current chapter, after being kicked, the disk scale heights $z_0$ of all models increase from $0.22$ kpc to $\gtrsim{1}$ kpc, while $\sigma_r$ is in the range $5$--$7$~kpc. However, we find that in all kicked cases, except for MWb with kicks occurring at the beginning, the long bar behaves like a relatively thin disk component. We have $x_{\rm lb} \approx y_{\rm lb}$, $C_{\perp,\rm lb} \sim 2$ and $R_{\rm out} \gtrsim{7}$, resulting in a density $\sim \exp(-R/R_0)$ in $R$ for scalelength $R_0$. These scalelengths would then be between approximately $2.4$ kpc and $2.8$ kpc. For comparison, in Bland-Hawthorn and Gerhard \cite{Bland-Hawthorn2016} the Milky Way disk scalelength is reported as $2.6 \pm 0.5$ kpc. The exponential scaleheights of these ``long bars" range between about $0.5$--$0.6$ kpc. In Fig.~\ref{fig:MWa_1d_profiles_kick_at_beginning}, for the MWa kicked at the beginning case and in Figs.~\ref{fig:MWa_1d_profiles_uniform_kick_rate}, \ref{fig:MWb_1d_profiles_uniform_kick_rate}, \ref{fig:MWc0.8_1d_profiles_kick_at_beginning} and  \ref{fig:MWc0.8_1d_profiles_uniform_kick_rate} for the uniform kick rate and MWb, and MWc0.8 case,
there may be, to varying degrees, an excess of kicked particles over the model in the region of $2\mbox{ kpc} \lesssim \abs{x} \lesssim 4$ kpc. However these are far less prominent than the long bar of the data that has not been kicked.

Our main aim in this chapter was to estimate the effect of the MSP kicks on their distribution in the Milky Way Galactic bulge. However, comparing to Cao et al.~\cite{Cao:2013dwa}, none of our simulations had quite the right bulge parameters. But, there appears to be a linear relationship between the unkicked scale parameters $x_b$, $y_b$, and $z_b$ and their kicked counterparts. Similarly, there appears to be a linear relationship between $C_\perp$, $C_{\parallel}$, and their kicked counterparts.
Therefore, we were able to estimate the Milky Way Galactic bulge kicked parameters as shown in  Fig.~\ref{fig:simulation_parameters} and Table~\ref{tab:milky_way_prediction}.
As can be seen, there is more scatter in the $x_b$ parameter. This is not unexpected due to the already mentioned degeneracy with the long bar.
Also, as can be seen, estimating the Milky Way bulge kicked $x_b$ parameter did involve a reasonable amount of extrapolation and so future simulations which have a larger $x_b$ will be needed to check it. 
A made-to-measure \cite{SyerTremaine1996,deLorenzi2007} approach may be needed. This would also be advantageous as it could take into account the X/peanut shaped morphology of the bulge \cite{Nataf:2010,Mc10,Wegg2015} as done in ref.~\cite{PortailWeggGerhard2015}. This would be particularly beneficial as there is some preliminary evidence that the X-shape may improve the fit to the Fermi-LAT gamma-ray data \cite{Coleman19}.

In conclusion, we used $N$-body simulations to explore the effect of a Maxwell distributed kick on the distribution of pulsars in the Galactic Center. We find that while a $700$--$900$ pc Gaussian smoothing of the stellar mass would be too aggressive, the bulge distribution of the kicked particles is slightly broader and less boxy. From these results, we expect that the GCE would not be exactly correlated with the stellar mass in the Galactic Center. As can seen from Table~\ref{tab:milky_way_prediction}, we would not expect the GCE to appear spherically symmetric due to the MSP kicks as that would require $x_b=y_b=z_b$ and $C_\perp=C_\parallel=2$ which are far from our inferred points relative to their error bars.

The amount  of spatial smoothing of the bulge MSPs will depend on the proportion of MSPs in the bulge that are made from the recycling channel and the proportion that are made from the 
 accretion induced collapse channel. Motivated by similarities between the bulge and disk population seen in Chapter \ref{ch:msp_pop} we have assumed this mixture is the same as the disk MSPs. 
 If the GCE is due to bulge MSPs, its morphology could be used to
 check our smoothing prediction by comparing if there are any deviations between the GCE morphology and the stellar spatial distribution.
 A complication to this approach would be the possibility of some smearing of the GCE due to cosmic ray electron diffusion \cite{Song2019,Macias_2021}.
 An additional complication is that if the MSP is spun up by a captured star then the MSP spatial distribution would be proportional to the stellar density squared \cite{Eckner2018, Macias19}.
 We have been assuming that the MSPs formed in a binary system and so have a density proportional to the stellar density. 
 Eventually, once the bulge MSPs are resolved \cite[Chapter \ref{ch:msp_pop}]{Calore2016}, comparing their spatial distribution to the stellar distribution should provide independent information to more robustly estimate the natal kick distribution.
 
\chapter{Summary and Future Work}

\graphicspath{{summary/Figs/}}

In this thesis we have explored the possibility that the GCE, an extended gamma-ray source found in the Fermi-LAT data, is produced by a bulge population of unresolved MSPs. Resolved MSPs have gamma-ray spectra that peak at a few GeV, like the GCE, and they have a luminosity distribution where they would be unlikely to be resolved at the distance of the Galactic Center. Additionally, the GCE now appears to have a spatial distribution similar to that of the Galactic bulge, indicating the source of the GCE is a population of unresolved astrophysical point sources. This disfavors the self-annihilating WIMPs scenario in which the distribution would be spherically symmetric.

In Chapter \ref{ch:msp_pop}, we used MSPs with confirmed gamma-ray pulsations detected in the Fermi-LAT data to model the Milky Way MSP population. Combining data in the Fermi Large Area Telescope fourth source catalog data release 2 \citep[4FGL-DR2:][]{Ballet:2020hze} and the ATNF pulsar catalog \citep{Manchester:2004bp}, we used MCMC to fit a model to their positions in galactic longitude $l$, latitude $b$, distance $d$, period $P$, period derivative $\dot{P}$, proper motions $\mu_{l}$ and $\mu_{b}$, gamma-ray flux $F$, spectral energy cut-off $E_{\rm cut}$, and  spectral index $\Gamma$. Our model consisted of a disk, boxy bulge and nuclear bulge, where the latter two components were responsible for a simulated GCE. We performed all fits both with and without the GCE.

We tried various models of the luminosity distribution. Ranking models using the WAIC, there was a clear preference for a dependence of luminosity on $E_{\rm cut}$, the magnetic field strength $B$ and the spin down power $\dot{E}$. For the form $L \propto E_{\rm cut}^{a_{\gamma}} B^{b_{\gamma}} \dot{E}^{d_{\gamma}}$, we found $a_{\gamma} = 1.2 \pm 0.3$, $b_{\gamma} = 0.1 \pm 0.4$ and $d_{\gamma} = 0.5 \pm 0.1$. This was consistent with the prediction for curvature radiation of Kalapotharakos et al.~\cite{Kalapotharakos_2019} where $a_{\gamma} = 4/3$, $b_{\gamma} = 1/6$ and $d_{\gamma} = 5/12$. It is inconsistent with their prediction for synchrotron radiation of $a_{\gamma} = 1$, $b_{\gamma} = 0$ and $d_{\gamma} = 1$. The worst model, with $\Delta{\rm WAIC}=42$ relative to the best model when the GCE was included in the fit (and $\Delta{\rm WAIC}=47$ otherwise), was one in which the luminosity was log-normally distributed and independent of any other property of an MSP.

Using a formula relating magnetic field strength, initial period and age to current period, we can model the present day period and period derivative distribution. We modelled MSPs as having log-normally distributed magnetic field strengths and initial periods, these distributions were fitted to the data. For our best model, we found for the magnetic field strength distribution that $\log_{10}(B/\textrm{G})$ had a mean of $8.21\substack{+0.05 \\ -0.06}$ with a standard deviation of $0.21\substack{+0.03 \\ -0.02}$ where $B$ has units of Gauss. For the initial periods with units of seconds, $\log_{10}(P_{i}/\textrm{s})$ had a mean of $-2.61\substack{+0.05 \\ -0.04}$ and a standard deviation of $0.13\substack{+0.02 \\ -0.02}$. Aside from their spatial distributions, the MSP populations making up each of the three components differed only in their SFR. Fitting a five bin DTD shared between the components, where the delay time is the time between star and MSP formation, we attempted to determine the age distribution of MSPs in each component. As an alternative age distribution, we had a uniform birth rate between the present and $10$ Gyr ago for all three MSP populations. Comparing these cases with the WAIC, we found that the case where the age distributions of the disk, boxy bulge and nuclear bulge were different was not significantly better than the uniform birth rate case. The fitted age distribution was poorly constrained and not clearly inconsistent with a uniform birth rate.

For our best model with $L \propto E_{\rm cut}^{a_{\gamma}} B^{b_{\gamma}} \dot{E}^{d_{\gamma}}$ and with the fitted DTD, we found the GCE could be produced by a bulge population of $N_{\rm bulge} = 34\substack{+17 \\ -10}$ thousand MSPs. The probability that none of these bulge MSPs have been resolved at present was $0.16$ with a $68\%$ interval of $1$--$4$. We identified three resolved MSPs with probability greater than $0.05$ of being bulge MSPs: PSR J1747-4036 with probability $0.4$, PSR J1811-2405 with probability $0.5$ and PSR J1855-1436 with probability $0.1$. However, these probabilities require that the distances of these pulsars really are large enough to locate them in a region of space where the bulge density is significant. That density is determined by the bulge spatial model and by the assumption that MSPs are the source of the GCE. Switching to a $r^{-2.4}$ bulge profile with a cutoff at $r=3.1$~kpc changed the previously mentioned probabilities to $0.5$, $0.2$, and $4\times 10^{-4}$ respectively.

In future, as the number of gamma-ray MSPs detected by Fermi-LAT increases over time, the constraints on the model parameters may be significantly improved. However, our model of the Galactic MSP population could also be improved. We included a randomly distributed magnetic axis angle $\alpha$ relative to the rotation axis as this affects the rate at which a pulsar spins down for a given magnetic field strength. Although we have a parameter $\eta$ that allows for random variation in pulsar luminosity, we do not explicitly account for the possibility that the flux may vary due to $\alpha$ as well as the viewing angle relative to rotation axis. This may mean the resolved MSPs are biased towards particular values of $\alpha$. This could be accounted for by including a model of the gamma-ray emission geometry.
Another improvement would be to include radio detected MSPs in the fit. This would increase the number of MSPs involved in constraining the model parameters. They would require a model of the radio luminosity and of the radio flux threshold. Almost all of the gamma-ray MSPs we have used in our analysis were discovered in radio before they were found in the Fermi-LAT data. As many may have been detected by Fermi-LAT only because of that previous discovery, including a radio model may allow explicit inclusion of a radio detection threshold component to the gamma-ray flux detection threshold. Currently, to model the possibility that MSPs with gamma-ray flux near the threshold may or may not be resolved for this reason, we simply allow the galactic longitude and latitude dependent flux threshold to vary log-normally, finding $\sigma_{\rm th} = 0.28\substack{+0.05 \\ -0.04}$. However, it may be that the radio threshold depends very differently on position in the sky.

As part of the fitted model in Chapter \ref{ch:msp_pop} we included a simple model of pulsar velocities in which they were travelling on circular orbits plus a Maxwell distributed peculiar velocity in a random direction. We found for the scale parameter of this peculiar velocity $\sigma_{v} = 77 \pm 6$ km s$^{-1}$. In Chapter \ref{ch:nbody} we investigated the effect of pulsar kicks on the structure of the bulge. We did this by running $N$-body simulations of a galaxy intended be similar to the Milky Way, adding in particles which were given a kick of a magnitude selected to produce a peculiar velocity distribution close to that of the resolved MSPs.

We ran $N$-body simulations using three different sets of initial conditions, and with kicks either occurring all at the beginning or at a uniform rate during the $10$ Gyr simulation. We found initial Maxwell distributed kicks of around $80$--$100$ km s$^{-1}$ depending on the initial conditions and kick rate. We fitted a parametric model of the Galaxy to the particle distributions. This model consisted of a disk, spherically symmetric bulge, long bar and a bar of the same form as in Cao et al.~\cite{Cao:2013dwa}. Eckner et al.~\cite{Eckner2018} argued using the virial theorem that MSP kicks where $\langle v^2 \rangle \lesssim (70 {\rm km~s^{-1}})^2$ would lead to a smoothing scale of the spatial distribution of $700$--$900$ pc. This would make the bulge distribution relatively spherical. We found the kicks broadened the bar structure and made it less boxy, but they did not render it spherically symmetric. The kicks did increase the spherically symmetric component at the galactic center from around $\sim 1\%$ to $5$--$8\%$ of the model, however the bar was the more significant structure in that region. Taking the fitted central values for the bar scale parameters $x_b$, $y_b$, and $z_b$ and separately the shape parameters $C_\perp$ and $C_{\parallel}$, we found they lie nearly on a line when plotting the values for the case with no kick against the values for the kicked particle distribution. Using a linear fit we showed the kicked scale parameters are increased by around $10\%$. From this we suggested that for the Cao et al.~\cite{Cao:2013dwa} model of the Milky Way boxy bulge fitted to the red clump giant distribution, the parameters $\left(x_b,y_b,z_b,C_\perp,C_{\parallel}\right) = \left(0.67,0.29,0.27,2,4\right)$ could, for a kicked version, become $\left(0.79,0.36,0.35,1.93,3.05\right)$.
We expect that the GCE, if it is produced by a population of unresolved MSPs, will differ from the stellar mass distribution in the Galactic Center. However, pulsar birth kicks should not result in a spherically symmetric distribution of bulge MSPs. Assuming a similar mixture of MSP formation channels applies to the bulge as the disk, so that the natal kick distributions in velocity and in time are not significantly different, future studies into the morphology of the GCE may find templates produced with a slightly larger and less boxy bar provide a better fit to the data.

\begin{spacing}{1.1}

\bibliographystyle{JHEP}
\cleardoublepage
\bibliography{References/references} %

\end{spacing}

\begin{appendices} %

\chapter{Likelihood probability density function of resolved MSPs} 
\label{app:probability}

In this appendix we implicitly assume that all
probabilities are conditioned on the parameters
($\thetab$). 
Also, as all of our expression here are also for an individual MSP we leave the subscript on each observational quantity as implicit.
As an example of these conventions, $p({\rm obs} \vert l, b, F)$ is equivalent to $p({\rm obs} \vert l_i, b_i, F_i,\thetab)$.

We assume the likelihood of observed MSPs with/without a parallax distance measurement depends only on distance $d$:
\begin{equation}
    \begin{multlined}
        p({\rm obs}, l, b, d, P, \dot{P}, \mu_l, \mu_b, F, E_{\rm cut}, \Gamma, \textrm{parallax/not parallax}) = p({\rm obs}, l, b, d, P, \dot{P}, \mu_l, \mu_b, F, E_{\rm cut}, \Gamma) \\ \times p(\textrm{parallax/not parallax} \vert d)
    \end{multlined}
\end{equation}
where the probability of a parallax measurement given distance $d$ is given by Eq.~\ref{eq:parallax_model}.
In terms of the various components of the MSP model, the probability density function of resolved MSPs at $l$, $b$, $d$, $P$, $\dot{P}$, $\mu_l$, $\mu_b$, $F$, $E_{\rm cut}$ and $\Gamma$ is: 
\begin{equation}
\label{eq:prob_density_int_init_period_alpha}
    p({\rm obs}, l, b, d, P, \dot{P}, \mu_l, \mu_b, F, E_{\rm cut}, \Gamma) =  \int \int   p({\rm obs}, l, b, d, P, \dot{P}, \mu_l, \mu_b, F, E_{\rm cut}, \Gamma, \alpha, P_I)\,\dd \alpha\, \dd P_I
\end{equation}
\noindent where we have integrated over the unknown magnetic axis angle and initial period. Then:
\begin{equation}
    \begin{multlined}
        p({\rm obs}, l, b, d, P, \dot{P}, \mu_l, \mu_b, F, E_{\rm cut}, \Gamma, \alpha, P_I) = p({\rm obs} \vert l, b, F) p(F \vert l, b, d, P, \dot{P}, \mu_l, \mu_b, E_{\rm cut}, \alpha) \\ \times p(E_{\rm cut}, \Gamma \vert l, b, d, P, \dot{P}, \mu_l, \mu_b) p(P, \dot{P} \vert l, b, d, \mu_l, \mu_b, \alpha, P_I) p(\mu_l, \mu_b \vert l, b, d) \\ \times p(l, b, d) p(\alpha) p(P_I)
    \end{multlined}
\end{equation}
\noindent The probability of observing an MSP with flux $F$ at $l$ and $b$ is given by Eq.\ \ref{eq:detection_probability}, so:
\begin{equation}
    p({\rm obs} \vert l, b, F) = p(F_{\rm th} \leq F \vert l, b)
\end{equation}

The probability density function of an MSP having flux $F$ conditional upon its other parameters is, given $\log(F) = \log(\eta) + f(...)$ where $f(...)$ is some function which doesn't depend on $\eta$:
\begin{equation}
\begin{split}
    p(F \vert l, b, d, P, \dot{P}, \mu_l, \mu_b, E_{\rm cut}, \alpha) &= p(\eta \vert l, b, d, P, \dot{P}, \mu_l, \mu_b, E_{\rm cut}, \alpha) \frac{\partial \eta}{\partial F} \\ &= p(\eta \vert l, b, d, P, \dot{P}, \mu_l, \mu_b, E_{\rm cut}, \alpha) \frac{\eta}{F}
\end{split}
\end{equation}
\noindent 
After a change of variables from period $P$ and observed period derivative $\dot{P}$ to magnetic field strength $B$ and age $t$, we get:
\begin{equation}
\label{eq:PPPdot}
    p(P, \dot{P} \vert l, b, d, \mu_l, \mu_b, \alpha, P_I) = p(B, t \vert l, b, d, \alpha, P_I) 
    \left\lvert \frac{\partial B}{\partial P} \frac{\partial t}{\partial \dot{P}} - \frac{\partial B}{\partial \dot{P}} \frac{\partial t}{\partial P} \right\rvert
\end{equation}
\noindent
where to evaluate the Jacobian in the above equation we
rewrite Eq.~\ref{eq:Shklovskii} as
\begin{equation}
\label{eq:Shklovskii1}
    \dot{P}_{\rm Shklovskii} = C_1 P
\end{equation}
where $C_1$ is a term that is independent of $P$ and $\dot{P}$. We also rewrite Eq.~\ref{eq:period_derivative_Galactic_correction} as
\begin{equation}
\label{eq:period_derivative_Galactic_correction1}
    \dot{P}_{\rm Galactic} = C_2 P
\end{equation}
where $C_2$ is also independent of $P$ and $\dot{P}$. 
We then obtain an equation for $B$ in terms of $P$ and $\dot{P}$ using
the above two equations with Eqs.~\ref{eq:period_derivative_observed_corrections} and  \ref{eq:magnetic_field_strength} to get 
\begin{equation}
\label{eq:B}
B^2=\frac{c^3 I P \left(- P({C}_1+{C}_2)+\dot{P}\right)}{\pi ^2 R^6 \left(\sin ^2(\alpha )+1\right)}
\end{equation}
Next we obtain an equation for $t$ in terms of $P$ and $\dot{P}$ by substituting the above equation into Eq.~\ref{eq:current_period} and solving $t$ to get
\begin{equation}
    t= \frac{P_I^2-P^2}{2 P \left(\left(C_1+{C}_2\right) P-\dot{P}\right)}
\end{equation}
Using the above two equations we can then solve for the Jacobian term in Eq.~\ref{eq:PPPdot} to get 
\begin{equation}
  \left\lvert \frac{\partial B}{\partial P} \frac{\partial t}{\partial \dot{P}} - \frac{\partial B}{\partial \dot{P}} \frac{\partial t}{\partial P} \right\rvert = \frac{c^6 I^2 P^2}{2 \pi^4 R^{12} (1 + \sin^2(\alpha))^2 B^3}
\end{equation}
where we have used Eq.~\ref{eq:B} to eliminate $C_1$ and $C_2$.
\noindent Similarly, using Eq.\ \ref{eq:proper_motion_to_linear_velocity}, $\partial v / \partial \mu \propto d$, so we find for proper motion:
\begin{equation}
   p(\mu_l, \mu_b \vert l, b, d) \propto p(v_l, v_b \vert l, b, d)  d^2
\end{equation}
\noindent where the proportionality constant is independent of our parameters and data and so does not affect our results. For position:
\begin{equation}
  p(l, b, d) \propto p(x, y, z) d^2 \cos(b)
\end{equation}
\noindent where $d^2 \cos(b)$ is proportional to the Jacobian of the change of variables from $x$, $y$ and $z$ to $l$, $b$ and $d$:
\begin{equation}
    \begin{split}
        x &= -R_0 + d \cos(l) \cos(b) \\
        y &= d \sin(l) \cos(b) \\
        z &= d \sin(b) \\
    \end{split}
\end{equation}
The density of resolved MSPs is the sum of the density in the disk, boxy bulge and nuclear bulge populations: 
\begin{equation}
\label{eq:density_sum_over_model_components}
    \rho(\textrm{obs}, ...) = N_{\rm disk} p_{\rm disk}(\textrm{obs}, ...) + N_{\rm bb} p_{\rm bb}(\textrm{obs}, ...) + N_{\rm nb} p_{\rm nb}(\textrm{obs}, ...)
\end{equation}
\noindent where we can calculate the number in each population using the parameters $\lambda_{\rm res}$, $\log_{10}(N_{\rm disk} / N_{\rm bulge})$ and $\log_{10}(N_{\rm nb} / N_{\rm bb})$ and solving with:
\begin{equation}
    \lambda_{\rm res} = N_{\rm disk} p_{\rm disk}(\textrm{obs}) + N_{\rm bb} p_{\rm bb}(\textrm{obs}) + N_{\rm nb} p_{\rm nb}(\textrm{obs})
\end{equation}
\noindent where each $p({\rm obs})$ is evaluated using Eq.~\ref{eq:pobs} for the corresponding spatial distribution.

In evaluating the likelihood for a given set of parameters, we used importance sampling to estimate integrals. 
As an example, in the case of the integrals over $P_I$ and $\alpha$ in Eq.~\ref{eq:prob_density_int_init_period_alpha} this integral becomes:
\begin{equation}
    \begin{split}
    p({\rm obs}, l, b, d, P, \dot{P}, \mu_l, \mu_b, F, E_{\rm cut}, \Gamma) &=  \int \int p({\rm obs}, l, b, d, P, \dot{P}, \mu_l, \mu_b, F, E_{\rm cut}, \Gamma, \alpha, P_I)\,\dd \alpha\, \dd P_I \\
    &=  \int \int p({\rm obs}, l, b, d, P, \dot{P}, \mu_l, \mu_b, F, E_{\rm cut}, \Gamma \vert \alpha, P_I)  p(\alpha) p(P_I) \,\dd \alpha\, \dd P_I \\
    &\approx \frac{1}{N} \sum_{i=1}^N p({\rm obs}, l, b, d, P, \dot{P}, \mu_l, \mu_b, F, E_{\rm cut}, \Gamma \vert \alpha_i, P_{Ii})
    \end{split}
\end{equation}
\noindent where we sum over $N$ samples $\alpha_i$ and $P_{Ii}$ from the probability distributions $p(\alpha)$ and $p(P_I)$.

\chapter{Measurement Uncertainties}
\label{appendix:uncertainties}

For the measurements of $F$, $E_{\rm cut}$ and $\Gamma$ we use the 4FGL covariance matrices\footnote{Kindly provided to us by Dr Jean Ballet of the Fermi-LAT collaboration.}
for the uncertainty in $N_0$, $\gamma_1$ and $a$, with the spectrum of the form:\footnote{"PLSuperExpCutoff2" at https://fermi.gsfc.nasa.gov/ssc/data/analysis/scitools/source\_models.html}
\begin{equation}
\label{eq:4fgl_xml_file_spectrum}
    \frac{\dd N}{\dd E} = N_0 \left(\frac{E}{E_0} \right)^{\gamma_1} \exp(-a E^{\gamma_2})
\end{equation}
\noindent where $\gamma_2=2/3$, and $E_0$ is fixed at a different value for each MSP.

For the small number of MSPs in 4FGL-DR2 that were not fitted with spectra in the form of Eq.\ \ref{eq:4fgl_xml_file_spectrum}, we simply use the reported flux and its error, treating $E_{\rm cut}$ and $\Gamma$ as missing. Since the catalog required $\Gamma > 0$, we also treat $E_{\rm cut}$ and $\Gamma$ as missing in the case where some MSPs had $\Gamma$ fixed near $0$. If, based on central estimates of $\dot{P}$, $\mu$ and $d$, $\dot{P}_{\rm int}$ appears to be negative for a particular MSP (i.e., the MSP is apparently spinning up) we treat $\dot{P}$ as missing. 
An apparently negative intrinsic period derivatives can result 
in the case that the true proper motion and/or distance is lower than the value we use, therefore causing the Shklovskii effect to be overestimated. An example of an updated proper motion measurement fixing this issue is given for PSR J1231-1411 in Abdo et al.~\cite{TheFermi-LAT:2013ssa}. A second potential cause of negative period derivative is radial acceleration in the direction of the sun significantly in excess of that accounted for by the $\dot{P}_{\rm Galactic}$ term in equations \ref{eq:period_derivative_observed_corrections} and \ref{eq:period_derivative_Galactic_correction}.

In the case of distance uncertainties, for the dispersion measure we use a method similar to that of Bartels et al.~\cite{Bartels2018}: The relative uncertainty in the dispersion measure is typically very small, so we assume that it is measured exactly. The main source of distance uncertainty is, therefore, associated with the model of the free electron density along the line of sight for each MSP. We use the YMW16 model of Yao et al.~\cite{Yao2017} and integrate over the uncertainty in the model parameters $\theta_{\rm YMW16}$ so that:
\begin{equation}
\begin{split}
    p(..., \textrm{DM}_j,...) &= \int \dd \theta_{\rm YMW16} p(\theta_{\rm YMW16}) p(..., \textrm{DM}_j, ... \vert \theta_{\rm YMW16}) \\
    &=  \!\begin{multlined}[t]
    \int \dd \theta_{\rm YMW16} p(\theta_{\rm YMW16}) \\ \times \left\lvert \frac{\partial}{\partial \textrm{DM}} d(\textrm{DM}_j, \theta_{\rm YMW16}) \right\rvert p(..., d(\textrm{DM}_j, \theta_{\rm YMW16}), ...) 
    \end{multlined} \\
    &= \int \dd \theta_{\rm YMW16} \frac{p(\theta_{\rm YMW16})}{n_e(d(\textrm{DM}_j, \theta_{\rm YMW16}), \theta_{\rm YMW16})} p(..., d(\textrm{DM}_j, \theta_{\rm YMW16}), ...)
\end{split}
\end{equation}
\noindent where $d(\textrm{DM}, \theta_{\rm YMW16})$ is distance as a function of dispersion measure and YMW16 model parameters. This can be found by solving for $d$ in $\textrm{DM}(\theta_{\rm YMW16}) = \int_0^d n_e(s,\theta_{\rm YMW16}) \dd s$ where $n_e(s,\theta_{\rm YMW16})$ is the free electron density at distance $s$ for model parameters $\theta_{\rm YMW16}$.

From Eqs.~\ref{eq:rhoD}, \ref{eq:likelihood_integral_over_true_vals} and \ref{eq:density_sum_over_model_components} we can work out the fraction of the likelihood contributed by the boxy bulge and nuclear bulge components for resolved  MSP $i$ given the model parameters and data uncertainties:
\begin{equation}
    p(\textrm{Bulge MSP}_i) = \frac{\rho_{\rm bb}(\D{i}) + \rho_{\rm nb}(\D{i})}{\rho(\D{i})}
    \label{eq:bulgelikelihood}
\end{equation}
\noindent where $\rho_{\rm bb}(\D{i}) = N_{\rm bb} p_{\rm bb}(\textrm{obs}, \D{i})$ and $\rho_{\rm nb}(\D{i}) = N_{\rm nb} p_{\rm nb}(\textrm{obs}, \D{i})$.

\chapter{Sampling methods}
\label{appendix:sampling}

The disk model, given in Eq.~\ref{eq:rho_disk}, can be sampled from by sampling in cylindrical coordinates a random $R$, $z$ and $\phi$. The radial coordinate $R$ is drawn from:
\begin{equation}
    p(R) = \frac{1}{\sigma_r^2} \exp(-R^2 / 2 \sigma_r^2) R
\end{equation}
\noindent for which, using the inverse of the cumulative distribution function of $R$, if $u$ is a uniformly random draw from $\left[0, 1 \right]$:
\begin{equation}
    R = \sigma_r \sqrt{-2 \log(1 - u)}
\end{equation}
\noindent The height $z$ is drawn from:
\begin{equation}
    p(z) = \frac{1}{2 z_0} \exp(-\abs{z} / z_0)
\end{equation}
\noindent which can be done by drawing $\abs{z}$ from an exponential distribution and choosing either a positive or negative sign each with $0.5$ probability. Finally, $\phi$ is drawn from a uniform distribution on $\left[0, 2 \pi \right]$.
For the boxy bulge and nuclear bulge distributions, we sampled the density using MCMC.

To sample from the age distribution, we used MCMC. We used standard library functions to sample from the various Gaussian distributions. To sample $\alpha$, again using inverse transform sampling:
\begin{equation}
    \alpha = \arccos(1 - 2 u)
\end{equation}
\noindent for $u$ uniformly drawn from $\left[0, 1 \right]$.

\chapter{Watanabe-Akaike Information Criterion (WAIC)}
\label{appendix:WAIC}

In order to rank the various models of the Galactic MSP population in Chapter \ref{ch:msp_pop}, correcting for the varying number of parameters, we use the WAIC. In this appendix we define the WAIC, then derive the contribution to the WAIC associated with the resolved MSPs component of the likelihood.

The WAIC is defined in terms of the log pointwise predictive density (lppd) and effective number of parameters ($p_\text{WAIC}$) as \citep{Gelman2013}:
\begin{equation}
    \label{eq:WAIC_definition}
    {\rm WAIC} = -2 ({\rm lppd} - p_\text{WAIC})
\end{equation}
\noindent where for data ${\bf y}_1,...,{\bf y}_N$ and $S$ parameter sets $\thetab^s$ in our Markov chain:
\begin{equation}
    \label{eq:lppd_definition}
    {\rm lppd} = \sum_{i=1}^N \log(\frac{1}{S} \sum_{s=1}^S p\left({\bf y}_i \;\middle\vert\; \thetab^s\right))
\end{equation}
\noindent where $p\left({\bf y}_i \;\middle\vert\; \thetab^s\right)$ is the predictive density of ${\bf y}_i$ given model parameters $\thetab^s$. Using the WAIC1 option  from  Gelman et al.~\cite{Gelman2013} 
\begin{equation}
    \label{eq:pwaic_definition}
    p_\text{WAIC} = 2 \sum_{i=1}^N \left( \log(\frac{1}{S} \sum_{s=1}^S p\left({\bf y}_i \;\middle\vert\; \thetab^s\right)) - \frac{1}{S} \sum_{s=1}^S \log(p\left({\bf y}_i \;\middle\vert\; \thetab^{s}\right)) \right)
\end{equation}

We can write the WAIC as the sum of two components, $\text{WAIC}_\text{res}$ and $\text{WAIC}_\text{GCE}$.
For the GCE contribution, $\text{WAIC}_\text{GCE}$, we use the Gaussian likelihood for each bin as in Eq.\ \ref{eq:likelihood_GCE}.
If we use, for $\text{WAIC}_\text{res}$, the contribution of resolved MSPs, a Poisson distribution for bins in the several dimensions in which we have data, then define $\lambda_{i,s}$ as the expectation value for bin $i$ for parameter set $s$ and $n_i$ the number of observations in bin $i$, then:
\begin{equation}
    \label{eq:waic_poisson}
    \begin{split}
        p\left(n_i \;\middle\vert\; \lambda_{i,s}\right) &= \frac{\exp(-\lambda_{i,s}) \lambda_i^{n_i}}{n_i !} \\
        &= \frac{\exp(-\delta \rho_{i,s}) (\delta \rho_{i,s})^{n_i}}{n_i !}
    \end{split}
\end{equation}
\noindent where $\lambda_{i,s} = \delta \rho_{i,s}$ with $\delta$ the bin volume and $\rho_{i,s}$ the average density within bin $i$. Then if we choose $\delta$ small enough such that $\delta \rho_{i,s} \ll 1$ and $n_i$ is either $0$ or $1$, we can derive the contribution to the ${\rm lppd}$ from the resolved MSP data:
\begin{equation}
    \label{eq:lppd_res_msps1}
    \begin{split}
        {\rm lppd}_{\rm res} &= \sum_{i=1}^N \log(\frac{1}{S} \sum_{s=1}^S p\left(n_i \;\middle\vert\; \lambda_{i,s}\right)), \mbox{ using Eq.~\ref{eq:lppd_definition}}  \\
        &= \!\begin{multlined}[t]
        -N \log(S) \\ + \sum_{i=1}^N \log(\sum_{s=1}^S \exp(-\delta \rho_{i,s}) (\delta \rho_{i,s})^{n_i}),\mbox{ using Eq.~\ref{eq:waic_poisson} and that $n_i$ is either $0$ or $1$ }
        \end{multlined} \\
        &\approx -N \log(S) + \sum_{i=1}^N \log(\sum_{s=1}^S (1 - \delta \rho_{i,s}) (\delta \rho_{i,s})^{n_i}), \mbox{ using $\delta \rho_{i,s} \ll 1$} \\
        &= -N \log(S) + \sum_{i=1}^N
       \begin{cases}
              \log(\sum_{s=1}^S (1 - \delta \rho_{i,s})) & n_i = 0 \\
              \log(\sum_{s=1}^S (1 - \delta \rho_{i,s}) (\delta \rho_{i,s})) & n_i = 1 
       \end{cases} \\
       &\approx -N \log(S) + \sum_{i=1}^N
       \begin{cases}
              \log(S (1 - \frac{1}{S} \sum_{s=1}^S \delta \rho_{i,s})) & n_i = 0 \\
              \log(\sum_{s=1}^S \delta \rho_{i,s}) & n_i = 1,  \mbox{ using $\delta \rho_{i,s} \ll 1$} 
       \end{cases} \\
       &= -N \log(S) + \sum_{i=1}^N
       \begin{cases}
              \left(\log(S) + \log(1 - \frac{1}{S} \sum_{s=1}^S \delta \rho_{i,s})\right) & n_i = 0 \\
              \left(\log(\delta) + \log(\sum_{s=1}^S \rho_{i,s})\right) & n_i = 1 
       \end{cases} \\
       &= -N \log(S) + N_{n_i=0} \log(S) + N_{n_i=1} \log(\delta) + \sum_{i=1}^N
       \begin{cases}
              \log(1 - \frac{1}{S} \sum_{s=1}^S \delta \rho_{i,s}) & n_i = 0 \\
              \log(\sum_{s=1}^S \rho_{i,s}) & n_i = 1 
       \end{cases}
       \end{split}
       \end{equation}
       where $N_{n_i=0}$ is the number of bin with zero counts and $N_{n_i=1}$ for the number of bins with one count.
       
       Using $-N  + N_{n_i=0}=-N_{n_i=1}$ in Eq.~\ref{eq:lppd_res_msps1} gives
       \begin{equation}
       \begin{split}
       {\rm lppd}_{\rm res} &= -N_{n_i=1} \log(S) + N_{n_i=1} \log(\delta) + \sum_{i=1}^N
       \begin{cases}
              \log(1 - \frac{1}{S} \sum_{s=1}^S \delta \rho_{i,s}) & n_i = 0 \\
              \log(S \frac{1}{S} \sum_{s=1}^S \rho_{i,s}) & n_i = 1 
       \end{cases} \\
       &\approx \!\begin{multlined}[t]
       -N_{n_i=1} \log(S) \\ + N_{n_i=1} \log(\delta) + \sum_{i=1}^N
       \begin{cases}
              -\frac{1}{S} \sum_{s=1}^S \delta \rho_{i,s} & n_i = 0, \mbox{ using $\delta \rho_{i,s} \ll 1$}  \\
              \left(\log(S) + \log(\frac{1}{S} \sum_{s=1}^S \rho_{i,s})\right) & n_i = 1 
       \end{cases}
       \end{multlined} \\
       &= N_{n_i=1} \log(\delta) + \sum_{i=1}^N
       \begin{cases}
              -\frac{1}{S} \sum_{s=1}^S \delta \rho_{i,s} & n_i = 0 \\
              \log(\frac{1}{S} \sum_{s=1}^S \rho_{i,s}) & n_i = 1 
       \end{cases} \\
       \end{split}
       \end{equation}
As most voxels will have $n_i=0$ we can approximate $\lambda_s=\sum_{n_i=0,n_i=1} \delta \rho_{i,s}\approx\sum_{n_i=0} \delta \rho_{i,s}$
     where $\lambda_s$ is simply the total expected number of resolved MSPs parameter.
     Therefore,
     \begin{equation}
     \begin{split}
     \label{eq:lppd_res_msps2}
    {\rm lppd}_{\rm res}   &\approx N_{n_i=1} \log(\delta) - \frac{1}{S} \sum_{s=1}^S \lambda_s + \sum_{i=1}^N
       \begin{cases}
              0 & n_i = 0 \\
              \log(\frac{1}{S} \sum_{s=1}^S \rho_{i,s}) & n_i = 1 
       \end{cases} \\
       &= N_{n_i=1} \log(\delta) - \frac{1}{S} \sum_{s=1}^S \lambda_s + \sum_{j=1}^{N_{\rm res}} \log(\frac{1}{S} \sum_{s=1}^S N_{{\rm tot},s} p\left({\rm obs},\D{j} \;\middle\vert\; \thetab^s \right)) \\
    \end{split}
\end{equation}
\noindent where in the last line we have used Eq.~\ref{eq:rhoD}.
The first term in the above equation can be ignored as long as we are comparing models fitted using the same data. 

The other term we need to evaluate is given by Eq.~\ref{eq:pwaic_definition}
which we write in our notation as
\begin{equation}
      p_\text{WAIC,res} = 2 \sum_{i=1}^N \left( \log(\frac{1}{S} \sum_{s=1}^S p\left(n_i \;\middle\vert\; \lambda_{i,s}\right)) - \frac{1}{S} \sum_{s=1}^S \log(p\left(n_i \;\middle\vert\; \lambda_{i,s}\right)) \right)\,.
\end{equation}
We substitute Eq.~\ref{eq:waic_poisson}  to get
\begin{equation}
\begin{multlined}
   p_\text{WAIC,res} = -2 N \log(S) \\ + 2 \sum_{i=1}^N \left( \log(\sum_{s=1}^S \exp(-\delta \rho_{i,s}) (\delta \rho_{i,s})^{n_i}) - \frac{1}{S} \sum_{s=1}^S \log(\exp(-\delta \rho_{i,s}) (\delta \rho_{i,s})^{n_i}) \right) 
\end{multlined}
\end{equation}
where we have used the fact that $n_i=0$ or $n_i=1$ so $n_i!=1$ in either case. Next we separate out the two possible values for $n_i$ and make use of $\delta\rho_{i,s}\ll1$ as follows:
\begin{equation}
    \begin{split}
      p_\text{WAIC,res} &= \!\begin{multlined}[t]
      -2 N \log(S) \\ + 2 \sum_{i=1}^N
       \begin{cases}
              \left( \log(\sum_{s=1}^S \exp(-\delta \rho_{i,s})) - \frac{1}{S} \sum_{s=1}^S \log(\exp(-\delta \rho_{i,s})) \right) & n_i = 0 \\
              \left( \log(\sum_{s=1}^S \exp(-\delta \rho_{i,s}) (\delta \rho_{i,s})) - \frac{1}{S} \sum_{s=1}^S \log(\exp(-\delta \rho_{i,s}) (\delta \rho_{i,s})) \right) & n_i = 1 
       \end{cases}
       \end{multlined} \\
       &\approx  \!\begin{multlined}[t]
       -2 N \log(S) \\ + 2 \sum_{i=1}^N
       \begin{cases}
              \left( \log(\sum_{s=1}^S (1 - \delta \rho_{i,s})) - \frac{1}{S} \sum_{s=1}^S (-\delta \rho_{i,s}) \right) & n_i = 0 \\
              \left( \log(\sum_{s=1}^S (1 - \delta \rho_{i,s}) (\delta \rho_{i,s})) - \frac{1}{S} \sum_{s=1}^S \log((1 - \delta \rho_{i,s}) (\delta \rho_{i,s})) \right) & n_i = 1 
       \end{cases} 
       \end{multlined} \\
       &\approx -2 N \log(S) + 2 \sum_{i=1}^N
       \begin{cases}
              \left( \log(S (1 - \frac{1}{S} \sum_{s=1}^S \delta \rho_{i,s})) - \frac{1}{S} \sum_{s=1}^S (-\delta \rho_{i,s}) \right) & n_i = 0 \\
              \left( \log(S \frac{1}{S} \sum_{s=1}^S \delta \rho_{i,s}) - \frac{1}{S} \sum_{s=1}^S \log(\delta \rho_{i,s}) \right) & n_i = 1 
       \end{cases} \\
       &= -2 N \log(S) + 2 \sum_{i=1}^N
       \begin{cases}
              \left( \log(S) + \log(1 - \frac{1}{S} \sum_{s=1}^S \delta \rho_{i,s}) + \frac{1}{S} \sum_{s=1}^S \delta \rho_{i,s} \right) & n_i = 0 \\
              \left( \log(S) + \log(\frac{1}{S} \sum_{s=1}^S \delta \rho_{i,s}) - \frac{1}{S} \sum_{s=1}^S \log(\delta \rho_{i,s}) \right) & n_i = 1 
       \end{cases} \\
       &\approx 2 \sum_{i=1}^N
       \begin{cases}
              \left(-\frac{1}{S} \sum_{s=1}^S \delta \rho_{i,s} + \frac{1}{S} \sum_{s=1}^S \delta \rho_{i,s} \right) & n_i = 0 \\
              \left(\log(\frac{1}{S} \sum_{s=1}^S \delta \rho_{i,s}) - \frac{1}{S} \sum_{s=1}^S \log(\delta \rho_{i,s}) \right) & n_i = 1 
       \end{cases} \\
       &= 2 \sum_{i=1}^N
       \begin{cases}
              0 & n_i = 0 \\
              \left(\log(\frac{1}{S} \sum_{s=1}^S \rho_{i,s}) - \frac{1}{S} \sum_{s=1}^S \log(\rho_{i,s}) \right) & n_i = 1\,. 
       \end{cases} 
    \end{split}
\end{equation}
Substituting Eq.~\ref{eq:rhoD} into the above equation gives
\begin{equation}
\label{eq:p_waic1_res_msps}
     p_\text{WAIC,res}=  
       2 \sum_{j=1}^{N_{\rm res}} \left(\log(\frac{1}{S} \sum_{s=1}^S N_{{\rm tot},s} p\left({\rm obs}, \D{j} \;\middle\vert\; \thetab^s \right))  - \frac{1}{S} \sum_{s=1}^S \log(N_{{\rm tot},s} p\left({\rm obs},  \D{j} \;\middle\vert\; \thetab^s \right)) \right)\, .
\end{equation}
Substituting Eqs.~\ref{eq:p_waic1_res_msps} and \ref{eq:lppd_res_msps2} into Eq.~\ref{eq:WAIC_definition} gives
\begin{equation}
\begin{split}
    {\rm WAIC}_{\rm res}=&2\left[
     \frac{1}{S} \sum_{s=1}^S \lambda_s - \sum_{j=1}^{N_{\rm res}} \log(\frac{1}{S} \sum_{s=1}^S N_{{\rm tot},s} p\left({\rm obs},\D{j} \;\middle\vert\; \thetab^s \right))\right]\\
    &
    +4\left\{ \sum_{j=1}^{N_{\rm res}} \left[\log(\frac{1}{S} \sum_{s=1}^S N_{{\rm tot},s} p\left({\rm obs}, \D{j} \;\middle\vert\; \thetab^s \right))  - \frac{1}{S} \sum_{s=1}^S \log(N_{{\rm tot},s} p\left({\rm obs},  \D{j} \;\middle\vert\; \thetab^s \right)) \right]
    \right\}\,.
    \end{split}
\end{equation}

\chapter{MWb and MWc0.8 $N$-body Simulation Figures}
\label{appendix:extra_figures_nbody}

\graphicspath{{nbody/Figs/}}

As the results of the MWb and MWc0.8 $N$-body simulations in Chapter \ref{ch:nbody} are similar to the MWa case, we have moved many figures specific to those models to this appendix.

\begin{figure}
    \centering
    \includegraphics[width=0.99\linewidth]{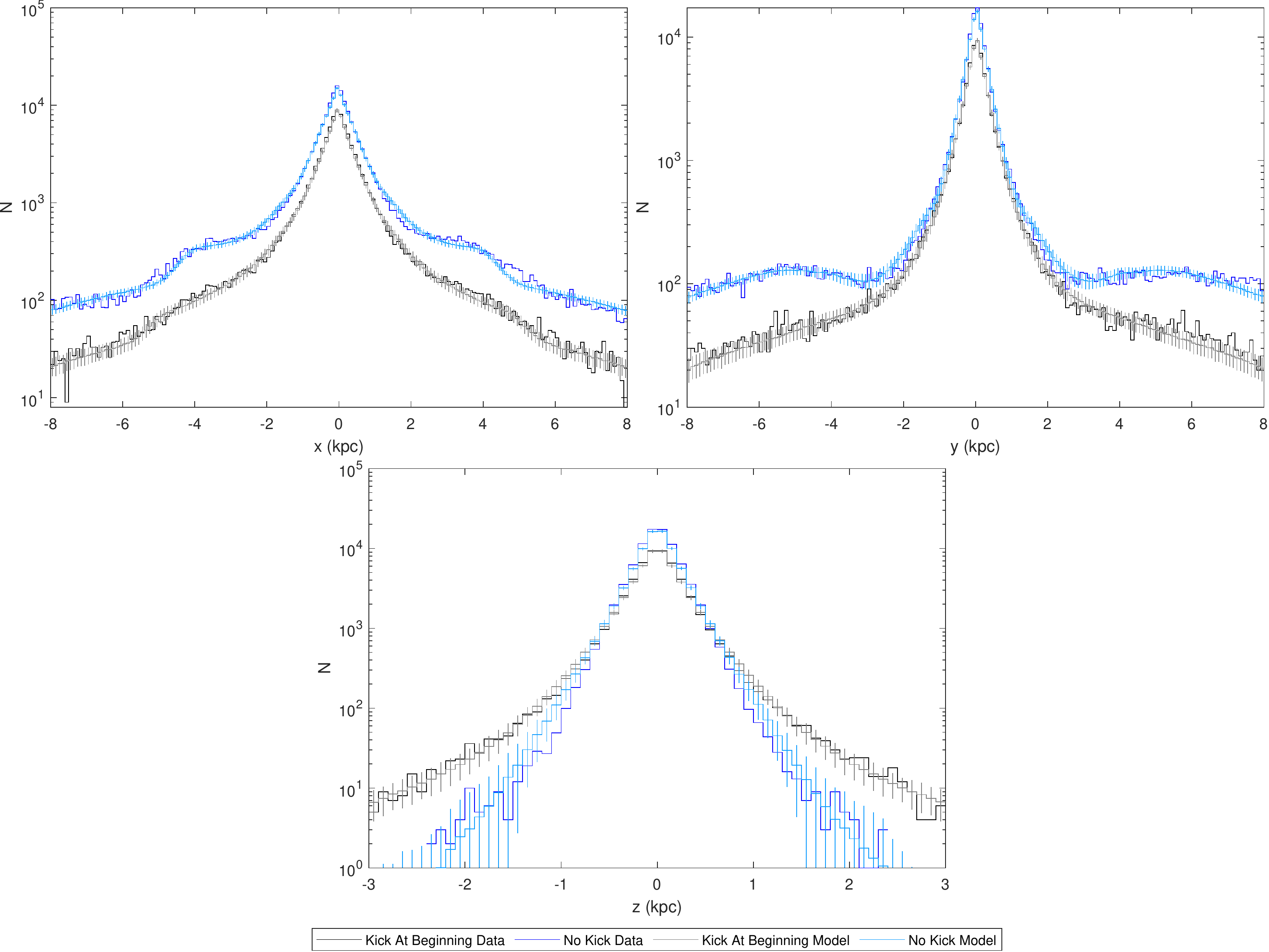}
    \caption[MWb $x$, $y$ and $z$ profiles with kicks at beginning.]{MWb profile along $x$, $y$ and $z$ axes with kicks occuring at the beginning. We show both $N$-body simulation data and data simulated using the fitted model. For the fitted model we show the mean number of particles in each bin and the standard deviation. }
    \label{fig:MWb_1d_profiles_kick_at_beginning}
\end{figure}

\begin{figure}
    \centering
    \includegraphics[width=0.99\linewidth]{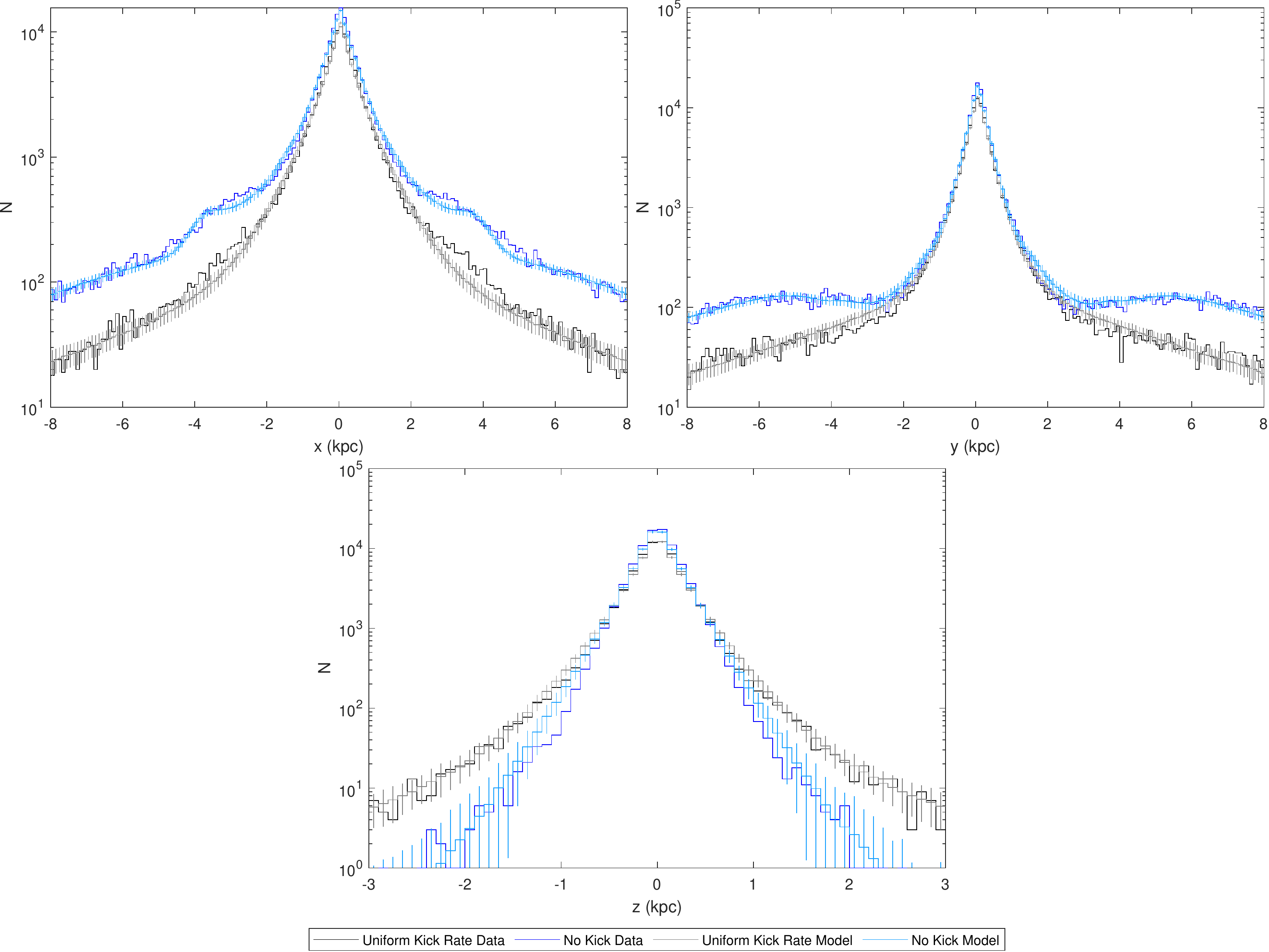}
    \caption[MWb $x$, $y$ and $z$ profiles with a uniform kick rate.]{MWb profile along $x$, $y$ and $z$ axes with a uniform kick rate. We show both $N$-body simulation data and data simulated using the fitted model. For the fitted model we show the mean number of particles in each bin and the standard deviation. }
    \label{fig:MWb_1d_profiles_uniform_kick_rate}
\end{figure}

\begin{figure}
    \centering
    \includegraphics[width=0.99\linewidth]{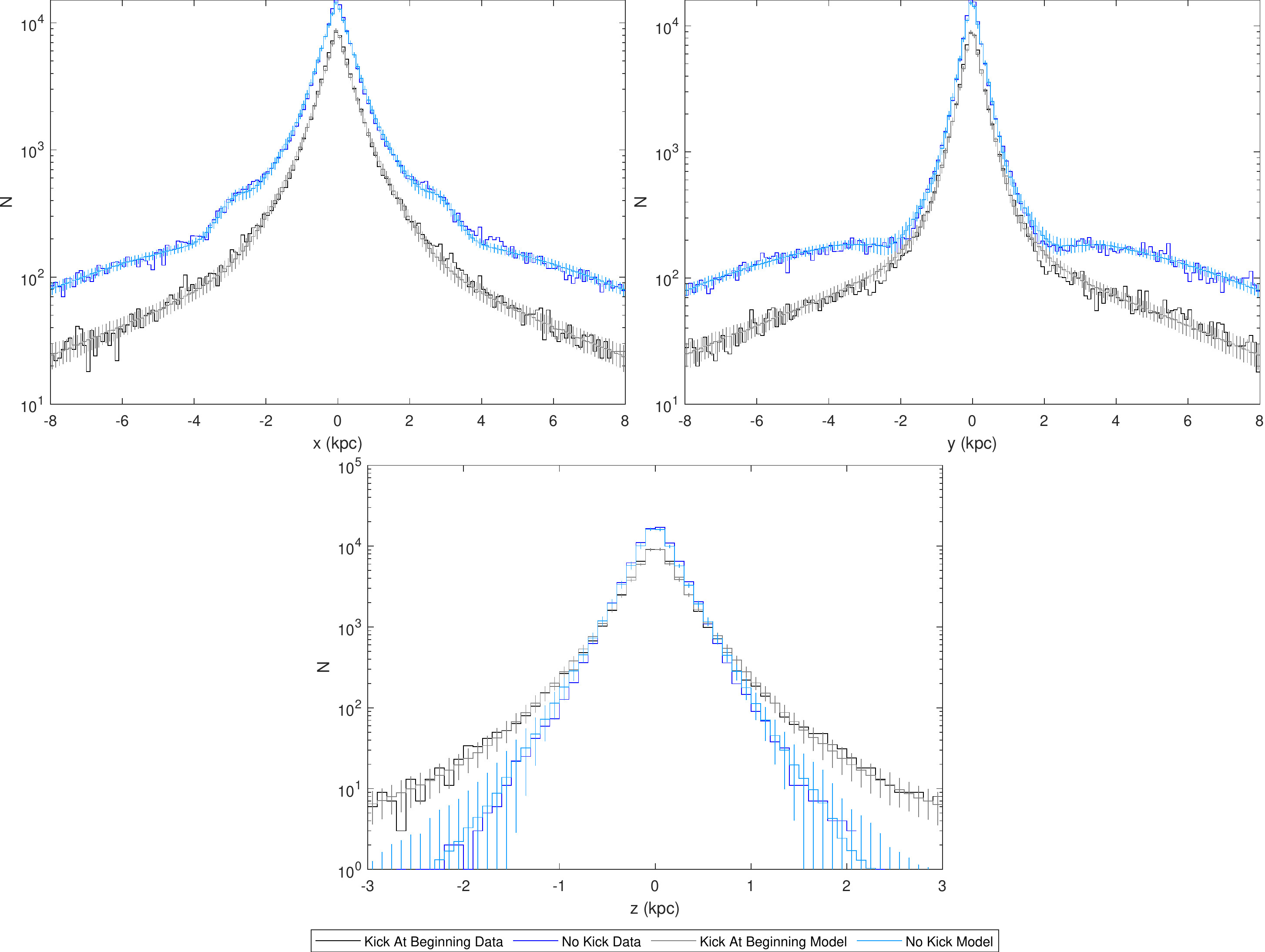}
    \caption[MWc0.8 $x$, $y$ and $z$ profiles with kicks at beginning.]{MWc0.8 profile along $x$, $y$ and $z$ axes with kicks occuring at the beginning. We show both $N$-body simulation data and data simulated using the fitted model. For the fitted model we show the mean number of particles in each bin and the standard deviation. }
    \label{fig:MWc0.8_1d_profiles_kick_at_beginning}
\end{figure}

\begin{figure}
    \centering
    \includegraphics[width=0.99\linewidth]{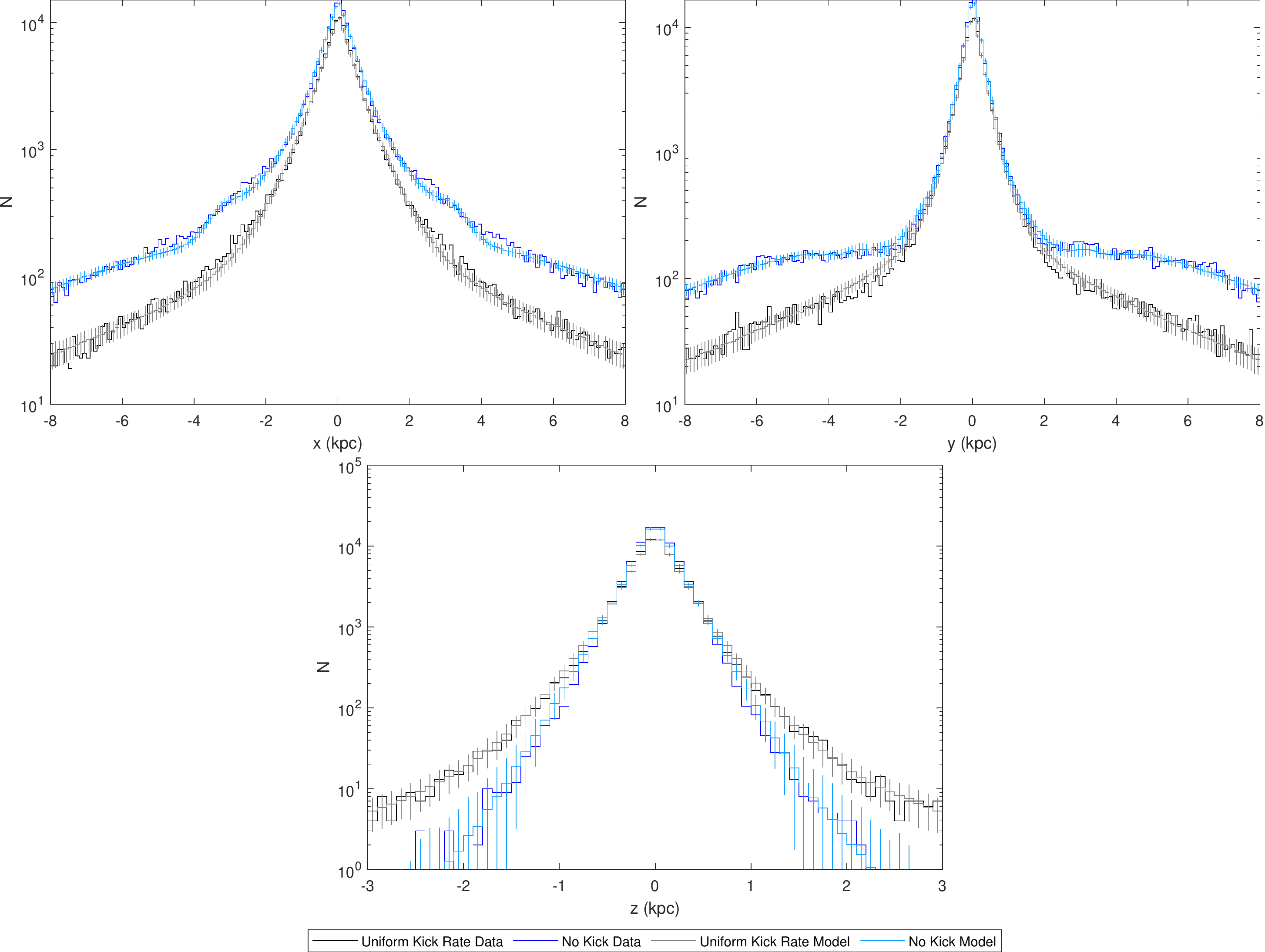}
    \caption[MWc0.8 $x$, $y$ and $z$ profiles with a uniform kick rate.]{MWc0.8 profile along $x$, $y$ and $z$ axes with a uniform kick rate. We show both $N$-body simulation data and data simulated using the fitted model. For the fitted model we show the mean number of particles in each bin and the standard deviation. }
    \label{fig:MWc0.8_1d_profiles_uniform_kick_rate}
\end{figure}

\begin{figure}
    \centering
    \includegraphics[width=0.99\linewidth]{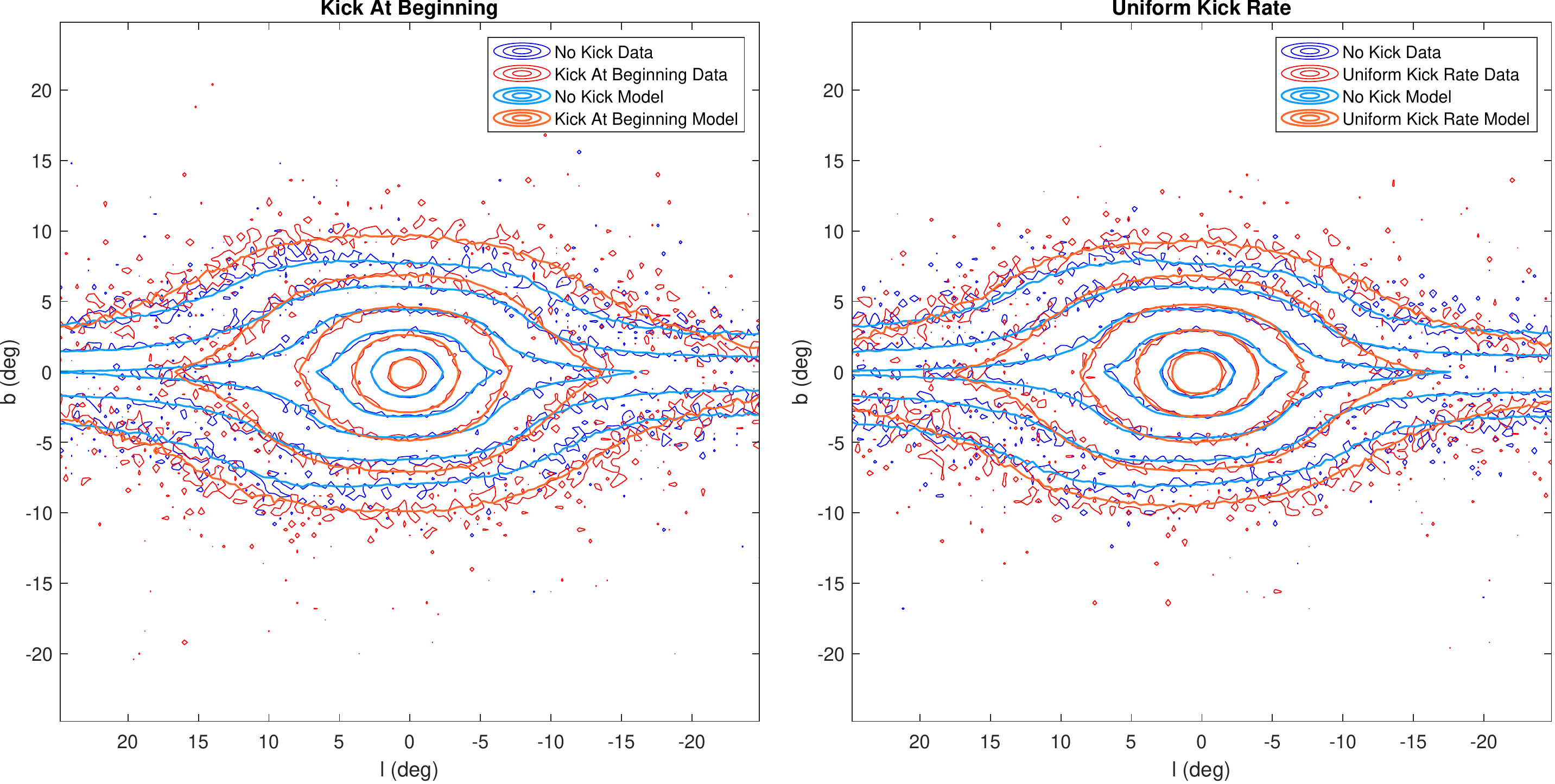}
    \caption[MWb flux distribution.]{MWb flux distribution in Galactic coordinates. The contours for each distribution are at $1$, $2$, $4$, $8$ and $16$ times the mean in this region. The Sun is placed at a distance of $7.9$ kpc, at an angle relative to the bar of $20^\circ$ and at a height of $15$ pc.  }
    \label{fig:MWb_data_los}
\end{figure}

\begin{figure}
    \centering
    \includegraphics[width=0.99\linewidth]{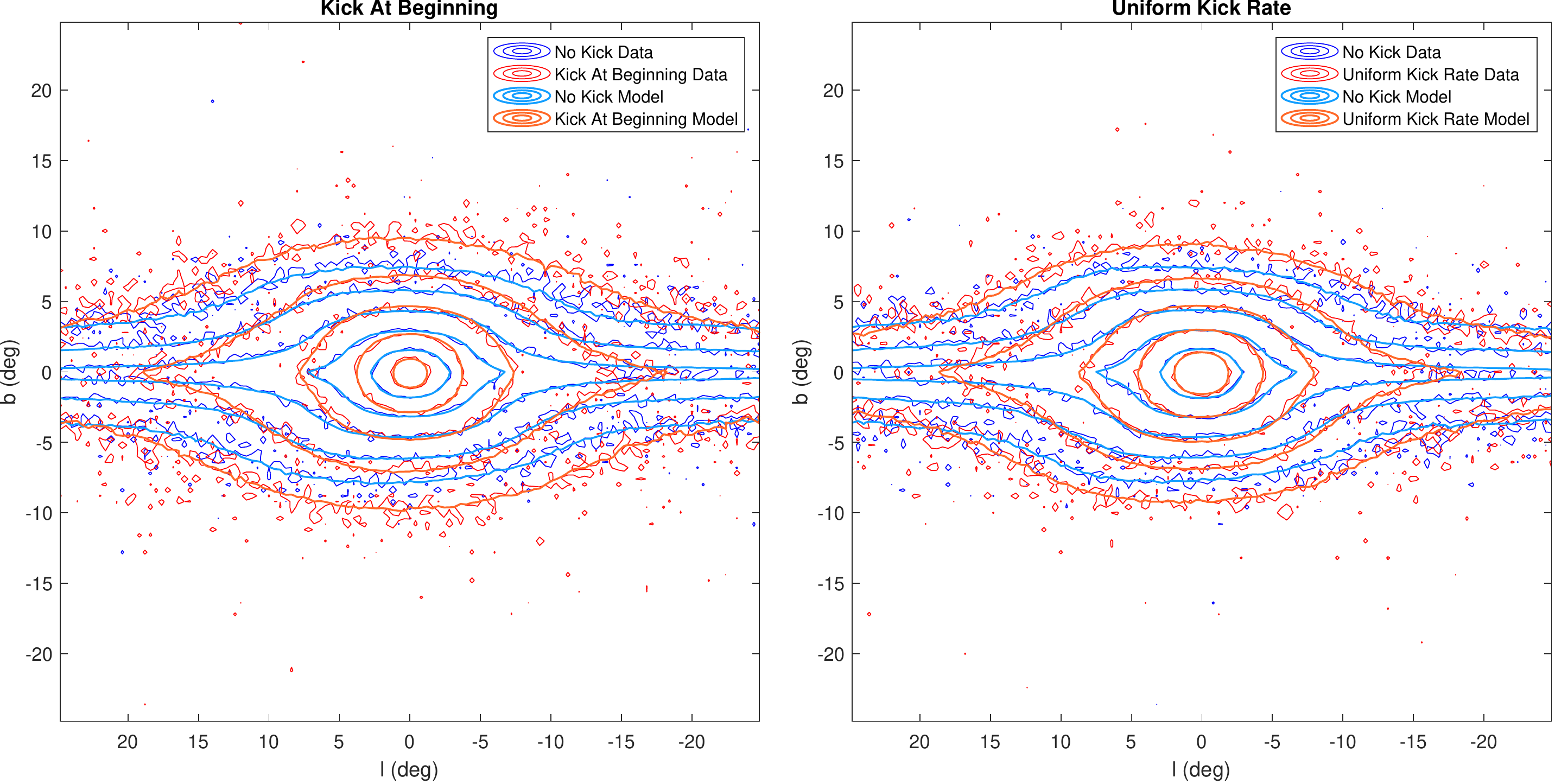}
    \caption[MWc0.8 flux distribution.]{MWc0.8 flux distribution in Galactic coordinates. The contours for each distribution are at $1$, $2$, $4$, $8$ and $16$ times the mean in this region. The Sun is placed at a distance of $7.9$ kpc, at an angle relative to the bar of $20^\circ$ and at a height of $15$ pc.  }
    \label{fig:MWc0.8_data_los}
\end{figure}

\end{appendices}

\printthesisindex %

\end{document}